\documentclass[11pt, a4paper,twoside]{memoir}   
\usepackage[top=2cm,bottom=2.5cm,right=3cm,left=2.5cm]{geometry} 
\usepackage{color}
\usepackage{graphicx}
\usepackage{subfig}
\usepackage{adjustbox}
\usepackage{titlesec, titletoc} 
\usepackage[T1]{fontenc}
\usepackage{blindtext}
\usepackage{enumitem}
\usepackage{amsmath,amsfonts,amssymb}
\usepackage{amsthm}
\usepackage{siunitx}
\usepackage{multirow}
\usepackage[colorlinks=true,linkcolor=orange,filecolor=cyan, citecolor=cyan,  urlcolor=magenta,linktocpage=true,anchorcolor=red]{hyperref} 
\usepackage{tensor}
\usepackage{empheq}
\usepackage{slashed}
\usepackage{wasysym} 
\usepackage{multicol}
\usepackage{acronym}[nohyperlinks] 
\usepackage[figurename=Figure,labelfont=bf]{caption}
\DeclareSymbolFont{extraup}{U}{zavm}{m}{n}
\DeclareMathSymbol{\varheart}{\mathalpha}{extraup}{86}
\DeclareMathSymbol{\vardiamond}{\mathalpha}{extraup}{87}
\usepackage{tikz}
\usetikzlibrary{shapes.misc,calc}
\usepackage{lipsum,fancyhdr}
\usepackage{bbm} 

\usepackage[
   backend=biber,        
   sorting=nyt,          
   citestyle=numeric-comp, 
]{biblatex}
\newcommand{\poubelle}[1]{} 

\newcommand{\myeq}[1]{Eq.~\eqref{#1}}

\newcommand{\YLMS}[3]{{_{#3}\tensor{Y}{_{#1,#2}}}}
\newcommand{\YLMSStar}[3]{{_{#3}Y^\star_{#1,#2}}}
\newcommand{\YLM}[2]{\tensor{Y}{_{#1,#2}}}
\newcommand{\YLMStar}[2]{Y^\star_{#1,#2}}
\newcommand{\sd}[0]{\slashed{\partial}}
\newcommand{\sds}[0]{{\slashed{\partial}^\star}}
\newcommand{\rll}[0]{\nu_{\ell}}
\newcommand{\rlla}[1]{\nu_{#1}}
\newcommand{\almphi}[2]{\phi_{#1,#2}}
\newcommand{\almphistar}[2]{\phi^{\star}_{#1,#2}}
\newcommand{\Clphi}[1]{C^\phi_{#1}}
\newcommand{\Fantome}[1]{
\phantom{\subfloat[\centering LABEL 2]{{\includegraphics[width=#1\textwidth]{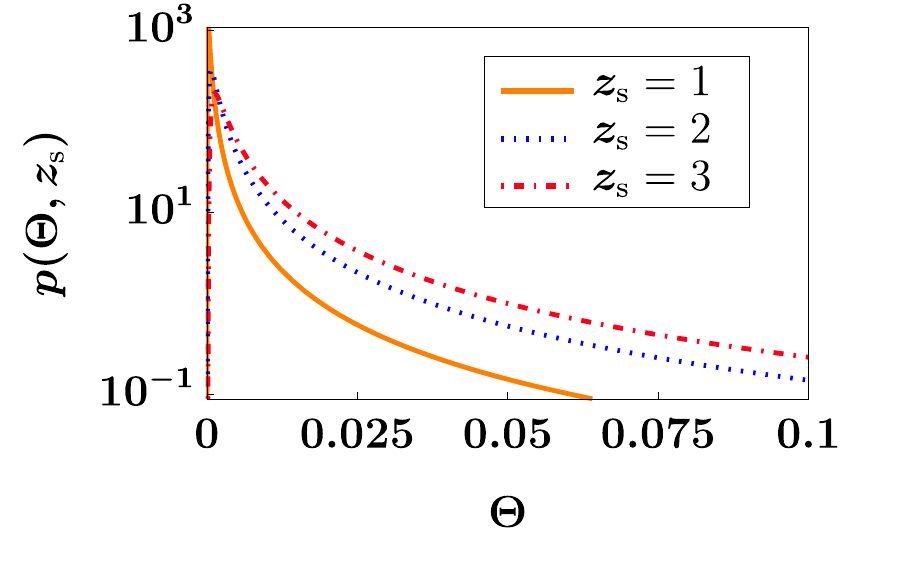} }}
}
}

\newcommand{\startchap}[0]{
\startcontents[mainsections]
\printcontents[mainsections]{l}{1}{\section*{Menu of the Chapter}
\setcounter{tocdepth}{2}}
} 
\newcommand{\stopchap}[0]{\stopcontents[mainsections]}

\newcommand{\MYdef}[2]{
\begin{mydef}[\textbf{#1}]
#2
\end{mydef}
\noindent
}

\newcommand{\MYrem}[2]{
\begin{myrem}[\textbf{#1}]
#2
\end{myrem}
\noindent
}

\newcommand{\MYthm}[2]{
\begin{mythm}[\textbf{#1}]
#2
\end{mythm}
\noindent
}

\newcommand{\boxemph}[1]{
	\begin{empheq}[box=\fbox]{align}
		#1
	\end{empheq}
}

\newcommand{\mapubli}[4]{\textbf{#1}, #2\\
#3, (#4)
}

\newcommand{\phiL}{{\phi}}
\newcommand{\ellRJ}{\tilde{\ell}_{\mathrm{R}}}
\newcommand{\ellRE}{{\ell}_{\mathrm{R}}}
\newcommand{\bds}{\boldsymbol}
\newcommand{\GN}{G_{\mathrm{N}}}
\newcommand{\pt}{\partial}
\newcommand{\bpt}{\boldsymbol{\partial}}
\newcommand{\dd}{\mathrm{d}}
\newcommand{\bdd}{\boldsymbol{\mathrm{d}}}
\newcommand{\ii}{\mathrm{i}}
\newcommand{\eee}{\mathrm{e}}
\newcommand{\LL}{{\mathrm{L}}}
\newcommand{\TT}{{\mathrm{T}}}
\newcommand{\MM}{{\mathrm{M}}}
\newcommand{\mc}{{\mathrm{c}}}
\newcommand{\me}{{\mathrm{e}}}
\newcommand{\mf}{{\mathrm{f}}}
\newcommand{\mo}{{\mathrm{o}}}
\newcommand{\mr}{{\mathrm{r}}}
\newcommand{\mpp}{{\mathrm{p}}}

\newcommand{\ms}{{\mathrm{s}}}

\newcommand{\mW}{{\mathrm{W}}}

\newcommand{\DA}{D_{\mathrm{A}}}
\newcommand{\DLum}{D_{\mathrm{L}}}
\newcommand{\DS}[0]{D_{\textrm{s}}}
\newcommand{\DL}[0]{D_{\ell}}
\newcommand{\DLS}[0]{D_{\ell \textrm{s}}}
\newcommand{\Mcal}{{\mathcal{M}}}

\newcommand{\Xcal}{{\mathcal{X}}}
\newcommand{\jacD}{{\bds{D}}}
\newcommand{\Ucal}{{\mathcal{U}}}
\newcommand{\Hcal}{{\mathcal{H}}}
\newcommand{\Lcal}{{\mathcal{L}}}
\newcommand{\Ecal}{{\mathcal{E}}}
\newcommand{\LEH}{\mathcal{L}_{\mathrm{EH}}}
\newcommand{\Mpl}{M_{\mathrm{Pl}}}
\newcommand{\Tpl}{T_{\mathrm{Pl}}}
\newcommand{\Lpl}{L_{\mathrm{Pl}}}
\newcommand{\MBH}{M_{\mathrm{BH}}}
\newcommand{\Msun}{M_{\astrosun}}
\newcommand{\bolde}{\bds{e}}
\newcommand{\zp}[0]{\zeta_{\mathrm{p}} }
\newcommand{\zc}[0]{\zeta_{\mathrm{c}} }
\newcommand{\Ng}[0]{N_{\mathrm{g}} }
\newcommand{\Ne}[0]{N_{\mathrm{e}} }
\newcommand{\Np}[0]{N_{\mathrm{p}} }
\newcommand{\Nc}[0]{N_{\mathrm{c}} }
\newcommand{\fsky}[0]{f_{\mathrm{sky}} }
\newcommand{\rg}[0]{r_{\mathrm{g}} }
\newcommand{\rH}[0]{r_{\mathrm{H}} }
\newcommand{\gS}[0]{\bds{g}_{\mathrm{S}} }
\newcommand{\cS}{c_{\mathrm{s}}}
\newcommand{\cSI}{c_{\mathrm{si}}}
\newcommand{\hSI}{\hbar_{\mathrm{si}}}
\newcommand{\eSI}{e_{\mathrm{si}}}
\newcommand{\GSI}{G_{N,\mathrm{si}}}
\newcommand{\Gnn}{G_{N,\mathrm{nat}}}
\newcommand{\MSI}{M_{\mathrm{si}}}
\newcommand{\Mn}{M_{\mathrm{nat}}}
\newcommand{\RSI}{R_{\mathrm{si}}}
\newcommand{\Rnn}{R_{\mathrm{nat}}}

\newcommand{\xSI}{x_{\mathrm{si}}}
\newcommand{\xn}{x_{\mathrm{nat}}}
\newcommand{\alSI}{\alpha_{\mathrm{si}}}
\newcommand{\aln}{\alpha_{\mathrm{nat}}}

\def\be{\begin{equation}}
\def\ee{\end{equation}}
\DeclareSIUnit\meterJ{\tilde{m}_J}
\DeclareSIUnit\nanometer{nm}
\DeclareSIUnit\meterE{m_E}
\DeclareSIUnit\year{yrs}
\DeclareSIUnit\kilom{km}
\DeclareSIUnit\Gyear{Gyrs}
\DeclareSIUnit\Megapc{Mpc}
\DeclareSIUnit\dam{dam}
\DeclareSIUnit\dm{dm}
\DeclareSIUnit\Ghz{GHz}

\newtheoremstyle{defstyle}
  {\topsep}
  {\topsep}
  {\itshape}
  {0pt}
  {\bfseries}
  {\ --}
  { }
  {\thmname{#1}\thmnumber{ #2}\textnormal{\thmnote{ (#3)}}}
\theoremstyle{defstyle}
\newtheorem{mydef}{Definition}[part]
\newtheorem{myrem}{Remark}[part]
\newtheorem{mythm}{Theorem}[part]

\definecolor{orange}{rgb}{1,0.5,0}
\definecolor{orangef}{rgb}{1,0.29,0}
\definecolor{orangec}{rgb}{1,0.728,0}
\definecolor{dgreen}{rgb}{0,0.6,0.1}

\titleclass{\part}{top} 
\titleformat{\part} 
[display] 
{\centering\normalfont\Huge\bfseries} 
{\MakeUppercase{\partname} \thepart} 
{10pt} 
{\titlerule[1pt]\vspace{1pc}\huge\MakeUppercase  } 
[ ] 
\titlespacing*{\part}{0pt}{0pt}{15pt}
\setsecnumdepth{subsection}

\begin{document}
\frontmatter

\thispagestyle{empty}
\newgeometry{left=2cm}
\newgeometry{top=1.5cm}
\newgeometry{bottom=2cm}
\Large 
\noindent\textsc{Universit\'e de Gen\`eve} \hfill \textsc{Facult\'e des Sciences}%

\noindent {D\'ept de Physique Th\'eorique} \hfill {Professeure Ruth \textsc{Durrer}}
\vspace{0.1 cm}
\noindent\rule{\textwidth}{0.4pt}
\vspace{1.3 cm}

\begin{center}
{\Huge \textbf{Observables In Cosmology:}
		\vspace{0.4 cm}
		
\textbf{Three Astronomical Perspectives}} 
\end{center}
\vspace{0.9cm}

\begin{center}
{\Large   \textsc{Th\`ese} }
\vspace{0.6 cm}

{\Large pr\'esent\'ee \`a la Facult\'e des Sciences pour obtenir le grade de\\ Docteur \`es Sciences, mention Physique}
\vspace{1.3 cm}

par
\vspace{0.4 cm}

\textbf{
{J\'er\'emie \textsc{Francfort}}
}
\vspace{0.4 cm}

de 
\vspace{0.4 cm}

Gen\`eve (GE) et Begnins (VD)
\vspace{1.5 cm}

\textsc{Th\`ese N\textsuperscript{0} ...}
\vspace{1.5 cm}

\textsc{Gen\`eve}

Impression ...

2022
\end{center}

\restoregeometry
\normalsize 
\newpage
\thispagestyle{empty}
\newgeometry{top=5cm,left=0.5 cm}

\Large 
\noindent \textbf{Cette th\`ese fut pr\'esent\'ee et soutenue publiquement le 24 juin 2022.} 
\vspace{0.5 cm}

\noindent\textbf{Membres du jury~:}
\vspace{0.5 cm}

\begin{itemize}
\item Prof. Ruth \textsc{Durrer}, superviseure\\
        {\large{Universit\'e de Gen\`eve}}
\item Prof. Camille \textsc{Bonvin}, membre du jury\\
        {\large{Universit\'e de Gen\`eve}}
\item Prof. Claudia de \textsc{Rham}, membre du jury\\
        {\large{Imperial College London}}
\item Dr Jean-Philippe \textsc{Uzan}, membre du jury\\
        {\large{Institut d'Astrophysique de Paris}}
\end{itemize}

\vspace{1.5 cm}
\noindent\Large\textbf{Avec le soutien de}
\vspace{1 cm}

\hspace{-0.5 cm} \includegraphics[width = 0.4\textwidth]{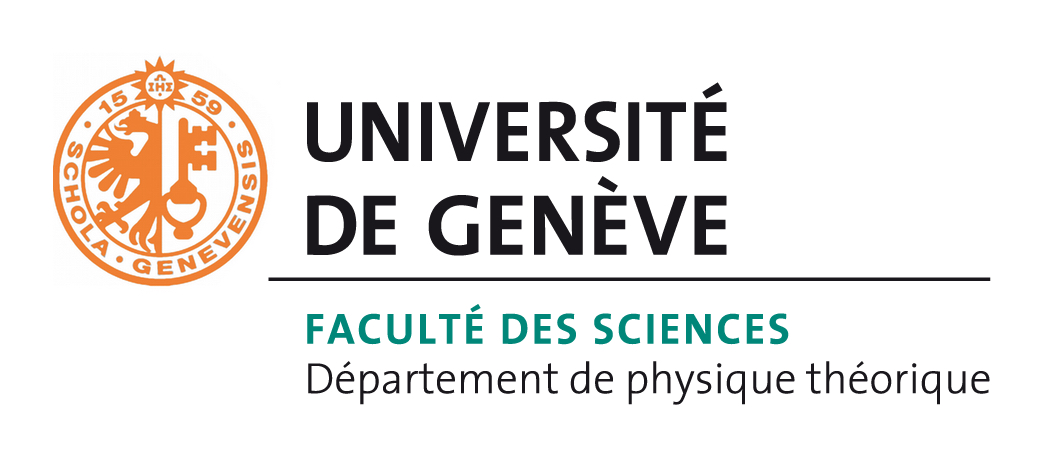}
\vspace{0.5 cm}

\includegraphics[width = 0.4\textwidth]{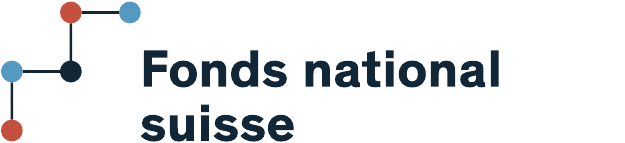}
\vspace{0.5 cm}

\hspace{-0.6 cm} \includegraphics[width = 0.4\textwidth]{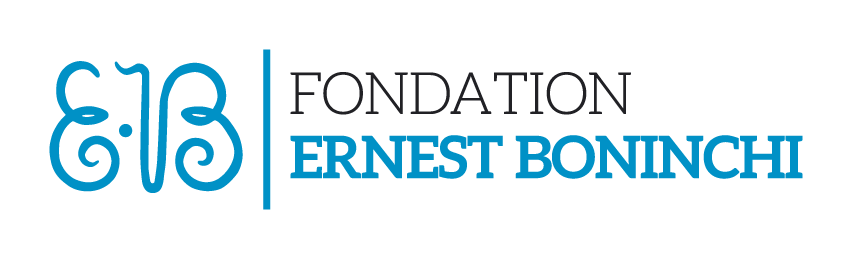}

\restoregeometry
\normalsize
\newpage 
\newgeometry{left=2.3cm,top=2cm,bottom=2.5cm,right=3cm}
\addcontentsline{toc}{part}{R\'esum\'e}
\begin{center}
    \LARGE \textbf {R\'esum\'e}
\end{center}
\noindent \emph{Summary in English below}
\vspace{0.3cm}

\noindent Durant ma th\`ese, j'ai r\'ealis\'e trois projets distincts, donnant lieu \`a quatre publications scientifiques~\cite{francfort_2019,Francfort:2020,Francfort:2021,Francfort:2022}. Ces projets sont ind\'ependants les uns des autres et portent sur des sujets divers de la cosmologie moderne. Un fil rouge, e\^ut-il fallu en trouver un, serait la d\'efinition d'observables physiques nous permettant de tester le paradigme de la relativit\'e g\'en\'erale et de le confronter aux mod\`eles dits de gravit\'e modifi\'ee, bien que ma th\`ese ne consid\`ere pas de tels mod\`eles en soi. Le mod\`ele standard de la cosmologie d\'ecrit un Univers homog\`ene et isotrope en expansion~: l'Univers FRLW. Le mod\`ele standard de la cosmologie pr\'edit une relation de proportionnalit\'e entre la distance d'une source lumineuse proche et son d\'ecalage vers le rouge. Cette pr\'ediction th\'eorique est confirm\'ee par des r\'esultats observationnels. De plus, le mod\`ele FRLW décrit aussi l'existence du fond diffus cosmologique (CMB)~: un ensemble de photons remplissant tout l'Univers et dont la distribution d'\'energie suit une loi du corps noir \`a une temp\'erature de $T_0=\SI{2.7}{\kelvin}$. Cette pr\'ediction th\'eorique est aussi observ\'ee exp\'erimentalement et fournit une preuve solide du bien-fond\'e du mod\`ele. 

Ce dernier peut \^etre raffin\'e en utilisant la th\'eorie des perturbations et en consid\'erant des inhomog\'en\'eit\'es, par exemple de densit\'e de mati\`ere. L'analyse quantitative de la distribution de ces inhomog\'en\'eit\'es est un champ de recherche tr\`es actif. L'id\'ee g\'en\'erale est que les propri\'et\'es statistiques de ces quantit\'es, par exemple leurs fonctions de corr\'elation, permettent d'estimer les divers param\`etres cosmologiques et de tester la validit\'e de la relativit\'e g\'en\'erale. 

Cette th\`ese est divis\'ee en quatre parties principales.

Dans la premi\`ere partie, j'introduis les notions th\'eoriques utilis\'es dans les suivantes. Je pr\'esente les outils de la g\'eom\'etrie diff\'erentielle et les concepts fondamentaux de la relativit\'e g\'en\'erale. Je discute le mod\`ele standard de la cosmologie en mettant un accent particulier sur la th\'eorie des perturbations et sur le lentillage.

La deuxi\`eme partie est bas\'ee sur "Cosmological Number Counts in Einstein and Jordan frames"~\cite{francfort_2019}. Cet article traite de l'invariance du comptage des galaxies sous transformation conforme. Je pr\'esente tout d'abord le concept de r\'ef\'erentiels conformes ainsi que l'interpr\'etation physique associ\'ee, ce qui me permet ensuite d'argumenter que les observables physiques ne doivent pas d\'ependre du r\'ef\'erentiel choisi. Finalement, je me concentre sur une observable en particulier~: le comptage des galaxies. Comme le nom l'indique, cette observable quantifie les fluctuations angulaires du nombre de galaxies observ\'ees. Je montre explicitement que le comptage des galaxies ne d\'epend pas du r\'ef\'erentiel, ce qui motive l'hypoth\`ese de d\'epart.

La troisi\`eme partie est bas\'ee sur "Image Rotation from weak Lensing" et "A new observable for cosmic shear"~\cite{Francfort:2021,Francfort:2022}. Ces deux articles traitent de l'effet de lentillage qui d\'ecrit comment une distribution de masse (ou d'\'energie) d\'evie les rayons lumineux et d\'eforme les images des galaxies que nous observons. Je pr\'esente d'abord les outils math\'ematiques utilis\'es pour quantifier le lentillage, et m'int\'eresse particuli\`erement au cisaillement. J'explique comment il peut, sous certaines conditions, induire une rotation des axes principaux des images des galaxies. J'utilise ce r\'esultat pour construire un estimateur des fonctions de corr\'elation du cisaillement cosmique. J'argumente enfin que, si le nombre de galaxies est assez grand, le rapport du signal sur bruit peut \^etre r\'eduit significativement, ce qui rend cette m\'ethode comp\'etitive.

La quatri\`eme partie est bas\'ee sur "Black Hole Gravitational Waves in the Effective Field Theory of Gravity"~\cite{Francfort:2020}. Dans cet article, j'\'etudie un trou noir de Schwarzschild dans une th\'eorie effective de gravit\'e modifi\'ee. Dans une telle th\'eorie, de nouveaux termes sont ajout\'es \`a l'action de Einstein-Hilbert de mani\`ere syst\'ematique mais agnostique, c'est-\`a-dire sans se soucier de l'origine desdits termes. Je calcule ensuite la m\'etrique du trou noir de Schwarzschild dans cette th\'eorie, perturbativement et \`a l'ordre lin\'eaire. Je m'int\'eresse ensuite aux ondes gravitationnelles produites par un tel trou noir. Finalement, je calcule la correction \`a leur vitesse de propagation et montre que cette vitesse peut diff\'erer de l'unit\'e, et je d\'etermine les corrections des fr\'equences quasinormales dans cette th\'eorie.

\newpage

\addcontentsline{toc}{part}{Summary}
\begin{center}
    \LARGE \textbf {Summary}
\end{center}
\emph{R\'esum\'e en fran\c cais au-dessus}
\vspace{0.3cm}

\noindent During my thesis, I carried out three distinct projects, giving rise to four scientific publications~\cite{francfort_2019,Francfort:2020,Francfort:2021,Francfort:2022}. These projects are independent from each other and investigate various topics of modern Cosmology. A common thread, should one be found, would be the definition of physical observables. This would allow us to test the predictions of General Relativity and to confront it to the so-called Modified Gravity models, although my thesis does not consider such models per se. The standard model of Cosmology describes a homogeneous and isotropic expanding Universe: The FRLW universe. The standard model of Cosmology predicts a proportional relationship between the distance to a nearby light source and its redshift. This theoretical prediction is confirmed by observational results. Moreover, the FRLW model also describes the existence of the Cosmological Microwave Background (CMB): A set of photons filling the whole Universe, whose energy distribution follows a black body law at a temperature of $T_0=\SI{2.7}{\kelvin}$. This theoretical prediction is also observed experimentally and provides a strong proof of the validity of the model. 

The model can be refined using perturbation theory and considering inhomogeneities, for example of the matter density. The quantitative analysis of the distribution of these inhomogeneities is a particularly active research field. The general idea is that the statistical properties of these quantities, for example their correlation functions, allows us to estimate the various cosmological parameters and to test the validity of general relativity. 

This thesis is divided into four main parts.

In the first part, I introduce the theoretical notions used in the following ones. I present the tools of differential geometry and the fundamental concepts of General Relativity. I discuss the standard model of Cosmology with special emphasis on perturbation theory and lensing.

The second part is based on "Cosmological Number Counts in Einstein and Jordan frames"~\cite{francfort_2019}. This paper deals with the invariance of Galaxy Number Counts under conformal transformation. I first present the concept of conformal reference frames and the associated physical interpretation. With this in mind, I argue that the physical observables should not depend on the chosen frame. Finally, I focus on one observable in particular: The Galaxy Number Counts. As the name suggests, this observable quantifies the angular fluctuations of the number of observed galaxies. I show explicitly that the Galaxy Number Counts is frame-independent, which supports the hypothesis I stated.

The third part is based on "Image Rotation from weak Lensing" et "A new observable for cosmic shear"~\cite{Francfort:2021,Francfort:2022}. These two papers deal with the lensing effect which describes how a mass (or energy) distribution deflects light rays and distorts the images of observed galaxies. I first present the mathematical tools used to quantify lensing with a focus my attention on the shear. I explain how it can, under certain conditions, induce a rotation of the principal axes of galaxy images. I use this result to build an estimator of the cosmic shear correlation functions. I finally argue that, if the number of galaxies is large enough, the signal-to-noise ratio can be significantly reduced, which makes this method a competitive one.

The fourth part is based on "Black Hole Gravitational Waves in the Effective Field Theory of Gravity"~\cite{Francfort:2020}. In this paper, I study a Schwarzschild Black Hole in a modified effective theory of gravity. In such a theory, new terms are added to the Einstein-Hilbert action in a systematic but agnostic way, i.e. without taking the origin of these terms into consideration. I then compute the Schwarzschild Black Hole metric in this theory, perturbatively and at linear order. I then focus on the gravitational waves produced by such a Black Hole. Finally, I compute the correction to their propagation speed and show that this speed can differ from unity, and I determine the corrections to the Quasinormal Modes in this theory.

\restoregeometry
\newpage
\addcontentsline{toc}{part}{Remerciements}
\begin{center}
    \LARGE \textbf {Remerciements}
\end{center}
Cette partie est l'une des plus difficiles \`a \'ecrire. Comme dirait Willy Wonka: \emph{Si peu de gens \`a remercier et tant de place}.
\vspace{0.2cm}

Je remercie tout d'abord les quatre membes du jury pour le temps qu'il et elles prendront afin de me permettre, je l'esp\`ere, de devenir Docteur J\'erem.
\vspace{0.2cm}

Je remercie aussi Basundhara, Bea, Charles, Clo\'e, Flann, Francesca, Giulia, Juanma, Lila, Manon, Ruth, William et Yamini pour les conseils et relectures qui m'ont sans aucun doute aid\'e enorm\'ement dans la r\'edaction de cette th\`ese.
\vspace{0.2cm}

Je tiens aussi \`a mentionner les trois mois merveilleux que j'ai pass\'es \`a Londres en 2019, et \`a ce titre je veux montrer toute ma gratitude \`a Claudia, Jun, Mat, Victor, et tout le reste du groupe, qui ont fait de cette exp\'erience ce qu'elle a \'et\'e. Je remercie aussi chaleureusement la Fondation Boninchi pour son soutien financier lors de cette visite.
\vspace{0.2cm}

Ces quatres ann\'ees dans le groupe de cosmologie auraient \'et\'e bien fades sans la compagnie de mes sympathiques coll\`egues. Charles, Giulia, Joyce, Louis, Michael, Mona, Nastassia, Sveva, Viraj, William, et les autres, ont contribu\'e \`a au plaisir que j'ai ressenti lors de cette aventure~!
\vspace{0.2cm}

Je suis aussi tr\`es reconnaissant envers Martin Kunz et Ruth Durrer, qui m'ont tout deux permis de donner quelques cours \`a leur place et ainsi de me laisser pratiquer cette activit\'e qu'est l'enseignement et qui me passionne.
\vspace{0.2cm}

Je tiens aussi \`a mentionner Bernard Vuilleumier, Francesco Huber et Philippe Pitteloud. Ils ont tous trois r\'eussi \`a attiser ma passion pour les math\'ematiques et la physique, lors d'une p\'eriode de la vie o\`u il est grandement important de se trouver. Ma pr\'esente et future carri\`ere leur doit beaucoup.
\vspace{0.2cm}

Je suis impressionn\'e par les dessins que Bea, Jessica et Juanma ont faits pour illustrer cette th\`ese. Cela me touche et, j'en suis s\^ur, rend ce travail plus beau~!
\vspace{0.2cm} 

Je salue avec reconnaissance l'aide pr\'ecieuse de Angela, Francine, Jacques, Lionel, Nathalie et Sandro. Ils et elles ont toujours \'et\'e disponibles pour r\'egler mes soucis techniques, et \`a chaque fois avec une rapidit\'e impressionante. 
\vspace{0.2cm}

Sur un autre plan, je tiens aussi  \`a remercier le FNS, et plus particuli\`erement tous et toutes les contribuables anonymes qui, par leur travail quotidien, ont financ\'e ce doctorat. Sans leur aide, on va pas se mentir, ça n'aurait pas \'et\'e possible.
\vspace{0.2cm}

Finalement, quelques mots pour ma superviseuse de th\`ese, Ruth. Durant quatre ans, j'ai toujours senti de la bienveillance et de la patience de sa part. Elle a toujours cru en moi et en mes capacit\'es pendant ce doctorat, et je la remercie pour cela. Je ne l'oublierai pas.
\vspace{1cm}

Sur le plan plus personnel, je n'aurais jamais termin\'e ce doctorat sans la pr\'esence de plusieurs personnes importantes.
\vspace{0.2cm}

Anne, Basundhara, Ben, Francesca, Jon, les Rassines, Seb, je suis reconnaissant pour tous les moments pass\'es ensemble, des balades \`a v\'elo aux sessions CG, en passant par les parties de squash matinales et revigorantes, les \'ecoles d'\'et\'e en Sardaigne et les \emph{focaccias}.
\vspace{0.2cm}

Les PSIons, Cheryne, Jessica, J\'ulia, Felipe, Kat, Lila, Niamh, que j'ai bien trop peu vus et vues mais dont la pr\'esence a toujours \'et\'e une source de plaisir et de r\'econfort.
\vspace{0.2cm}

Et finalement, je ne peux omettre Bea, Kahi et la petite L\'ea, mes $3$ \emph{bestahs}. Les nombreuses \emph{pyjamas party}, petit-d\'ej et news, CG, et soir\'ees votations ont \'et\'e autant de moments qui m'ont accompagn\'e et repos\'e ces quatre ann\'ees, merci~! $\varheart$ \vspace{0.2cm}

Je remercie aussi tous les \emph{caf\'es gentrifi\'es} du Canton de Gen\`eve pour l'atmosph\`ere qu'ils cr\'eent et qui m'a permis de r\'ealiser ce doctorat et d'\'ecrire cette th\`ese.
\vspace{0.2cm}

Quelques phrases pour ma famille. Mes parents m'ont toujours soutenu dans mes \'etudes et dans ma carri\`ere professionnelle. Je les remercie pour leur pr\'esence et leur patience, et tout simplement pour m'avoir fabriqu\'e. Un petit mot aussi pour Nicolas, Nonna, Ulla et Val\'erie, et pour leur soutien depuis toujours. Merci \`a vous tous et toutes !
\vspace{0.2cm}

Finalement, un mot pour mon \emph{charmant}, Juanma. Il a \'et\'e \`a mes c\^ot\'es depuis le d\'ebut de cette th\`ese. Il a su me motiver, me faire avancer et m'aider quand j'en avais besoin, confronter mes id\'es quand c'\'etait n\'ecessaire et partager mes moments de joie lors de cette aventure. Je le remercie infiniment~!

\restoregeometry
\newpage
\newgeometry{left=2cm}

\addcontentsline{toc}{part}{Publications}
\begin{center}
    \LARGE \textbf {Publications}
\end{center}

\normalsize
\noindent This thesis is based on the following publications.
\begin{enumerate}
\item \mapubli{Cosmological Number Counts in Einstein and Jordan frames~\cite{francfort_2019}}{\textbf{J\'er\'emie~Francfort}, Basundhara Ghosh and Ruth Durrer}{JCAP}{2019}
\item \mapubli{Black Hole Gravitational Waves in the Effective Field Theory of Gravity~\cite{Francfort:2020}}{Claudia de Rham, \textbf{J\'er\'emie Francfort} and Jun Zhang}{PRD}{2020}
\item \mapubli{Image rotation from lensing~\cite{Francfort:2021}}{\textbf{J\'er\'emie Francfort}, Giulia Cusin and Ruth Durrer}{CQG}{2021}
\item \mapubli{A new observable for cosmic shear~\cite{Francfort:2022}}{\textbf{J\'er\'emie Francfort}, Giulia Cusin and Ruth Durrer}{JCAP}{2022}
\end{enumerate}
\noindent I also contributed to the following two publications during my thesis.
\begin{enumerate}[resume]
\item \mapubli{RASSINE: Interactive tool for normalising stellar spectra~\cite{Cretignier:2020}}{Michael Cr\'etignier, \textbf{J\'er\'emie Francfort}, Xavier Dumusque, Romain Allart and Franscesco Pepe}{AA}{2020}
\item \mapubli{Cosmological Number Counts under Disformal Transformations~\cite{Ghosh:2022}}{Basundhara Ghosh, \textbf{J\'er\'emie Francfort} and Rajeev Kumar Jain}{Submitted PRD}{2022}
\end{enumerate}

\normalsize
\restoregeometry
\newpage
\addcontentsline{toc}{part}{Abbreviations and Conventions}
\begin{center}
    \LARGE \textbf {Abbreviations and Conventions}
\end{center}

\noindent\textbf{\Large List of Abbreviations}
\vspace{0.3 cm}
\begin{acronym}[MPC] 
\acro{CMB}{Cosmic Microwave Background}
\acro{FRLW}{Friedmann-Robertson-Lema\^itre-Walker}
\acro{GR}{General Relativity}
\acro{GNC}{Galaxy Number Counts}
\acro{GW}{Gravitational Waves}
\acro{QNM}{Quasinormal Modes}
\acro{SNR}{signal-to-noise ratio}
\acro{SSSS}{Static Spherically Symmetric Spacetime}
\acro{ST}{Scalar-Tensor}
\acro{LCDM}[$\Lambda$CDM]{Lambda Cold Dark Matter}
\end{acronym}
\vspace{0.7 cm}
\noindent\textbf{\Large List of Conventions}
\vspace{0.5 cm}

\noindent \textbf{General Conventions}
\vspace{0.5 cm}

\noindent We use the signature $(-,+,+,+)$ and the Einstein summation convention. Tensors, vectors and forms are written in bold, e.g.
\begin{align*}
\bds{A} &= \tensor{A}{_\mu ^\nu} \, \bdd x^\mu \otimes \bpt_\nu\,, \\
\bpt_\nu &\equiv  \frac{\pt}{\pt x^{\mu}}\,.
\end{align*}
Symmetric tensor products of forms are written
$$
\bdd x^\mu \bdd x^\nu = \frac{\bdd x^\mu \otimes \bdd x^\nu+\bdd x^\nu \otimes \bdd x^\mu}{2}\,.
$$
Scalar products are denoted 
$$
\bds u \cdot \bds v \equiv u^\mu v^\nu g_{\mu \nu}\,.
$$
The components of the Riemann tensor read
$$
\tensor{R}{^\mu_{\nu \rho \sigma}} =
\tensor{\Gamma}{^\mu_{\rho \alpha}}\tensor{\Gamma}{^\alpha_{\nu \sigma}}
-\tensor{\Gamma}{^\mu_{\sigma \alpha}}\tensor{\Gamma}{^\alpha_{\nu \rho}}
+\pt_\rho \tensor{\Gamma}{^\mu_{\nu \sigma}}
-\pt_\sigma \tensor{\Gamma}{^\mu_{\nu \rho}}\,,
$$
where $\tensor{\Gamma}{^\mu_{\nu \rho}}$ are the Christoffel symbols. In coordinate-free notation, the Riemann tensor is
$$
\bds{\mathcal{R}} = \tensor{R}{^\mu_{\nu \rho \sigma}} \, \bpt_\mu \otimes \bdd x^\nu \otimes \bdd x^\rho \otimes \bdd x^\sigma\,,
$$
while the Ricci tensor is
$$
\bds{R} = \tensor{R}{_{\mu \nu}}\, \bdd x^\mu \otimes \bdd x^\nu\,.
$$
\noindent Newton's constant is denoted $\GN$ while the Einstein tensor is
$$
\bds{G}=\tensor{G}{_\mu_\nu}\,  \bdd x^\mu \otimes \bdd x^\nu\,.
$$
Greek indices ($\mu, \nu, \dots$) run from $0$ to $4$, Latin indices at the beginning of the alphabet ($a,b,\dots$) run from $1$ to $2$ and Latin indices in the middle of the alphabet ($i,j,k$) run from $1$ to $3$.

\noindent We choose units such that $c=\hbar=1$, hence lengths, times and inverse masses have the same dimension (see Chapter~\ref{chapintro:units}).
\vspace{0.5 cm}

\noindent \textbf{Cosmological Setup}
\vspace{0.5 cm}

\noindent The cosmic time is $t$ while the conformal time is $\eta$. They are related as $$a\,  \dd \eta = \dd t\,,$$ where $a$ is the scale factor. The comoving radial coordinates is $r$, and hence physical distances are of the form $D_\mathrm{p} = ar$. The usual spherical angles are $(\theta, \varphi)$, and the angle element is 
$$
\bdd \Omega^2 = \bdd \theta^2 + \sin^2 \theta \bdd \varphi^2\,.
$$ 

\noindent A dot is a derivative with respect to cosmic time while a prime is a derivative with respect to conformal time. Background quantities will be denotes either with an index $0$ or a bar, depending on the context.

\noindent The metric perturbations are $(\phi, \psi, \xi, \zeta)$, the Bardeen potentials are $(\Phi, \Psi)$ and the Weyl potential is $\Psi_\mW$. In particular, note the difference between the coordinate $\varphi$ and the metric perturbation $\phi$. 

\noindent In the sections about conformal frames, $\phi$ denotes the scalar field, while in the sections about lensing, $\phi$ denotes the scalar field.

\noindent \textbf{Important:} The variable $\phi$ denotes three different objects in this thesis. However, they never appear at the same time and the distinction is, hopefully, clear from the context.

\noindent The Jacobi matrix and the lensing potential $\phi$ are parametrised by $(\DA,\kappa,\gamma,\chi)$ representing the angular diameter distance, the convergence, the shear and its orientation respectively. The components of the shear are
\begin{align*}
    \gamma_1 &= \gamma \cos 2 \chi\,, \\
    \gamma_2 &= \gamma \sin 2 \chi\,, \\
    \gamma^{\pm} &= \gamma_1 \pm \ii \gamma_2\,.
\end{align*}
The lensing potential satisfies
\begin{align*}
\langle\almphi{\ell}{m}(z) \rangle  &= 0\,, \\
\langle \almphi{\ell_1}{m_1}(z_1) \almphistar{\ell_2}{m_2}(z_2)\rangle  &=
\Clphi{\ell_1}(z_1,z_2) \delta_{\ell_1,\ell_2} \delta_{m_1,m_2}\,,
\end{align*}
where the average represents different statistical realisations of the Universe.

\noindent The angle between two unit vectors $\bds n_1$ and $\bds n_2$ is given by
$$
\mu \equiv \cos \varphi  = \bds n_1 \cdot \bds n_2\,.
$$
The convention for the Spin Spherical Harmonics is
\begin{align*}
\YLMS{\ell}{-m}{-s} &= (-1)^{s+m}\YLMSStar{s}{\ell}{m}\,. \\
\YLMS{\ell}{m}{0} &\equiv \YLM{\ell}{m}\,. 
\end{align*}
\noindent \textbf{Schwarzschild setup}
\vspace{0.5 cm}

\noindent The cosmic time is $t$ but here $r$ is the usual radial distance. The spherical coordinates follow the same convention as in the Cosmological setup. A prime denotes here derivatives with respect to $r$. Again there should hopefully not be any confusion as this different topics are independent.
\newpage
\begin{KeepFromToc} 
	\tableofcontents
\end{KeepFromToc}
\mainmatter\pagenumbering{arabic}
\part{Introduction and Theoretical Tools}\label{Part:Intro}
\begin{figure}[ht!]
	\centering
	\includegraphics[width = 0.9\textwidth]{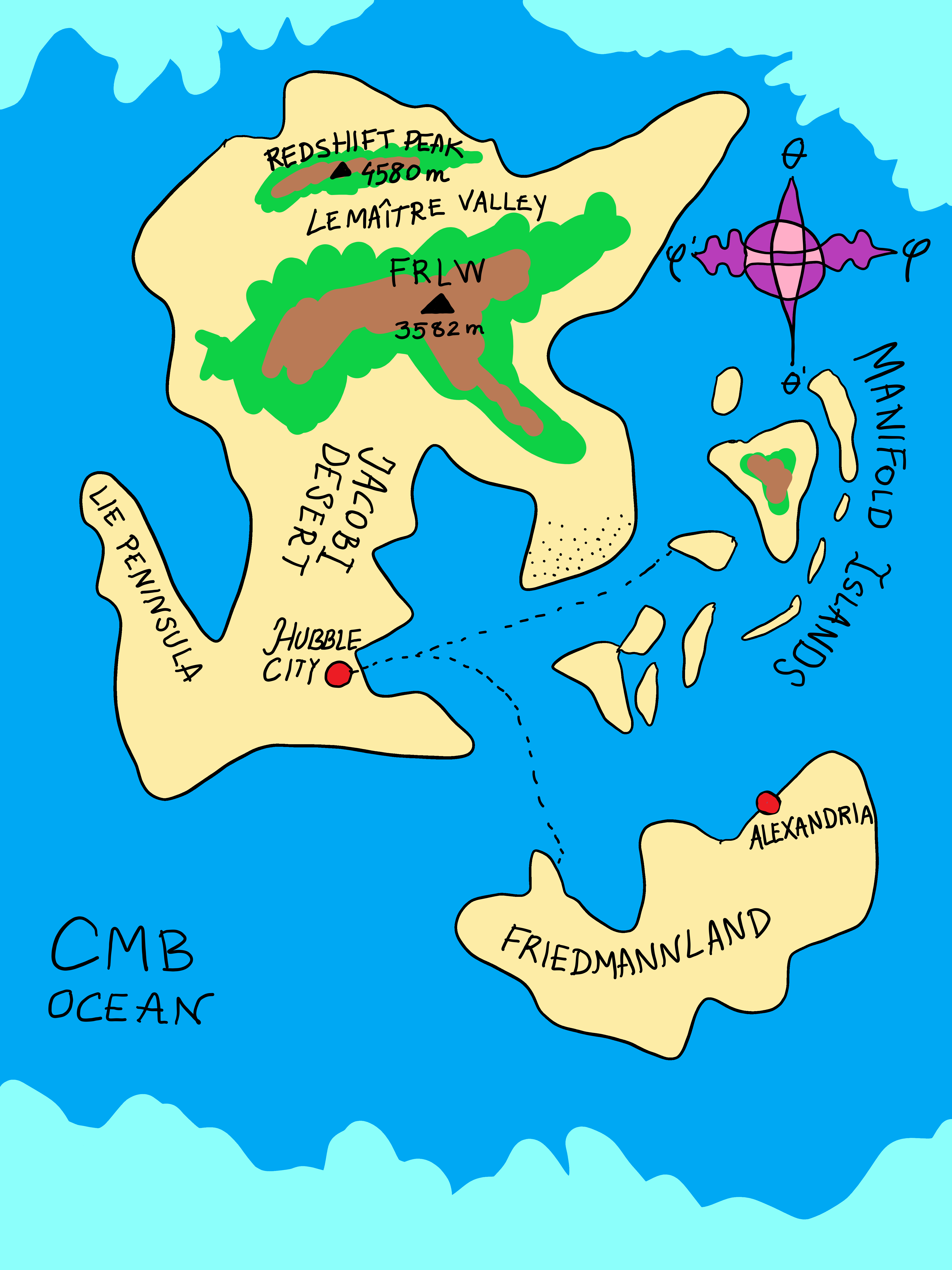} 
	\caption*{\textbf{A Journey into Differential Geometry}, Juan Manuel Garc\'ia Arcos}
\end{figure}

\chapter{General and Historical Introduction} \label{chap:historical_intro}
When you tell people you study Cosmology, they often become very interested. The Sun, the stars, the Solar System or the cosmos, they have always fascinated mankind. In the past, Astronomy, Astrology or Cosmogony were only one topic at the border between science and spirituality. Nowadays, Astronomy and Cosmology are established topics in Physics, in which a lot of research is carried out. In this Chapter, first we introduce \ac{GR} and Cosmology from a generic context, having \emph{guest stars} with us. Then, we present the current work more precisely. This Chapter is mostly qualitative. Its purpose is to introduce the subject of my thesis to a broad audience. Each of the following parts will be introduced and concluded more precisely.

Retracing the history of Physics and  Cosmology is no easy task. There is definitely not a \emph{first} physicist. However, it is still possible to identify some milestones in the development of physical theories. The first person we consider is Galileo Galilei (1564-1642). He was, among other activities, an Italian physicist and he is often considered as the father of modern Physics. He developed the scientific method through his astronomical observations. His work was extremely useful in discovering and understanding the Solar System. For example, he was one of the first people to describe the topography of the Moon and he discovered some of Jupiter's moons. One of his most famous works (if not \emph{the}) is \emph{Dialogue Concerning the Two Chief World Systems} (1632) \cite{galilee:1632} where he, through his characters Sagredo, Salviati and Simplicio, defends heliocentrism, a system where the Sun is in the center of the Solar System. He also argues that physical experiments should yield the same results if made in different frames, moving at constant speed relative to each other. This is nothing else but Galilean Relativity, the ancestor of \ac{GR}!

One year after Galilei's death, baby Isaac Newton (1643-1727) was born\footnote{Newton actually was not born on the same year as Galilei's death. The birth's year of Newton is indeed 1642, but in the \emph{Julian} calendar!} in England. Newton's master piece was definitely \emph{Philosophiæ Naturalis Principia Mathematica} (1688) \cite{newton:1688}, where he introduced the three laws that any teenager studying Physics is taught. Newton observed and studied both the mechanical phenomena on Earth, and the trajectories of astronomical objects, such as the Earth or the planets. His genius was to understand that these effects are actually only one and can be described mathematically using only one formalism. This is how he came to write his \emph{Principia}. This is the canonical example of unification in Physics: Merging two, apparently distinct, fields into one coherent entity, described by one set of physical rules and mathematical equations. This process is what physicists have done historically, and what we still try to achieve in Physics today.

Our third and last guest is no one else than Albert Einstein (1879-1955). More than a century had passed since Newton's death, and a lot of work had been done in Mathematics and Physics. This allowed Einstein to write several articles about mechanics and gravity (1905, 1915) \cite{Einstein:1905,Einstein:19052,Einstein:1915}. This work is, without any doubt, the foundation of \ac{GR} and modern Cosmology.

Today, Cosmology is a science per se. The kick-off of this field can be considered to be the development of the \ac{FRLW} metric in the 1920's by Friedmann, Robertson, Lema\^itre and Walker independently. Soon afterwards, observations showed that the Universe was expanding, a feature easily accommodated by the \ac{FRLW} solution. Another prediction of the \ac{FRLW} metric is the existence of the \ac{CMB}: The Universe should be entirely filled with light, whose spectrum follows a black-body law at a temperature today of $T \approx \SI{2.7}{\kelvin}$. This light was observed in the 60's by Arno Allan Penzias and Robert Woodrow Wilson. This observation is probably the strongest proof for the cosmological model: The Universe has been expanding since the Big Bang and is, on large scales, homogeneous and isotropic. 

The story could end now if it were not for three major dramas: The Hubble tension, dark matter and dark energy.

The first problem is the Hubble Tension. It is related with Hubble's law, which states that the receding speed of galaxies with respect to us is proportional to their distance from the Milky Way. The coefficient of proportionality is the Hubble factor. As a matter of fact, we can measure the Hubble factor using two different methods, either by directly measuring the receding velocities of the nearest galaxies, or extracting it from the inhomogeneous properties of the \ac{CMB}. The problem is the following: These two methods do not lead to the same values (including the experimental errors). The source of this discrepancy is today not known, and several hypothesis are on the table, see \cite{DiValentino:2021} for examples.

The second problem is dark matter. It started with the early theory of Fritz Zwicky in the 1930's which was at first not taken into consideration. The idea was later resurrected by Vera Rubin, Kent Ford and Ken Freeman in the 1970's when they measured the rotation curves of galaxies (i.e. the rotation velocity of objects in galaxies as a function of their distance to the galactic center). The observations suggest that most of the mass of a galaxy is located near its center. Using Newtonian physics (or even including \ac{GR} correction), the rotation velocity should decrease as the distance to the galactic center increases. This is not what they observed. Rather they realised that the rotation speed reaches a plateau at large distance from the galactic center. To explain this observation, we postulate the existence of dark matter, which would interact with the regular matter through gravity, but which would not emit light nor interact with it. Until today, dark matter has never been observed directly. For a more detailed review of the historical development of the topic see \cite{Bertone:2018,Carr:1994}.

The third problem is dark energy. In the Big Bang model, the content of the Universe is not predicted. Let us assume that it is only filled with regular matter, dark matter and light. Using the machinery of \ac{GR} and the assumptions of the model, we would expect that the expansion of the Universe would decelerate, or at most reach a constant rate. Based on the seminal work of Henrietta Swan Leavitt, astronomers were able to measure distances to distant supernovae. Combining this information with their luminosity provides the history of the expansion of the Universe. The work of Adam Riess, Brian Paul Schmidt and Saul Perlmutter in 1998 \cite{Riess:1998} showed that the Universe is indeed in accelerated expansion. To explain such an effect, it is necessary to introduce a new component in the Universe: dark energy. Its properties must be quantitatively different from regular matter and dark matter. Roughly speaking, dark energy should have a \emph{negative} gravity, which could trigger this repulsion and accelerated expansion. As for the two other problems, no satisfactory explanation is known.

Research in Cosmology does not focus only on these three problems, but these are sufficiently important to be put forward. As aforementioned, no satisfactory answer is known for any of them. The existence of dark matter and dark energy can be taken into account in the so-called \ac{LCDM} model. This overarching model assumes that \ac{GR} (and Cosmology) correctly describes gravity and that the Universe is homogeneous, isotropic and filled with matter, cold dark matter and dark energy. Observations suggest that dark energy, dark matter and regular matter are responsible for, $70\%$, $25\%$ and $5\%$ of the energy content of the Universe respectively.

This model is not yet entirely satisfactory, as it is only an effective theory and does not specify the nature of the two dark components. Alternatively, it is also possible that \ac{GR} is not the best physical description of gravity. In this case, a theory of \emph{Modified Gravity} is necessary. This theory should however be able to reproduce all the results predicted by \ac{GR}, which makes the establishment of such a theory not an easy task. See \cite{Clifton:2012,Mannheim:2006,Nojiri:2017} for examples of what has been done in the past.

Whether the \ac{LCDM} model is correct, or whether Modified Gravity is the answer, it appears necessary to make as many and as precise as possible independent cosmological observations. Combining them with experimental results allows us to constrain and eliminate different models. In this thesis, I mostly focus on two major topics: Galaxy surveys and \ac{GW} observations.

First, galaxy surveys is a vast field in which astronomers observe galaxies in the sky together with as many properties as possible. This topic includes two different projects. First, I present the \ac{GNC} and argue why this quantity is frame-independent. Second, I show how interesting properties of galaxies such as their position (parametrised by the redshift and their angle in the sky), their orientation and their polarisation can be used and combined to extract information about cosmological models.

Second, \ac{GW} astronomy has become a hot topic since $2015$ and the first observation of \ac{GW} by LIGO. \ac{GW} open a totally new window of observations, and hence they are the second direction I took in my thesis. More specifically, I investigate the correction to the frequencies of \ac{GW} around a Schwarzschild Black Hole in a theory of Modified Gravity.

To summarise, the goal of the present work is to present several ways to either test \ac{GR}, or provide good candidates for future observations. Special effort will be made to convince the reader that the various results we present are good physical observables, in a sense which will be discussed below.

This thesis is structured as follows. Part~\ref{Part:Intro} is the current part, in which we introduce the work and the useful tools in \ac{GR} and Cosmology which will be used in this thesis. Part~\ref{Part:Frames} is based on \cite{francfort_2019}. We present the \ac{GNC}, which is a true observable in galaxy surveys. We show that this quantity is independent of the frame, focusing in the theory of conformal frames. Part~\ref{Part:RotLensing} is based on \cite{Francfort:2021,Francfort:2022}. We show that the rotation of the main-axes of galaxies with respect to the light polarisation is a good observable sensitive to shear. We explain how to apply this result to a Schwarzschild and a cosmological setup to probe cosmic shear. Part~\ref{Part:BH} is based on \cite{Francfort:2020}. We present an effective theory of Modified Gravity, where generic corrections to \ac{GR} are considered in an agnostic manner. We focus our study on Schwarzschild-like metrics, and we specifically study two observables: The speed of \ac{GW} and their \ac{QNM}. Part~\ref{Part:Conc} is the conclusion of the present work, where we summarized what has been done and propose future directions of investigation. Part~\ref{Part:App} is a dumping ground of appendices where we provide some expressions and explanations that we omit in the main text.

Let us start this journey together, I hope you will enjoy the ride!

\poubelle{

In this chapter we discuss blabla.
hahahahaha blanc bec 

\textbf{conventions and abbreviations}
signature metric ; convention riemann tensor. \\
List of symbols \\
List of abbreviations
$$
dxa dxb = dx \otimes dx /2
$$
point egal time derivative. prime egal conformal itme. hubble fsctor. comma egal derivartive, semicolon nabla\\
mu nu pour 1,4 et 0 pour temps et ij pour espace\\
ab the angles \\
Einstein whtout tilde Jordan with.\\
phi : lensing potential

chi : orientation of the shear

Bold = tensors vectors but not spatial.

Mettre co2 emissions these

L'element de volume : $\bds d \Omega^2$

\textbf{A FAIR}
\begin{itemize}
    \item Les majuscules dans les titres trouver une coherence. Je pense dans les titres oui mais pas dns le texte.
    \item en grad ou pas nabla a decider.
    \item legendes des imagesen  gras a la ligne etc..
    \item Appendix, Section, Chapter, Fig Tab, Rem. Definition 
    \item Mettre les arguments des fonctions et l'ordre : d'abord z etc.
    \item dans chaque chapitre mettre le meme nom pur intro et onclusion.
    \item faire les chapeauxdes intros dans les bons endroits.
    \item encvadrer les eqs
    \item observables : i dont wanna chooosee
    \item categorie de importantes rmq ??????
\end{itemize}

Images : inflation money prix du gaz

}

\chapter{Differential Geometry and General Relativity \label{chap:DGGR}} 
In this Chapter, we first briefly present some useful concepts of Differential Geometry to study \ac{GR}. Differential Geometry is a beautiful topic well-adapted to describe curved spaces or spacetimes. Its use goes much further than \ac{GR}. To give only one example, the physics of membrane in living organism can be described by Differential Geometry. Here we give only a brief introduction, but much more can be found in the literature, for example in the very nice toolkit of Poisson~\cite{poisson_2004}, or the Lecture Notes of Gourgoulhon~\cite{gourgoulhon_2012}. These are very nice overviews of the Mathematics of Differential Geometry, together with an introduction to the $3+1$ formalism.

After this introduction, we present the most important ideas and tools of \ac{GR}, which is, at least today, the best mathematical and physical theory to describe Gravity and Cosmology. Again, this chapter is not aimed to be used as Lecture Notes, but rather the goal is to put into context the definitions and the vocabulary relevant to study the topic. Plenty of references exist in the literature, for example the famous book written by Wald~\cite{wald_1984}, or the nicely written \emph{nutshell} by Zee~\cite{zee_2013}.

\startchap

\section{Differential Geometry in a Nutshell} \label{sec:DGGR}
\subsection{Manifolds and Charts}
When you walk in the street, you somehow feel like you are moving in $\mathbb R^2$. Differential Geometry is the mathematical framework allowing us to formalise this idea. We will provide here the main ideas and concepts used in \ac{GR}. Differential Geometry is definitely more complicated and interesting that what is presented here. The interested reader can find more information in \cite{Straumann:2013,zee_2013}. In Differential Geometry, the main objects are manifolds, which we define now.

\begin{mydef}[\textbf{Manifold}]
Let $\Mcal$ be a topological set and $\{ \Ucal_\alpha\}_\alpha$ a collection of open sets such that for every $x\in \Mcal$ there exists (at least) one $\alpha$ with $x\in \Ucal_\alpha$. Moreover, for each open set $\Ucal_\alpha$ there exist a $C^\infty$-diffeomorphism $\psi_\alpha : \Ucal_\alpha \rightarrow  \mathbb{R}^n $.\\
With these properties, the topological set $\Mcal$ is a $n$-dimensional manifold. 
\end{mydef}
The functions $\psi_\alpha$ are called the charts of the manifold, and the image $\psi_\alpha(x)\in \mathbb{R}^n$ are the coordinates of the point $x$, which we will, in general, not distinguish from the points $x$ itself. The coordinates of a point $x$ are generally written $x^\mu$, with $\mu = 1, \dots, n$.

\begin{figure}[ht!]
	\centering
	\includegraphics[width = 0.95\textwidth]{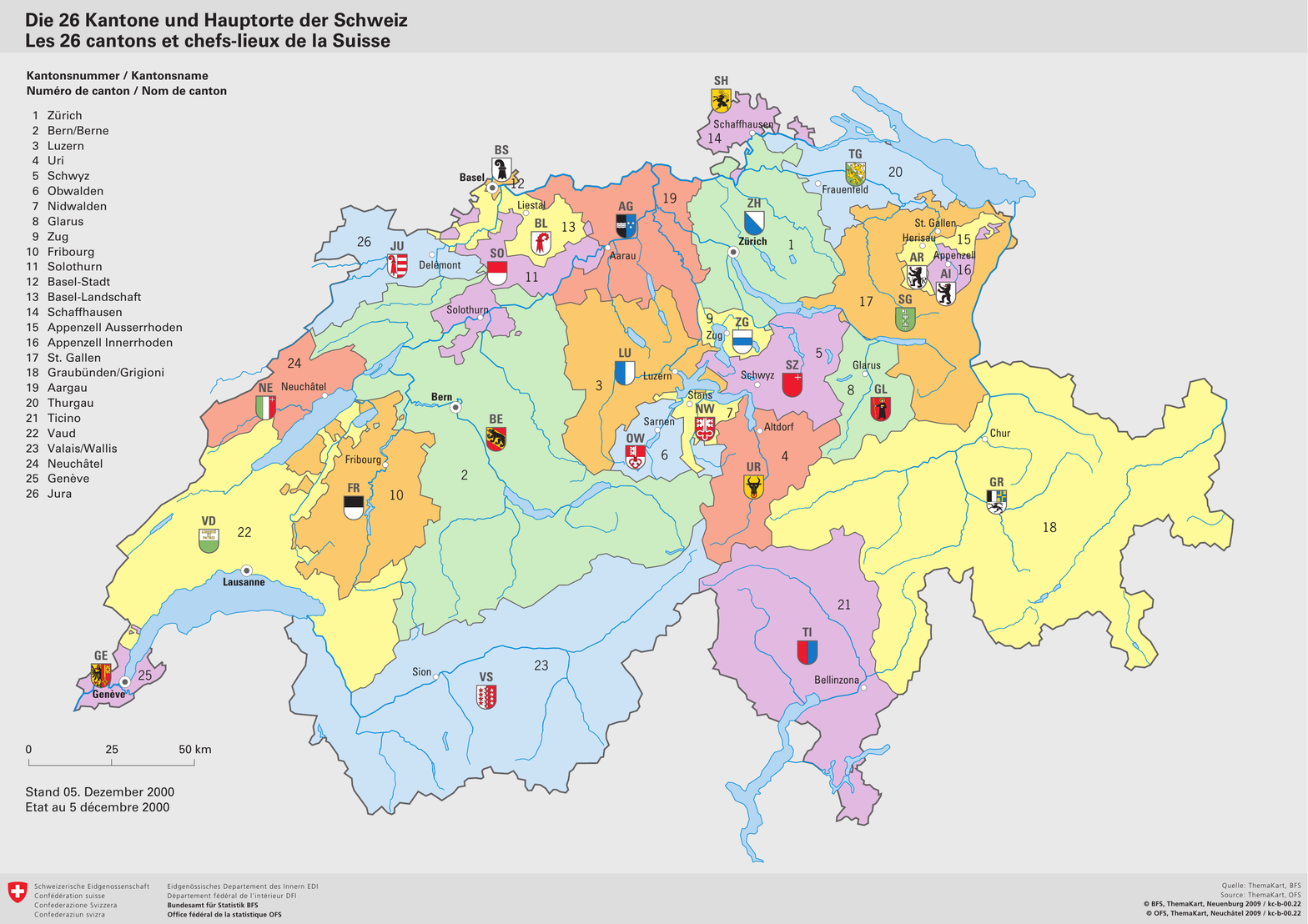} 
	\caption[Les 26 cantons et chefs-lieux de la Suisse]{\textbf{Les 26 cantons et chefs-lieux de la Suisse}\\
	Switzerland is a good example of a manifold: The country is divided into cantons, corresponding to the sets $\Ucal_\alpha$. Federalism being what it is, each of them is free to choose any set of coordinates on its own territory. For example, Geneva can use the usual latitude and longitude coordinates, Valais can use the distance to the Rh\^one, as well as the position of the projection on the river, while Ticino may define a Cylindrical-like system centered around Bellinzona. Note that each of this patch should be slightly extended above the boundaries of the respective canton, allowing overlaps and proper coordinate transformations, when one goes from one canton to another.\\
	Conf\'ed\'eration Suisse, D\'epartement F\'ed\'eral de l'Int\'erieur (DFI), Office F\'ed\'eral de la Statistique (OFS) \cite{cantons:2022}}
\end{figure}

\MYrem{Physics does not depend on coordinates}{For a given point $x$, there may be several $\alpha$ such that $x\in \Ucal_\alpha$, say $\alpha_1$ and $\alpha_2$. This leads to some ambiguity regarding the definition of the coordinates of $x$: Shall we take $\psi_{\alpha_1}(x)$ or $\psi_{\alpha_2}(x)$?

The answer is that both choices are correct, and this is one of the main idea of relativity: Physics\footnote{Here, we intentionally use the word \emph{Physics} in a very vague way.} should not depend on the coordinates used to describe and analyse the situation. Hence, while intermediate computations are often made in a particular system, any physical observable we measure in an experiment should not depend, in fine, on the coordinate system.}

\subsection{Vectors, Forms and Tensors}
In this section, we present the Mathematics of tensors and we see why it is relevant to do Physics in a coordinate-independent fashion. In this section, we consider a generic $n$-dimensional differentiable manifold $\Mcal$. Moreover, we do not distinguish between points of a manifold and their coordinates $x^\mu \in \mathbb{R}^n$. The first step is to define Vectors, and the space they belong to, the Tangent Space.
 
\begin{mydef}[\textbf{Tangent Space}]
At each point $x\in \Mcal$, we define the Tangent Space $T_x(\Mcal)$ as the set of derivatives. A derivative is an linear operator acting on a scalar function satisfying the Leibniz property.
\end{mydef}
\noindent
Roughly speaking, given a function $f: \mathcal{M} \rightarrow \mathbb{R}$, we can define a derivative at a given point $x$ by choosing $n$ numbers $v^\mu$ and setting $\boldsymbol{v}(f) = v^\mu \partial_\mu f$. It is straightforward to show that $\boldsymbol{v}(f)$ is a linear operator and satisfies the Leibniz property. This simple example leads us to the following definition.
\begin{mydef}[\textbf{Basis Vectors}]
At each point $x\in \Mcal$, a basis of the tangent space $T_x(\Mcal)$ are the set of partial derivatives $\{\boldsymbol{\partial}_\mu \}_{\mu=1}^n$.
\end{mydef}
Using this basis, we construct a vector at a point $x$ as $\boldsymbol v = v^\mu \boldsymbol{\partial}_\mu$. A vector field on $\Mcal$ (or on an open set $\mathcal{N} \subset  \Mcal$) is built by assigning at each point $x$ a set of components $v^\mu(x)$, i.e.
\be 
\boldsymbol v(x) = v^\mu(x) \boldsymbol{\partial}_\mu\,.
\ee 

The partial derivatives being defined with respect to a coordinate system $x^\mu$, we expect the components $v^\mu(x)$ of the vector field to depend on the coordinate system. The important point is that the vector, as a mathematical entity, is invariant. More precisely, given two coordinate systems $(x^\mu)$ and $(\tilde{x}^{\tilde{\mu}})$\footnote{Note that we put a tilde also on the index, as the abstract indices may be different from one system to another, e.g. if we go from the Cartesian coordinates $(x,y)$ to the Cylindrical ones $(r,\varphi)$.}, we have (at a point $x\in \Mcal$)
\be 
\boldsymbol{v}(x) = v^{\mu}(x) \bpt_\mu = \tilde{v}^{\tilde{\nu}}(\tilde x) \tilde{\bpt}_{\tilde{\nu}}\,.
\ee 
From this equality, and using the chain rule to convert the partial derivatives, we get the transformation rule for Vectors
\boxemph{ \label{eqgr:vecrule}
\tilde{v}^{\tilde \nu}(\tilde x) = v^\mu(x) \frac{\pt \tilde{x}^{\tilde{\nu}}}{\pt x^\mu}\,,
}
where it is implied that $x$ and $\tilde x$ are the coordinates of the same point and are related by a given coordinate transformation.

As mentioned before, at a given point $x\in \Mcal$, the set of basis tangent vectors forms the vector space $T_x(\Mcal)$. We want to build objects allowed to \emph{eat} vectors. This is possible with the help of the two following definitions.
\begin{mydef}[\textbf{Cotangent Space}]
At each point $x\in \Mcal$, the cotangent space $T_x^\star(\Mcal)$ is the dual space of the tangent space $T_x(\Mcal)$. In other words, it is the space of linear real functions on the tangent space $f:T_x(\Mcal)\rightarrow  \mathbb{R}$.
\end{mydef}
\begin{mydef}[\textbf{Coordinate $1$-Forms}]
At each point $x\in \Mcal$, and for a given system of coordinates, the coordinate differentials $\{ \bdd x^\mu\}_{\mu=1}^{n}$ form a basis of the cotangent space $T_x^\star(\Mcal)$ such that 
\be 
\bdd x^\mu(\bpt_{\nu}) = \tensor{\delta}{^\mu_\nu}\,.
\ee
\end{mydef}
From this definition, we construct an arbitrary form as a linear combination of the coordinate form, namely
\be
\boldsymbol{\omega}(x) = \omega_\mu(x)\, \bdd x^\mu\,.
\ee 
Using again the change rule, we get the coordinate transformation rule for forms
\boxemph{ \label{eqgr:formrule}
\tilde{\omega}_{\tilde \nu}(\tilde x) = \omega_\mu(x) \frac{\pt x^\mu}{\pt \tilde{x}^{\tilde \nu}}\,.
}
\MYrem{Dimension of the tangent and cotangent spaces}{From the two definitions above, it is clear that the dimension of the manifold $n$ is also the dimension of the tangent space and of the cotangent space.}

Finally, we describe fundamental objects in Physics: Tensors. A Tensor is, roughly speaking, a combination of vectors and forms. More precisely, a tensor $\boldsymbol{T}$ of type $(p,q)$ is an object of the form
\be
\boldsymbol{T} = \tensor{T}{^{\mu_1 \dots \mu_p}_{\nu_1 \dots \nu_q}} \bpt_{\mu_1} \otimes \dots \otimes \bpt_{\mu_p} \otimes \bdd x^{\nu_1} \otimes \dots \otimes \bdd x^{\nu_q}\,.
\ee 
Under a change of coordinates, the coordinates of the tensor change combining the \emph{simple} rules for tensor and forms given by Eq.~\eqref{eqgr:vecrule} and Eq.~\eqref{eqgr:formrule} for each single index, namely
\boxemph{
\tensor{\tilde{T}}{^{\tilde{\mu}_1 \dots \tilde{\mu}_p}_{\tilde{\nu}_1 \dots \tilde{\nu}_q}}(\tilde x)
=
\tensor{T}{^{\mu_1 \dots \mu_p}_{\nu_1 \dots \nu_q}}(x)
\frac{\pt \tilde{x}^{\tilde{\mu}_1}}{\pt x^{\mu_1}}
\dots 
\frac{\pt \tilde{x}^{\tilde{\mu}_p}}{\pt x^{\mu_p}}
\frac{\pt x^{\nu_1}}{\pt \tilde{x}^{\tilde{\nu}_1}}
\dots 
\frac{\pt x^{\nu_q}}{\pt \tilde{x}^{\tilde{\nu}_q}}\,.
}

\MYdef{Covariant and contravariant}{
Often, upper indices are called \emph{contravariant indices} and lower indices \emph{covariant indices}. We will avoid as often as possible to use these expressions.}

\subsection{The Metric}
We present in this Section one of the most (if not the) important tensor in \ac{GR}: The Metric.
\begin{mydef}[\textbf{Metric}]
A metric $\boldsymbol{g}$ is a tensor of type $(0,2)$ defined at each point, namely 
\boxemph{
\boldsymbol{g}=\bdd s^2 = g_{\mu \nu}\; \bdd x^\mu \bdd x^\nu\,,
}
satisfying the following properties.
\begin{enumerate}
    \item The metric must be symmetric
    \be 
    g_{\mu \nu}= g_{\nu \mu}\,.
    \ee
    \item The metric must be non-degenerate, i.e. the components $g_{\mu \nu}$ seen as a matrix should be invertible. The inverse metric is denoted
    \be 
        \bds g = g^{\mu \nu} \bpt_\mu \bpt_\nu\,,
    \ee
    i.e.
    \be 
    g_{\mu \nu}(x) g^{\nu \rho}(x) = \tensor{\delta}{_\mu^\rho}\,.
    \ee
    \item The metric should have the signature $(-,+,+,+)$, i.e. there is an orthonormal system of $4$ vectors $\bds{v}_a$ satisfying
    \be 
    \bds{v}_a \cdot \bds{v}_b = \tensor{v}{_a^\mu} g_{\mu \nu}\tensor{v}{_b^\nu}=\eta_{ab}\,,
    \ee 
    where $\eta_{ab}$ is the Minkowski metric
    \be  \label{eqGR:eta}
    \bds \eta = 
    \begin{pmatrix}
    -1&0&0&0\\ 0 & 1 &0&0 \\ 0 &0 &1 &0 \\ 0 &0&0&1
    \end{pmatrix}\,.
    \ee
\end{enumerate}
\end{mydef}
\MYrem{Signature}{
The negative sign in the Minkowski metric Eq.~\eqref{eqGR:eta} corresponds to timelike intervals. This makes time a \emph{primus inter pares}: Even if time and space are set on equal footage in \ac{GR}, they still differ by this sign, which actually leads to very interesting aspects, e.g. regarding causality.
}

\section{General Relativity}
\subsection{A Collection of Tensors \label{secGR:collec}}
With the tools presented in the previous section in our hand, we can now build more objects which are very useful in \ac{GR}. The first objects we construct are the Christoffel symbols.
\begin{mydef}[\textbf{Christoffel symbols}]
The Christoffel symbols $\tensor{\Gamma}{^\rho_{\mu \nu}}$ are defined through the metric as
\boxemph{ \label{eqGR:Christoffel}
\tensor{\Gamma}{^\rho_{\mu \nu}} = \frac 12 g^{\rho \sigma} 
\left( \pt_\nu g_{\mu \sigma} + \pt_\mu g_{\sigma \nu} - \pt_\sigma g_{\mu \nu} \right)\,.
}
\end{mydef}
The Christoffel symbols are symmetric in their lower indices and, despite the notation, do not form a tensor. From these objects, we can construct the Covariant derivative.
\begin{mydef}[\textbf{Covariant derivative}]
The covariant derivative $\boldsymbol{\nabla}$ is an operator whose action on scalar functions gives the differential, namely
\be
\boldsymbol{\nabla} f \equiv \bdd f = \nabla_\mu f\,  \bdd x^\mu\,,
\ee
with $\nabla_\mu f= \pt_\mu f$. For a tensor of type $(1,1)$ given by $\boldsymbol{T}=\tensor{T}{^\mu_\nu} \bpt_\mu \otimes \bdd x^\nu$, its covariant derivative is given by
\be
\boldsymbol{\nabla} \boldsymbol{T}
=
\nabla_\rho \tensor{T}{^\mu_\nu}\,  \bdd x^\rho \otimes \bpt_\mu \otimes \bdd x^\nu\,,
\ee 
with
\boxemph{
\nabla_\rho \tensor{T}{^\mu_\nu}  = 
\partial_\rho \tensor{T}{^\mu_\nu} 
+ \tensor{\Gamma}{^\mu_{\rho \alpha}} \tensor{T}{^\alpha_\nu}
- \tensor{\Gamma}{^\alpha_{\rho \nu}} \tensor{T}{^\mu_\alpha}\,.
}
The generalisation to more indices is straightforward. Moreover, note that often the following notation is used
\be
\nabla_\rho \tensor{T}{^\mu_\nu}  = \tensor{T}{^\mu_\nu_{;\rho}}\,.
\ee 
\end{mydef}
As it names suggests, the covariant derivative is the generalisation of the derivative for curved spacetime. From the Christoffel symbols, we can also build the various curvature tensors.
\begin{mydef}[\textbf{Riemann curvature tensor}]
The Riemann curvature tensor (or Riemann tensor) $\tensor{R}{^\mu_{\nu \rho \sigma}}$ is a tensor of type $(1,3)$ whose components are given by
\boxemph{
\bds{\mathcal{R}} &= \tensor{R}{^\mu_{\nu \rho \sigma}} \, \bpt_\mu \otimes \bdd x^\nu \otimes \bdd x^\rho \otimes \bdd x^\sigma\,, \\
\tensor{R}{^\mu_{\nu \rho \sigma}} &=
\tensor{\Gamma}{^\mu_{\rho \alpha}}\tensor{\Gamma}{^\alpha_{\nu \sigma}}
-\tensor{\Gamma}{^\mu_{\sigma \alpha}}\tensor{\Gamma}{^\alpha_{\nu \rho}}
+\pt_\rho \tensor{\Gamma}{^\mu_{\nu \sigma}}
-\pt_\sigma \tensor{\Gamma}{^\mu_{\nu \rho}}\,.
}
\end{mydef}
\begin{mydef}[\textbf{Ricci curvature tensor}]
The Ricci curvature tensor (or Ricci tensor) $\tensor{R}{^\mu_{\mu \nu}}$ is a tensor of type $(0,2)$ whose components are constructed via the contraction of the Riemann tensor via
\boxemph{
\bds{R} &= \tensor{R}{_{\mu \nu}}\, \bdd x^\mu \otimes \bdd x^\nu\,, \\
\tensor{R}{_{\mu \nu}} &=
\tensor{\delta}{_\alpha ^\beta}\tensor{R}{^\alpha_{\mu \beta \nu}}
=
\tensor{R}{^\alpha_{\mu \alpha \nu}}\,.
}
\end{mydef}
\begin{mydef}[\textbf{Ricci curvature scalar}]
The Ricci curvature scalar (or Ricci scalar) $R$ is the trace or the Ricci tensor, namely
\boxemph{
R =
g^{\mu \nu}\tensor{R}{_{\mu \nu}}
=
\tensor{R}{^\mu_{\mu}}\,.
}
\end{mydef}
\MYrem{Identities of the curvature tensors}{
The Riemann curvature tensor satisfy the following identities
\begin{align}
    \label{eqGR:riemid1}
    R_{\mu \nu \rho \sigma}&= -R_{\nu \mu \rho \sigma} = -R_{\mu \nu \sigma \rho} \,, \\
    \label{eqGR:riemid2}
    R_{\mu \nu \rho \sigma}&= R_{\rho \sigma\mu \nu} \,, \\
    \label{eqGR:riemid3}
    R_{\mu \nu \rho \sigma} +R_{\mu \sigma\nu \rho } +  R_{\mu \rho\sigma\nu  } &= 0 \,, \\
        \label{eqGR:riemid4}
        R_{\mu \nu \rho \sigma;\alpha} + R_{\mu \nu \alpha\rho; \sigma}
        +R_{\mu \nu  \sigma\alpha;\rho}&=0\,.
\end{align}
The last two formulas are the Bianchi identities. Moreover, the Ricci curvature tensor is symmetric, i.e.
\be
R_{\mu \nu} = R_{\nu \mu}\,.
\ee
}

The covariant derivative is very useful and allows us to define the concept of parallel transport.
\begin{mydef}[\textbf{Parallel transport}]
A vector $\boldsymbol{u}=u^\mu \bpt_\mu$ is parallel transported along a vector $\boldsymbol{v}=v^\mu \bpt_\mu$ if
\be  \label{eqgr:partra}
\boldsymbol{\nabla}_{\boldsymbol{v}} \boldsymbol{u}
=0\,,
\ee 
or in components
\be 
\nabla_{\bds v} u ^\nu =  v^\mu \nabla_{\mu} u^\nu  =0\,.
\ee 
\end{mydef}
A trajectory $x^\mu(\tau)$, where $\tau$ is an affine parameter, is a geodesics if the 4-velocity $u^\mu = \dd x^\mu / \dd \tau$ is parallel transported along itself. This concept is central if one works in curved spacetime, as it generalizes the notion of straight line (or the one of the shortest path between two point). It is also a key concept in the theory of \ac{GR} as we will explain below.

\subsection{Einstein Field Equations}
We defined in the previous section the metric, and derived various tensors from it. However, one question still remains: For a given physical situation, how can we determine the tensor $\bds{g}(x)$ at any spacetime point? 

We know from Newtonian Mechanics that the matter density $\rho$ determines the gravitational potential $\Phi$ through the Poisson equation. We would like to find the generalisation of this relation in \ac{GR}. This can be done starting with an action and using the variational principle. The Einstein-Hilbert action is
\boxemph{
\label{greq:EH}
\mathcal{S}_{\mathrm{EH}}[\bds{g} ,\psi] = \int \; \left(\frac{1}{16\pi \GN} R + \mathcal{L}_{\mathrm{m}}[\bds{g},\psi]\right) \sqrt{-g}\, \mathrm{d}x^4 x\,,
}
where $\psi$ denotes collectively all the \emph{matter} fields which we included in the Lagrangian $\mathcal{L}_{\mathrm{m}}[\bds{g},\psi]$, $R$ is the Ricci scalar built from $\bds g$ and $g$ its determinant, where $\bds g$ seen as a matrix. The matter Lagrangian depends on the metric only through minimal coupling with matter (e.g. kinetic terms are expressed with the covariant derivatives). We do not explain here why we take this Ansatz as the correct action describing gravity. Rather, we take it as a \emph{Deus ex machina}, and derive its consequences.

Varying the Einstein-Hilbert action given by \myeq{greq:EH} with respect to the metric $\bds g$ leads to the Einstein Field Equations (or Einstein's Equations)
\boxemph{
\label{eqgr:EFE1}
\bds{G} &= 8 \pi \GN \bds{T}\,, \\
\label{eqgr:EFE}
G_{\mu \nu} &= 8 \pi \GN T_{\mu \nu}\,,
}
where 
\begin{align} 
\label{eqGR:Gmn}
\bds{G} &\equiv \bds{R} - \frac 12 R \bds{g} \,, \\
G_{\mu \nu} &\equiv R_{\mu \nu} - \frac 12 R g_{\mu \nu} \,.
\end{align}
is the Einstein tensor, $\GN$ is Newton's constant, and 
\be 
\label{eqGR:emt}
T_{\mu \nu} = - \frac{2}{\sqrt{-g}} \frac{\delta (\sqrt{-g} \mathcal{L}_{\mathrm{m}})}{\delta g^{\mu \nu}}\,,
\ee
is the energy-momentum tensor quantifying the matter and energy content of the spacetime. 

\MYrem{Einstein's Equations in vacuum}{In vacuum, with $\bds T=0$, the Einstein's Equations read $\bds{G}=0$. The trace of the Einstein tensor Eq.~\eqref{eqGR:Gmn} is\footnote{Recall that the Newton constant is $\GN\neq G$.}
\be 
G = -R\,,
\ee 
which implies
\be 
R_{\mu \nu} = G_{\mu \nu} - \frac 12 G g_{\mu \nu}\,.
\ee 
Hence, in vacuum, the condition $\bds G=0$ is equivalent to $\bds R=0$.
}

One last comment regarding the consistency of this equation. The Einstein tensor is identically divergence free
\be  \label{eqgr:bianchiG}
\nabla_\mu \tensor{G}{_\nu ^\mu} =0\,.
\ee 
This condition is also sometimes called the Bianchi identity. From the Einstein's Equations, it implies 
\begin{align}  \label{eqgr:conseq}
\nabla_\mu \tensor{T}{_\nu ^\mu} &=0\,, \\
\bds \nabla \cdot \bds T &=0\,.
\end{align}
This last condition is nothing else as the energy-momentum conservation equation! This is a safety check for the validity of the Eintein's Equations, as we know from Special Relativity and from Electrodynamics that the energy-momentum tensor should satisfy such a condition.

\subsection{Motion of Particles}
The Einstein's Equations are the machinery to determine the metric $\bds g$ if the distribution of matter and energy is known. However, the dynamics of the particle is itself governed by the metric structure of spacetime. We would like to derive the equations of motion for particles moving inside a gravitational field. 

The starting point is, as usual, an action. We consider a particle whose trajectory is given by $x^\mu(\lambda)$, where $\lambda$ is an affine parameter. This means that we can reparametrise it as $\tilde{\lambda} = \tilde{\lambda}(\lambda)$, provided the relation is a bijection. From this trajectory, we define the $4$-velocity through
\begin{align}
    \bds u &= u^\mu \bpt_\mu \,,
    u^\mu &= \frac{\dd x^\mu}{\dd \lambda}\,.
\end{align}
The $4$-velocity together with the metric allows us to compute the infinitesimal time (or distance) between $\lambda$ and $\lambda + \dd \lambda$ as
\be 
\dd \tau^2 = g_{\mu \nu} \dd x^\mu \dd x^\nu = g_{\mu \nu} u^\mu u^\nu \, \dd \lambda^2 = \bds u^2 \, \dd \lambda^2\,.
\ee
Moreover, we know from Relativistic Mechanics that this quantity is $0$ if the particle is massless, and negative if it is massive. In addition, we can, in the case of a massive particle, adjust the affine parameter such that $\bds u^2 = -1$, which we will assume from now on.

We claim now that the relativistic action for a particle is simple the total proper time of the trajectory
\be  \label{eqGR:actionpart}
\mathcal{S} \propto - \int\; \dd \tau = - \int\; \sqrt{-g_{\mu \nu} u^\mu u^\nu }\, \mathrm{d}\lambda\,.
\ee
We do not specify here the correct prefactor as it can differ between the massless and the massive case.

Varying the action Eq.~\eqref{eqGR:actionpart} with respect to the trajectory $x^\mu$ leads to the equations of motion
\begin{align}
    \ddot{x}^\mu + \Gamma^\mu_{\alpha \beta} \dot{x}^\alpha \dot{x}^\alpha =0\,.
\end{align}
These equations are equivalent to the geodesics condition
\be 
\bds{\nabla}_{\bds{u}} \bds{u}=0
\ee 
mentioned in Section~\ref{secGR:collec} which is the final result of this Chapter: Particles, whether they are massive or massless, follow geodesic when moving on a curved spacetime. Note that this does not hold in some theories of Modified Gravity when the action for matter particles can be different (we will see such an example in Part~\ref{Part:Frames}).

\stopchap
\chapter{Units and Observables}
In this chapter we discuss the concepts of units and observables. Even if this topics can seem boring or not worth your time, I do think that a thorough discussion is always enlightening. We first discuss the natural system of units from a more conceptual point of view as what is usually done, and then we try to give a (qualitative) definition of the notion of observables.

\startchap
\section{Units\label{chapintro:units}}
\subsection{Introduction}
After replying something like \emph{The speed of the ball is $28$}, Many students have heard their Physics teacher shouting \emph{$28$ what? Oranges? Mangos?} In Science, and often in general in life, it does not make sense to describe quantities without units. The only exception being dimensionless number, when separable objects can be counted. For many people, scientists included, units are a nightmare not worth spending time. I remember the nightmare it was, in middle school, when I was asked to draw tables to convert $\SI{119.94}{\dm^3}$ to $\SI{0.0011994}{\dam^3}$...

Historically speaking, several unit systems have been created, in different epochs and countries, and it is difficult to see any rational behind it. Anyone who has been confronted with the differences between, e.g., the Anglo-Saxon and the SI system is familiar with these issues. The problem comes from the fact that Humans defined units using arbitrary rationales. For example, it seems that the Babylonians divided hours in minutes in $60$ as they were counting in this basis. Together with astronomical considerations, this provides an empirical definition of the second. Later, a definition based on some physical process was defined: The second is defined such that the frequency of the emission of the atomic transition between the two hyperfine ground states of Caesium $133$ is exactly $\SI{9 192  631 770}{\second^{-1}}$. This number was obviously chosen such that the final result is close enough to the historical value. When the value of one second is defined, the meter can be defined as the distance travelled by light in vacuum in a given amount of time. Finally, the kilogram can be defined imposing the mass of a meter cube of water at its maximal density. Interestingly enough, there was for several decades a standard mass defining the kilogram and kept in Paris. It was clear that this solution was not perfect, as this object had to bee frequently renewed. The other units were defined in a similar fashion, but we will restrict ourselves to these three for the present discussion. 

In $2018$ basis of the units system were redefined. The philosophy is now different: Rather than defining units with reference objects, the value of various physical constants is defined to be \emph{fixed} in the SI system. This constrains all the other units. For example, the frequency of the caesium $133$, the value of the speed of light $c$ and of the reduced Planck constant $\hbar$ are kept fixed in the SI system, which uniquely defined the second, the meter and the kilogram. This is the philosophy of the natural system, which we present below.

The point of this Section is to show that units are much more than boring tables. For example, they are the basis of dimensional analysis, which has saved me in the past and will save me in the future. They also provide a natural system of units in Theoretical Physics as we will discuss later. Finally, and more important, they are the first step towards the more general concept of observables, which we present in Section~\ref{secintro:Obs}.

\subsection{The Natural System\label{secintro:natural}}
We want to motivate the definition the natural system of units. As explained before, in this system, the values of some natural constants are kept fixed. We will restrict ourselves to time, lenghts and masses, and their respective SI units, second, meter and kilogram. To define precisely these three units, we need first the speed of light and the Planck constant. We introduce the following notation
\begin{align}\label{eqintro:defc}
        c &= \cSI \, \mathrm{m} \times \mathrm{s}^{-1}\,, \\
        \label{eqintro:defhbar}
        \hbar &= \hSI\, \mathrm{kg} \times \mathrm{m}^{2} \mathrm{s}^{-1}\,. \\
\end{align}
Here, the values $\cSI$ and $\hSI$ are the numerical dimensionless values of these constants when expressed in the SI system. Using the new definition of $2019$, they read
\begin{align}
    \cSI &\equiv \num{299 792 458}\,, \\
    \hSI &\equiv \frac{1}{2\pi} \num{6.62607015e-34}\,.
\end{align}
We need one more constant: The frequency of the caesium transition. However, this is not what is customarily done in the literature. We take the more usual approach where the proton charge is defined to be constant
\be 
e =\eSI \, \mathrm{C}\,,
\ee
with
\be 
\eSI \equiv \num{1.602176634e-19}\,.
\ee 
We stress here that the quantities $\cSI$, $\hSI$ and $\eSI$ are \emph{exact} and are not subjected to any experimental errors. To close the system, we need a third condition which is generally taken to be the definition of the electronvolt
\be  \label{eqintro:defeV}
1\, \mathrm{eV} \equiv  \eSI\, \mathrm{J}=\eSI \, \mathrm{kg} \times \mathrm{m}^{2} \times \mathrm{s}^{-2}\,.
\ee 

The three equations Eq.~\eqref{eqintro:defc}, Eq.~\eqref{eqintro:defhbar} and Eq.~\eqref{eqintro:defeV} can be inverted to express the SI units as
\boxemph{ \label{eqintro:meter_eV}
        1\, \mathrm{m} &= \frac{\eSI}{\cSI \hSI}\,  c \hbar\,  \frac{1}{\mathrm{eV}} \,, \\
        \label{eqintro:second_eV}
        1\, \mathrm{s} &= \frac{\eSI}{ \hSI}\,   \hbar\,  \frac{1}{\mathrm{eV}} \,, \\
         \label{eqintro:kg_eV}
        1\, \mathrm{kg} &= \frac{\cSI^2}{\eSI}\,  \frac{1}{c^2} \,  \mathrm{eV} \,.
}

These relations form the basis of the natural system of units. They can be used to convert quantities from the SI system to the natural system. For example, Newton's constant is given by 
\be
\GN= \GSI\, \mathrm{m}^3 \times \mathrm{kg}^{-1} \times \mathrm{s}^{-2}
= \frac{\GSI \eSI^2 }{\cSI^5 \hSI}\, c^5 \hbar \, \frac{1}{\mathrm{eV}^2}\,.
\ee 
Usually (if not always), people are lazy and just say \emph{We set $c=\hbar=1$}. What they actually mean is that physical quantities are written using powers of $c$, $\hbar$ and $\mathrm{eV}$, and that at the end they artificially set $c=\hbar=1$. In the example above, this would give (using $\GSI\approx 6.67\times 10^{-11}$)
\be  \label{equnits:Gnat}
G = 6.7 \times 10^{-57} \, \frac{1}{\mathrm{eV}^2}\,.
\ee

We also understand from Eq.~\eqref{eqintro:meter_eV}, Eq.~\eqref{eqintro:second_eV}, and Eq.~\eqref{eqintro:kg_eV} why inverse times, inverse lengths and masses are said to have the inverse unit of an energy, as the powers of $\mathrm{eV}$ in their expression agree.

Let us show how this formalism simplify the physical equations. We take as an example the equation for the Schwarzschild radius
\be \label{eqintro:rsFull}
R = 2 \frac{\GN M}{c^2}\,.
\ee
We stress that this equation, in this form, holds in \emph{every} unit systems. If we apply this in the SI system we get
\be 
\RSI \, \mathrm{m} =
\left(2  \GSI \, \mathrm{m}^3 \times \mathrm{kg}^{-1} \times \mathrm{s}^{-2}\right)\, 
\left(\MSI \, \mathrm{kg} \, \cSI^{-2}\right)\,
\left( \mathrm{m}^{-2} \times \mathrm{s}^2\right)\,,
\ee 
or
\be  \label{eqintro:rsSI}
\RSI  =2  \frac{\GSI \MSI}{\cSI^2}\,.
\ee 
This is exactly what you would do to compute the Schwarzschild radius with a simple calculator: Plug the dimensionless number and you know that the final result should be expressed in meters.

We describe now how to rewrite this equation in the natural system. First, we want to see what is the dimensionless value for, say, $R$ in the natural system. We know that $R$, as a physical quantity, does not depend on the unit system, i.e.
\be 
R = \RSI \, \mathrm{m}  = \Rnn \, \frac{c \hbar}{\mathrm{eV}}\,,
\ee 
where $\Rnn$ is the dimensionless value of $R$ in the natural system and we used Eq.~\eqref{eqintro:meter_eV} to know the form of a length in this system. Using again Eq.~\eqref{eqintro:meter_eV}, we get
\be 
\Rnn = \RSI \frac{\eSI}{\cSI \hSI}\,.
\ee 
Similarly, we get for Newton's constant and the mass
\begin{align}
    \Gnn &= \GSI \frac{\eSI^2}{\cSI^5 \hSI}\,, \\
    \Mn &= \MSI\, \frac{\cSI^2}{\eSI}\,.
\end{align}
Using these relations and the fact that, in the SI system, the relation is given by Eq.~\eqref{eqintro:rsSI}, it is direct to show that
\be  \label{eqintro:rsNatural}
\Rnn = 2 \Gnn \Mn\,,
\ee 
or in other words: In the natural system, the physical equation are obtained setting $c_{\mathrm{nat}}=1$! It is very important however to note that the true physical equation is Eq.~\eqref{eqintro:rsFull} involving dimensional quantities, while Eq.~\eqref{eqintro:rsSI} or Eq.~\eqref{eqintro:rsNatural} are simply its translation in one system or another.

\MYrem{Our convention}{In this work, we will be lazy and simply \emph{set} $c=\hbar=1$, without always mentioning it, but we encourage the reader to have the discussion of this Section in mind at all times! Hence, times, length and inverse masses have the same units.}
\MYrem{Planck units}{
	Using only $c$, $\hbar$ and $G$, one can define (the factor of $8\pi$ is a mere convention)
	\begin{align}
	\Lpl &= \sqrt{\frac{\hbar \GN}{c^3}} \,, \\
	\Tpl &= \sqrt{\frac{\hbar \GN}{c^5}} \,, \\
	\Mpl &= \sqrt{\frac{c\hbar}{8\pi\GN}} \,.
	\end{align}
It is direct to check that these quantities are respectively a length, time and mass. They are the Planks units. Their value in the SI system can be computed directly using the SI values of the three constant. The exercise is more interesting in the natural system. Using the definition of Newton's constant in the natural system Eq~\eqref{equnits:Gnat}, we get
\begin{align}
	\Lpl &\approx 8\times 10^{-29} \, \frac{1}{\mathrm{eV}}\sqrt{\frac{\hbar}{c^3}}\,, \\
	\Tpl &\approx 8\times 10^{-29} \, \frac{1}{\mathrm{eV}}\sqrt{\frac{\hbar}{c^5}}\,, \\
	\Mpl &\approx 2\times 10^{27}\,  \mathrm{eV}\sqrt{c \hbar}\,.
\end{align}
And, again, usually the factors $c$ and $\hbar$ are neglected.
}

\subsection{The Vector Space of Physical Values}
This section is inspired by the excellent reference \cite{Ansmann_2015}.  We will follow the ideas presented there, with slight differences in the notations. Similar ideas were presented in~\cite{Uzan:2002}. Again, we only consider here physical quantities which can be expressed as powers of length, time and mass. This includes most of the quantities we want to study in this work (mass and energy density, pressure, frequency, \dots) and the generalisation to quantities with other units (for example temperature) is straightforward.

Any physical quantity $X$ can be expressed as
\be 
X = 10^{\alpha_X} L^{\alpha_\LL} T^{\alpha_\TT} M^{\alpha_\MM}\,.
\label{units:def_X}
\ee 
where $\alpha_X$ is a dimensionless number, $L$, $T$ and $M$ are given units of length, time and mass (such as the meter, the second and the kilogram) and $\alpha_\LL\,, \alpha_\TT\,, \alpha_\MM \in \mathbb{Z}$ (or even $\mathbb{Q}$). The dimensionless number is written as a power of $10$ for reasons that will become clear.

For example, if I use the SI system, my height $h$ corresponds to
\be 
\left(\alpha_h,\alpha_\LL, \alpha_\TT,\alpha_\MM \right) = (0, 1,0,0)
\ee 
and my weight $w$ corresponds to
\be 
(\alpha_w,\alpha_\LL, \alpha_\TT,\alpha_\MM) = (3, 1,-2,1)\,.
\ee 
In these two examples, the first number is to be understood as a power of $10$, i.e. $h\sim 10^0\, \mathrm{m}$ and $w\sim 10^{3}\, \mathrm{N}$. These two examples, yet simple, are very instructive: we understand that any physical quantity, provided a system of units has been chosen, can be fully determined through a $4$-dimensional vector.

In this Vector Space of Physical Values as they call it in the article, the addition and the multiplication of two physical quantities is defined by
\boxemph{
(\alpha_X,\alpha_\LL, \alpha_\TT,\alpha_\MM) \oplus
(\alpha_Y,\alpha_\LL, \alpha_\TT,\alpha_\MM)
&\equiv (\log_{10}(10^{\alpha_X } + 10^{\alpha_Y}),\alpha_\LL, \alpha_\TT,\alpha_\MM)\,, \\
(\alpha_X,\alpha_\LL, \alpha_\TT,\alpha_\MM) \odot 
(\alpha_Y,\beta_\LL, \beta_\TT,\beta_\MM)
&\equiv
(\alpha_X+ \alpha_Y,\alpha_\LL + \beta_\LL, \alpha_\TT + \beta_\TT,\alpha_\MM+\beta_\MM)\,.
}
Note that the first relation holds only if all the $\alpha$'s are equal, or as the old saying goes \emph{You cannot add eggplants to chocolate!}. More details can be found in the article (e.g. the discussion about this space being a vector space).

We want to describe how to go from one system to another. As before, we will consider the International System and the natural system, as describe above. A physical quantity can be written as
\begin{align}
    X&= 10^{\alSI} \, \mathrm{m}^{\alpha_\mathrm{m}} \times \mathrm{s}^{\alpha_\mathrm{s}} \times \mathrm{kg}^{\alpha_{\mathrm{kg}}}\,, \\
    X&= 10^{\aln} \, \mathrm{eV}^{\alpha_{\mathrm{eV}}} \times c^{\alpha_c} \times \hbar^{\alpha_\hbar}\,.
\end{align}
The knowledge of the four quantities $(\xSI, \alpha_m,\alpha_s,\alpha_kg)$ or $(\xn,\alpha_{\mathrm{eV}}, \alpha_c, \alpha_\hbar)$ completely the physical quantity $X$, which, itself, does not depend on the coordinate system. Using the results of the previous sections (especially Eq.~\eqref{eqintro:meter_eV}, Eq.~\eqref{eqintro:second_eV}, and Eq.~\eqref{eqintro:kg_eV}), the transformation rule is given by
\boxemph{
\begin{pmatrix}
\aln \\ 
\alpha_{\mathrm{eV}} \\ 
\alpha_c \\ 
\alpha_\hbar
\end{pmatrix}
=
\begin{pmatrix}
1 &  \log_{10}\left( \frac{\eSI}{\cSI \hSI } \right ) & \log_{10}\left( \frac{\eSI}{ \hSI } \right ) &\log_{10}\left( \frac{\cSI^2}{\eSI } \right )  \\ 
0 &-1  &-1  &1 \\ 
0 &  1& 0 &-2 \\ 
0 &1  &1  &0 
\end{pmatrix}
\begin{pmatrix}
\alSI \\ 
\alpha_{\mathrm m} \\ 
\alpha_{\mathrm s} \\ 
\alpha_{\mathrm{kg}}
\end{pmatrix}\,.
}
This shows that a change of unit system is nothing else as a usual linear coordinate transformation in the vector space of units!

\subsection{Dimensional Analysis for Tensors}
We end this Section by briefly discussing dimensional analysis for tensors. This discussion is based on the nice article by P.G.L. Mana~\cite{Mana:2021}, which goes much deeper as what we are going to discuss now. In this Section, we do not set $c=\hbar=1$ which we keep explicit.

To understand why the question of dimensional analysis, as usual, a simple example is illuminating. We consider the flat metric in the plane, in two coordinates systems 
\be 
\bdd s^2  = \bdd x^2 + \bdd y^2 = \bdd r^2 + r^2 \bdd \theta^2\,.
\ee 
From this example, it is clear that the components of the metric such as $g_{xx}$ or $g_{\theta \theta}$ do not have the same dimension. Hence, and as usual when working with tensors, working with components is not a good idea. On the other side, it is clear that $\bdd{d}s^2$ always has the dimension of a length squared, as it should be as it is the job of the metric tensor. This motivates the following definitions.
\MYdef{Dimensions of the coordinates}{
Let $x^\mu$ be a coordinate system. We define $[x^\mu]$ as the unit of the coordinate $x^\mu$. 
}
For example, using the usual spherical coordinates $x^\mu=(t,r,\theta,\varphi)$, we have
\be 
[x^\mu] = (\mathrm{T}, \mathrm{L},1,1)\,,
\ee 
where $\mathrm{T}$ and $\mathrm{L}$ represent respectively a time and a length.
\MYdef{Intrinsic dimension of a tensor}{
Let $\bds T$ be a generic tensor given in coordinates by
\be 
\boldsymbol{T} = \tensor{T}{^{\mu_1 \dots \mu_p}_{\nu_1 \dots \nu_q}} \bpt_{\mu_1} \otimes \dots \otimes \bpt_{\mu_p} \otimes \bdd x^{\nu_1} \otimes \dots \otimes \bdd x^{\nu_q}\,.
\ee 
All the terms in this sum have the same dimension, called the \emph{intrinsic dimension of the tensor $\bds T$} and denoted $[\bds T]$. In particular, using 
\begin{align} 
    [\bpt_{\mu}] &= \frac{1}{[x^\mu]}\,, \\
    [\bdd x^\nu] &= [x^\nu]\,,
\end{align}
we have
\be 
[\bds{T}] =[\tensor{T}{^{\mu_1 \dots \mu_p}_{\nu_1 \dots \nu_q}}] \,
\frac{ [x^{\nu_1}] \dots [x^{\nu_q}] }{[x^{\mu_1}] \dots [x^{\mu_p}]}\,.
\ee 
}
From these definitions, it is straightforward to derive the intrinsic dimension of the various tensors presented in Chapter~\ref{chap:DGGR}. 

We take the convention such that 
\be 
[\bds{g}]=\mathrm{L}^2\,,
\ee 
or in components
\be 
g_{\mu \nu} = \frac{\mathrm{L}^2}{[x^\mu][x^\nu]} \,.
\ee 
Note that other choices are possible, see \cite{Mana:2021} for more details. From this, the units of the components of the inverse metric are
\be 
[g^{\mu \nu}] = \frac{[x^{\mu}][x^{\nu}]}{L^2}\,
\ee 
which ensures that
\be 
[g^{\mu \alpha}] [g_{\alpha \nu}] = 1\,.
\ee 
Using the definition of the Christoffel symbols and of the various tensors in Section~\ref{secGR:collec}, we get
\begin{align}
[\Gamma^\mu_{\nu \rho}] &= \frac{[x^\mu]}{[x^\nu][x^\rho]}\,, \\
[\tensor{R}{^\mu_{\nu \rho \sigma}} ] &= \frac{[x^\mu]}{[x^\nu][x^\rho][x^\sigma]} \,, \\
[R_{\mu \nu}] &= \frac{1}{[x^\mu][x^\nu]}\,, \\
[R] &= \mathrm{L}^2\,.
\end{align}
These relations imply 
\be 
[\bds{\mathcal{R}}]=[\bds{R}]=[\bds{G}]=1\,.
\ee 

We want now to determine the dimension of the energy-momentum tensor. Again, there is some freedom in the exact definition is the speed of light is not set to $1$. Let $\bds u$ be a $4$-velocity with intrinsic dimension
\be 
[\bds u] = \mathrm{T}\,.
\ee
We postulate that the quantity
\be 
[\bds u \cdot \bds T \cdot \bds u] = \frac{\mathrm{M}}{\mathrm{L}\mathrm{T}^2}\,,
\ee 
which is a spatial energy density. This implies
\begin{align}
    [T_{\mu \nu}] &= \frac{\mathrm{M}}{\mathrm{L}} \frac{1}{[x^\mu][x^\nu]}\,, \\
    [\bds T] &= \frac{\mathrm{M}}{\mathrm{L}}\,.
\end{align}
As mentioned, several conventions are possible, each of them with a different power of $c$ in the final Einstein's Equations. Recalling that $[\bds{G}]=1$ and using the usual dimensions for the Newton's constant, the full Einstein's equations read
\be 
\bds{G} = \frac{8\pi \GN}{c^2} \bds{T}\,.
\ee 
We could also make this analysis from the definition of the energy-momentum tensor Eq.~\eqref{eqGR:emt}, but we would need in this case to divide by an extra factor of $c$. Finally, note that the Einstein-Hilbert Lagrangian reads, with our convention,
\be 
\mathcal{L}_{\mathrm{EH}} = \frac{c^3 R}{16\pi \GN}\,.
\ee

\section{Observables\label{secintro:Obs}}
\subsection{General considerations}
We discussed in the previous Section the important concept of units and presented two different systems used in Physics. A natural question arises: Which system is \emph{the best}? This somehow implies that there is one choice of units which is better than any other. The answer is that such a system does not exist: Any problem can be solved in any unit system. You \emph{could} study General Relativity using miles, quarantines and ounces as basic units of length, time and masses and you \emph{could} use the Planck system in your daily life. 

The only thing that matters, in the end, is that physical observable should not depend on your choice. What is a physical observable is a more subtle issue. We will come back to this below. The important point is that there is no good choice for the units. Some of them may make computations more convenient, as it was the case in the Section~\ref{secintro:natural}, where the expression for the Schwarzschild radius is simpler in the natural system. However, the physical observable, in this case $R$, does \emph{not} depend on this choice.

In the next Section, we will show that the choice of units is not the only choice on which physical observables should not depend, and we will give a (non-exhaustive) list of transformations under which physical observables should not change.

\subsection{Examples}
We know already that in Electrodynamics and Quantum Mechanics there is the freedom to choose the gauge or the phase of the wave function. Final results, e.g. the force felt by a charged particle in an electromagnetic field or the probability to observe a particle, should not depend on this choice. We present in this Section other of these examples which will be relevant in the present work.
\begin{description}
\item[Choice of units:] As explained in the previous Sections, the description of physical systems should not, in fine , depend on the system of units. In practise, final answers do contain units, but an equality as $L=1.5\, \mathrm{m}$ is indeed unit-independent, as both the number $1.5$ and the unit (meter here) both change under a unit transformation. We will study in more details in Part~\ref{Part:Frames} how this invariance can actually be even more general to include \emph{spacetime dependent} unit transformations, and we will apply this to the Galaxy Number Counts, which is expected to be a genuine physical observable.
\item[Choice of coordinates:] Undergraduate students are allowed to solve problems in the coordinate system of their choice, but advised to work in the most convenient one. This reflects the fact that physical observables should not depend on this choice. We will discuss this in Part~\ref{Part:RotLensing}, where we will define several observables which are independent on the choice on coordinates (in this example spherical coordinates on the celestial sphere). We will encounter a more interesting example in Part~\ref{Part:BH} where we will see that even if the Master Equation describing waves around Black Holes is not unique and depend on a choice of coordinates, physical quantities as the speed of the gravitational waves or the frequencies of such waves do not depend on this choice.
\item[Rescaling in Cosmology:] In Chapter~\ref{chap:cosmo} we will see that in Cosmology, the scale factor, the radial coordinate and the curvature are not uniquely defined, and hence are not physical observables. We will still be able to build physical observables from these quantities, such as the Hubble factor.
\item[Choice of gauge:] Also in Chapter~\ref{chap:cosmo}, we will discuss the perturbations in Cosmology and around a \ac{SSSS} and see that there is again some freedom to parametrise these quantities, on which final observables should not depend. Note that this gauge choices are actually related to a freedom in the choice of coordinates.
\end{description}

\subsection{Definition}
We are now able to define physical observables in physics (or more generally in science).
\MYdef{Physical Observables}{
We consider a set of physical quantities $\bds{\mathcal{Q}} = \{\mathcal{Q}_i \}$ defined on a manifold $\Mcal$. We define a set of transformations $\{\mathcal{S} \}$ from $\Mcal$ to itself. Let $\mathcal{O}(\bds{\mathcal Q})$ be a quantity built from the physical quantities. This quantity is a physical observable if
\boxemph{
\mathcal{O}(\bds{\mathcal Q}) = \mathcal{O}(  \mathcal{T}(\bds{\mathcal Q})  )
}
for every $\mathcal{T} \in \mathcal{S}$. In other words, the value of a physical observable should be invariant under all the transformations.
}
There is one caveat with this definition: It is (purposely) very vague. Defining precisely what is a physical quantity, or making an exhaustive list of all the transformations in the set $\mathcal{S}$ is a difficult task that I am not pretending to perform. Moreover, there is a slight loophole: We can, first, impose which quantities should be physical observables, and then only keep the transformations $\mathcal{T}$ which leaves the observables invariant, or vice-versa (first choosing the transformations and the defining the observables). We do not have a fully satisfactory answer for now, and we will just keep this ideas in mind in this work. Note also that, these details being said, we will not mention them all the time. We will also sometimes work with quantities that are, stricto sensu, not physical observables, but the context should be always clear to avoid any ambiguity.

\stopchap
\chapter{The Standard Cosmological Model\label{chap:cosmo}}
In this Chapter, we present in the first Section the Standard Cosmological Model. In Cosmology, the Universe is assumed to be homogeneous and isotropic on large scales. Moreover, we also know from observations that the Universe is expanding. We present several important concepts, particularly the scale factor, the Hubble constant and the redshift, and we briefly discuss the Cosmic Microwave Background. For a more thorough introduction to Cosmology, see the book by Jean-Philippe Uzan et Patrick Peter \cite{peter_2013}. Obviously, much more references can be found for an introduction to Cosmology.

In the second Section, we discuss the theory of perturbations applied to Cosmology. The model of a homogeneous and isotropic Universe is not realistic: The Universe contains inhomogeneities, for example the galaxies, and this should be taken into account. For example, I found the Lecture Notes of Hannu Kurki-Suonio \cite{kurki_2020} and the review of Malik \cite{Malik:2008} very useful to learn the basis of perturbation theory.

The last Section of the Chapter is devoted to lensing, the general theory of light deflection as it travels through the Universe. We discuss the lens map and the Jacobi formalism which are the mathematical tools to describe quantitatively lensing of bundles of light rays. We briefly present the Mathematics of the Spherical Harmonics, and their generalisation to Spin Weighted Spherical Harmonics and discuss the statistics of lensing. The main reference for this part is the book about the \ac{CMB} written by Ruth Durrer \cite{durrer_2021}.

\startchap

\section{The FRLW Universe}
\subsection{The FRLW Metric}
We know that General Relativity describes gravitational phenomena extremely well. Examples of such include trajectories of planets around stars, behaviour of galaxies and even Black Holes. So let us be bold and ask the following question: Why do not we apply the machinery of General Relativity to describe the behaviour of the entire Universe? This is the main task of Cosmology!

We want to find, in fine, the equations governing the evolution of the Universe at large scales. To do so, we need to make some assumptions about the general form of the metric.

First, we require the Universe to be static, which means that the metric cannot depend on time and that the time-space elements $g_{0i}$ vanish. Second, we want the Universe to be, at any instant in time, homogeneous and isotropic. This assumptions is the cosmological principle. Based on this assumptions, we can build the \ac{FRLW} metric.
\MYdef{FRLW metric}{
The \ac{FRLW} metric is defined, using spherical coordinates $(t,r,\theta,\varphi)$, as
\boxemph{ \label{eqcosmo:FRLW}
\boldsymbol{g}  = - \bdd t^2 + a(t)^2 \left( \frac{\bdd r^2}{1-k r^2}  + r^2 \bdd \theta^2 + r^2 \sin^2 \theta  \bdd \varphi^2\right)\,,
}
where $a(t)$ is the scale factor depending on time only, $k$ the spatial curvature of the constant-time hypersurfaces, whose sign can be either positive, negative or zero.
}
\MYrem{Normalisation of time}{
A prefactor in front of the time-time part of the metric would be possible. It would then read
\be 
\boldsymbol{g} = - N(\tilde{t})^2\bdd {\tilde{t}}^2 +\dots\,.
\ee 
However, this can be removed defining the cosmic time $t$ via (provided $N(\tilde{t} )>0$)
\be 
N(\tilde{t} )\dd \tilde{t} = \dd t \,.
\ee 
}
Another more important comment: There is some freedom in the definitions of the radial coordinate. Indeed, requiring the line-element to be invariant, we define
\begin{align} \label{eqcosmo:atransfo}
    \tilde{a}  &= \lambda a \,, \\
    \tilde{r}  &= \lambda^{-1} r \,, \\
    \tilde{k} &= \lambda^2 k\,.
\end{align}
It can easily be checked that the line-element is invariant is we use the tildes quantities. This freedom in the definition of the quantities tells us that $r$, $a$ and $k$ are not observable per se. One should consider invariant quantities such as
\boxemph{
\mathcal{Q}_1&= ar\,,\\
\mathcal{Q}_2&= k a^{-2\,.}}
The first quantity $\mathcal{Q}_1$ is a physical distance and has the units of a length, while $\mathcal{Q}_2$ is a physical measure of the spatial curvature and has the units of a inverse length squared, as it should be.
\MYrem{Spatial curvature}{
In this thesis, we will only consider flat spatial surfaces with $k=0$.
}
\MYrem{Conformal time}{\label{remcosmo:conftime}
Sometimes, it is more convenient to work in conformal time $\eta$ defined implicitly as $\dd t = a(t) \dd \eta$. Using this time coordinate, the \ac{FRLW} metric reads
\be
\boldsymbol{g} = a(\eta)^2 (-\bdd \eta^2 + \bdd r^2 + r^2 \bdd \Omega^2 )\,,
\ee 
where we set $k=0$. Moreover, under the transformation Eq.~\eqref{eqcosmo:atransfo}, the conformal time becomes
\be 
\dd \tilde{\eta} = \lambda^{-1} \dd \eta\,,
\ee 
which implies that any product of the form $\mathcal{Q}_3 = a \Delta \eta$ is invariant.
}
\MYrem{Today}{\label{remcosmo:today}
	By convention, quantities today are denoted with the index $0$, e.g. $t_0$ or $a_0$. It is common to set $a_0=1$ in the literature, but we will never do this in this thesis.
}

\subsection{The Hubble factor}
In this Section, we want to define physical quantities relevant in Cosmology. The first choice would be the scale factor $a(t)$. However, as discussed above, it is not a physical observable per se. Rather, we should consider for a physical quantity a typical physical distance $D_\mathrm{p}(t)=a(t) R$, where $R$ is a fixed radial coordinates.

The second quantity one would like to measure is the speed of expansion of the Universe. A naive answer would be to consider simply $v_{\mathrm{exp}} = \dot a$. This is however a very bad choice... Why? The reason is again that this is not an observable per se: The value of the scale factor $a(t)$ can be rescaled arbitrarily by a positive constant $\lambda$, as explained in the previous section. A better definition of the expansion rate of the Universe is the Hubble factor defined as
\boxemph{
H(t)=\frac{\dot a(t)}{a(t)}\,.
}
Indeed, if one defines $\tilde{a} = \lambda a$ (while keeping the physical time $t$ unchanged), the Hubble factor is not modified. Hence, this definition is a good candidate to quantify the expansion rate of the Universe. 
\MYrem{Hubble factor in conformal time}{
Using the conformal time defined in Rem.~\ref{remcosmo:conftime}, we define the Hubble factor in conformal time as
\be
\mathcal{H} = \frac 1a \frac{\dd a}{\dd \eta}  = aH\,.
\ee
}
Note that the Hubble factor is not constant! Its value today is the Hubble constant $H_0$ given by
\be 
H_0 = 100h\,  \frac{\mathrm{km}}{\mathrm{s}}\, \mathrm{Mpc}^{-1}\,,
\ee 
with $h\approx 0.6/0.7$, see Section~\ref{seccosmo:Fried} for more details about the time evolution of the Hubble factor. The physical interpretation of the Hubble factor goes as follows. Consider two comoving friends, one located at coordinate $r=0$ and one located at constant coordinate $r\neq 0$ (and at a fixed values of the angles $(\theta, \varphi)$ which is not relevant here). The physical distance $D_{\mathrm p}$ between these two friends can be computed using the \ac{FRLW} metric given by Eq.~\eqref{eqcosmo:FRLW} and is 
\be  
D_{\mathrm p }(t) = a(t) r\,.
\ee 
Note that this distance depends on time precisely because of the expansion of the Universe! The recession speed of the second friend (with respect to the first one) is the time derivative of this quantity given by
\be \label{eqcosmo:hubble_law}
\dot D_{\mathrm p} = \dot a r = H(t) a r = H(t) D_{\mathrm{p}}\,.
\ee 
This is the \emph{Hubble's law}! It gives an interpretation of the Hubble constant: It quantifies the relative recession speed of any distance object in the Universe! This also explains its interesting units: An object (say a galaxy) located at $1\, \mathrm{Mpc}$ away from us will recede with a speed of order $100\, \mathrm{km\times s^{-1}}$.

Let us finish this section by briefly discuss the Hubble tension. Today, we have two different experimental ways to measure the Hubble constant. 

The first method is the \emph{close method}: Roughly speaking, it relies on the principle described by the Hubble's law \myeq{eqcosmo:hubble_law}. The general idea is to observe several galaxies, and measure together their receding speed and their distance. Fitting those two quantities together can provide an estimation of the Hubble constant.

The second method is the \emph{far method}: Without going into the details, the actual value of the Hubble constant can be deduced from the observed power spectrum of the \ac{CMB}.

The story could end now if these two methods did not give different values (taking into account the error bars). This difference is called the Hubble tension and we do not have any satisfactory answer yet, even if several hypothesis have been done, see for example \cite{DiValentino:2021}.

The two estimable videos (in French) \cite{scienceset:1,scienceset:2} of the Channel \emph{ScienceEtonnante} present an excellent description of the problem, even for Physicists!

\subsection{Redshift in Cosmology}
In this Section, we introduce another major concept in Cosmology (and in general in Physics):~The redshift. Let us give first a qualitative explanation. When a photon travels in the Universe, it is somehow \emph{attached} to the expanding space, which makes it expand too, or more precisely, which increases its wavelength. We know from Electrodynamics and Quantum Mechanics that the energy of such a photon will decrease. Hence, in the Universe, travelling photons lose energy. If those photons would be in the visible range, they would then start \emph{blue-ish} with a high energy, and then become \emph{red-ish}, with less energy. This is the origin of the term \emph{redshift}.

We want to show this in a more formal way. We consider a radially travelling photon whose 4-velocity is given by
\begin{align}
\boldsymbol{k} &= k^\mu \bpt_\mu\,, \\
k^\mu &= \omega(t) (1, a(t)^{-1},0,0)\,.
\end{align}
The vector is chosen such that $\boldsymbol{k}^2 =0$, but the prefactor outside can in principle depend on time. Photon, as any particle, must follow geodesics, hence the 4-vector $\boldsymbol{k}$ must satisfy the parallel transport condition given by Eq.~\eqref{eqgr:partra} (with $\bds u = \bds b = \bds k$). Using the \ac{FRLW} metric and the formalism presented in Section~\ref{sec:DGGR}, we find that the only condition is 
\be 
(a(t) \omega(t))^{\bullet} =0\,,
\ee
or in other words
\be 
\omega(t) = \omega_0 \frac{a_0}{a(t)}\,,
\ee 
where $\omega_0$ and $a_0$ are some initial values.

We need now to explain the interpretation of $\omega(t)$. In order to do so, we introduce a stationary observer, for example on Earth. Her 4-velocity is given by 
\be 
\boldsymbol{u} = \bpt_t\,,
\ee 
which satisfies $\boldsymbol{u}^2 = -1$. She observes the photon arriving at her with 4-velocity $\boldsymbol{k}$. The frequency she receives is given by
\be 
\omega_{\mathrm{r}} \equiv  - \boldsymbol{k} \cdot \boldsymbol{u} = \omega(t)\,.
\ee
This definition is motivated by the fact that it is a coordinate invariant quantity and it leads to the correct answer in Special Relativity, e.g. by taking into account the Doppler effect.

Then, if she knows the physical process from which the photon was emitted, she can reproduce this experiment in her lab and will measure the emitted frequency $\omega_\me$, that she will compare with the observed frequency $\omega_\mr$. This is done by defining the redshift
\boxemph{ \label{eqcosmo:redshift}
1+z 
\equiv \frac{\omega_{\mathrm{e}}}{\omega_{\mathrm{r}}} 
= \frac{
a(t_{\mathrm{r}})
}
{
a(t_{\mathrm{e}})
}\,,
}
where the subscripts means \emph{emission} and \emph{reception}. This relation is fundamental in Cosmology: Assuming that $a(t_{\mathrm r}) = a_0$ is the scale factor today (and hence is constant), this relation translates the scale factor at emission into a redshift. Hence, the redshift can be used as a \emph{clock} for the history of the Universe (assuming that the scale factor is an increasing function of $t$). One should note that this clock goes backward: at the Big-Bang, we had $a(t_\mathrm{e})=0$ and $z=\infty$, and today we have $a(t_{\mathrm e}) = a_0$ and $z=0$.

\subsection{The Friedmann Equations\label{seccosmo:Fried}}
In this Section, we derive the Friedmann equations. These equations describe the dynamics of the \ac{FRLW} Universe. They are obtained through the Enstein's Equations Eq.~\eqref{eqgr:EFE}. The left hand side of the equations, the Einstein tensor, can be computed using the machinery of Differential Geometry and the \ac{FRLW} metric. However, we have not defined yetthe right hand side. We explain here how to define a reasonable energy-momentum tensor in Cosmology. 

We assume that the Universe is filled with several different fluids, with total energy density $\rho$ and pressure $P$. These quantities, because of the homogeneity assumption, can only depend on time. Moreover, we assume that the fluid has a constant 4-velocity $\bds{u} = \bpt_t$, which is the only possible choice because of the isotropy assumption. From this 4-velocity, we can build the spatial metric 
\be
\bds{h} = \bds{g} + \bds{u} \otimes \bds{u}\,,
\ee 
satisfying $\boldsymbol{h} \cdot \boldsymbol{u}=0$ (as $\boldsymbol{u}^2 =-1$). With these considerations, we can introduce the energy-momentum tensor.
\MYdef{Energy-momentum tensor}{ \label{defcosmo:Tmunu}
The energy-momentum tensor for a perfect fluid is
\boxemph{  \label{eqcosmo:tmunu}
\bds{T} &= P \bds{h} + \rho \bds{u} \otimes \bds{u}\,, \\
T_{\mu \nu} &= P(g_{\mu \nu} + u_{\mu} u_{\nu}) + \rho u_{\mu} u_{\nu}\,,
}
where $\rho$ is its energy density, $P$ its pressure and $\bds u$ its $4$-velocity.
}
This definition can be justified recalling that the energy density $\rho$ is defined to be (minus) the eigenvalue of the energy-momentum tensor associated with the timelike eigenvector $\bds{u}$, i.e. $\rho = - u^\mu T_{\mu \nu} u^{\nu}$. This definition fixes the second term. The first term can be interpreted as the definition of pressure, as being the spatial part of the energy-momentum tensor.

In Definition~\ref{defcosmo:Tmunu}, we assumed that the fluid has only one component. We could consider several fluids, and the total energy-momentum tensor would read
\be 
\boldsymbol{T}_{\mathrm{tot.}} = \sum_i \boldsymbol{T}_{i}\,,
\ee 
where the fluid $i$ has energy density $\rho_i$ and pressure $P_i$. Moreover, we will assume that the pressure is proportional to the energy density
\be 
P_i = w_i \rho_i\,,
\ee 
where $w_i$ is the barotropic index of the fluid. The three most common examples are $w_{\mathrm{m}} =0$ for cold matter, $w_{\mathrm{r}} = 1/3$ for radiation and $w_\Lambda = -1$ for a cosmological constant.

We have now all the ingredients to derive the Friedmann equations. Using the energy-momentum tensor given by Eq.~\eqref{eqcosmo:tmunu}, and computing the Einstein tensor from the \ac{FRLW} metric given by Eq.~\eqref{eqcosmo:FRLW}, we get two independent Friedmann equations : One corresponding to $G_{00}$ and one to the spatial trace $\tensor{G}{_i^i}$
\boxemph{ \label{eqcosmo:fri1}
H^2  &= \frac{8\pi \GN}{3} \rho - \frac{k}{a^2}\,, \\
\frac{\ddot a}{a} &= - \frac{4\pi\GN}{3} (\rho + 3P)\,.
}
From these two equations, or equivalently using the conservation equation given by Eq.~\eqref{eqgr:conseq}, we get
\be 
\dot{\rho} + 3H (\rho + P) =0\,.
\ee 
If we assume that the fluids are not interacting with each other, this relation holds for each of them independently
\be  \label{eqcosmo:rhoidot}
\dot{\rho}_i + 3H (\rho_i + P_i) =0\,.
\ee 
Moreover, if we set $P_i = w_i \rho_i$, and using Eq.~\eqref{eqcosmo:rhoidot}, we can relate the energy density $\rho_i(t)$ with the scale factor $a(t) as$
\be  \label{eqcosmo:rhoi}
\rho_i(t) = \rho_{i0} \left(\frac{a(t)}{a_0}\right)^{-3(1+w_i)}\,,
\ee 
where $\rho_{i0}$ and $a_0$ are arbitrary initial conditions. We can then rewrite the first Friedmann Equation Eq.~\eqref{eqcosmo:fri1}, getting
\begin{align}
H^2 &= \frac{8\pi \GN}{3} \rho(t) \\
&=  \frac{8\pi \GN}{3} \sum_i \rho_{i}(t) \\
&= \frac{8\pi \GN}{3}  \rho_0 \sum_i \frac{\rho_i(t)}{\rho_0} \\
&= H_0^2 \sum_i\frac{\rho_{i0}}{\rho_0} \left(\frac{a(t)}{a_0}\right)^{-3(1+w_i)}\,,
\end{align}
where we used the first Friedmann Equation applied today (represented by the index $0$) and the solution for each species given by Eq.~\eqref{eqcosmo:rhoi}. We introduce the cosmological parameters today as
\be 
\Omega_i \equiv \frac{\rho_{i0}}{\rho_0}\,,
\ee 
satisfying $\sum_i \Omega_i=1$. The Friedmann equation reads then
\boxemph{ \label{eq:}
\frac{H^2}{H_0^2} = \sum_i \Omega_i \left(\frac{a(t)}{a_0}\right)^{-3(1+w_i)}\,.
}
Using the definition of the redshift given by Eq.~\eqref{eqcosmo:redshift}, we can also write this relation as
\be 
\frac{H^2}{H_0^2} = \sum_i \Omega_i  (1+z)^{3(1+w_i)}\,.
\ee
Defining the function 
\be 
E(z) = \sqrt{\sum_i \Omega_i (1+z)^{3(1+w_i)}}\,,
\ee 
we get
\boxemph{ 
		\label{eqcosmo:Hz}
		H(z) = H_0 E(z)\,,
}
which relates the Hubble factor at arbitrary redshift $z$ with the Hubble constant today.

\subsection{Timeline of the Universe}
\begin{figure}[h!t]
	\begin{center}
	\includegraphics[width = 0.49\textwidth]{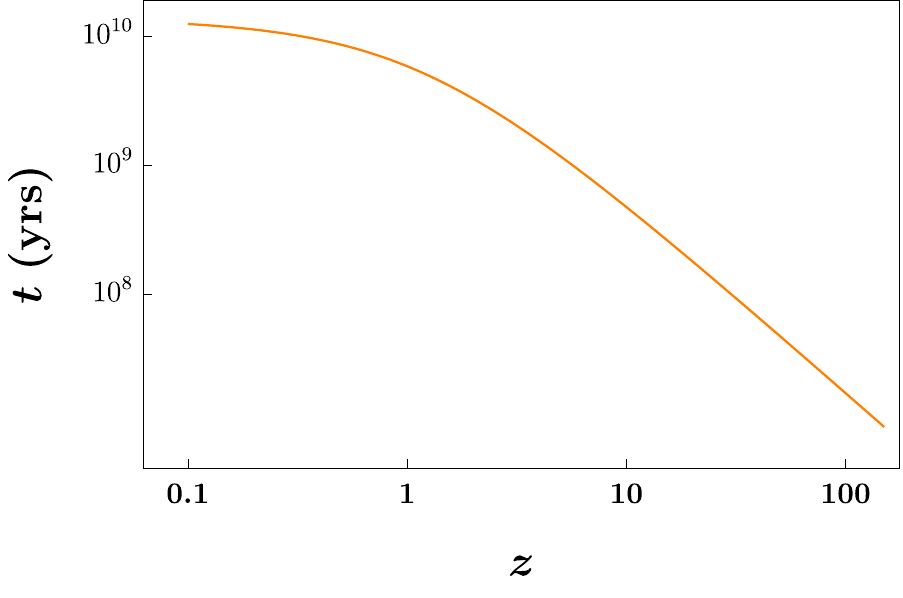} 
	\includegraphics[width = 0.49\textwidth]{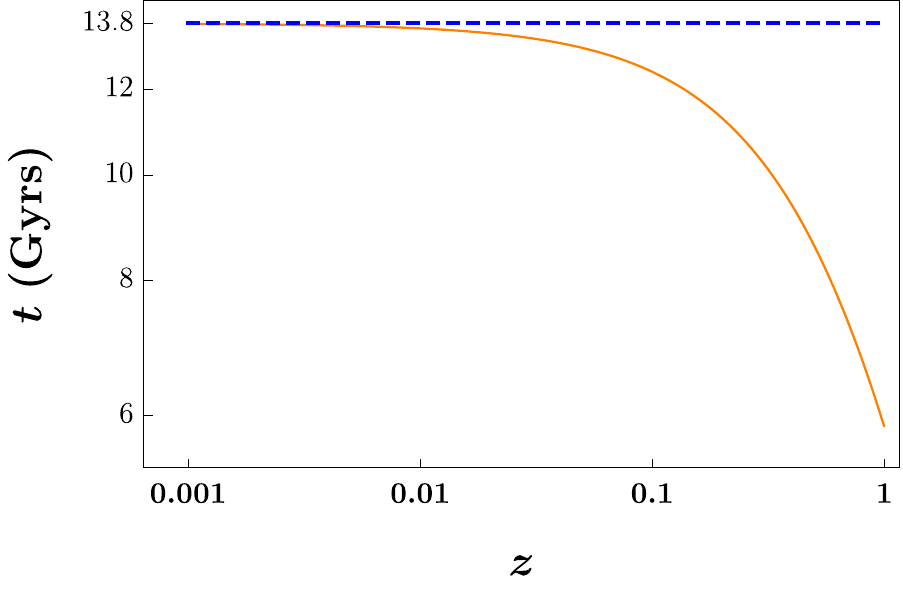} 
	\end{center}
	\caption[Timeline of the Universe]{\label{figcosmo:tVSz}
	\textbf{Timeline of the Universe}\\
	Relationship between the redshift of an event and the \emph{age} of the Universe when this event happened. We can read off, for example, the age of the Universe by looking at $z=0$.}
\end{figure}

In this Section, we want to present the \emph{biography} of the Universe. In other words, we would like to describe several events, together with the time at which they happened. We assume that the Universe was born during the Big-Bang at $t=0$, and we denote the present time $t_0$. We consider an event which happened at time $t$ such that $0\leq t < t_0$. The time separating us from this event can be computed as
\begin{align}
    t_0 - t  &= \int_{t}^{t_0}\; \mathrm{d}t \\
    &=\int_{a}^{a_0}\; \frac{1}{\dot{a}}\, \mathrm{d}a \\
    &= \int_{a}^{a_0}\; \frac{1}{Ha}\, \mathrm{d}a \\
    &=\int_{0}^{z}\; \frac{1}{H(\tilde{z}) (1+\tilde{z})}\, \mathrm{d}\tilde{z} \\
    \label{eqcosmo:tfunz}
    &= \frac{1}{H_0}\int_{0}^{z}\; \frac{1}{E(\tilde{z}) (1+\tilde{z})}\, \mathrm{d}\tilde{z}\,,
\end{align}
where we used the relation between the redshift and the scale factor given by Eq.~\eqref{eqcosmo:redshift} and the expressed Hubble factor as function of the redshift with Eq.~\eqref{eqcosmo:Hz}. We can integrate this relation numerically to obtain pairs $(t,z)$ linking between time and redshift. In particular, setting $z=\infty$, we can compute the age of the Universe. Using the Cosmological \ac{LCDM} parameters (Planck \cite{Planck:VI})
\begin{align}
    h&\approx 0.7 \,,\\
    \Omega_{\mathrm m} &\approx 0.31\,, \\
    \Omega_\Lambda &\approx 0.69\,,
\end{align}
and neglecting radiation, we find that age of the Universe is 
\be 
t_{\mathrm{U}} \approx 13.8\, \mathrm{Gyrs}\,.
\ee 
In Fig.~\ref{figcosmo:tVSz}, we show this relationship between time and redshift, where the time is computed since the Big-Bang until the considered event. For example, \emph{Homo Sapiens} appeared roughly $500'000$ years ago. A photon emitted by the first Human would be seen today (by an alien) with a redshift of $z\approx 0.003$, and a photon emitted by a galaxy $\SI{6}{\Gyear}$ after the Big-Bang would be seen today with a redshift of $z\approx 1$.
\begin{table}[!ht]
\centering
\begin{tabular}{lcc}
        Event & $z$ & Time after Big-Bang\\ \hline \hline 
        CMB formation & $1100$ & $ \SI{380000}{\year}$  \\ \hline
        First stars & $ 1000$ & $ \SI{500000}{\year} $  \\ \hline
        Formation of the Sun & $0.4$ & $ \SI{9.3}{\Gyear}$ \\ \hline 
        Dark energy dominates & $0.3$ & $ \SI{10.3}{\Gyear}$  \\ \hline
        Homo Sapiens & $0.0035$ & $  \SI{13.5}{\Gyear}$ \\ \hline
        My Thesis defense! & $0$ & $  \SI{13.8}{\Gyear}$ \\ \hline
\end{tabular}
\caption{\label{tabcosmo:timeline}Timeline of the Universe: Several cosmological events, their redshift today, and the time after the Big-Bang at which they happened (approximate values).}
\end{table}

Anyone studying Cosmology should always remember the typical time scales, which are in general much longer that any time scale a human can think of. To give you a little anecdote, when I was taking my first Cosmology class, the professor said that dark energy started to dominate \emph{recently}. I thought that the first Humans witnessed this transition! As one can see in Tab.~\ref{tabcosmo:timeline}, I was quite far-off, as humans appeared almost $3$ billion years after the transition...

\subsection{Distances\label{secintro:distances}}
In Cosmology, because of the large scales involved, the concept of distance is not well-defined. We present here various definitions of distance with different interpretations. Let us consider a photon received on Earth with redshift $z$. Its time of emission is $t$, where the explicit relation with the redshift is given by Eq.~\eqref{eqcosmo:tfunz}. We assume that the observer is located at $r=0$.
\begin{enumerate}
\item First, let us compute the comoving distance $r$ of the source. As the photons travels along null geodesic, using the \ac{FRLW} metric yields
\be
\dd r = - \frac 1a \dd t\,,
\ee 
where the negative sign comes from the fact that the photon travels inwards. Using the same steps as before, we get
\boxemph{ \label{eqcosmo:rint}
r = \frac{1}{a_0 H_0} \int_0^z \; \frac{1}{E(\tilde z)}\, \dd \tilde{z}\,.
}
Note that this distance does \emph{not} depend on time: The source is located at fixed $r$ and does not move. Moreover, using the definition of the conformal time, it is clear that $\dd r = - \dd \eta$, which implies
\be  \label{eqcosmo:etaz}
\eta_0 - \eta = r\,,
\ee 
where $\eta_0$ (resp. $\eta$) is the conformal time today (resp. at emission).

\MYdef{Conformal distance between two sources}{
We can generalise this to show that the conformal distance between two sources observed at redshift $z_1$ and $z_2$ (assuming the sources are aligned and $z_1<z_2$) as
\be 
\chi(z_1,z_2) \equiv 
\frac{1}{a_0 H_0} \int_{z_1}^{z_2} \; \frac{1}{E(\tilde z)}\, \dd \tilde{z}\,.
\ee 
}
\item We can also compute the distance today $D_0$. This distance would correspond to an \emph{instantaneous} measurement, which is indeed not possible. However, this distance can be mathematically computed. The scale factor being $a_0$ today and the comoving distance being $r$ given above, the distance today is
\boxemph{
D_0 = a_0  r = \frac{1}{ H_0} \int_0^z \; \frac{1}{E(\tilde z)}\, \dd \tilde{z}\,.
}
\item We can perform the same computation, but at the time of emission to get the instantaneous distance when the photon was emitted $D_{\me}$. As the conformal distance is constant, and using the definition of the redshift, we get
\boxemph{
D_\me = a_\me r = \frac{a_0}{1+z} r = \frac{1}{ (1+z)H_0} \int_0^z \; \frac{1}{E(\tilde z)}\, \dd \tilde{z}\,.
}
\item The penultimate distance we consider is the angular diameter distance $\DA$. If we observe an object under a solid angle $\Delta \Omega$, and we know its  physical area $A_{\mathrm p}$, we define the angular diameter distance $\DA$ through the usual trigonometric relation
\be 
A_{\mathrm p} = \Delta \Omega \DA^2\,.
\ee  
The solid angle is not modified when the photon travels ($\theta$ and $\varphi$ are constant). Hence, at the emission time and position, the physical area of the object was given by
\be 
A_{\mathrm p}  = a^2 r^2 \Delta \Omega \,,
\ee 
from which we infer
\boxemph{ \label{eqintro:Dang}
\DA= a r = \frac{a_0}{1+z} r = \frac{1}{ (1+z)H_0} \int_0^z \; \frac{1}{E(\tilde z)}\, \dd \tilde{z}\,,
}
which is actually equal to $D_\me$.
\item Finally, we define the luminosity distance. If an object with intrinsic luminosity $L_{\mathrm{int.}}$ emits light, and a flux of energy $F$ is received on Earth, we define the luminosity distance $\DLum$ via\footnote{We consider here that we receive the full flux, i.e. from all around the source. To be more realistic, we would need to consider the flux received in a small solid angle $\Delta \Omega$, but the final result would not be different.} 
\be 
F = \frac{  L_{\mathrm{int.}} }{4\pi \DLum^2}\,.
\ee 
The light of the source has been redshifted, so the energy we observe is 
\be 
E_{\mo} = \frac{E_{\mathrm{int.}}}{1+z}\,,
\ee
where $E_{\mathrm{int.}}$ is the intrinsic energy emitted by the source. Moreover, there is a time dilatation. Indeed, imagine that the source emits light during a small interval of time $\Delta t_\me$. During this time, the light travels a conformal distance $\Delta r$ given by $\Delta t_\me = a \Delta r$, where $a$ is the scale factor as emission. The conformal distance $\Delta r$ is constant during the light's journey. As the scale factor expands, the time it takes for this light ray to pass increases. More precisely, the time at the observer position is given by
\be
\frac{\Delta t_\mo}{ a_0} = \frac{\Delta t_\me}{ a}\,,
\ee 
or equivalently
\be
\Delta t_\mo = (1+z)\Delta t_\me\,.
\ee
The observed luminosity is then given by 
\be
L_\me = \frac{E_{\mo}}{\Delta t_\mo} = \frac{1}{(1+z)^2} L_{\mathrm{int.}}\,.
\ee 
This energy is spread on a sphere of radius $a_0 r$, hence the observed flux is
\be 
F_\me = \frac{L_\me}{4\pi a_0^2 r^2 } = \frac{1}{(1+z)^2} \frac{L_{\mathrm{int.}}}{4\pi a_0^2 r^2}\,.
\ee 
From this, we infer that the luminosity distance is given by
\boxemph{  \label{eqintro:Dlum}
\DLum = (1+z) a_0 r = \frac{1+z}{H_0} \int_0^z \; \frac{1}{E(\tilde z)}\, \dd \tilde{z}
}
\end{enumerate}
We defined $5$ different distances. In practice, only two are observables: The angular diameter distance $\DA$ and the luminosity distance $\DLum$. The first former is an observable is the angular size of the object is known, we call such objects \emph{standard rulers}. The latter is an observable is the intrinsic luminosity of the object is known, which we call \emph{standard candels}.

Note that from these definitions Eq.~\eqref{eqintro:Dang} and Eq.~\eqref{eqintro:Dlum}, we get the relation
\boxemph{
\DLum = (1+z)^2 \DA\,.
}
This relation is very interesting as it only rely on a few assumptions. We assumed that photons follow null geodesics and travel across a \ac{FRLW} Universe. Moreover, in the derivation of the luminosity distance, we implicitly assumed that the number of photons is conserved during the process.

\subsection{The Cosmic Microwave Background}
Right after the Big Bang, the Universe was a mere hot soup of elementary particles. Within the first half hour, the quarks combined to form protons and neutrons, which themselves merged to form deuterium and helium nuclei. However, at that time, the photons travelling in the Universe were energetic enough to break any neutral atom that would have dared to form. Another consequence is that light in this primordial Universe could not travel freely. The Universe is said to be opaque, with a thick \emph{mist} penetrating it.

This situation lasted for quite a long time. The Universe had to wait no more than $380^\prime 000$ years to form neutral hydrogen atoms. By that time, because of the expansion of the Universe, a majority of the photons were not hot enough anymore and could not \emph{break} neutral atoms. Hence, light was not constantly stopped and could travel freely in the Universe. These light is the origin of the Cosmic Microwave Background which we still observe today!

\begin{figure}[ht!]
	\begin{center}
	\includegraphics[width = 0.49\textwidth]{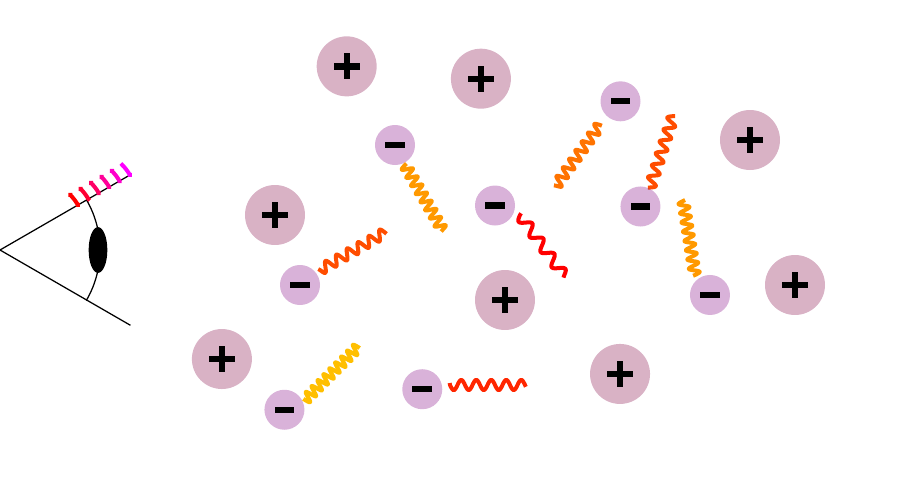} 
	\includegraphics[width = 0.49\textwidth]{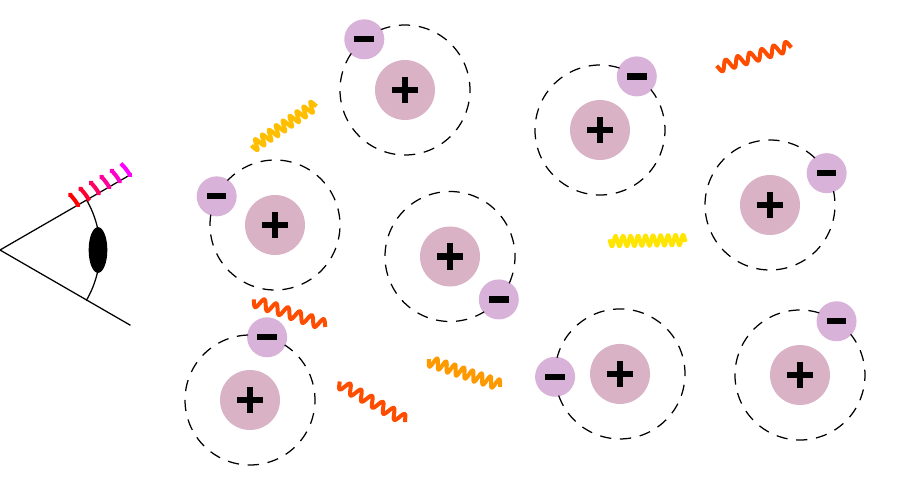}
	\end{center}
	\caption[Formation of the CMB]{\label{figcosmo:CMB}
		\textbf{Formation of the CMB}\\
		\textbf{Left:} Before decoupling, electrons and protons are not bounded. The free photons scatter against them and cannot travel freely.\\
	 \textbf{Right:} After decoupling, neutral atom are formed. The photos do not scatter and can travel freely across the Universe.}
\end{figure}

The \ac{CMB} is a relic of this primordial light which was in thermal equilibrium with the other particles. Hence, the energy distribution of these photons should follow a perfect black body law, whose temperature would be $T_0 = 2.7\, \mathrm{K}$ today. This is exactly what we observe! The observed \ac{CMB} is then a very good evidence for the Big Bang model used in Cosmology.

\section{Perturbation Theory\label{seccosmo:PT}} 
\subsection{General Idea}
In this Section, we present perturbation theory. This is a mathematical and physical topic in its own right, and one could definitely write a full thesis about it. I do not have the pretension to be thorough and complete here. Our goal is to present the idea from a general point of view, and then describe the relevant cases for the present work: Cosmological perturbations and perturbations around \ac{SSSS}. The basic idea of perturbation theory can be summarized as follows:
\begin{center}
\emph{Small problems require small solutions.}
\end{center}
This is a general advice to follow in life. For example, if your noodles are slightly hot, and if you slightly blow on them, I'm quite sure you can eat them. Sometimes, however, small problems require big solutions: If you have a small spiky rock in your shoes, you should not slightly shake your shoes to solve the issue, rather you should stop walking and take care of it.

The Mathematical idea is similar: Assume you have a problem, say an equation (the heat in the example), that is very close from a problem whose solution is known. You can fairly expect the solution you are seeking to be very close from the known solution. The small difference between the two (the slight blow in the example) can be found using perturbation theory.

However, there are problems that seem to be very close from each other, but whose solutions are far apart, as in the example of the rock. For such examples, perturbation theory does not work and more advanced methods should be used.

Perturbation theory is actually very useful in Physics: In general, we always consider situations that are close to ideal models, from which the solutions can be easily computed. Perturbation theory helps us to compute approximation of the solutions. In Cosmology, for example, we assume that the Universe is isotropic and homogeneous. This is an idealized situation from which several things can be computed. The reality is a bit different: The Universe is not fully homogeneous, it contains some small \emph{ripples} here and there, e.g. galaxies or dark matter halos. Cosmological Perturbation Theory is extremely useful to describe those so-called inhomogeneities and extract very interesting predictions therefrom.

\subsection{Lie Derivative\label{seccosmo:Lie}} 
\subsubsection{Motivation}
There is one caveat one should be careful about when using perturbation theory in General Relativity. We start with a simple example to exhibit the problem and generalize it to General Relativity in the next section.

Suppose we study a $2D$ scalar function $f(t,x)$. We will assume a Cosmological background, where the $0$\textsuperscript{th}-order only depends on $t$, and consider a $1$\textsuperscript{st}-order generic perturbation, i.e.
\be 
f(t,x)=\bar{f}(t) + \varepsilon\, \delta f(t,x)\,,
\ee
where $\varepsilon$ is a parameter representing the perturbation and considered to be small. We will work only at first order in $\varepsilon$.

Suppose we perform a change of coordinates given by
\be 
\left\{\begin{matrix}
\tilde{t} &=& t-\varepsilon\, x \\
\tilde{x} &=  & x \,.
\end{matrix}\right.
\ee 
This transformation is also called a \emph{gauge transformation}. The function $f$ being a scalar, its transformation rule is given by
\be 
\tilde{f}\left(\tilde t, \tilde x \right) = f\left(
t
,x
\right)\,,
\ee
where it is implied that $t$ and $x$ depend on $\tilde t$ and $\tilde x$. This yields
\be 
\tilde{f}(\tilde t, \tilde x) = f\left(\tilde t+\varepsilon\, \tilde{x}, \tilde{x} \right)
= \bar{f} \left(\tilde{t}\right) + \varepsilon \left[ \delta f(\tilde t, \tilde x) + x \dot{\bar{f}}(\tilde t) \right]\,.
\ee 
This means that if we use the $(\tilde t, \tilde x)$ system of coordinates, we could think that the new perturbation is
\be 
\delta \tilde{f} (\tilde t, \tilde x) =  \delta f(\tilde t, \tilde x) + x \dot{\bar{f}}(\tilde t) \,.
\ee
However, the new term only comes from the fact that we perform a slight coordinate transformations, and does not have any physical meaning. Hence, we need a way to determine the true physical perturbation. Note also that this problem only arises at the perturbed level: We want to keep the Cosmological background $\bar{f}$ in all coordinate systems, which prevents us from doing a change of coordinate at this order.

\subsubsection{Scalar Field}
We now present the relevant mathematical tool to formalise this issue: The Lie Derivative. It describes quantitatively the transformation of perturbations under change of coordinates. We will explicitly explain how to define it for scalar functions, vectors and forms, and will generalize it to tensors afterwards. In this Section, we consider generic coordinates $x^\mu$ and $\tilde{x}^\mu$ related by (at first order)
\begin{align} \label{eqcosmo:xtilfun}
\tilde{x}^\mu  &= x^\mu + \varepsilon\, \xi^\mu ( x)\,, \\
\label{eqcosmo:xfun}
x^\mu &= \tilde{x}^\mu - \varepsilon\, \xi^\mu (\tilde{ x})\,.
\end{align}
Note that here the function $\boldsymbol{\xi}({x})$ is the same for both transformations (up to its sign), as we work only at first order in $\varepsilon$. Under this assumption, it is straightforward to show that the transformation rules Eq.~\eqref{eqcosmo:xtilfun} and Eq.~\eqref{eqcosmo:xfun} are inverse of each other. 

We consider a scalar function $F( x)$ separated into a background part and a perturbation as
\be  \label{eqcosmo:Fdef}
F( x) = \bar{F}( x) + \varepsilon\, \delta F( x)\,,
\ee 
where we do not make any assumption about the background function, e.g. we do not impose a Cosmological setup. The transformation rule for scalar functions is given by
\be  \label{eqcosmo:scalarF}
\tilde F (\tilde{ x}) = F( x)\,,
\ee 
where it is implied here that $x^\mu$ is given in terms of $\tilde{x}^\mu$ through Eq.~\eqref{eqcosmo:xfun} and vice-versa. This yields
\begin{align}  \label{eqcosmo:Ftildextilde}
\tilde{F}(\tilde{  x }) 
&= F( x - \varepsilon\, \bds \xi) \\
&= \bar{F}(\tilde{ x}) 
+ \varepsilon
\left[ 
\delta F(\tilde{ x}) - \xi^\mu(\tilde x) \partial_\mu \bar{F}(\tilde{ x})
\right]\,,
\end{align}
where in the last term we replaced $F$ by $\bar F$ as this quantities are equal at first order. We would like now to define a way to compare the functions $F$ and $\tilde{F}$. There are two ways to do this. The first one would be to compute this difference at the same physical point, i.e. between $F( x)$ and  $\tilde{F}(\tilde{ x})$. However, as the transformation rule for a scalar function given by Eq.~\eqref{eqcosmo:scalarF}, we know that this difference would be $0$, and hence would not be very interesting. What we could do instead is choose a given coordinate $\tilde{\boldsymbol{x}}$ and compute the difference between $F(\tilde{\boldsymbol{x}})$ and $\tilde{F}(\tilde{\boldsymbol{x}})$, which can be done directly through Eq.~\eqref{eqcosmo:Fdef} and Eq.~\eqref{eqcosmo:Ftildextilde}
\be  
\tilde{F}(\tilde{x}) - F(\tilde{x})
=
- \varepsilon \xi^\mu(\tilde{x}) \partial_\mu \bar{F}(\tilde{x})\,.
\ee 
Moreover, we define the perturbations in the new system as
\be  \label{eqcosmo:deltaFtilde}
\tilde{F}(\tilde{ x}) = \bar{F}(\tilde{ x})+ \varepsilon \,
\delta \tilde{F}(\tilde{ x})
\ee 
In other words, we assume that the background function is the same in both coordinate systems. From the definition of the scalar function Eq.~\eqref{eqcosmo:Fdef} and Eq.~\eqref{eqcosmo:deltaFtilde}, it is clear that 
\be  \label{eqcosmo:deltaF1}
 \tilde{F}(\tilde{x})-F(\tilde{x})
=
\delta \tilde{F}(\tilde{x})-\delta F(\tilde{x}) \,.
\ee 
Combining now Eq.~\eqref{eqcosmo:Ftildextilde} and Eq.~\eqref{eqcosmo:deltaF1} gives
\be
\delta \tilde{F}(\tilde{x})-\delta F(\tilde{x})
=
-  \xi^\mu(\tilde x) \partial_\mu \bar{F}(\tilde{x})\,.
\ee 
From this, we can define the Lie Derivative of a scalar field.
\MYdef{Lie Derivative - scalar field}{
The Lie derivative of a scalar field ${F}( x)$ under the change of coordinates given by Eq.~\eqref{eqcosmo:xtilfun} is defined as
\boxemph{
\pounds_{\boldsymbol \xi} {F} =  - \bds \xi \cdot \bpt F\,,
}
or in coordinates
\boxemph{
\pounds_{\boldsymbol \xi} {F} = -  \xi^\mu \partial_\mu {F}\,,
}
}
In other words, the Lie Derivative of the background function $\bar F$ gives the variation of the perturbations $\delta \tilde{F}(\tilde{x})-\delta F(\tilde{x})$ under the infinitesimal change of coordinates Eq.~\eqref{eqcosmo:xtilfun}.
\MYrem{Coordinate dependence}{As this expression is at first order, it can be expressed at the coordinate $ x$ or $\tilde{ x}$. Moreover, the $\varepsilon$ term is implicit as the function $\boldsymbol{\xi}$ is perturbative.}
\MYrem{Convention}{Some references define the Lie Derivative with the opposite sign. Here we use the rule \emph{New field minus the Old field} to define the Lie Derivative.}

\subsubsection{Vector Field}
We now turn our attention to vector fields. Let $\bds V(x)$ be a vector field decomposed into a background part and a perturbation as
\be  \label{eqcosmo:vectorfield}
\bds{V}(\bds{x}) = \bar{\bds{V}}(\bds{x}) + \varepsilon \, \delta \bds{V}(\bds{x})\,.
\ee 
Under the change of coordinates given by Eq.~\eqref{eqcosmo:xtilfun}, and using the transformation rule for vectors given by Eq.~\eqref{eqgr:vecrule}, we get
\begin{align}
\tilde{V}^\mu (\tilde{\bds{x}}) &=
V^\nu (\bds x) \frac{\partial \tilde{x}^\mu}{\partial x^\nu}\\
&= V^\nu (\bds x) (\tensor{\delta}{^\mu_\nu} + \varepsilon\, \pt_\nu\tensor{\xi}{^{\mu}})\,.
\end{align}
As before, we want to compare $\bds V(\tilde {\bds x})$ with $\tilde{\bds V}(\tilde {\bds x})$. Using the coordinate change Eq.~\eqref{eqcosmo:xtilfun} yields
\be 
V^\mu(\bds x) = V^\mu(\tilde{\bds x}- \varepsilon\, \bds{\xi})
= V^\mu(\tilde{\bds x}) - \varepsilon\, \xi^\nu \pt_\nu \tensor{\bar{V}}{^\mu}\,,
\ee 
where again we used that the last term is a first order quantity. From this, we get that the variation of the vector field keeping fixed the name of the coordinate is
\be  \label{eqcosmo:diffV}
\tilde{V}^\mu(\tilde{ x}) - V^\mu(\tilde{ x}) =
\tensor{\bar{V}}{^\nu} \pt_\nu \tensor{\xi}{^\mu}
- \xi^\nu \pt_\nu \tensor{\bar{V}}{^\mu}\,.
\ee 
As before, we define the perturbation of the vector field in the new coordinate system as
\be  \label{eqcosmo:vectorfieldprime}
\tilde{\bds V}(\tilde{ x})= \bar{\bds{V}}(\tilde{ x}) + \varepsilon\, \delta \tilde{{V}} (\tilde{ x})\,.
\ee 
From the decomposition of the vector field Eq.~\eqref{eqcosmo:vectorfield} and Eq.~\eqref{eqcosmo:vectorfieldprime}, we get
\be \label{eqcosmo:deltaV1}
\tilde{\bds V}(\tilde{ x}) - \bds{V}(\tilde{ x}) =
\delta \tilde{\bds V}(\tilde{ x}) - \delta \bds{V}(\tilde{ x})\,,
\ee 
Finally, combining Eq.~\eqref{eqcosmo:diffV} and Eq.~\eqref{eqcosmo:deltaV1} yields
\be 
\delta \tilde{V}^\mu(\tilde{ x}) - \delta V^\mu(\tilde{ x}) 
=
\tensor{\bar{V}}{^\nu} \pt_\nu \tensor{\xi}{^\mu}
- \xi^\nu \pt_\nu \tensor{\bar{V}}{^\mu}\,,
\ee 
which motivates the following definition.
\MYdef{Lie Derivative - vector field}{
The Lie derivative of a vector field $\bds{V}( x)$ under the change of coordinates given by Eq.~\eqref{eqcosmo:xtilfun} is defined as
\boxemph{
\pounds_{\boldsymbol \xi} \bar{\bds{V}} = 
[\bds{V} \cdot \bpt] \bds \xi - [\bds{\xi} \cdot \bpt] \bds{V}\,,
}
or in coordinates
\boxemph{  \label{eqcosmo:lievectorcoord}
\pounds_{\boldsymbol \xi} \tensor{V}{^\mu} = 
\tensor{V}{^\nu} \pt_\nu \tensor{\xi}{^\mu}
- \xi^\nu \pt_\nu \tensor{V}{^\mu}\,.
}
}
\subsubsection{Form Field}
The generalisation to $1$-forms is straight forward and follows the same steps. The only difference comes from the transformation rule given by Eq.~\eqref{eqgr:formrule}, which bring the opposite sign for the corresponding term (namely the term with the gradient of $\bds \xi$). Assuming we have a form given by 
\be
\bds{W}(\bds x) = \bar{\bds{W}}( x) + \varepsilon\, \delta \bds{W}( x)\,,
\ee
and we perform a change of coordinates, we obtain
\be 
\tilde{W}_\mu(\tilde{ x}) - W_\mu(\tilde{ x}) =
-\tensor{\bar{W}}{_\nu} \pt_\mu\tensor{\xi}{^\nu}
- \xi^\nu \pt_\nu \tensor{\bar{W}}{_\mu}\,.
\ee 
If we define the perturbation in the new system of coordinates as
\be \label{eqcosmo:deltaW1}
\tilde{\bds{W}}(\tilde{ x}) = \bds{\bar{W}}(\tilde{ x}) + \varepsilon\, \delta \tilde{\bds{W}}(\tilde{ x})\,,
\ee
we get
\be \label{eqcosmo:deltaW2}
\tilde{\bds W}(\tilde{ x}) - \bds{W}(\tilde{ x}) =
\delta \tilde{\bds W}(\tilde{ x}) - \delta \bds{W}(\tilde{ x})\,.
\ee 
These relations imply
\be 
\delta \tilde{W}_\mu(\tilde{ x}) - \delta W_\mu(\tilde{ x}) 
=
-\tensor{\bar{W}}{_\nu} \pt_\mu\tensor{\xi}{^\nu}
- \xi^\nu \pt_\nu \tensor{\bar{W}}{_\mu}\,.
\ee 
The definition of the Lie Derivative for a form then comes naturally.
\MYdef{Lie Derivative - $1$-form field}{
The Lie derivative of a form field $W( x)$ under the change of coordinates given by Eq.~\eqref{eqcosmo:xtilfun} is defined as
\boxemph{ 
\pounds_{\boldsymbol \xi} \boldsymbol{W} = 
- \bpt \bds \xi \cdot \bds W - [\bds \xi \cdot \bpt] \bds W\,,
}
or in coordinates
\boxemph{ \label{eqcosmo:lieformcoord}
\pounds_{\boldsymbol \xi} \tensor{W}{_\mu} = 
-\tensor{W}{_\nu} \pt_\mu\tensor{\xi}{^\nu}
- \xi^\nu \pt_\nu \tensor{W}{_\mu}\,.
}
}
The generalisation to tensors is straight forward. Basically, the second term of Eq.~\eqref{eqcosmo:lievectorcoord} or Eq.~\eqref{eqcosmo:lieformcoord} is always present, as it comes from the transformation from $x^\mu$ to $\tilde{x}^\mu$. On the other side, there is a term similar to the first one in Eq.~\eqref{eqcosmo:lievectorcoord} for every \emph{contravariant index} and one similar to the first on in Eq.~\eqref{eqcosmo:lieformcoord} for every \emph{covariant index}.

We just mention the Lie Derivative of the metric (or a $2$-form) as it is the one that will be useful. The metric $\bds{g}$ is a tensor of type $(0,2)$. Its Lie Derivative under the change of coordinates Eq.~\eqref{eqcosmo:xtilfun} is given, in coordinates, by 
\boxemph{  \label{eqcosmo:liedermetric}
\pounds_{\boldsymbol \xi} \tensor{g}{_{\mu \nu}} = 
- g_{\rho \nu} \partial_\mu \xi^\rho - g_{\mu \rho} \partial_\nu \xi^\rho
- \xi^{\rho} \partial_\rho g_{\mu \nu}
}

To conclude this Section, let us stress why the concept of Lie Derivative is important. We consider a perturbed tensor, typically the metric, where the functional form of the background is the same in all the coordinate frames, typically the \ac{FRLW} metric. We use the fact that, under a change of coordinates, the difference between the perturbations in both frames is given by the Lie Derivative of the background tensor, which can be seen through Eq.~\eqref{eqcosmo:deltaF1}, Eq.~\eqref{eqcosmo:deltaV1} and Eq.~\eqref{eqcosmo:deltaW2}: The definition of the Lie Derivative corresponds exactly to this difference. With this tool in our hand, we derive transformations laws of perturbations under gauge transformation. We also build gauge invariant perturbations which truly represent perturbation and are not a mere artifact of the chosen coordinates.

\subsection{Three Important Examples}
\subsubsection{Cosmological Perturbations}
We discuss here the formalism used in the framework of Cosmological Perturbation Theory. We consider the unperturbed \ac{FRLW} metric\footnote{We work in conformal time $\eta$ as it makes the computations simpler and we assume that the spatial curvature vanishes.}
\be 
\bar{\bds{g}} = a^2 \left( - \bdd {\eta}^2 + \bdd{\bds{x}^2}\right)\,.
\ee 
We add some perturbations and consider the full metric
\be  \label{eqcosmo:pertmetric}
\bds{g} = \bar{\bds{g}} + a^2 \bds{h}\,,
\ee 
where $\bds{h}$ represents the perturbation and is understood to be small.

We need now to parametrise the tensor $\bds h$ in a smart way. Obviously, we could say that the variables of interests are the individual components $h_{\mu \nu}$, and consider the $10$ of them. This would not be very convenient in the end, as the Einstein's Equations are not very nice in this case.

A better way to consider such perturbations goes as follows. The first thing to notice is that the \ac{FRLW} background is, by construction, invariant under spatial rotation. In such a case, the correct method is to consider objects that behave well under such transformations. In the case of rotations, these objects are the scalars, the vectors and the tensors, whose transformation rules correspond to the one of a field with spin $s=0$, $1$ and $2$ respectively. This process is the so-called \emph{SVT decomposition}. 

In this work, we will only consider scalar perturbations, hence we will only present them here. The generic case, including vectors and tensors, can be found in virtually any textbook on cosmological perturbations, see for example \cite{kurki_2020}. In this case, the various components of the perturbation tensor are defined as
\boxemph{ \label{eqcosmo:h00}
h_{00} &= - 2 \psi\,, \\
\label{eqcosmo:h0i}
h_{0i} &= \pt_i \xi\,, \\
\label{eqcosmo:hij}
h_{ij} &= - 2 \delta_{ij} \phi + \left(\pt_i \pt_j - \frac 13 \delta_{ij} \Delta \right) \zeta\,.
}
The four scalar perturbations are then given by the $4$ functions $\psi$, $\phi$, $\xi$ and $\zeta$.

We perform now a gauge transformation given by
\be  \label{eqcosmo:changecoord}
\tilde{\bds x} = \bds{x} + \bds{\chi}\,,
\ee
where the vector field $\bds \chi$ is also parametrised in terms of $2$ scalars, $\tau$ and $\sigma$, as
\begin{align}
 \chi^0 &= \tau \,, \\
 \chi^i &= \tensor{\delta}{^{ij}}\pt_j \sigma\,.
\end{align}
Using the result of the next section, and especially Eq.~\eqref{eqcosmo:liedermetric}, we get the gauge transformation of the scalar functions under the change of coordinates given by Eq.~\eqref{eqcosmo:changecoord}
\boxemph{
\label{eqcosmo:psitra}
\tilde{\psi} &= \psi - \Hcal \tau - \tau^\prime\,, \\
\label{eqcosmo:phitra}
\tilde{\phi} &= \phi + \Hcal \tau + \frac 13 \Delta \sigma\,, \\
\label{eqcosmo:xitra}
\tilde{\xi} &= \xi + \tau - \sigma^\prime\,, \\
\label{eqcosmo:zetatra}
\tilde{\zeta} &= \zeta - 2 \sigma\,,
}
where a prime denotes here a derivative with respect to the conformal time $\eta$. Any gauge transformation is parametrised by $2$ scalar $\tau$ and $\sigma$ and we have $4$ functions. This means that we expect $4-2=2$ physical degrees of freedom which cannot be removed by any gauge transformation. It is very funny to build these degrees of freedom from the transformation rules given by Eqs.~\eqref{eqcosmo:psitra}-\eqref{eqcosmo:zetatra}.

For example, if we starts with $\psi$, we need to remove the $\tau^\prime$ part, which can be done by adding a term proportional to $\xi^\prime$. Starting with the higher derivatives and doing this method recursively, we define the gauge independent variables as
\boxemph{
\Psi &= \psi + \left(\xi - \frac{\zeta^\prime}{2} \right)^\prime + \Hcal \left(\xi -\frac{\zeta^\prime}{2} \right)\,, \\
\Phi &= \phi + \frac 16 \Delta \zeta - \Hcal\left(\xi - \frac{\zeta^\prime}{2} \right)\,.
}
Indeed, using the transformation rules given by Eqs.~\eqref{eqcosmo:psitra}-\eqref{eqcosmo:zetatra}, it is straightforward to show that $\tilde{\Psi} = \Psi$ and $\tilde{\Phi} = \Phi$. Those quantities are the \emph{Bardeen potentials}, see \cite{Bardeen:1980} for the seminal work. In other words, these two quantities are gauge independent and correspond to physical degrees of freedom. Hence, they are good observables. For example, if they both vanish in one system of coordinates, we can be sure that there is no physical perturbations and that the perturbations of the metric are spurious and are caused by a mere change of coordinates.

In general, one chooses a specific gauge and works in it. For example, it is clear that $\zeta$ can be set to $0$ by specifically setting $\sigma$, and then $\xi$ can be set to $0$ by specifically setting $\tau$. 
\MYdef{Longitudinal gauge}{This choice corresponds to the longitudinal gauge and in this case $\Psi = \psi$ and $\Phi = \phi$.}

\subsubsection{Perturbations to the Energy-Momentum Tensor \label{eqcosmo:pertEMT}}
In this section, we briefly present the perturbations to the energy-~momentum tensor. We use the expression of the energy-~momentum tensor Eq.~\eqref{eqcosmo:tmunu} but here we consider that the energy density, the pressure and the $4$-velocity of the fluids are perturbed as
\begin{align}
\label{eqcosmo:deltarho}
\rho &= \bar{\rho} + \delta \rho\,, \\
\label{eqcosmo:deltaP}
P &= \bar{P} + \delta P\,, \\
\label{eqcosmo:deltav}
\bds{u}&= \bar{\bds{u}} + \delta \bds{u}\,.
\end{align}
Here, the barred quantities are the time-dependent background quantities of the homogeneous and isotropic Universe, and the perturbed quantities depend either on space and time. The normalisation condition 
\be 
\bds{u}^2 =0\,,
\ee
with the perturbed metric given by Eq.~\eqref{eqcosmo:pertmetric} imposes the perturbed velocity to be of the form
\be
\bds{u} = \frac{1}{a}\left( (1-\psi) \bpt_t + \pt_i v \bpt_i\right)\,.
\ee
where we introduce the perturbed velocity $v$, and where we considered only scalar perturbations. Moreover, we could also consider the non-diagonal part of the spatial tensor, i.e. a term corresponding to a non-isotropic shear. This physical term is not relevant for us and we neglect it here, i.e. we assume the fluid to be perfect. 
\MYrem{Relative perturbation}{
Usually, in a cosmological background with coordinates $(\eta, \boldsymbol{x})$ one defines the relative density perturbation as
\be 
\delta_\rho(\eta, \bds{x}) \equiv \frac{\rho(\eta, \bds{x})- \bar \rho(\eta)}{\bar \rho(\eta)}\,,
\ee
and likewise for the pressure.
}
Under the change of coordinates given by Eq.~\eqref{eqcosmo:changecoord}, the perturbed quantities of the energy-momentum tensor transform as
\boxemph{
\delta \tilde{\rho} &= \delta \rho - \tau {\bar{\rho}}^\prime \,, \\
\delta \tilde{P} &= \delta P - \tau \bar{P}^\prime\,, \\
\tilde{v} &= v + {\sigma}^\prime \,,
}
from which it is direct to build gauge independent variables
\boxemph{
D_\rho &= \delta \rho + {\bar{\rho}}^\prime\left( \xi - \frac 12  {\zeta}^\prime\right) \,, \\
D_P &= \delta P + {\bar{P}}^\prime\left( \xi - \frac 12  {\zeta}^\prime\right)  \,, \\
V &= v + \frac 12 \zeta^\prime\,,
}
and gauge independent equations
\boxemph{
\Phi &= \Psi\,, \\
\Delta \Phi -3 \Hcal (\Hcal \Phi +  \Phi^\prime) &= 4 \pi \GN a^2 D_\rho \,, \\
 \Phi^\prime + \Phi \Hcal &= V \left(  \frac{ a^{\prime \prime}}{a} - 2 \Hcal^2\right)\,, \\
\Phi^{\prime \prime} + 3 \Hcal \Phi^\prime - \Hcal^2 \Phi + 2 \frac{a^{\prime \prime}}{a}\Phi &= 4 \pi \GN a^2 D_P\,.
}
These equations govern the evolution of perturbations. Note that the first two equations are constraints, while the last two are evolution equations. Moreover, the first condition $\Phi = \Psi$ is only valid because we neglected the non-isotropie stress. The more general expressions can be found e.g. in~\cite{durrer_2021,peter_2013}.

\subsubsection{Spherically Symmetric Static Spacetimes \label{seccosmo:ssss}}
In this Section, we present the last important example: Perturbations around Spherically Symmetric Static Spacetimes. This topic has been covered in details in several references. The seminal papers by Regge and Zerilli \cite{Regge:1957,Zerilli:1970} introduce the perturbations around a Schwarzschild Black Hole, and more recent reviews discuss this topic in great details, for example \cite{Nollert:1999, Pani:2013}. We only present here the relevant results without going into any proof.

In the usual spherical coordinates, such spacetimes are given by the metric
\be 
\label{eqcosmo:metricAB}
\bds{\bar{g}} = 
-A(r) \bds \dd t^2
+B(r)^{-1} \bds \dd r^2 
+ r^2 \bds \dd  \Omega^2\,,
\ee
where $A(r)$ and $B(r)$ are arbitrary radial functions. We want to consider perturbations of the form $\bds g = \bds{\bar{g}} + \bds h$. Using the  spherical symmetry of the background metric, we can decompose the perturbations as
\boxemph{
\bds h = \sum_{\ell,m} \sum_{n=1}^{10} F_{\ell,m}^{n}(t,r) \bds{Y}_{\ell,m}^{n}(\theta, \varphi).
}
In this formula, $\bds{Y}_{\ell,m}^{n}(\theta, \varphi)$ are $10$ tensors on the sphere related to the Spherical Harmonics. Without going into the details, they are linked with the spin-$2$ irreducible representations of the rotation group. Each of this individual perturbation is parametrised by a function $F_{\ell,m}^{n}(t,r)$ which quantifies this specific perturbation.

There is one more symmetry we can use: The metric is invariant under parity, or explicitly under the change of coordinates given by $\tilde \theta = \pi - \theta$. This allows us to perform one more decomposition of the tensor $\bds h $ as
\be
\bds h = \bds h_{\mo} + \bds h_{\me}\,
\ee 
where $\bds{h}_{\mo} $ (respectively $\bds{h}_{\me}$) represents the odd (respectively even) perturbations. 

The general idea behind these decomposition comes from the fact that, at first order in perturbation theory, the different modes decouple. To solve the equations, we need to consider a fixed value of $\ell$ and $m$ and either the odd or the even perturbations. Then, the equations can be solved perturbatively.

In coordinates, these two tensors (for fixed values of $\ell$ and $m$) are given by\footnote{The matrices are separated in two pieces for purely aesthetic reasons.}
{\footnotesize
\boxemph{
 (h_{\me})_{\mu \nu}^{\ell,m}&=
        \begin{pmatrix}
        -  A(r)\,  H_0\,  Y_{\ell,m}&
         H_1\,  Y_{\ell,m} & 
        H_3 \, \partial_\theta Y_{\ell,m} & 
        H_3 \, \partial_\varphi Y_{\ell,m} \\
        H_1\,  Y_{\ell,m}  &  
        \frac{H_2\, Y_{\ell,m}}{B(r)} & 
        H_4 \, \partial_\theta Y_{\ell,m} &
        H_4 \, \partial_\varphi Y_{\ell,m} \\
        H_3 \, \partial_\theta Y_{\ell,m}  &
        H_4 \, \partial_\theta Y_{\ell,m}  &  
        0& 
        0\\
        H_3 \, \partial_\varphi Y_{\ell,m}  &
        H_4 \, \partial_\varphi Y_{\ell,m}  & 
        0& 
        0\\
        \end{pmatrix} \\
         \nonumber
&+ 
        \begin{pmatrix}
        0&0&0&0 \\
        0&0&0&0 \\
        0 &
        0 &  
        r^2 (K_0\,  Y_{\ell,m} +K_1\, \partial_\theta^2  Y_{\ell,m} )& 
        r^2 K_1 X(\theta, \varphi)\\
        0 &
        0 & 
        r^2 K_1 X(\theta, \varphi)& 
        r^2 (K_0 \sin^2 \theta\, Y_{\ell,m} + K_1 W(\theta, \varphi)) \\
        \end{pmatrix}\,, \\ 
 (h_{\mo})_{\mu \nu}^{\ell,m}&=
        \begin{pmatrix}
        0 &0  &  - h_0 \sin^{-1}\theta \partial_\varphi Y_{\ell,m}&  h_0 \sin \theta  \partial_\theta Y_{\ell,m} \\
        0 & 0 & - h_1 \sin^{-1}\theta \partial_\varphi Y_{\ell,m} &  h_1 \sin \theta  \partial_\theta Y_{\ell,m}\\
        - h_0 \sin^{-1}\theta \partial_\varphi Y_{\ell,m}&  - h_1 \sin^{-1} \theta  \partial_\varphi Y_{\ell,m} &0&0\\
        h_0 \sin \theta \partial_\theta Y_{\ell,m}&  h_1 \sin \theta  \partial_\theta Y_{\ell,m} & 0&0 
        \end{pmatrix}  \\
         \nonumber
&+
        \begin{pmatrix}
        0 &0  &  0& 0\\
        0 & 0 & 0 & 0\\
         0&  0 & -2r^2 h_2 \sin^{-1}\theta X(\theta, \varphi) &  r^2 h_2 Z(\theta, \varphi)\\
         0&  0&  r^2 h_2 Z(\theta, \varphi) & 2r^2 h_2\sin \theta X(\theta, \varphi)  \\
        \end{pmatrix} \,,\\
    \nonumber
&\mathrm{with} \\
X(\theta, \varphi)&\equiv (\partial_\theta \partial_\varphi  Y_{\ell,m} - \cot \theta \,  \partial_\varphi  Y_{\ell,m})\,, \\
W(\theta, \varphi)&\equiv \left( \partial_\varphi^2 Y_{\ell,m} +\cos \theta \sin \theta \,  \partial_\theta Y_{\ell,m} \right)\,, \\
W(\theta, \varphi)&\equiv \sin\theta\, \partial_\theta^2Y_{\ell,m} - \sin^{-1} \theta \partial_\varphi^2 Y_{\ell,m} - \cos \theta \partial_\theta \partial_\varphi Y_{\ell,m}\,.
}
}
Here, the functions $Y_{\ell,m}$ are the Spherical Harmonics, see Section~\ref{seccosmo:SH}.

These expressions can be written in a covariant form, see for example \cite{Nollert:1999}. In total, there are $3$ odd variables ($h_0$, $h_1$ and $h_2$) and $7$ even variables ($H_0$, $H_1$, $H_2$, $H_3$, $H_5$, $K_0$ and $K_1$), corresponding to the $10$ generic functions $F_{\ell,m}^{n}$.

We now study the transformations rules for these perturbations under an infinitesimal change of coordinates. We consider a given pair $(\ell,m)$ which we do not write explicitly, and we use the relation for the Lie Derivative of the metric given by Eq.~\eqref{eqcosmo:liedermetric}.

For the odd perturbations, we take for the infinitesimal change
\be 
\bds \xi = \Lambda(t,r) \left( \frac{\partial_\varphi Y_{\ell,m}}{\sin \theta} \bpt_\theta -  \frac{\partial_\theta Y_{\ell,m}}{\sin \theta}\bpt_\varphi  \right)\,.
\ee 
The gauge transformations are given by
\boxemph{
\tilde{h}_0 &= h_0 + r^2 \dot \Lambda\,, \\
\tilde{h}_1 &= h_1 + r^2  \Lambda^\prime\,, \\
\tilde{h}_2 &= h_2 +   \Lambda\,,
}
where a dot (resp. prime) represents a derivative with respect to conformal time $\eta$ (resp. conformal distance $r$). With $1$ parameter at hand, it is possible to build $2$ gauge independent quantities, namely
\boxemph{
k_0 &= h_0 - r^2 \dot{h}_2\,, \\
k_1 &= h_1 - r^2 {h}^\prime_2\,.
}
The simplest choice is to choose $h_2=0$, the so-called \emph{Regge-Wheeler} gauge, in which $k_0=h_0$ and $k_1=h_1$.

For the even perturbations, the infinitesimal change reads
\be
\bds \xi = M_0  Y_{\ell,m} \bpt_t  + M_1 Y_{\ell,m} \bpt_r
+ M_2  \partial_\theta Y_{\ell,m} \bpt_\theta 
+ M_2  \partial_\varphi Y_{\ell,m} \bpt_\varphi\,, 
\ee 
and the transformations are given by
\boxemph{
\tilde{H}_0 &= H_0 - 2 \dot M_0  - \frac{A^\prime}{A} M_1 \,, \\
\tilde{H}_1 &= H_1 - \frac{1}{B} \dot M_1 + A M_0^\prime  \,, \\
\tilde{H}_2 &= H_2 + \frac{B^\prime}{B} M_1 - 2  M_1^\prime\,, \\
\tilde{H}_3 &= H_3 + A M_0- r^2 \dot M_2 \,, \\
\tilde{H}_4 &= H_4 - \frac{1}{B} M_1 - r^2  M_2^\prime\,, \\
\tilde{K}_0 &= K_0 - \frac{2}{r} M_1 \,, \\
\tilde{K}_1 &= K_1 -2 M_2\,.
}
Again, with $3$ parameters at hand, it is possible to build $4$ gauge independent variables, for example
\boxemph{
\Gamma_0 &= H_0  + \frac{2}{A} \dot{H}_3- B \frac{A^\prime}{A} H_4 + \frac{r^2}{2} B \frac{A^\prime}{A} K_1^\prime - \frac{r^2}{A} \ddot K_1\\
\Gamma_1 &=  H_1 - A \left( \frac{1}{A} H_3\right)^\prime -\dot H_4 + \frac 12 A \left( \frac{r^2}{A} K_1\right)^\prime + \frac{r^2}{2}  \dot{K}_1^\prime\,, \\
\Gamma_2 &= H_2 + B^\prime H_4 - 2 (B H_4)^\prime - \frac{r^2}{2} B^\prime K_1^\prime + (r^2 B K_1^\prime)^\prime  \,, \\
\Gamma_3 &= K_0 - \frac{2}{r}B H_4 +r B  K_1^\prime\,,
}
In the Regge-Wheeler gauge, we have $H_3=H_4=K_1=0$, hence $\Gamma_0=H_0$, $\Gamma_1=H_1$, $\Gamma_2=H_2$ and $\Gamma_3 = K_0$.

One last comment about the time dependence of the perturbations. As the background is static, we can again perform a Fourier decomposition 
\be
F_{\ell,m}^{n}(t,r) = \int\; \eee^{-\ii \omega t} F_{\ell,m}^{n}(\omega,r)\, \dd \omega\,.
\ee 
and for the same reason as explained before, perturbations with different frequencies $\omega$ do not interact at first order. In general, as the equations with different $\ell$, $m$ and $\omega$ decouple, we will simply write
\be 
F_{\ell,m}^{n}(\omega,r) = F^{n}(r)\,.
\ee 

\section{Lensing}
\subsection{General Principle}
A very important effect of General Relativity is lensing. As the name suggests, lensing describes the deflection of light as it travels through non-homogeneous space, because of the gravitational field. Lensing has two main effects.

\begin{figure}[ht!]
	\begin{center}
	\includegraphics[width = 0.49\textwidth]{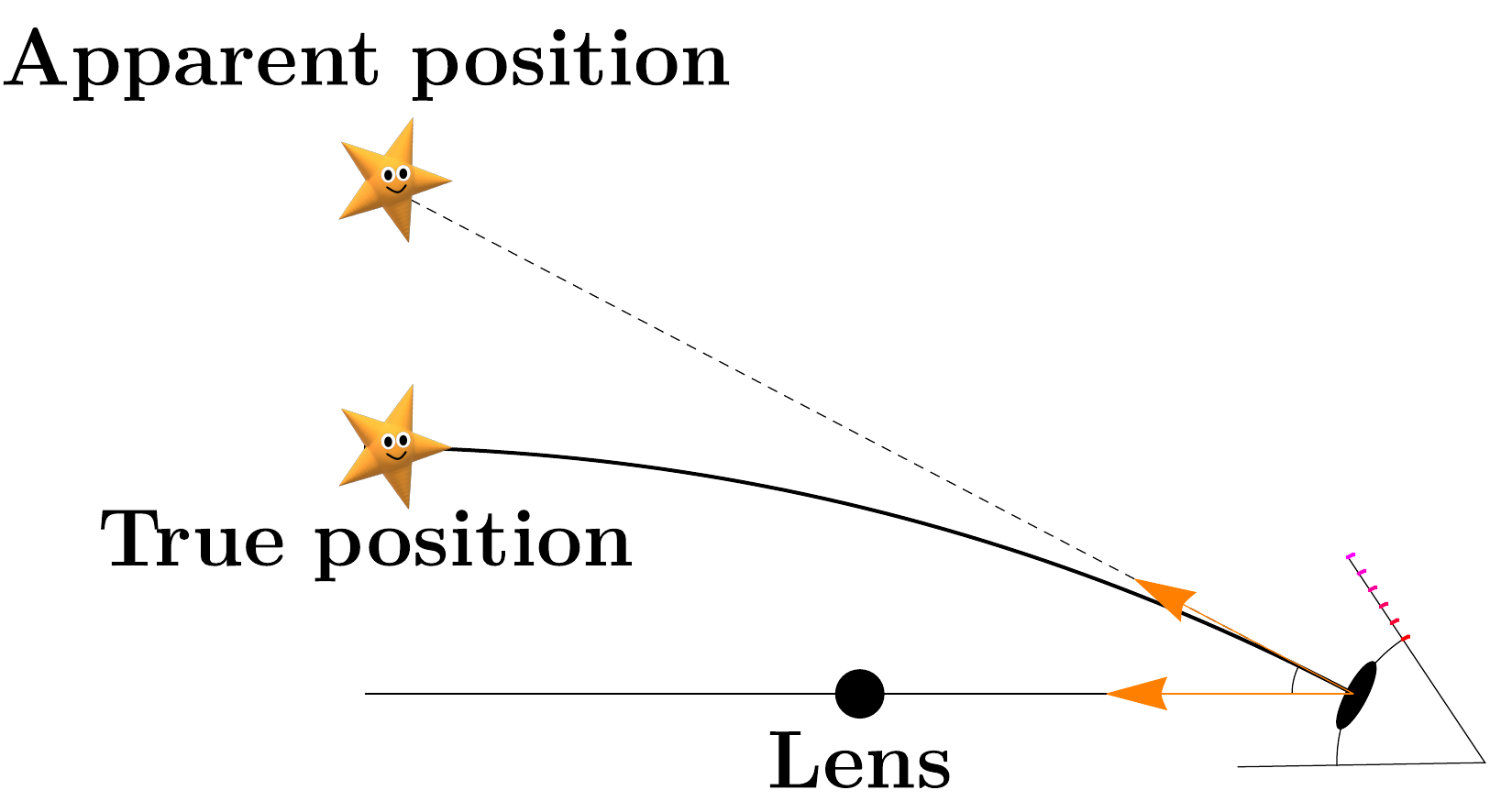} 
	\includegraphics[width = 0.49\textwidth]{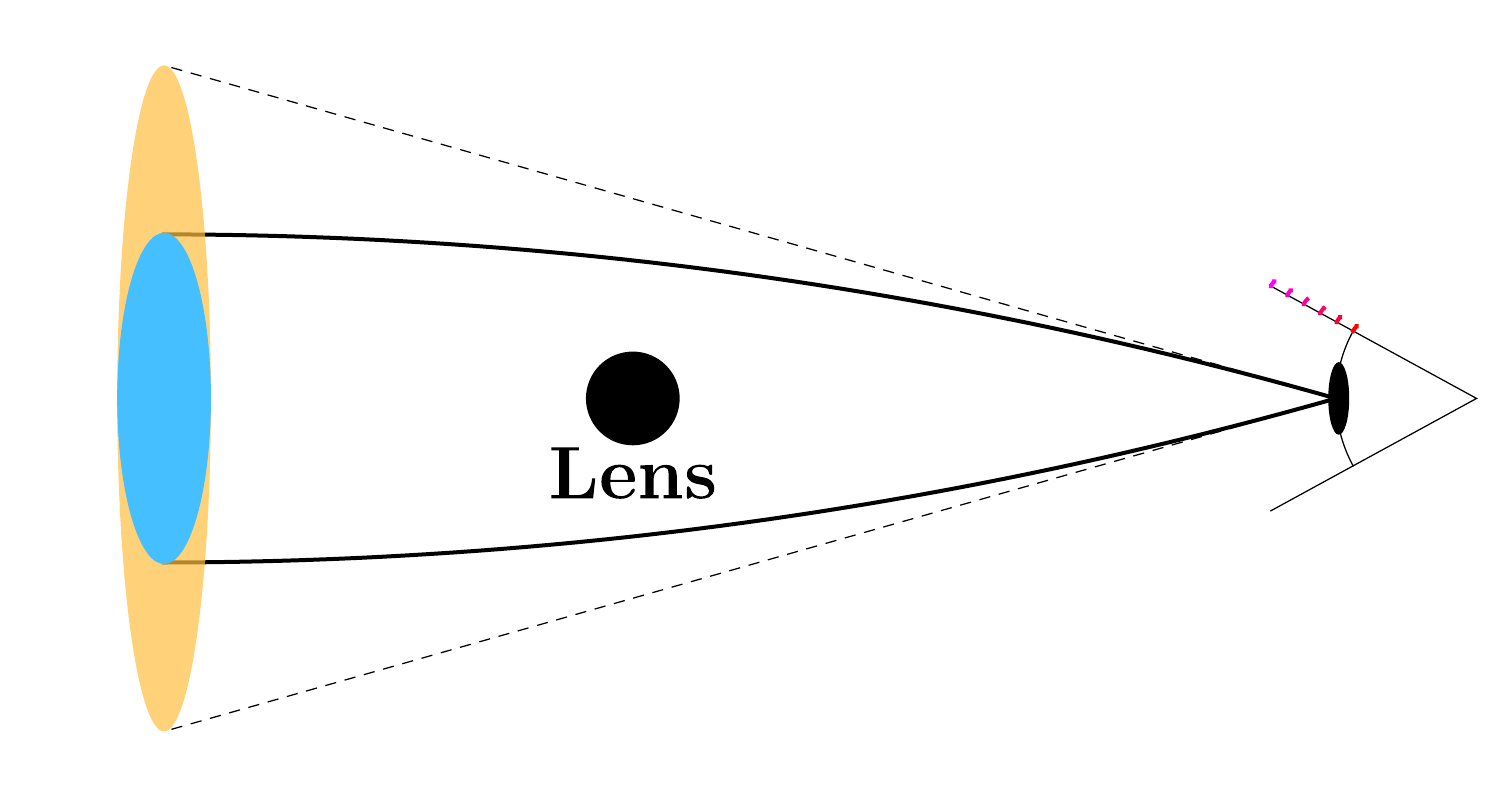}
	\end{center}
	\caption[Effects of lensing]{\label{figcosmo:lensing1}
	\textbf{Effects of lensing}\\
	\textbf{Left:} Image are displaced. \\ 
	\textbf{Right:} Images are distorted}
\end{figure}

The first effect is the displacement of the images, as it happens in the hot desert when you see a mirage. The light being bent, the apparition position (of a star or a galaxy for example) on the sky, as seen from the Earth, can be displaced with respect to its true position, see Fig.~\ref{figcosmo:lensing1} (left).

The second effect happens when we observe a bundle of light instead of a single ray. Indeed, the effect of lensing on neighbouring rays is slightly different as they do not travel on the same spacetime points. This effect can be understood easily as follow. Schematically, lensing is a map from the plane to itself (to make things simpler) given by 
\be 
\bds{L}(\bds \alpha) = \bds \alpha + \varepsilon\, \bds{\Lambda}(\bds \alpha)\,,
\ee 
where $\varepsilon$ is a small parameter and $\bds \alpha$ is a vector on the plane. The function $\bds \Lambda$ is the deflection angle and quantifies lensing. We will study this function more in details below. We set our coordinate system such that $\bds L(\bds 0)=\bds 0$. We consider a light ray starting at $\bds x$. Under lensing, and at first order in $\bds x$, this vector is mapped to
\be 
L^i(\bds x) = x^j (\tensor{\delta}{^i_j} + \varepsilon\, \partial_j {\Lambda}^i).
\ee
This shows that distorsion effects are given by the gradients of the deflection angle. This will become clear in the next Section. This second effect is shown in Fig.~\ref{figcosmo:lensing1} (right).

\subsection{The Lens Map \label{seccosmo:lens_map}}
In this Section, we want to formalise the concepts explained previously. Much more can be found about lensing in \cite{durrer_2021} or in \cite{fleury_2015}. We present here the main concepts which are relevant for our work.

Mathematically speaking, the study of light propagation across the Universe is equivalent to study a map from the sphere $S^2$ to itself. Indeed, when you observe the sky during the night and you see a star, you can measure its observed position on the celestial sphere $\bds{n}_\mo = (\theta_\mo, \varphi_\mo)$, where $\theta$ and $\varphi$ represent the usual spherical angles. The space between you and this star is not homogeneous but filled with randomly distributed galaxies and dark matter. The light ray is deviated from its emitted position $\bds{n}_\me=(\theta_\me, \varphi_\me)$. This position is not an observable per se on Earth as we have no way to access it. However, we still want to relate those two positions, as this can help us interpret the observations and from this we can learn a lot about the matter distribution in the Universe. This map relating $\bds{n}_\mo$ and $\bds{n}_\me$ is the Lens map that we want to discuss here. A lot of work has been done about this in the past and we present the main results.

Photons travelling through a curved background follow null geodesics. Defining the Lens map is nothing more than studying the propagation of null geodesics in curved spacetime. Let us consider the perturbed \ac{FRLW} spacetime in spherical coordinates and conformal time (and using the longitudinal gauge, i.e. setting $\xi=0$ and $\zeta=0$ in Eq.~\eqref{eqcosmo:h0i} and Eq.~\eqref{eqcosmo:hij})
\be 
\bds{g} = a^2 \left(-(1+2\Psi)\bdd \eta^2 + 
(1-2\Phi)(\bdd r^2 + r^2 \bdd \Omega^2) \right)\,.
\ee 
Here $\Psi$ and $\Phi$ correspond to the gauge independent variables. We now state a theorem of Differential Geometry about null geodesics, without proof.
\MYthm{Null geodesics of conformally related metrics}{\label{thmcosmo:nullgeo}
Let $\bds g$ and $\tilde{\bds g}$ be two conformally related metrics, i.e. there exist a (positive) function $F(\bds x)$ such that
\be 
\tilde{\bds{g}} = F(\bds x) \bds{g}\,.
\ee 
Then, the null geodesics associated with the metric $\bds{g}$ and $\tilde{\bds{g}}$ are the same.
}
This theorem is very useful as it allows us to change the metric we consider it with multiplying by a global arbitrary function. This simplifies the study of null geodesics. In our case, we consider the metric
\begin{align}
\tilde{\bds g} &= a^{-2} (1+2\Phi) \bds{g}\\
&=
-\big(1+2(\Psi+\Phi)\big)\bdd \eta^2 + 
\bdd r^2 + r^2 \bdd \Omega^2\\
&\equiv 
-\left(1+4 \Psi_{\mW}\right)\bdd \eta^2 + 
\bdd r^2 + r^2 \bdd \Omega^2\,,
\end{align}
where we introduced the Weyl potential
\be 
\Psi_\mW \equiv \frac 12 (\Psi+\Phi)\,,
\ee 
and where we worked at first order in the perturbations $\Psi$ and $\Phi$. From Theorem~\ref{thmcosmo:nullgeo}, we know that the geodesics of light in the perturbed \ac{FRLW} spacetime will only depend on the Weyl potential, i.e. on the sum of the Bardeen potentials.

In the next step, we should compute perturbatively the geodesics of the metric $\tilde{\bds g}$. We only present here the final results. We consider a photon emitted at $\bds{x}_\me = (\eta_\me, r_\me, \theta_\me, \varphi_\me)$ and observed at position $\bds{x}_\mo = (\eta_\mo, r_\mo, \theta_\mo, \varphi_\mo)$. At the background level $\Psi_\mW=0$, the equations of motion are easily solved and give
\boxemph{
r(\eta ) &= \eta_\mo -\eta\,, \\
\theta_\me&= \theta_\mo\,, \\
\varphi_\me&= \varphi_\mo\,. 
}
At the perturbed level, we are only interested in the angular deflection. The final solution gives the emission angles as function of the observed angles and the emission time
\boxemph{\label{eqcosmo:thetae}
\theta_\me &= \theta_\mo + \pt_\theta \phiL(\eta_\me, \theta_\mo, \varphi_\mo)\,, \\
\label{eqcosmo:phie}
\varphi_\me &= \varphi_\mo + \frac{1}{\sin^2 \theta_\mo} \pt_\varphi \phiL(\eta_\me, \theta_\mo, \varphi_\mo)\,, \quad \mathrm{with} \\
\phiL(\eta_\me, \theta_\mo, \varphi_\mo) &= -2 \int_{\eta_\me}^{\eta_\mo}\; \frac{\eta-\eta_\me}{(\eta_\mo-\eta_\me)(\eta_\mo-\eta)} \Psi_\mW(\eta,\eta_\mo-\eta,\theta_\mo, \varphi_\mo)\, \dd \eta\ \,,
}
where we have introduced the lensing potential $\phiL(\eta_\me, \theta_\mo, \varphi_\mo)$. 
\MYrem{Lensing potential}{Note that in the lensing potential, we integrated along the unperturbed path as $\Psi_\mW$ is already a first order quantity (Born's rule). Moreover, it seems a bit fishy to have a dependence on the emission (conformal) time $\eta_\me$. However this is not a problem, as in general we can measure the redshift of a source, and express the lensing potential in terms of this redshift, instead of the emission time using the relationship between time and redshift given by Eq.~\eqref{eqcosmo:tfunz}. It can be shown that under a transformation of the form Eq.~\eqref{eqcosmo:atransfo}, the lensing potential is indeed invariant, which makes it a good physical observable.
}
\MYrem{Apparent divergence}{The integral defining the lensing potential seems to be divergent for $\eta \rightarrow \eta_\mo$. However, this divergence is not observable per se. Indeed, the observable is the deviation angle, given by the spherical gradient of the lensing potential. Hence, if inside the integral we perform the replacement
\be 
\Psi_\mW(\eta, \eta_\mo - \eta, \theta_\mo, \varphi_\mo) \rightarrow \Psi_\mW(\eta, \eta_\mo - \eta, \theta_\mo, \varphi_\mo)-
\Psi_\mW(\eta_\mo, 0, \theta_\mo, \varphi_\mo)\,,
\ee 
the deviation angle is not affected. Indeed, $\Psi_\mW(\eta_\mo, 0, \theta_\mo, \varphi_\mo)$ is simply $\Psi_\mW(\eta_\mo, 0)$ and does not depend on the angles. Moreover, this term is of the form
\be 
\Psi_\mW(\eta, \eta_\mo - \eta, \theta_\mo, \varphi_\mo)-
\Psi_\mW(\eta_\mo, 0, \theta_\mo, \varphi_\mo) \sim \partial \Psi_\mW \times  (\eta_\mo - \eta)\,,
\ee
where $\partial \Psi_\mW$ schematically represents a linear combination of derivatives of $\Psi_\mW$, which are well-defined at $\eta=\eta_\mo$. Hence, the terms $\eta_\mo-\eta$ cancel out, and the final result is not divergent.

Note that adding a constant term to $\Psi_\mW$ is equivalent to changing the monopole term corresponding to $\ell=0$, which is not observable in lensing experiments, see Section~\ref{seccosmo:stats} and Rem~\ref{remcosmo:monopole} below.
}
Let us rewrite Eq.~\eqref{eqcosmo:thetae} and Eq.~\eqref{eqcosmo:phie} in a more convenient form. We assume that we observe a small patch of the celestial sphere around arbitrary angles $\bar{\theta}$ and $\bar{\varphi}$ which can be different from $\theta_\mo$ and $\varphi_\mo$. Around these, we introduce the local coordinates
\begin{align}
    \alpha^1 &= \theta - \bar{\theta}\,, \\ 
    \alpha^2 &= \sin \bar{\theta}(\varphi-\bar{\varphi})\,,
\end{align}
which is equivalent to work in the basis
\be
(\bds e_+, \bds e_2 )= (\bpt_\theta , \sin^{-1} \bar \theta \bpt_\varphi)\,.
\ee 
The metric in this patch in is then given by
\be 
\bdd \Omega^2 = \bdd \theta^2 + \sin^2 \bar{\theta} \bdd \varphi^2 =
\delta_{ab} \bdd \alpha^i \bdd \alpha^j\,.
\ee
The deflection relations Eq.~\eqref{eqcosmo:thetae} and Eq.~\eqref{eqcosmo:phie} become in this system
\begin{align} \label{eqcosmo:alphae}
    \bds{\alpha}_\me = \bds{\alpha}_\mo + \bds{\nabla} \phiL(\bds{\alpha}_\mo)\,,
\end{align}
where $\bds \nabla$ is here the gradient on a flat $2D$-space. Note that in the expression for $\varphi_\me$, we can replace $\sin \theta_\mo$ by $\sin \bar \theta$ as the lensing potential is a first order quantity. The relation Eq.~\eqref{eqcosmo:alphae} is simply a map from the observed plane to the emission plane, where we recall that we consider a small neighbourhood around a point on the sphere. We would like now to study how neighbouring points are deflected. To do so, we consider an arbitrary position $\bds \alpha_\mo$ close to $\bds \alpha=0$, corresponding to $(\bar \theta, \bar \varphi)$. We can expand the relation Eq.~\eqref{eqcosmo:alphae} at first order in $\bds \alpha_\mo$ to get 
\begin{align} \label{eqcosmo:alphae2}
\bds \alpha_\me &= \bds{\alpha}_\mo + \bds \nabla \phi (0) + 
\bds \alpha_\mo \bds \nabla \bds \nabla \phi(0)\,.
\end{align}
We can without loss of generality define 
\be 
\tilde{\bds \alpha}_\me =  \bds{\alpha}_\me - \bds{\alpha}_\me(0) = \alpha_\me-\bds \nabla \phi (0) \,,
\ee 
which is a mere translation of the coordinates in the emission plane. Finally, the map given by Eq.~\eqref{eqcosmo:alphae2} reads\footnote{We drop the tilde on the emission plane for clarity.}
\boxemph{ \label{eqcosmo:lensingmap}
\bds{\alpha}_\me &= \bds{\alpha}_\mo + (\bds{\alpha_\mo}\cdot \bds{\nabla}) \bds{\nabla}\phiL \equiv \bds{A} \bds{\alpha}_\mo \,, \\
\label{eqcosmo:lensingmapcoord}
\alpha_\me^a &= (\tensor{\delta}{^a_b} + \nabla^a \nabla_b \phiL) \alpha_\mo^b \equiv \tensor{A}{^a_b} \alpha_\mo^b\,.
}
Where we defined the \emph{Jacobi map} given by the matrix $\bds{A}$. This relationship is fundamental when one studies lensing of light across the Universe. Note that the important point here is that the deflection angle given by $ + \bds{\nabla} \phiL(\bds{\alpha}_\mo)$ does depend on the position $\alpha_\mo$. If this was not the case, the cosmological lensing would be a mere translation and would not be very interesting, as it would not lead to any change in the shape of the images. We now describe quantitatively the change of shape of images undergoing lensing.
 
\MYrem{Dependence on the reference angle}{
We defined reference angles $(\bar \theta, \bar \varphi)$ around which we defined a patch of flat coordinates and expanded the Lens map. The final result involving the Jacobi map does not depend on this choice and is very general: The matrix $\bds A$ correctly describes lensing, and in particular, how small bundles of light are deformed when they travel across the Universe.
}

\MYrem{CMB lensing}{
The lensing formalism finds a great application to the CMB. Indeed, the CMB is itself lensed. In this work, we will only analyse lensing of galaxies, but we turn the interested reader to \cite{Cayuso:2022,Lewis:2009} for nice references about the topic.
}
\subsection{Jacobi Formalism \label{seccosmo:jacobi}}
\begin{figure}[h!t]
	\begin{center}
	\includegraphics[width =0.55 \textwidth]{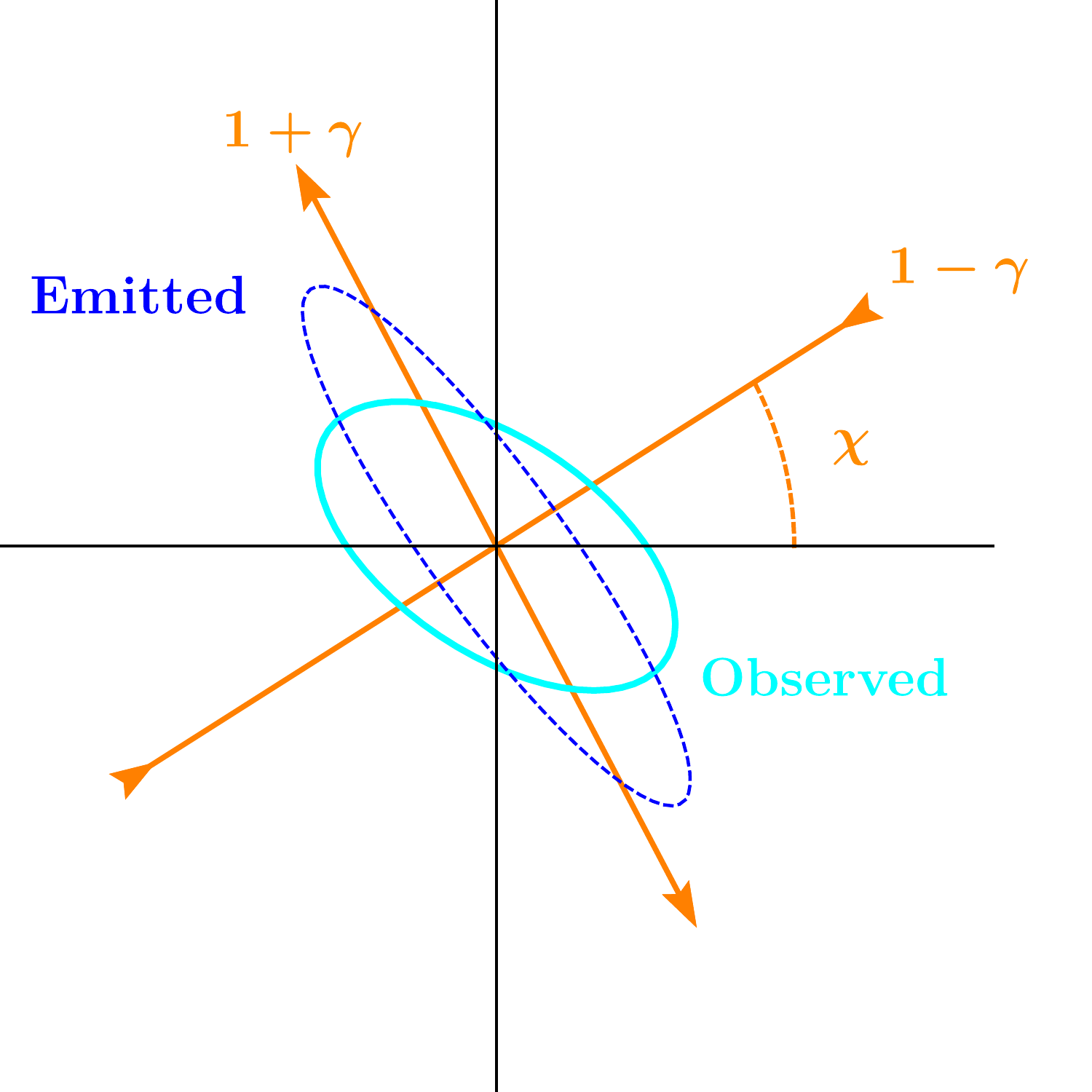} 
	\end{center}
	\caption[Effect of the Jacobi map]{\label{figcosmo:shear}
	Effect of the Jacobi map: The observed shape is shown in blue. The principal axes are rotated by angle $\chi$ with respect to the original basis. The first principal direction is shrunk while the second is stretched. This gives the shape of the image at the emission. Note that there is also a recalling given by $1-\kappa$ that we do not show here.}
\end{figure}

In this Section, we would like to understand from a qualitative point of view the action of the lensing map given by Eq.~\eqref{eqcosmo:lensingmapcoord}. This transformation tells us the transformation of a bundle of light. Indeed, if one observes say a galaxy around a small patch of the sky, one can measure several points $\bds{\alpha}_\mo$ and compute the shape of the emitted galaxy by computing the set of points corresponding to $\bds{\alpha}_\me$. The most generic expression is
\boxemph{ \label{eqcosmo:Lensing_map}
\bds{A} &= \begin{pmatrix}
1-\kappa - \gamma_1 & - \gamma_2  \\
-\gamma_2 &   1-\kappa + \gamma_1 
\end{pmatrix}\quad  \mathrm{with} \\
\kappa &= - \frac 12 \Delta \phi\,, \\
\label{eqcosmo:gamma1expr}
\gamma_1 &= - \frac 12 (\nabla_1 \nabla_1 - \nabla_2 \nabla_2) \phi\,, \\
\label{eqcosmo:gamma2expr}
\gamma_2 &= - \nabla_1 \nabla_2 \phi\,.
}
Note that as the lensing potential $\phi$ is a scalar, the order of the covariant derivatives does not matter and as the metric in the $(\bds e_1, \bds e_2)$ system is flat, neither does the position of the indices. This justifies also that the matrix $\bds A$ was taken symmetric. This decomposition of $\bds{A}$ in terms of $\kappa$, $\gamma_1$ and $\gamma_2$ is typical in lensing and is called the Jacobi formalism. We give now an interpretation to these terms. Defining $\chi$ and $\gamma>0$ through
\begin{align}
    \gamma_1 &= \gamma \cos 2 \chi\,, \\
    \gamma_2 &= \gamma \sin 2 \chi\,,
\end{align}
it is direct to show that, at first order in $\kappa$ and $\gamma$, the transformation map becomes
\boxemph{
\label{eqcosmo:Adecompos}
\bds{A}=
\begin{pmatrix}
1-\kappa  & 0  \\
0 &   1-\kappa 
\end{pmatrix}
\begin{pmatrix}
\cos \chi & -\sin \chi  \\
\sin \chi &  \cos \chi
\end{pmatrix}
\begin{pmatrix}
1-\gamma & 0 \\
0 &  1+\gamma
\end{pmatrix}
\begin{pmatrix}
\cos \chi & \sin \chi  \\
-\sin \chi &  \cos \chi
\end{pmatrix}\,.
}
The decomposition of $\bds{A}$ given by Eq.\eqref{eqcosmo:Adecompos} has a vivid interpretation.It goes as follows (recall the definition $\bds{\alpha}_\me = \bds{A} \bds{\alpha}_\mo$).
\begin{enumerate}
    \item The first matrix rotates the observed shape by an angle $\chi$ clockwise (assuming $\chi>0$).
    \item The second matrix stretches the first axis by a factor $1-\gamma$ and expands the second axis by a factor $1+\gamma$ (recall $\gamma>0$).
    \item The third matrix rotates back the shape by an angle $\chi$ counterclockwise.
    \item The last matrix is a global dilatation of the shape by a factor $1-\kappa$.
\end{enumerate}
In other words, there is a shear in direction $(\bds{e}_-, \bds{e}_+)$ called \emph{principal axes}, which is rotated by an angle $\chi$ \emph{counterclockwise} with respect to the basis $(\bds{e}_1, \bds{e}_2)$. The direction $\bds{e}_-$ is shrank while the direction $\bds{e}_+$ is stretched. Finally, there is a global rescaling by a factor $1-\kappa$. For these reasons, $\gamma$ is called the \emph{shear} and $\kappa$ the \emph{convergence}. The effect of these various transformations are shown in Fig.~\ref{figcosmo:shear}.

\MYrem{Jacobi map}{\label{remcosmo:jacobi_map}
Note that here we worked at first order in $\kappa$ and $\gamma$. A more general definition is also possible, when one considers also a net rotation of the final image at the end, corresponding to the antisymmetric part of the Jacobi map. This quantity vanishes at first order in perturbation theory and we do not mention it here. Moreover, one can be interested to map observed \emph{angles} to physical \emph{distances} at emission. This is also taken into account by a global prefactor in front, which is related to the angular distance. We will present these details in Section~\ref{seclensing:jacobi} as they will be useful when we discuss the Schwarzschild setup, but they are irrelevant in a Cosmological context.
}

\subsection{Spherical Harmonics \label{seccosmo:SH}}
We defined the various lensing parameters. The next step is study their statistics. Those parameters are fields of various nature on the sphere $S^2$. Hence, we make here a small mathematical interlude where we present briefly the Spherical Harmonics and the Spin Weighted Spherical Harmonics.

The usual Spherical Harmonics are scalar functions on the sphere $\YLM{\ell}{m}(\theta, \varphi)$ satisfying the Laplace's equation
\be 
\Delta_\Omega \YLM{\ell}{m} = - \ell (\ell+1) \YLM{\ell}{m}\,,
\ee 
where $\Delta_\Omega$ is the angular Laplacian. They are defined for $\ell\in \mathbb N$ and $\vert m \vert \leq \ell$. They also form a complete set of orthonormal functions on the sphere. We also choose the convention such that 
\be \label{eqcosmo:conjYLM}
\YLMStar{\ell}{m} = (-1)^m \YLM{\ell}{-m}\,.
\ee
Explicit expressions for $\ell=1,2$ are given in Appendix.~\ref{sec:appYLM}.

We also need to study tensors on the sphere. This is done with the Spin Weighted Spherical Harmonics. Much more information can be found for example in Refs.~\cite{durrer_2021,seibert_2018}. The second is a very nice book written by Seibert covering much more that what is presented here.

We want to define the Spin Weighted Spherical Harmonics, $ _sY_{\ell,m}$, where $s$ represents the Weight. The $s=0$ case corresponds to the usual Spherical Harmonics functions $_0\YLM{\ell}{m}\equiv \YLM{\ell}{m}$ presented above. For a generic integer $s$, we define first the spin raising and spin lowering operations, $\sd $ and $\sd^\star$, 
on a  function $\YLMS{\ell}{m}{s}$ with Spin Weight $s$ as
\boxemph{
\label{eqcosmo:sd}
\sd \, (\, \YLMS{\ell}{m}{s}\, )  &=
\left( s \cot \theta - \partial_\theta - \frac{\ii}{\sin \theta} \partial_\varphi \right) (\,\YLMS{\ell}{m}{s} \, ) \,, \\
\label{eqcosmo:sds}
\sds \, (\, \YLMS{\ell}{m}{s} \, )  &=
\left( -s \cot \theta - \partial_\theta + \frac{\ii}{\sin \theta} \partial_\varphi \right) (\, \YLMS{\ell}{m}{s} \, ) \,.
}
Those derivatives are also called \emph{slashed derivatives}. The Spin Weighted Spherical Harmonics for generic $s\in \mathbb{Z}$ are obtained recursively with the spin raising and spin lowering operators given by Eq.~\eqref{eqcosmo:sd} and Eq.~\eqref{eqcosmo:sds} via
\boxemph{
\sd( \YLMS{\ell}{m}{s}  ) &= \sqrt{(\ell-s)(\ell+s+1)} \ 
{ \YLMS{\ell}{m}{s+1} }\,,  \\
\sds( \YLMS{\ell}{m}{s}  ) &=- \sqrt{(\ell+s)(\ell-s+1)}\ 
\YLMS{\ell}{m}{s-1} \,.
}
This justifies the name of \emph{Spin raising/lowering} operators for the slashed derivatives. In particular, for $s=\pm 2$, these definitions yield
\begin{align} 
\label{eqcosmo:ytwoplus}
\sd^2( {Y}{_{\ell,m}}  )  &= \rll \; \;  {_{2}\tensor{Y}{_{\ell,m}}}\,,  \\
\label{eqcosmo:ytwominus}
\sds^2( {Y}{_{\ell,m}}  ) &= \rll\;\;  {_{-2}\tensor{Y}{_{\ell,m}}}\,, \\
\rll &= \sqrt{\frac{(\ell+2)!}{(\ell-2)!}}\,.
\end{align}
The Spin Weighted Spherical Harmonics satisfy the orthogonality condition 
\be \label{eqcosmo:yortho}
\int\; \YLMS{\ell_1}{m_1}{s} \, \YLMSStar{\ell_2}{m_2}{s}  \, \mathrm{d} \Omega^2 = \delta_{\ell_1,\ell_2} \delta_{m_1,m_2}\,,
\ee
and the conjugation relation
\be
\label{eqcosmo:ylmconj}
\YLMS{\ell}{-m}{-s} = (-1)^{s+m}\YLMSStar{s}{\ell}{m}\,.
\ee 
Again, explicit expressions of the Spin Weighted Spherical Harmonics for $s=0,1,2$ and $\ell=0,1$ are given in Appendix.~\ref{sec:appYLM}.

\subsection{Statistics\label{seccosmo:stats}}
In this Section, we want to introduce a very important tool in Cosmology, and in Physics in general: The power spectrum. We recall the definition of the lensing potential
\be \label{eqcosmo:lensingpotential}
\phi(z, \bds{n}) = -2 \int_{\eta_\me}^{\eta_\mo}\; \frac{\eta-\eta_\me}{\left(\eta_\mo-\eta_\me \right)(\eta_\mo-\eta)} \Psi_\mW(\eta,\eta_\mo- \eta , \bds{n})\, \dd \eta\,,
\ee
where we expressed it as a function of the observed redshift of a given source $z$, the explicit relation between $z$ and $\eta$ being given by Eq.~\eqref{eqcosmo:rint} and Eq.~\eqref{eqcosmo:etaz}. Hence it is understood that $\eta_\mo$ and $\eta$ stand for $\eta_\mo(z)$ and $\eta(z)$ respectively, and that
\be 
\dd \eta = - \frac{\dd z}{a_0 H(z)}\,.
\ee 
Moreover, it is understood that $\bds{n}$ is a given direction in the sky.
The lensing potential depends on the Weyl potential $\Psi_\mW$, which depends, through the Einstein's Equation, on the fluctuations of matter distribution. Without going into much details, the distribution of homogeneities is a random distribution. Hence, one cannot predict, say, the exact value of $\phi(z, \bds{n})$ for a given value of $z$ and $\bds{n}$. However, we can study statistical properties of this random field.

To do so, we introduce the decomposition of the lensing potential in terms of the spherical harmonics
\be  \label{eqcosmo:phidecompos}
\phi(z, \bds{n}) =\sum_{\ell,m} \almphi{\ell}{m}(z) \YLM{\ell}{m}(\bds{n})\,.
\ee 
The complex coefficients $\almphi{\ell}{m}$ are equivalent to the Fourier coefficients of a \emph{usual} function $f(\bds x)$. Using the conjugation property of the Spherical Harmonics Eq.~\eqref{eqcosmo:conjYLM}, and the fact that the lensing potential $\phi$ is real, we obtain the same kind of condition for these coefficients
\be  \label{eqcosmo:alpphistar}
\almphistar{\ell}{m}(z) = (-1)^m \almphi{\ell}{-m}(z)\,.
\ee 
The main idea is the following. We assume that these coefficients have zero mean and follow a Gaussian distribution, i.e.
\begin{align} 
    \langle\almphi{\ell}{m}(z) \rangle  &= 0\,, \\
    \label{eqcosmo:Clphi}
    \langle \almphi{\ell_1}{m_1}(z_1) \almphistar{\ell_2}{m_2}(z_2)\rangle  &=
    \Clphi{\ell_1}(z_1,z_2) \delta_{\ell_1,\ell_2} \delta_{m_1,m_2}\,.
\end{align}
Here, the expectation value is to be understood as an average over all the \emph{random realisation of the Universe}. Of course, this is not possible in principle: One would need to be a God-like creature to create several realisations of the Universe and compute these expectation values. We will set this comment aside for now, and later we will explain how in practise one can use these relations. In the second condition, the numbers $\Clphi{\ell}$ are the power spectrum of the lensing potential and are fundamental in Cosmology: Their values can be predicted from various cosmological models and observations allow us to confront these different models. This second condition enforces that the random field $\phi$ is statistically isotropic. This is also seen computing the correlation function given by
\begin{align}
\zeta^\phi(z_1,\bds{n}_1 ; z_2, \bds{n}_2) &\equiv 
\langle \phi(z_1, \bds{n}_1)  \phi(z_2, \bds{n}_2) \rangle \\
&= \sum_{\ell_1, m_1} \sum_{\ell_1, m_2} \langle \almphi{\ell_1}{m_1} \almphi{\ell_2}{m_2} \rangle  \YLM{\ell_1}{m_1}(\bds{n}_1) \YLM{\ell_2}{m_2}(\bds{n}_2)\\  &= 
\sum_{\ell_1, m_1} \sum_{\ell_1, m_2} \langle \almphi{\ell_1}{m_1} \almphistar{\ell_2}{-m_2} \rangle  (-1)^{m_2} \YLM{\ell_1}{m_1}(\bds{n}_1) \YLM{\ell_2}{m_2}(\bds{n}_2)\\ 
&=
\sum_{\ell_1, m_1} \sum_{\ell_1, m_2} \Clphi{\ell_1}(z_1,z_2) \delta_{\ell_1, \ell_2}  \delta_{m_1,m_2} \YLM{\ell_1}{m_1}(\bds{n}_1) \YLMStar{\ell_2}{-m_2}(\bds{n}_2) \\
&= \sum_{\ell} \Clphi{\ell}(z_1,z_2) \left(\sum_m \YLM{\ell}{m}(\bds{n}_1) \YLMStar{\ell}{-m}(\bds{n}_2) \right) \\
&= \sum_{\ell} \frac{2\ell+1}{4\pi} \Clphi{\ell}(z_1,z_2) P_\ell(\mu)\,,
\end{align}
where $P_\ell$ are the Legendre polynomials. We used the conjugation properties of the Spherical Harmonics and the coefficients $\almphi{\ell}{m}$ and the addition property of the Spherical Harmonics, see Appendix~\ref{sec:appADD}. Moreover, we defined
\be 
\mu \equiv \bds{n}_1 \cdot \bds{n}_2 =  \cos \varphi\,,
\ee
where $\varphi$ is the angle between $\bds n_1$ and $\bds n_2$. In the end, we have that the correlation function of the lensing potential is given by 
\boxemph{ \label{eqintro:corr_fun_full_sky}
\zeta^\phi(z_1,z_2 ; \mu) =
\sum_{\ell} \frac{2\ell+1}{4\pi} \Clphi{\ell}(z_1,z_2) P_\ell(\mu)\,.
}
This relation shows that the correlation function only depends on the angle between the two direction $\bds{n}_1$ and $\bds{n}_2$, which is the definition of the statistical isotropy. This justifies, afterwards, the assumption that the coefficients $\almphi{\ell}{m}$ follow a Gaussian distribution.

Finally, we want to decompose the shear as a tensor field on the sphere. Recall that the relationship between the components of the shear and the lensing potential are given by Eq.~\eqref{eqcosmo:gamma1expr} and Eq.~\eqref{eqcosmo:gamma2expr}. We define the shear in the $(+,-)$ basis as
\be 
\gamma^{\pm} = \gamma_1 \pm \ii \gamma_2
\ee 
From these relations, it is straightforward to show that
\begin{align}
    \gamma^+ &= - \frac 12 {\sd}^2 \phi\,, \\
    \gamma^- &= -\frac 12 {\sds}^2 \phi\,.
\end{align}
More details can be found for example in \cite{Francfort:2022,durrer_2021}. Using the decomposition of the lensing potential in Spherical Harmonics Eq.~\eqref{eqcosmo:phidecompos} and the properties of the slashed derivatives Eq.~\eqref{eqcosmo:ytwoplus} and Eq.~\eqref{eqcosmo:ytwominus}, we can show that the decomposition of the components of the shear is
\boxemph{
\label{eq:cosmogammap}
\gamma^+(z,\boldsymbol n) &= -\frac 12 \sum_{\ell=2,m} \phi_{\ell,m}(z) \rll  \;  \YLMS{\ell}{m}{+2}(\boldsymbol n)\,, \\
\label{eq:cosmogammam}
\gamma^-(z,\boldsymbol n) &= -\frac 12 \sum_{\ell=2,m} \phi_{\ell,m}(z) \rll\;  \YLMS{\ell}{m}{-2}(\boldsymbol n)\,.
}
Note that the terms corresponding to $\ell=0,1$ vanish because of the definitions of the slashed derivatives Eq.~\eqref{eqcosmo:sd} and Eq.~\eqref{eqcosmo:sds}.

\MYrem{Monopole and dipole}{\label{remcosmo:monopole}
In the context of Spherical Harmonics expansion, terms corresponding to $\ell=0$ and $\ell=1$ are called respectively \emph{monopole} and \emph{dipole} terms. Generally, these terms are not used or neglected. We discuss here the reason.

A monopole term, as $Y_{0,0}(\theta, \varphi)$ is constant, corresponds to
\be
\phi(z,\boldsymbol n) = \phi_0(z)\,,
\ee
using the deflection relation Eq~\eqref{eqcosmo:alphae}, this implies
\be
\bds{\alpha}_\me = \bds{\alpha}_\mo\,.
\ee
In other words, a monopole term does not lead to any lensing and is hence not observable.

To understand why a dipole term is not observable, let us consider an observer moving at constant speed $v$ in direction $\tilde{z}$ in a pure \ac{FRLW} spacetime. Her $4$-velocity is (in cosmic time)
\be 
\tilde{\bds u} = \frac{1}{\sqrt{1-a^2 v^2}} (\bpt_t + v \bpt_{\tilde{z}})\,,
\ee and the metric is
\be 
\bds g = - \bdd t^2 + a^2 \bdd \tilde{\bds{x}}^2\,.
\ee 
Defining the coordinates
\begin{align}
    t &= \tilde t \,, \\
    x &= \tilde x \,, \\
    y &= \tilde y\,, \\
    z &= \tilde{z}-vt\,,
\end{align}
the $4$-velocity becomes 
\be 
\bds u =  \frac{1}{\sqrt{1-a^2 v^2}} \bpt_t\,,
\ee
i.e. the observer is now at rest.
The metric in the new coordinate system reads
\begin{align}
\bds g &=
-(1-a^2 v^2) \bdd t^2 + a^2 \bdd \bds{x}^2+ 2a^2 v \bdd t \bdd z \\
&= a^2 \big(
-(1-a^2 v^2) \bdd \eta^2 +  \bdd \bds{x}^2+ 2a v \bdd t \bdd z
\big) \,,
\end{align}
where the last equality is expressed in conformal time. Using the formalism presented in Section~\ref{seccosmo:PT}, we get
\begin{align} 
\psi &= - \frac 12 v^2 \,, \\
\phi &= 0 \,, \\
\xi &= a v z\,, \\
\zeta &= 0\,,
\end{align}
and the Weyl potential is given by
\be
\Psi_\mW  = - \frac 12 v^2 + a^\prime v z = - \frac 12 v^2 + a^\prime v r \cos \theta\,,
\ee
where we converted $z$ in Spherical Coordinates. The first term leads to a monopole term in the lensing potential and does not contribute as we discussed above. The contribution of the second term to the lensing potential is of the form
\be 
\phi(z,\bds n) = \phi_1(z) \cos \theta\,,
\ee 
which corresponds to a term with $\ell=1$ $m=0$ (terms with $m=\pm1$ can be obtained the same way with a velocity in different directions). The bottom line is the following:Even in a non-perturbed Universe, the motion of the observer with respect to the background Universe leads to a \emph{fake} dipole term with $\ell=1$. Hence, there is a degeneracy between an intrinsic dipole coming from the cosmological perturbations and from the velocity of the observer. Moreover, the shear (the quantity we want to study) is insensitive to any dipole term. This can be seen directly using the explicit formula Eq.~\eqref{eqcosmo:gamma1expr} and Eq.~\eqref{eqcosmo:gamma2expr} with $\phi \propto \YLM{1}{m}$ or noting that the sums in Eq.~\eqref{eq:cosmogammap} and Eq.~\eqref{eq:cosmogammam} start at $\ell=2$.
}

\stopchap
\part{Galaxy Number Counts in Conformal Frames}\label{Part:Frames}
\begin{figure}[h!]
	\centering
	\includegraphics[width = 0.9\textwidth]{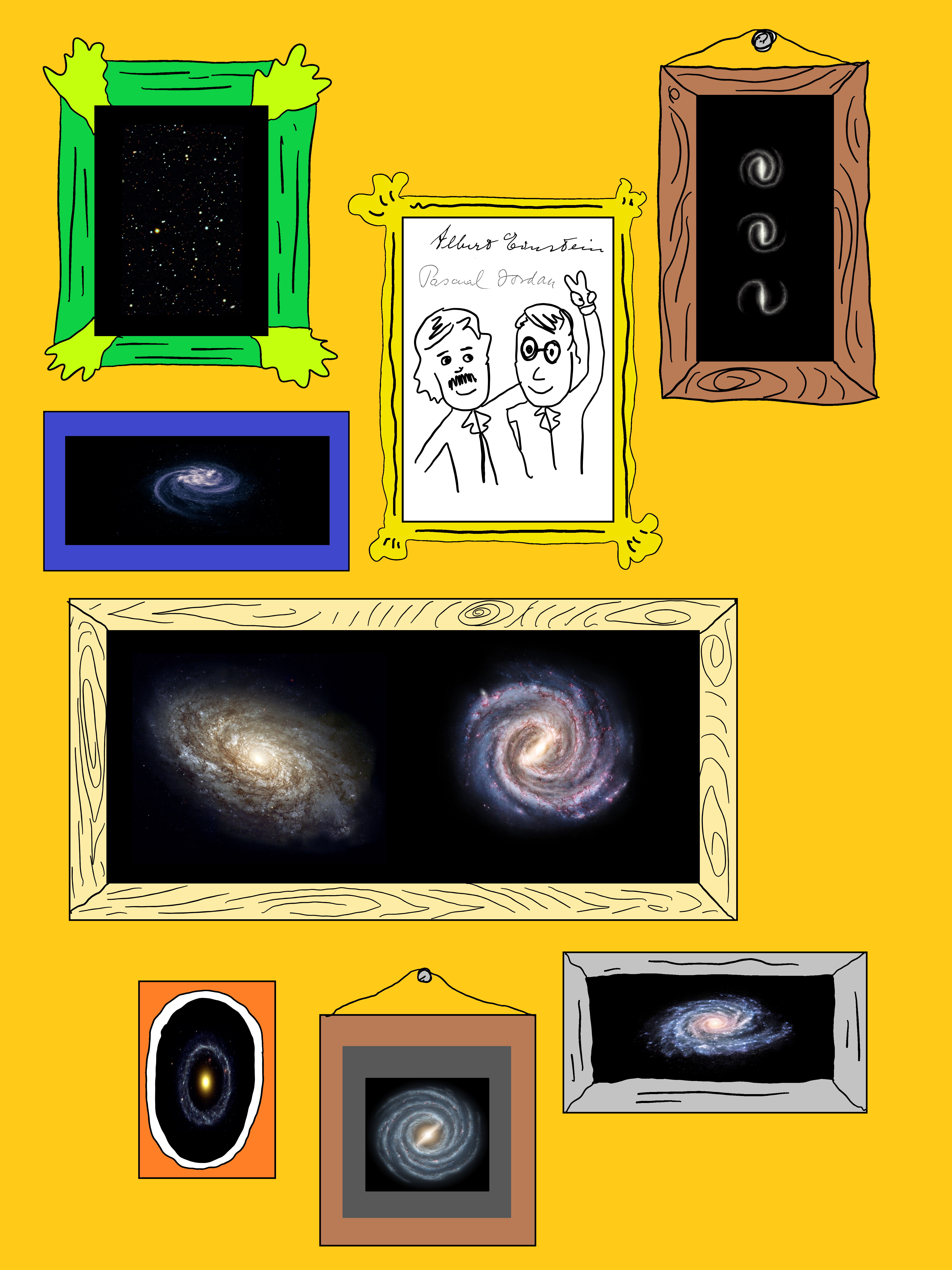} 
	\caption*{\textbf{The Einstein and Jordan Frames}, Juan Manuel Garc\'ia Arcos}
\end{figure}

\chapter{General Introduction\label{chapframes:intro}}
In this Part, I present the article Basundhara, Ruth and I wrote~\cite{francfort_2019}. This was the first article of my PhD, and I really enjoyed working on the topic with Basundhara. I remember the day I managed to prove the main relations, I was in a Unicorn-themed caf\'e in Paris, and I messaged Basundhara straight away!

One big puzzle in modern Cosmology is Dark Energy. Large scale observations suggest that the Universe has been recently accelerating, since roughly $5$ billions years ago. The problem is that the measured value of the cosmological constant, or equivalently, of the vacuum energy, is approximately $10^{-120}$ times smaller than the naive value expected from Quantum Field Theory. Tremendous attempts have been made to reconcile these two values. Modified Theories of Gravity form a broad topic in Cosmology. A famous example is Horndeski's theory, see for example \cite{Jana:2020,Dalang:2019,Clifton:2012,Nojiri:2017,Boehmer:2021} for generic reviews.

The simplest models of Modified Gravity are the so-called \ac{ST} theories, which we will consider here. In such theories, a scalar field is coupled to gravity (hence the name), see for example \cite{Fujii:2003,Sotiriou:2015lxa} for a nice introduction. Horndeski's theory mentioned above is, under some assumptions, the most general \ac{ST} theory.

\ac{ST} theories can be studied in different frames, the two most important ones being the Einstein and the Jordan frames. In the Jordan frame, matter is minimally coupled to gravity in the action, but there is an extra coupling term between the curvature and the scalar field, which introduces new terms in the Einstein's Equations. In the Einstein frame, there is a non-minimal coupling between matter and the metric. The matter particles do not follow geodesics in this frame. However, the scalar field is not directly coupled to the curvature, and the Einstein's Equations take their usual form.

People have long debated about the equivalence between these two frames, and if physical observables should be frame-independent, see for example \cite{Chiba:2013, Catena:2006bd,Faraoni:1998, Faraoni:1999, Rondeau:2017,Capozziello:1996xg,Dicke:1962,Deruelle:2010ht,Deruelle:2010ht,Rondeau:2017,Morris:2014,Karamitsos:2017elm} for different discussions. However, the answer is not consensual yet. Some authors claim that the equivalence is not established \cite{Banerjee:2016lco} and that observables depend on the frame \cite{Hyun:2018, Hyun:2019}. It has also been proposed that the equivalence breaks down at the quantum level \cite{Faraoni:2007}.

The objective of our work is to consider the \ac{GNC}, an observable introduced by Bonvin in \cite{Bonvin:2011bg} which quantifies the angular fluctuations of the number of observed galaxy. This quantity has been studied extensively recently, see \cite{DiDio:2013, Tansella:2017,Ghosh:2020,Durrer:2016}. As the \ac{GNC} is indeed an observable that astronomers can measure, we expect it to be frame-independent, which we show explicitly.

The Part is structured as follows. Chapter~\ref{chapframes:intro} is this Introduction. In Chapter~\ref{chapframes:CF} we discuss general concepts about Conformal Frames. In particular, we give a vivid physical interpretation to conformally related frames.  We also present the main ingredients to work with \ac{ST} theories, and discuss the specific example of the \ac{FRLW} spacetime. In Chapter~\ref{chapframes:NC}, we define the \ac{GNC} and show explicitly that this observable is frame-independent. We conclude in Chapter~\ref{chapframes:conc}.

\chapter{Conformal Frames\label{chapframes:CF}}
In this Chapter, we present the theory of conformal transformations and conformal frames. In the first Section, we present the definition of conformal transformations and discuss their physical interpretation. Conformal transformations were introduced by Brans and Dicke, see for example~\cite{Brans:1961, Dicke:1962,Dicke:1962, Lu:2019rck}. Physics of conformal frames has since then been broadly studied, for example by Catena~\cite{Catena:2006bd} or by Chiba~\cite{Chiba:2013}.

In the second Section, we introduce a coupling between gravity and a scalar field. We discuss the conformal transformations in this context and present our first \emph{dictionary}: A generic method to derive equivalent theories from one frame to another. The study of a scalar field coupled to conformally related metrics has been studied extensively, and has plenty of applications, e.g. multifield inflation theories, see \cite{Karamitsos:2017elm} for a good example.

In the third Section, we discuss the specific Cosmological example of the \ac{FRLW} metric in two conformally related metrics. Again, this topic has been studied in great details in the past, for example by Faraoni and Rondeau \cite{Faraoni:1998,Faraoni:1999,Faraoni:2007,Rondeau:2017}.

\startchap

\section{Definition}
\subsection{A Tale of Two Frames}
We present in this Section the concept of conformally related frames, old friends we already encountered in Theorem~\ref{thmcosmo:nullgeo}. We recall here the definition.
\MYdef{Conformally related frames}{
Let $\bds g$ and $\tilde{\bds g}$ two metric on the same manifold (with the same signature) with coordinates $ x^\mu$. They are said to be conformally related if there is a positive and smooth function $\Omega( x ) $ such that
\be  \label{eqframes:definitionconf}
\bds{g}( x) = \Omega^2( x) \tilde{\bds g}( x)\,.
\ee 
This relation is also called a conformal transformation, and the function $\Omega^2( x)$ is the conformal factor.
}
In other words, the two metrics differ only by a coordinate-dependent multiplication. We also stress the fact that a conformal transformation is not a coordinate transformation. The manifold on which both metrics act, together with its set of coordinates $x^\mu$, is not modified. Hence, a conformal transformation only changes the value of the metric tensor. We will call a manifold together with a specific metric \emph{a frame}. Hence, as we consider two metrics here, we have two different frames in our hands.
\MYdef{Jordan and Einstein frames}{The frame associated with $\tilde{\bds g}$ is the \emph{Jordan frame} while the frame associated with $\bds{g}$ is the \emph{Einstein frame}.
}
Under the transformation Eq.~\eqref{eqframes:definitionconf}, the various tensors of General Relativity in both frames can be related . The exact expressions for these transformations can be found e.g. in~\cite{Faraoni:1998, Fabiani:2017,Domenech:2016}. To only give one example, the Ricci scalars are related as (in $4$ dimensions)
\boxemph{
\label{eqframes:ricci1}
R&= \frac{1}{\Omega^2}\left( \tilde{R} - 6\frac{\tilde{\Box} \Omega }{\Omega}\right)\,, \\
\label{eqframes:ricci2}
\tilde{R}&= \Omega^2 \left( R + 6\frac{\Box \Omega}{\Omega} - 12 \frac{g^{\mu \nu}\nabla_\mu \Omega \nabla_\nu \Omega)}{\Omega^2}\right)\,.
}
Note that in these relations, the covariant derivatives (and the d'Alembertian) are not the same: The connection $\bds \nabla$ (resp. $\tilde{ \bds \nabla}$) is associated with $\bds g$ (resp. $\tilde {\bds g}$). The first term of these relations can be interpreted as follows: The relationship between the frames Eq.~\eqref{eqframes:definitionconf} means that dimensionless lengths are bigger in the Einstein frame (assuming $\Omega>1$). The curvature being the inverse of a length squared, it is smaller in the Einstein frame, hence the presence of $\Omega^2$ in the denominator.

\subsection{Physical Interpretation}
\label{secframes:phys_int}
The definition of a conformal transformation Eq.~\eqref{eqframes:definitionconf} together with the various transformation (e.g. Eq.~\eqref{eqframes:ricci1} and Eq.~\eqref{eqframes:ricci2}) do not provide much more physical insight. In this Section, we give a nice physical interpretation to better understand conformal transformations. This interpretation will follow us through this journey and we want to make it clear from the beginning. The discussion is inspired from \cite{Chiba:2013} and \cite{Dicke:1962}.

The relationship between the two frames Eq.~\eqref{eqframes:definitionconf} is nothing else as a proxy describing how lengths and times are measured in both frames. More precisely, let us assume that we are interested to measure a given length, say your own height, in both frames. To do so, we need to define a \emph{ruler}: An object (or a physical process) with which the length of another object will be compared. When this is done, the length of any given object can be measured in units of the ruler. 

In the Jordan frame, one specific ruler is used to define $\SI{1}{\meterJ}$, a \emph{Jordan-meter}. With this ruler, any given length can be expressed in Jordan-meter as
\be \label{eqframes:LJordan}
L = \SI[parse-numbers = false]{\tilde \ell}{\meterJ}\,,
\ee 
meaning that this objects measures $\ell$ units of length in this frame. Note that here, $L$ is the physical length of the object, as an invariant and abstract quantity, while $\ell$ is a dimensionless number measuring this length in a given frame with given units, see Section~\ref{chapintro:units}. The same reasoning can be done in the Einstein frame, defining $\SI{1}{\meterE}$, a \emph{Einstein-meter} and measuring the length of the same object as
\be  \label{eqframes:LEinstein}
L =  \SI[parse-numbers = false]{\ell}{\meterE}\,.
\ee 

We can now give a nice interpretation of the definition of conformal frames Eq.\eqref{eqframes:definitionconf}. It simply relates the two dimensionless numbers as (recall that $\bdd{s}^2={g}_{\mu \nu}\bdd{x}^\mu \bdd{x}^\nu$ measures a length \emph{squared}) 
\boxemph{  \label{eqframes:ell}
\ell(  x) = \Omega( x) \tilde{\ell}( x)\,.
} 
\MYrem{Relation between the rulers}{
The physical length of an object appearing in Eq.~\eqref{eqframes:LJordan} and Eq.~\eqref{eqframes:LEinstein} should be the same in both frames, which means that the units are related as
\be \label{eqframes:ratio}
\frac{\ell(  x)}{\tilde \ell(  x)} =
\frac{\SI[parse-numbers = false]{1}{ \meterJ}(  x)}{\SI[parse-numbers = false]{1}{\meterE}(  x)} = \Omega(x)\,.
\ee 
We can generalize the fundamental relation Eq.~\eqref{eqframes:ell} to other units. For example (setting $c=\hbar=1$), times follow the same relations, while the inverse relation holds for masses or energies.
This justifies the fact that the rulers (and the units) are also coordinate-dependent! To phrase it nicely: In the theory of conformal frames, you are allowed to change your ruler over time as well as when you travel! Note also that if the intrinsic value for $\SI[parse-numbers = false]{1 }{\meterE}$ increases, then $\ell$ has to decrease. In other words, my size in meters (roughly $1.7$) is much smaller than my size in centimeters (roughly $170$)! 
}
In this context, it can seem difficult to do Physics\footnote{But also very convenient: If, a couple of years ago, you would had added a little bit of mass to the reference kilogram in the \emph{Bureau international des poids et mesures}, you would have lost weight without any effort!} : How can I make precise experiments if every two minutes I change my ruler (and my clocks, scales, thermometers, and so on)? The answer is that one should define a \emph{reference length} by considering a given physical process (which can be of any kind: The wavelength of a given atomic transition, the distance between the Earth and the Sun, the mean size of an animal,...). In both frames, this reference process has dimensionless length $\ellRJ$ and $\ellRE$. When measuring another physical length, the key is two compare its dimensionless length $\tilde \ell$ and ${\ell}$ with the reference length of the considered frame, which is nothing else to say that in Physics, ratios are more important than absolute numbers, which is the essence of what we stated in Section~\ref{chapintro:units}!

\subsection{An Example: FRLW and Minkowski \label{secframes:FMin}}
We present here a toy-model of a conformal transformation: The \ac{FRLW} metric. This will be explained in more details in Section~\ref{secframes:FRLW}, but we find it insightful to provide a qualitative explanation first. 

Let us consider the two metrics\footnote{As there is only one scale factor, we omit the tilde for clarity.}
\begin{align}
    \bds{g} &= - \bdd \eta^2 + \bdd \bds{x}^2\,, \\
    \tilde{\bds{g}} &=  a(\eta)^2(-\bdd \eta^2 + \bdd \bds{x}^2)\,.
\end{align}
They are good friends of us: The Minkowski and the \ac{FRLW} metric, respectively. They are obviously related with the conformal factor $\Omega( x) = a(\eta)^{-1}$.

We will assume that the Jordan frame is our usual metric system, where $\SI{1}{\meterJ} = \SI{1}{\meter}$, while in the Einstein frame, people take as ruler the distance between the Milky Way and another given far away galaxy, let us call it the Rooibos Galaxy. They use then $\SI{1}{\meterE} = \SI[parse-numbers=false]{10^{23}}{\meter}$. In this frame, usual distances are really tiny, a trip between Geneva and St. Gallen back and forth is about $\SI[parse-numbers=false]{10^{-17}}{\meterE}$...

Using the property that physical distances are frame independent Eq.~\eqref{eqframes:ratio}, we get
\be 
\frac{\ell}{\tilde{\ell}} = \Omega=10^{-23}\,,
\ee
in other words, distances in the Einstein frames appear much smaller as the ruler is much bigger. Note however that this relation holds only at the present time: As in the Einstein frame, the ruler is a distance between the Milky Way and the Rooibos Galaxy, the physical value of $\SI[parse-numbers=false]{1}{\meterE}$ will change over time due to the expansion, and hence the dimensionless length ${\ell}$ of any measurement in this frame will decrease!

The question arising now is the following: How to make measurements in the Einstein frame? In the Jordan frame, it is \emph{business as usual}: The Universe is in expansion and everything goes as explained in Chapter~\ref{chap:cosmo}. In the Einstein frame, things are different: Astronomers measuring the distance between us and the Rooibos Galaxy always find $\SI{1}{\meterE}$, by definition, and conclude, for example, that the Hubble constant $H(t)$ vanishes at all times, which is indeed the case in the Minkowski metric. The key is to consider, as explained in Section~\ref{secframes:phys_int}, a given physical process, say the wavelength of a given atomic transition, which defines $\ellRE$. As the ruler in the Einstein frame keeps getting bigger, the dimensionless value of $\ellRE$ keeps decreasing! This will let the astronomers understand that the Universe is indeed in expansion.

In practice, the expansion of the Universe is inferred through redshift measurements. Assume again that an astronomer receives some light signal, from the Rooibos Galaxy, produced by an atomic transition. The astronomer will reproduce this transition in the lab, measure its wavelength, and compare it with the received wavelength. In this context, the lab wavelength takes the role of the reference length and there is in fine no need to use a ruler whatsoever, the received wavelength is bigger than the reference one anyway. 

One last important comment: Assume that this measurement has been performed, and that the received wavelength is indeed smaller than the wavelength produced in the laboratory. Does this necessarily mean that the Universe is expanding? The wavelength of a given atomic transition is given by $\lambda= C a_{\mathrm{B}} \alpha$, where $a_{\mathrm{B}}$ is the Bohr radius and $\alpha$ is the fine structure constant, and $C$ is an irrelevant constant number. If the Bohr radius decreases over time, or if the value fine structure constant gets smaller, the wavelength will intrinsically become smaller. In this case, the observer will measure a redshift, which won't be due to the expansion of the Universe! We will discuss this point in more details in Section~\ref{secframes:philo}.

\subsection{Disformal Transformations} \label{secframes:disformal}
In this Section, we briefly discuss disformal transformations, an more general version of conformal transformations. They were first introduced by Bekenstein~\cite{Bekenstein:1993}, and were later developed by several authors, see for example~\cite{Chiba:2020,Minamitsuji:2014,Fumagalli:2016,Domenech:2015} for various discussions and applications to Cosmology. We present here the main ideas and more details can be found in~\cite{Ghosh:2022}, an extension of the reference we present in this Part. Under a disformal transformation, both frames as related as
\boxemph{ \label{eqconf:disformal}
\bds g &= \Omega^2 \tilde{\bds{g}} + \Lambda \bds{\nabla} \phi \otimes \bds{\nabla}  \phi\,, \\
g_{\mu \nu} &= \Omega^2 \tilde{g}_{\mu \nu} + \Lambda \nabla_\mu \phi \nabla_\nu \phi\,.
}
Here, $\Omega$ and $\Lambda$ are smooth function of the scalar field $\phi$. A dependence on $(\bds{\nabla}\phi)^2$ would also be possible but we do not consider this case here. As in the case of conformal transformations, the metric $\tilde{\bds g}$ (resp. $\bds g$) is associated with the Jordan frame (resp. Einstein).

We assume that the metric in the Jordan frame is \ac{FRLW} (in cosmic time)
\be 
\tilde{\bds{g}} = - \bdd t^2 + a^2 \bdd \bds{x}^2\,.
\ee 
Using the transformation rule Eq.~\eqref{eqconf:disformal}, the metric in the Einstein frame is given by 
\be  \label{eqconf:geinstein}
\bds{g} = - (\Omega ^2- \Lambda \dot{\phi}^2) \bdd t^2 + \Omega ^2 a^2 \bdd \bds{x}^2\,,
\ee 
where we assumed a cosmological setup, i.e. the scalar field only depends on time. The main difference with conformal transformations is that times and lengths are not affected in a similar way, as can be seen directly from Eq.~\eqref{eqconf:geinstein}, as both coefficients are different. In this context, lengths $\ell$ and times $\tau$ are related as
\begin{align}
    \ell &= \Omega \tilde{\ell} \,, \\
    \tau &= \sqrt{\Omega^2 - \Lambda \dot \phi^2} \tilde{\tau}\,.
\end{align}
From these rules, other transformation rules can be derived. Let us detail the example for a mass. If we assume that the unit for action (e.g. $\hbar$ is kept constant in both frames, see Chapter~\ref{chapintro:units} for more details), a mass is given by
\be 
M = \frac{A T}{L}\,,
\ee
where $A$, $T$ and $L$ respectively represent an action, a time and a length. The transformation rule is given by 
\be 
m = \frac{\sqrt{\Omega^2 - \Lambda \dot \phi^2}}{\Omega} \tilde{m}\,.
\ee
More examples are provided in \cite{Ghosh:2022} (Section~$2.1$).

The bottom line is that, even in the case of disformal transformations, both frames are still equivalent. This should not be a surprise: If one is allowed to change lengths over space and time, why not changing time too? We will not go further into the details, as the logic is the same as for conformal transformations.

\subsection{A Philosophiconomical Reflexion\label{secframes:philo}}
\subsubsection{A varying reference length, the only explanation?}
We discussed the possibility of varying physical laws at the end of Section~\ref{secframes:FMin}. This topic has been discussed extensively in \cite{Uzan:2010,Olive:2009}, and experimental tests have been mentioned in~\cite{Berengut:2011,Garcia:2007,Bronnikov2006,Karshenboim:2003} and I am not claiming to state anything revolutionary. The goal of the present section is to warn the reader about other possibilities.

We stated that in Physics, one should consider dimensionless ratio, and avoid using absolute or dimensionful quantities. We also argued why the use of a reference length is important: A different reference length over time is a smoking gun for a change of the ruler over time. Note that we will here only discuss lengths (and rulers) and time variations, but the logic applies for any other physical quantity and spatial variations.

However, the physical interpretation of a varying reference length can be different: Maybe the ruler stays the same but the reference length itself varies over time! This would be a signature that the law of Physics themselves change over time. Assume that you take as reference length the wavelength of the first atomic transition of the hydrogen atom (the Lyman series with $n=2$). For the sake of the argument, let us also assume that you use some mechanical device to measure this wavelength, hence the gravity plays a role in the measurement. You make this measurement today and find $\lambda_{0} =\SI{121.57}{\nanometer}$, and when you redo it tomorrow, you obtain this time $\lambda_{0} =\SI{124.3}{\nanometer}$. Assuming your errors are well-controlled, what happened?

According to what we said above, we would have assumed that your ruler changed, maybe you modified the gravitational device: Did you change the length of one pendulum? Or the oscillating mass? We would like to investigate the other solutions to the problem, so we will assume that, as a diligent scientist, you did not change your setup between the different experiments.

The other assumption would be that the law of Physics are changing over time. For example, what if the energy of the photon is given by $E=\hbar \omega + \delta E(t)$, where $\delta E(t)$ is a  function of time, starting at $0$ and slowly increasing? Or what if the value of $\hbar$ is changing over time? Or even better, what if the dimensionless fine structure constant changed? All these effects would modify the wavelength of the transition. Another possibility is that the value of $\GN$ changes over time, which would modify the action of your mechanical device. In all these scenarios, the reference length would be modified, not because your ruler has changed but because the laws of Physics are different (see Section \ref{secframes:scalar} where we will see a simple example where this is due to the scalar field).

The question is now: In this context, how can we be sure that Physics is not changing over time? Checking every day the wavelength of all the atomic transition is not realistic, and this actually would not even help. Indeed, if Physics is changing you may not even notice it if the ruler you're using is changing at a specific rate, cancelling the effect of the varying laws. 

A very good is example is the \ac{FRLW} Universe: As we saw in Section~\ref{secframes:FMin}, we cannot distinguish between and expanding Universe and the Minkowski spacetime, where the rulers is varying and a scenario in which physical laws (e.g. the value of the fine structure constant) change over time. Both interpretations lead to the same observables and should be considered as equivalent (at least as long as we only consider observations at the background level).

The problem is clear: There is somehow a loophole, and it is very difficult (if not impossible) to prove that Physics does not vary with time. It is however reasonable to assume that this is the case: From a Human point of view, the Universe seems to be the same over time. I have never seen my cat getting suddenly bigger than me, and I can check every day when I bike that the laws of Mechanics and Gravity are the same (luckily for my safety!).

To conclude this discussion in a broader context, let me share maybe my personal experience. When I try to solve a Physics problem, I often find myself using tools and equations I know without questioning their use or relevance enough. The ideas presented in this Section should, according to my opinion, always accompany Physicists, especially theoreticians who may get lost in their ideas. What I discussed here can seem very basic but is actually very deep if one tries to think about it. I hope these words are not just a bottle in the sea but will resonate in some minds!

\subsubsection{Money money money}
I would like to give a short example taken from economics and daily life to show how the concepts presented in this section can be applied outside Physics. Before going into details, I will simplify things and only consider two countries without considering fees of any kind or other subtleties.

In economics, it is also meaningless to deal with \emph{dimensional} variables. For example, what is the intrinsic value of one Swiss Franc, or of one Euro? This question alone does not have any quantitative answer. However, the ratio between those two quantities is in general well-defined : $1\ \mathrm{EUR}/1\ \mathrm{CHF}$ has a specific value (in general between $1$ and $1.5$ in this example) which can vary in time. The money units $\mathrm{CHF}$ and $\mathrm{EUR}$ act as the rulers in both frames we discussed above!

However, one could still argue that this ratio is meaningless. Indeed, there is plenty of currencies in the world and they do not mean anything per se. What makes much more sense for a human is to know what they can get with such financial units, for example how many Unconstructed Mokas or Noodles plates they can buy. In this sense, the meaningful ratios are of the form
\be
\frac{1\ \mathrm{Noodles\ plate}}{1\ \mathrm{CHF}}\,.
\ee 
In the framework we described above, this ratio correspond exactly to the \emph{reference length}, which in this case is a \emph{reference price}, more generally called \emph{consumer price index}, indicating the price of a pool of goods. The fact that Physics should be the same in different frames translates here into the fact that you cannot get richer by changing your Swiss Francs to Euros (again, neglecting differences between countries, fees, speculation, ...). For a physical distance $L = \SI[parse-numbers = false]{\ell}{\meter}$, the comparison is shown in Tab.~\ref{tabframes:economy}
\begin{center}
\begin{table}
\centering
\begin{tabular}{c c   c   }
Noodles plate & $\rightarrow$ & $L$ \\
Currency & $\rightarrow$ & $1\mathrm{m}$ \\
Price & $\rightarrow$ & $\ell$
\end{tabular}
\caption{\label{tabframes:economy}Comparison between the physical formalism and an example taken from Economics.}
\end{table}
\end{center}

One last comment on this topic: When I was given my first Cosmology class, the professor mentioned inflation. By that time, I was still a kid and had a vague idea that inflation was something bad related to money. At first, I did not understand why one would talk about this in a theoretical Physics class, until I discovered what it really was. Since then, I put this idea I had inflation somewhere in my brain and never looked at it until the middle of my PhD, when I was working on the article with Basundhara and Ruth. It's only by that time that I understood that inflation is the same concept in Cosmology and in Economics: In both cases, the reference length/price suddenly increases very fast over time!

Much more can be discussed on this topic, and I did put too many elements under the carpet, as I do not want to crowd this thesis with unnecessary details... Let me just point out, to close this interlude, that I find very interesting and actually very insightful to think about Economics or any other field using physical concepts and I can only encourage everyone to do the same!

\section{Coupling with a Scalar Field \label{secframes:scalar}}
\subsection{Action of a Scalar Field}
In this Section, we introduce \ac{ST} theories: How can we study General Relativity together with a scalar field (as matter fields)? This has been done extensively in the past. The book by Fujii \cite{Fujii:2003} is a very good introduction to this topic and goes much further than this thesis. 

Theoretical physicists love actions. Let us make them happy and build an action for a scalar field. First, we would like to reproduce something similar to the classical formula: \emph{Kinetic minus Potential}. The kinetic energy is proportional to $\dot{\phi}^2$. Moreover, if you imagine a metal spring, the spatial deformation will act as a potential energy, so $\nabla_i \phi \nabla^i \phi$ is a good candidate. In a tensorial notation, and taking into account our sign convention, the Lagrangian for a scalar field is
\be 
\mathcal{L}_{\phi} = - \frac 12 k(\phi) (\bds{\nabla}\phi)^2 -V(\phi)\,,
\ee 
where we allow a generic kinetic term $k(\phi)$ and we consider an external potential $V(\phi)$. The full action of a \ac{ST} is obtained using the standard prescription and reads
\boxemph{ \label{eqframes:actionST}
\mathcal{S}_{\mathrm{ST}}[\bds{g},\phi] = 
\int \; \left(
\frac{1}{16\pi \GN} R  - \frac 12 k(\phi) (\bds{\nabla}\phi)^2 -V(\phi) 
\right) \sqrt{-g}\, \mathrm{d}^4 x\,.
}
The scalar field is said to be \emph{minimally coupled} to gravity as they only \emph{interact} through the square root of the metric determinant and the covariant derivative.

Much more can be said about \ac{ST} theories. More details are given by Sotiriou in \cite{Sotiriou:2015lxa}, where the general theory is presented together with implications for Cosmology

\subsection{Dictionary for the Scalar Field}
\subsubsection{The action \label{secframes:action}}
We would like to study how the action given by Eq.~\eqref{eqframes:actionST} transforms under a conformal transformation of the metric. More details can be found in the very nice thesis of Giammarco Fabiani \cite{Fabiani:2017} where they present the \ac{ST} theories in much details, in particular they consider the possibility of field redefinition. Any interested reader is encouraged to have a look at this work! The work of Morris provided also more details in this direction, see \cite{Morris:2014}.

Let us consider a slightly more general version of the \ac{ST} action Eq.~\eqref{eqframes:actionST} by adding a non-minimal coupling term to the curvature $A(\phi)$ and an interaction with the scalar field $\alpha(\phi)$ in the matter Lagrangian
\be
\mathcal{S}_{\mathrm{ST}}[{\bds{g}},{\phi}] = 
\int \; \left(
\frac{A(\phi)}{16\pi \GN} R  - \frac 12 k(\phi) (\bds{\nabla}\phi)^2 -V(\phi) 
+ \mathcal{L}_{\mathrm{m}}[ \alpha(\phi)\bds{g}, \psi]
\right) \sqrt{-g}\, \mathrm{d}^4 x\,,
\ee
where $\psi$ represent the matter fields, e.g. a cosmological fluid.
We perform the usual change of frame given by
\begin{align} \label{eqframes:deftransfo}
    \bds{g} &= \Omega^2 ( \phi ) \tilde{\bds{g}}\,,
\end{align}
where we slightly changed the definition of the conformal factor. Indeed, we assumed that it only depends on the coordinates through the scalar field.  Under these transformations the action can be written in the same functional form as
\be \label{eqframes:STtilde}
\tilde{\mathcal{S}}_{\mathrm{ST}}\left[\tilde{\bds{g}},{ \phi} \right] = 
\int \; \left(
\frac{\tilde A( \phi)}{16\pi \GN} \tilde R  - \frac 12 \tilde k( \phi) (\tilde{\bds{\nabla}} \phi)^2 -\tilde V( \phi) 
+ \mathcal{L}_{\mathrm{m}}[ \tilde{\alpha}(\phi)\tilde{\bds{g}}, \psi]
\right) \sqrt{-\tilde g}\, \mathrm{d}^4 x\,.
\ee
Using that the action is invariant, i.e.
\be 
\mathcal{S}_{\mathrm{ST}}[{\bds{g}},{\phi}]  = 
\tilde{\mathcal{S}}_{\mathrm{ST}}\left[\tilde{\bds{g}},{ \phi} \right] \,,
\ee 
the functions $\tilde A$, $\tilde k$ and $\tilde V$ are defined as
\boxemph{
\label{eqframes:Atilde}
\tilde A ( \phi)&= \Omega^2 A( \phi) \,,  \\
\label{eqframes:Vtilde}
\tilde V ( \phi) &= \Omega^4 V( \phi )\,, \\
\label{eqframes:alphatilde}
\tilde \alpha ( \phi) &= \alpha(\phi)\Omega^2\,, \\
\label{eqframes:ktilde}
\tilde k ( \phi) &=
\Omega^2
\left( 
k( \phi) -
\frac{3}{16\pi \GN} 
\left(\frac{\Omega^\prime}{2\Omega } \right)^2
\right)\,.
}
With this transformation, we can define classes of equivalent theories: For each quadruplet of functions $(A,k,V,\alpha)$, we perform the change of frame parametrised by $\Omega$ to obtain the same theory expressed with the tilde quantities. As this transformation is nothing but a mere change of frame, we expect the law of Physics to be the same in both frames, which is what interests us in this article, with the \ac{GNC} discussed hereafter.

Let us turn now to our specific example. We make the following assumptions.
\begin{itemize}
    \item In the Einstein frame, the scalar field is minimally coupled to gravity: ${A}(\phi) =1$.
    \item In the Jordan frame, the fluid is not coupled to the matter fields: $\tilde{\alpha} = 1$
\end{itemize}
Under these assumptions, the transformations Eq.\eqref{eqframes:Atilde}, Eq.~\eqref{eqframes:Vtilde}, Eq.~\eqref{eqframes:alphatilde} and Eq.~\eqref{eqframes:ktilde} become
\begin{align} 
    \label{eqframes:Afinal}
    \tilde{A} &= \Omega^{2}\,, \\
     \label{eqframes:Vfinal}
    V &= \Omega^{-4} \tilde{V}\,, \\
         \label{eqframes:alphafinal}
    \alpha &= \Omega^{-2}\,, \\
     \label{eqframes:kfinal}
    k&= \frac{\tilde k}{\Omega^2} + \frac{3}{16\pi\GN} \left(\frac{\Omega^\prime}{2\Omega } \right)^2\,.
\end{align}
and the actions in both frames read
\boxemph{ \label{eqframes:actionEfinal}
\tilde{\mathcal{S}}_{\mathrm{ST}}[{\bds{g}},{\phi}] &= 
\int \; \left(
\frac{1}{16\pi \GN} R  - \frac 12 k(\phi) (\bds{\nabla}\phi)^2 -V(\phi)
+ \mathcal{L}_{\mathrm{m}}[ \Omega^{-2}(\phi){\bds{g}}, \psi]
\right) \sqrt{-g}\, \mathrm{d}^4 x\,, \\
 \label{eqframes:actionJfinal}
\mathcal{S}_{\mathrm{ST}}\left[\tilde{\bds{g}},{ \phi} \right] &= 
\int \; \left(
\frac{ \Omega^2}{16\pi \GN}  \tilde R  - \frac 12 \tilde k( \phi) (\tilde{\bds{\nabla}} \phi)^2 -\tilde V( \phi) 
+ \mathcal{L}_{\mathrm{m}}[ \tilde{\bds{g}}, \psi]
\right) \sqrt{-\tilde g}\, \mathrm{d}^4 x\,,
}
with $k$, $\tilde k$, $V$ and $\tilde V$ arbitrary functions related by Eq.~\eqref{eqframes:kfinal} and Eq.~\eqref{eqframes:Vfinal}. Note that the condition $k,\tilde k>0$ ensure the stability of the theory.

\subsubsection{The energy-momentum tensor}
To conclude this section, we discuss the transformation rules of the energy-momentum tensor under the conformal transformation. This transformation being a (local) change of units, we expect the energy-momentum tensor to be modified accordingly, as it quantifies the energy density and pressure. Moreover, as there is a factor $\Omega^{2}$ in the matter Lagrangian in Eq.~\eqref{eqframes:actionJfinal}, we expect the usual conservation equation to be modified.

We present here the final results, but more explanations and derivations can be found in the article \cite{francfort_2019} or in the work of Santiago \cite{Santiago:2000}, with a thorough discussion about this topic.

Using the definition of the energy-momentum tensor given by Eq.~\eqref{eqGR:emt}, we get the dictionary for the energy-momentum tensor
\begin{align}
    \tilde{T}_{\mu \nu} &= \Omega^2 T_{\mu \nu} \,, \\ 
    \label{eqframes:tupdown}
    \tensor{\tilde{T}}{^\mu_\nu} &= \Omega^4 \tensor{{T}}{^\mu_\nu}\,, \\
    \tensor{\tilde{T}}{^\mu^\nu} &= \Omega^6 \tensor{{T}}{^\mu^\nu}\,. 
\end{align}
In the Jordan frame, matter is not coupled to the scalar field, hence the conservation equation still holds, while it is modified in the Einstein frame because of the non trivial interaction $\alpha\neq 1$ in Eq.~\eqref{eqframes:alphafinal}
\boxemph{ 
    \label{eqframes:consE}
    \bds \nabla \cdot \bds T &= - \bds \nabla \phi \frac{\pt_\phi \Omega}{\Omega} T\,, \\
    \label{eqframes:consJ}
    \tilde{\bds \nabla} \cdot \tilde{\bds T }&=0\,, 
}
or in components
\begin{align}
    \label{eqframes:consEcomp}
    \nabla^\mu T_{\mu \nu} &= - \nabla_\nu \phi \frac{\pt_\phi \Omega}{\Omega} T\,, \\
        \label{eqframes:consJcomp}
    \tilde{\nabla}^\mu \tilde{T}_{\mu \nu} &= 0 \,, 
\end{align}
with $T=\tensor{T}{^\mu_\mu}$. The last term represents a non-conservation term for the fluid, which we will explain below.

\section{FRLW in Two Frames \label{secframes:FRLW}}
\subsection{Background Universe}
\subsubsection{Scale factor, Hubble constant and Friedmann equations}
In this Section, we apply the \ac{ST} theory to a \ac{FRLW} background. Again, we present the results of the \cite{francfort_2019}. We consider two \ac{FRLW} metrics given by 
\begin{align}
    \bds{g} &= a^2 \left(- \bdd \eta^2 + \bdd \bds{x}^2 \right)\,, \\
    \tilde{\bds{g}} &=  \tilde{a}^2 \left(- \bdd \eta^2 + \bdd \bds{x}^2 \right)\,.
\end{align}
Both metrics are related by the conformal transformation Eq.~\eqref{eqframes:deftransfo}, which implies
\boxemph{ \label{eqframes:a}
a = \Omega_0 \tilde{a}\,.}
\MYrem{Background variables}{
Here and in what follows, we will denote background variables with the subscript $0$. It is then understood that all these variables only depend on the conformal time, e.g.
\be 
\Omega_0 = \Omega(\phi_0(\eta))\,,
\ee 
or
\be 
\Omega_0^\prime = \partial_\phi \Omega \phi_0^\prime\,.
\ee
}
\MYrem{Reference lengths}{
As explained in Section~\ref{secframes:phys_int}, the study of conformal frames is easily understood if one define references lengths, associated with a given physical process. Here, we will assume that the reference length does not change in the Jordan frame and equals $\tilde{\ell}=1$ (which is somehow equivalent to say that the ruler does not change in this frame). The conformal transformation Eq.~\eqref{eqframes:deftransfo} implies that the reference length in the Einstein frame is 
\be 
\ell = \Omega\,,
\ee 
where we drop the subscript $\mathrm{R}$ for clarity.
}
The relationship between the scale factors allows us to relate both Hubble constants as
\boxemph{  \label{eqframes:Hubble}
\tilde{\Hcal} = 
\frac{\tilde{a}^\prime}{\tilde a} 
= \frac{\Omega}{a} \left(  \frac{a}{\Omega}\right)^\prime 
= \Hcal - \frac{\Omega_0^\prime}{\Omega_0} 
= \Hcal - \frac{\ell_0^\prime}{\ell_0}\,.
}
This relation can be easily understood. Recall that in the \emph{usual} Jordan frame, the ruler is not modified while it is in the Einstein frame. Assume for example that $\tilde \Hcal=0$ and $\ell_0^\prime >0$, which means that the Universe is not expanding and that the ruler in Einstein is shrinking (hence the lengths are artificially increasing). Then we have 
\be 
\Hcal = \frac{\ell_0^\prime}{\ell_0}>0\,
\ee 
which is coherent with the assumptions: The lengths seem to be growing, which mimics an expanding Universe.

We present now the Einstein equations in both frames for completeness. Again, more details can be found in \cite{Morris:2014} or in the article. We only present the final results here. In the Einstein frame, the Friedmann equations read
\begin{align}
	3 \Hcal^2 &= 8\pi \GN\left(a^2 \rho_0 + \frac{{(\phi_0^\prime)}^2 k}{2} + a^2 V\right)\,,\\
	\Hcal^2 - 2 \frac{{a^{\prime \prime}}}{a}&= 8\pi \GN\left(a^2 P_0 + \frac{ (\phi_0^\prime)^2 k}{2} - a^2 V\right)\,.
\end{align}
They are simply the Friedmann equations with the usual energy density and pressure of a scalar field (and matter). In this frame, the scalar field is considered as matter like the fluid. In the Jordan frame, these equations become
\begin{align}
    \label{eqframes:friedmann_i}
	3 \tilde\Hcal^2 &= 
	\frac{8\pi \GN}{\Omega_0^2}
	\left(
	\tilde{a}^2 \tilde\rho_0 
	+ \frac{(\phi_0^\prime)^2 \tilde k}{2} 
	+ \tilde{a}^2 \tilde V
	\right) 
	-6 \tilde\Hcal\frac{ \phi^\prime_0 \Omega_0^\prime}{\Omega_0}\,,
	\\
	\label{eqframes:friedmann_ii}
	\tilde\Hcal^2 - 2 \frac{ \tilde a^{\prime \prime} }{\tilde a}
	&= \frac{8\pi \GN}{\Omega_0^2}
	\left(
	\tilde a^2 \tilde P_0
	+ \frac{(\phi_0^\prime)^2\tilde k}{2} 
	- \tilde a^2 \tilde V
	\right)\\ 
	&+ 2\tilde\Hcal \frac{\dot \phi_0 \Omega_0^\prime}{\Omega_0} 
	+ 2\frac{ \phi^{\prime \prime }_0 \Omega_0^\prime}{\Omega_0} 
	+ 2 (\phi_0^\prime)^2 \left(
	\left( \frac{ \Omega_0^{ \prime}}{\Omega_0}\right)^2  +  \frac{ \Omega_0^{ \prime \prime}}{\Omega_0}
	\right) \,, \nonumber
\end{align}
where we notice a modification with respect to their usual form: The scalar field is not considered as matter, and the non-minimal coupling introduces new term, which we can interpret as \emph{modified gravity}.

\subsubsection{The energy-momentum tenosr}
We now turn our attention to the energy-momentum tensor and to the energy conservation equation. Using its transformation rule Eq.~\eqref{eqframes:tupdown}, and the definition of the fluid energy-momentum tensor Eq.~\eqref{eqcosmo:tmunu}, we get
\begin{align}
\label{eqframes:rho0}
    \tilde{\rho_0} &= \Omega_0^4 \rho_0\,, \\
    \label{eqframes:P0}
    \tilde{P}_0 &= \Omega_0^4 P_0\,.
\end{align}
These relations are easily understood considering that pressure and energy density have mass dimension $+4$, and hence their transformation rule must be the opposite one as for lengths. Using the conservation equations Eq.~\eqref{eqframes:consJ} and Eq.~\eqref{eqframes:consE}, we get the energy conservation equations
\boxemph{
\tilde{\rho}_0^\prime &= - 3 \tilde{\Hcal} (\tilde \rho_0 + \tilde P_0)\,, \\
\rho_0^\prime &= - 3 {\Hcal} ( \rho_0 +  P_0) + \frac{\ell_0^\prime}{\ell_0} (3P_0 - \rho_0)\,.
}
It is easy to derive the last relation either using the general form Eq.~\eqref{eqframes:consE} or the explicit transformations of the energy density Eq.~\eqref{eqframes:rho0}, the pressure Eq.~\eqref{eqframes:P0} and the Hubble constant Eq.~\eqref{eqframes:Hubble}. Again, we can understand the last relationship considering two interesting examples. We will assume that $\tilde{\Hcal}>0$ is constant, which represents a physical expansion in the Jordan frame, and $\Hcal=0$ which means that the Einstein frame is artificially static. The relationship between those two quantities Eq.~\eqref{eqframes:Hubble} implies $\ell_0^\prime =- \tilde{\Hcal}\ell_0 $: In the Einstein frame, dimensionless lengths are (exponentially) decreasing, the ruler is expanding and dimensionless masses are (exponentially) growing. We consider two specific examples.
\begin{enumerate}
    \item If the Universe is filled with cold matter with $P_0=0$, we get in the Einstein frame
    \be 
    \rho_0^\prime = +\tilde{\Hcal} \rho_0\,.
    \ee 
    Moreover, the energy density in a given volume $V_0$ is defined as
    \be
    \rho_0 = \frac{m_0}{V_0}\,,
    \ee
    where $m_0$ is the mass inside the volume. To compute the evolution of the energy, we follow a given mass with time. In the physical Jordan frame, the physical volume $V$ becomes, after a time $\dd t$, due to the expansion of the Universe as
    \be
    V_{\mathrm p} = V(1+3 \tilde{\Hcal} \dd t)\,.
    \ee 
    However, in the Einstein frame, the ruler is expanding so dimensionless volume are decreasing with the rate
    \be 
    V^\prime = -3 \tilde{\Hcal} V\,,
    \ee 
    the factor $3$ coming from the fact that a volume is a length to the third power. Those two effects cancel exactly, and it appears in the Einstein frame that the dimensionless volume is kept constant. On the other side, dimensionless masses in the Einstein are increasing as
    \be 
    m^\prime = \tilde{\Hcal} m\,,
    \ee
    which is the only relevant effect whose consequence is that $\rho_0$ seems to be increasing in this frame.
    \item If the Universe is filled with Dark Energy with $P_0=-\rho_0$, we get 
    \be
    \rho_0^\prime = 4 \tilde{\Hcal} \rho_0\,.
    \ee
    Here, the dilution effect does not take place, but dimensionless masses are still growing while dimensionless volumes are still decreasing: The first effect brings a factor $1$ while the second a factor of $3$, which makes $1+3=4$ in total.
\end{enumerate}

\subsubsection{The redshift}
We end this discussion on the background \ac{FRLW} metric by presenting the correct expression for the redshift. Again, in the Jordan frame the ruler is kept constant ($\ell_{\mathrm{R}}=1$) the basis expression Eq.~\eqref{eqcosmo:redshift} still holds\footnote{Recall that the indices mean \emph{emission} and \emph{reception}.}
\be 
\label{eqframes:redshift}
1+z 
=
\frac{\tilde{\omega}_\me}{\tilde{\omega}_\mr} 
=
\frac{\tilde{a}_\mr}{\tilde{a}_\me}\,,
\ee
where the quantities are measured in the Jordan frame. To compute the redshift in the Einstein frame, we recall that frequencies scale as inverse times, and hence their transformation rules read
\be
\omega = \frac{1}{\ell_0} \tilde{\omega}\,.
\ee
Combining those elements, we obtain the redshift expressed in the Einstein frame as
\be \label{eqframes:redshiftEF}
1+z 
=
\frac{ \ell_{0,\me}\  \omega_\me }{ \ell_{0,\mr}\  \omega_\mr } 
=
\frac{\ell_{0,\me}\  {a}_\mr}{ \ell_{0,\mr}\ {a}_\me}\,.
\ee
Again, we can understand this relation with a simple example. Say that $\ell_{0,\mr} < \ell_{0,\me}$ and $ \omega_\mr =  \omega_\me$. This means that dimensionless lengths are getting smaller, so that the ruler is expanding exactly at the same rate as the expansion of the Universe: The dimensionless wavelengths and frequencies are not modified. Thanks to the correction in the formula, we still obtain $z>1$, which is coherent with a physical, frame-independent, redshift.

\subsection{Perturbed Universe}
\subsubsection{Bardeen potentials}
We would like to consider now perturbations to the \ac{FRLW} background. We use the same notations as in Section~\ref{seccosmo:PT}. We consider the perturbed metrics in the longitudinal gauge 
\begin{align}
\bds{g} &= a^2 \left(- (1+2\Psi)\bdd \eta^2 + (1-2 \Phi) \bdd \bds{x}^2  \right)\,, \\
\tilde{\bds{g}} &= \tilde{a}^2 \left(- (1+2\tilde{\Psi})\bdd \eta^2 + (1-2 \tilde{\Phi}) \bdd \bds{x}^2  \right)\,,
\end{align}
as well as the perturbed scalar field
\be 
\phi(\eta, \bds{x}) = \phi_0(\eta) + \delta \phi (\eta, \bds{x})\,,
\ee
and the perturbed reference length (in the Einstein frame)
\be 
\ell(\eta, \bds{x}) = \ell_0(\eta) + \delta \ell (\eta, \bds{x})\,,
\ee 
and recall that $\Omega = \ell$.
The interpretation of $\delta \ell$ is the following: Consider one constant time slice where $\ell_0(\eta)$ has some fixed value. If at some given spatial point $\bds{x}$ we have $\delta \ell (\eta, \bds{x})>0$. This means that at this specific point, dimensionless are slightly bigger than on average for this time slice, and equivalently that the ruler is slightly smaller. We will use this interpretation later on when we consider perturbed equations.

The dictionary for the Bardeen potentials can be easily found using the definitions and reads
\boxemph{
\label{eqframes:Phi}
\Phi &= \tilde{\Phi} - \frac{\delta \ell}{\ell_0}\,,  \\
\label{eqframes:Psi}
\Psi &= \tilde{\Psi} + \frac{\delta \ell}{\ell_0}\,.
}
This can be easily understood as follows. Consider a small coordinate displacement $\delta x$. In the Jordan frame, the measured distance is 
\be
L_{\mathrm{J}} = \tilde a \delta x (1-2 \tilde \Phi)\,,
\ee
while in the Einstein frame, the measured distance is
\be
L_{\mathrm{E}} =  a \delta x (1-2 \Phi)\,.
\ee
Recalling that those two distances are related thanks to the conformal transformation via
\be 
L_{\mathrm{E}}  = \ell L_{\mathrm{J}}  = (\ell_0 + \delta \ell) L_{\mathrm{J}} \,,
\ee 
it is straightforward to show that the relation Eq.~\eqref{eqframes:Phi} holds. The same logic can be applied for a time measurement, justifying the relation for $\Psi$ given by  Eq.~\eqref{eqframes:Psi}.

A direct consequence of Eq.~\eqref{eqframes:Phi} and Eq.~\eqref{eqframes:Psi} is that Weyl potential is frame independent
\be
\Psi_\mW = \Psi + \Phi = \tilde \Psi + \tilde \Phi = \tilde{\Psi}_\mW\,.
\ee 
This is not a surprise as the Weyl potential is used to compute the lensing potential, and we know from Theorem~\ref{thmcosmo:nullgeo} that lensing should not be affected by a conformal transformation. More details about invariance under conformal transformations in Cosmology can be found e.g. in~\cite{Li:2015}.

\subsubsection{Energy-momentum tensor}
We want now to describe the perturbations to the energy-momentum tensor. We use the same convention as in Eq.~\eqref{eqcosmo:deltarho}, Eq.~\eqref{eqcosmo:deltaP} and Eq.~\eqref{eqcosmo:deltav} to describe this perturbations in both frames. We want to relate the relative perturbations 
\be
\delta(\eta, \bds{x}) = \frac{\rho(\eta, \bds{x})- \rho_0(\eta)}{\rho_0(\eta)}\,.
\ee 
Using that
\be
\rho = \ell^{-4} \tilde \rho\,,
\ee
it is easy to see that at first order, both perturbations are related as
\boxemph{ \label{eqframes:deltarho}
\delta &= \tilde \delta - 4 \frac{\delta \ell}{\ell}\,,
}
and the same holds for the relative Pressure perturbation. Again, we can get a nice understanding of this formula considering a simple example: Assume that $\tilde \delta (\eta,  x) =0$ at some spacetime point. This means that in the Jordan frame, there is no energy perturbation at this specific point. If $\delta \ell>0$, lengths tend to be measured slightly bigger at this point, and masses slightly lower in the Einstein frame, which implies that the energy density are also measured slightly lower, with the usual factor of $4$, as explained in the previous section.

The perturbed $4$-velocity in both frame read
\begin{align}
\bds{u} &= \frac{1}{a}\left( (1-\Psi) \bpt_t + v^i \bpt_i\right)\,, \\
\tilde{\bds{u}} &= \frac{1}{\tilde{a}}\left( (1-\tilde{\Psi}) \bpt_t + v^i \bpt_i\right)\,, \\ 
\end{align}
where the velocity $\bds v$ is the same in both frames, which could be expected as it is a dimensionless quantity (and hence there is no need for a tilded version).

\subsubsection{Redshift}
\begin{figure}[h!t]
	\begin{center}
	\includegraphics[width =0.5\textwidth]{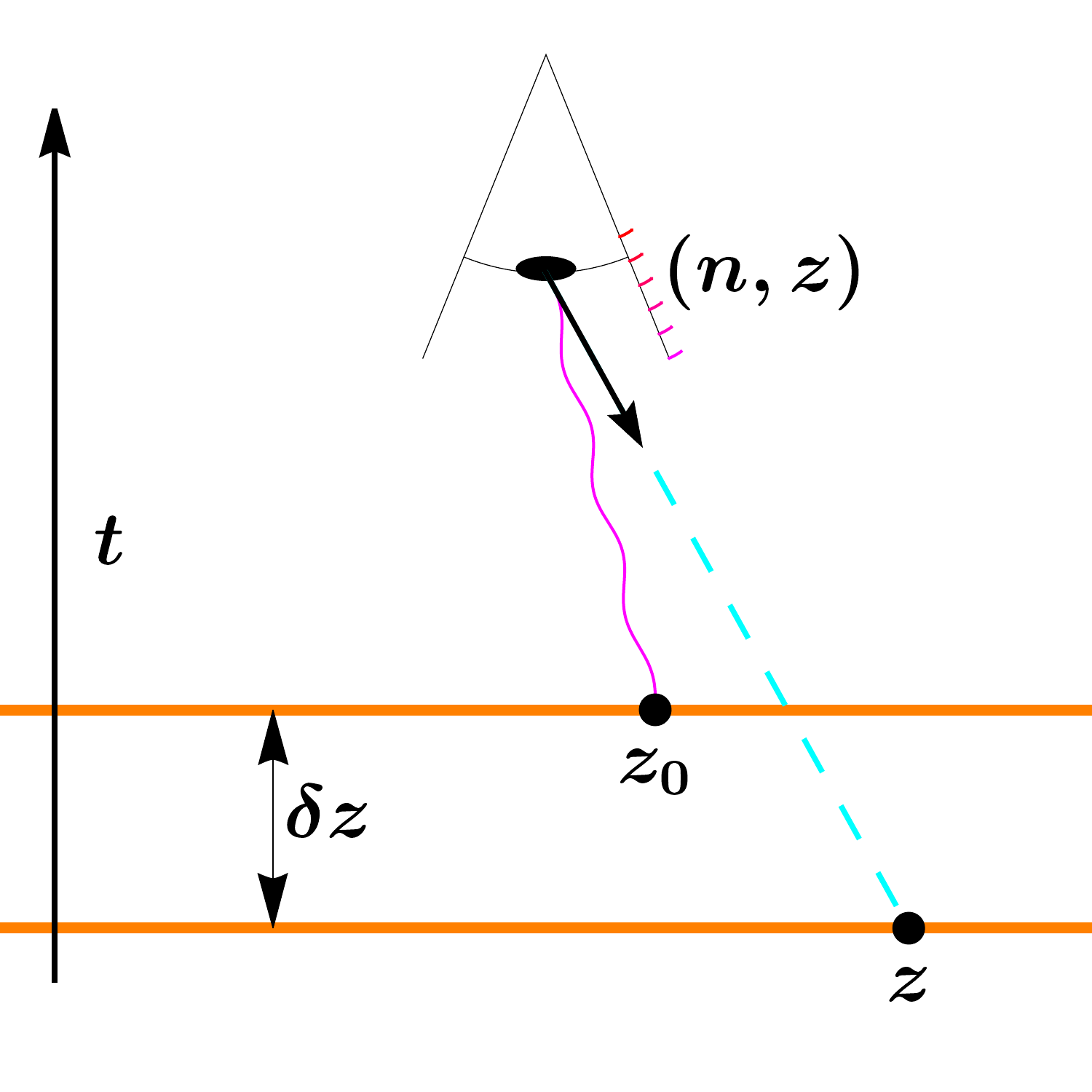}
	\end{center}
	\caption[Redshift perturbation]{\label{figframes:setupredshift}
	\textbf{Redshift perturbation}\\
	 The photon is emitted at a time slice corresponding to $z_0$, but the observed redshift is $z$. The difference betweeen these two quantities is $\delta z$. In this specific example, we have $z>z_0$ so $\delta z >0$. Note that the observed direction is also perturbed because of lensing.}
\end{figure}

We end this Section about perturbations with an important point: The perturbation of the redshift. Let us stress an important point: When observing a photon coming from a galaxy, we measure its direction in the sky $\boldsymbol n$ and its redshift $z$. Both quantities are in a sense \emph{perturbed} as the photon traveled across the Universe which is not homogeneous and isotropic. Mathematically speaking, we formalise this by assuming that the photon was emitted on some constant time slice corresponding to a background redshift $z_0\neq z$. This would be the observed redshift if the Universe was truly \ac{FRLW}, and the formula would be given by Eq.~\eqref{eqcosmo:redshift}. In reality, this background redshift and the observed one do not coincide and we define the redshift difference as
\be  \label{eqframes:diffz}
\delta z \equiv z-z_0\,.
\ee 
We need to stress here that $z_0$ and $\delta z$ are \emph{not} observables! They are mathematical tools we use in our computations, but any final, physical, observable results should not depend on it. The background redshift $z_0$ is not physical as, at the perturbed level, the background spacetime is not well-defined and can be redefined through an infinitesimal coordinate transformation, see Section~\ref{seccosmo:Lie} where we discussed this effect in more details. The situation is shown in Fig.~\ref{figframes:setupredshift}, where a photon is emitted and observed in a perturbed Universe.

The expression for $\delta z$ can be found e.g. in \cite{Bonvin:2011bg} and reads in the Jordan frame
\boxemph{ \label{eqframes:deltaz}
\delta z(z, \boldsymbol n) = - (1+z_0)
\left( \tilde \Psi(z_0, \boldsymbol n) - \boldsymbol{n} \cdot \boldsymbol{v} + \int_0^{r(z_0)}\; (\tilde \Psi^\prime + \tilde \Phi^\prime)\, \mathrm d r  \right)\,.
}
We can give a physical interpretation for each term in Eq.~\eqref{eqframes:deltaz}.
\begin{enumerate}
    \item The first term is a usual potential term: Let us assume that the Bardeen potential $\tilde \Psi>0$ at the emission position. While travelling towards us, the photon goes down this potential, and as any biker knows, its energy increases. Hence, the observed redshift is reduced and this implies $\delta z <0$.
    \item The second term is a Doppler effect, where $\boldsymbol n$ is the observed direction and $\boldsymbol v$ is the peculiar velocity of the galaxy. If $\boldsymbol n \cdot \boldsymbol v>0$, the galaxy is receding away from us, which increases the redshift because of the Doppler effect, hence $\delta z >0$ with this term. This effect is sometimes called the simple Sachs Wolfe effect, see \cite{Nishizawa:2014} for more explanations.
    \item The last term is the Integrated Sachs Wolfe effect. It can be understood as follows: If you are biking up hill and the ground is helping you by going up while you are biking, you will lose less energy that if the ground was not moving. This is what happens if the photons travel through a time-dependent potential field, and this reduces their observed redshift. Note that for this term, the \emph{spatial} potential $\Phi$ also contributes.
\end{enumerate}

\MYrem{Monopole terms}{In Eq.~\eqref{eqframes:deltaz}, the terms depending only on the observer position were neglected. These terms are constant and, for example, only contribute to the monopole term of the power spectrum, see Remark~\ref{remcosmo:monopole}.}

The expression for the redshift perturbation given by Eq.~\eqref{eqframes:deltaz} is valid in the Jordan frame, as there is no correction to make. In the Einstein frame, the only new term at the source position is $\ell_{\me}$, see Eq.~\eqref{eqframes:redshiftEF}. Using the definition of the perturbation of the reference length, the redshift perturbation in the Einstein frame reads
\boxemph{ \label{eqframes:deltazE}
\delta z(z, \boldsymbol n) = - (1+z_0)
\left(  \Psi(z_0, \boldsymbol n)  - \frac{\delta \ell (z_0, \bds n)}{\ell_0(z_0)}- \boldsymbol{n} \cdot \boldsymbol{v} + \int_0^{r(z_0)}\; ( \Psi^\prime +  \Phi^\prime)\, \mathrm d r  \right)\,.
}
Note that, using the dictionary for the Bardeen potentials Eq.~\eqref{eqframes:Psi}, it is direct to show that $\delta z$ is frame invariant, i.e. that  Eq.~\eqref{eqframes:deltaz} and Eq.~\eqref{eqframes:deltazE} are equal.

\stopchap

\chapter{Galaxy Number Counts\label{chapframes:NC}}
In this Chapter, we present the concept of \ac{GNC}. In the first Section, we introduce the \ac{GNC} as a Cosmological observable introduced by Bonvin et al. in \cite{Bonvin:2011bg}. The key idea is that, when we perform galaxy surveys, we observe galaxies, together with their angular position and observed redshift. The \ac{GNC} is a physical observable, contrary to, for example, the energy perturbation density. Hence, we expect its value to be frame-independent. In the second Section, we prove this explicitly, at the perturbation level, in a second section.

\startchap

\section{Introduction}
\subsection{General Idea}
We present here the concept of \ac{GNC}. This quantity answers to the question asked by Bonvin et al.~\cite{Bonvin:2011bg}. To rephrase them,
\begin{center}
\emph{What do galaxy surveys really measure?}
\end{center}
When astronomers observe a galaxy, as mentioned in the previous Section, they measure two quantities: Its redshift $z$ and its direction on the celestial sphere $\boldsymbol n$. Both quantities cannot be measured exactly, hence the galaxy are grouped together into redshift and angular \emph{bins}. If you are provided with a galaxy survey (assuming there is sufficiently enough galaxies), you can perform the following steps.
\begin{enumerate}
    \item Choose a redshift bin around $z$ and an angular bin around the direction $\boldsymbol n$.
    \item Count \emph{all} the galaxies observed in the redshift bin $z$ (considering the full sky). 
    \item Compute the average number of galaxies $\langle N \rangle (z)$ per angular bin.
    \item Count the number of galaxies $N(z, \boldsymbol n)$ in the specific bin around $z$ and $\boldsymbol n$.
    \item Compute $N(z, \boldsymbol n)- \langle N \rangle (z)$ and normalize it by $\langle N \rangle (z)$ to determine if, in this specific direction, there is an over- or under- galaxy density, compared with the average galaxy density at constant redshift $z$.
\end{enumerate}
This process makes sense from an observational point of view: In fine, we only observe galaxies, not energy density nor the full matter distribution.

Following the steps above, we define the \emph{Galaxy Number Counts} (\ac{GNC})
\boxemph{
\Delta (z, \boldsymbol n) \equiv \frac{N(z, \boldsymbol n)- \langle N \rangle (z)}{\langle N \rangle (z)}\,.
}
\begin{figure}[h!t]
	\begin{center}
		\includegraphics[width =0.75\textwidth,height=0.51\textheight]{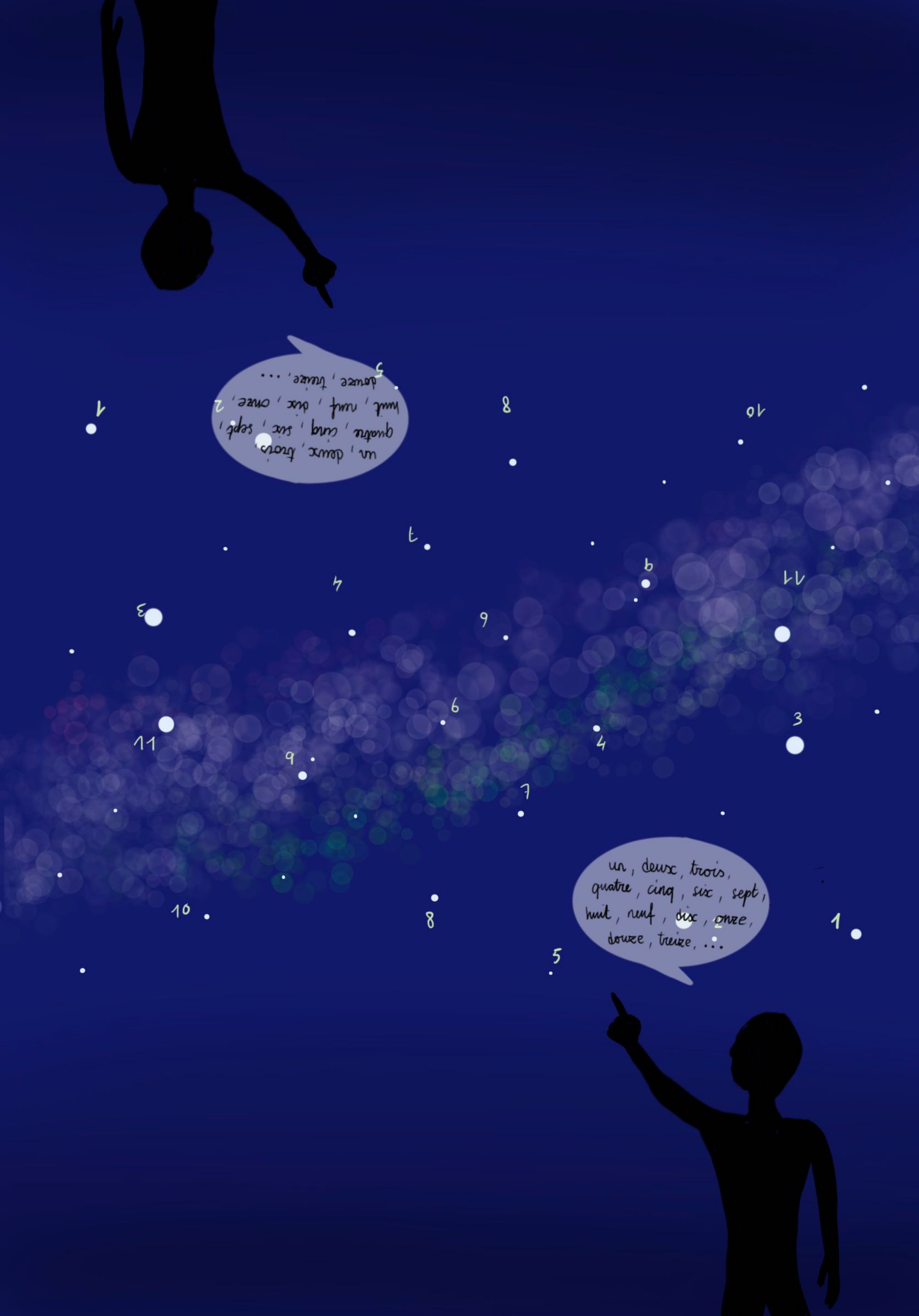} 
	\end{center}
	\caption*{\label{figframes:Comptage}
		\textbf{Compter les \'etoiles des deux c\^ot\'es}, Jessica Mascellino
	} 
\end{figure}

\MYrem{Bias}{If you really want to be picky, you could argue that we do not observe galaxies, but the photons they emit. There is a bias problem: Only bright galaxies can be observed, or some galaxies will be hidden by some foreground matter. We neglect these subtleties here and consider the \ac{GNC} from a theoretical perspective, assuming that we are able to observe all the galaxies.}

\subsection{Expression of the Galaxy Number Counts}
\begin{figure}[h!t]
	\begin{center}
	\includegraphics[width =0.5\textwidth]{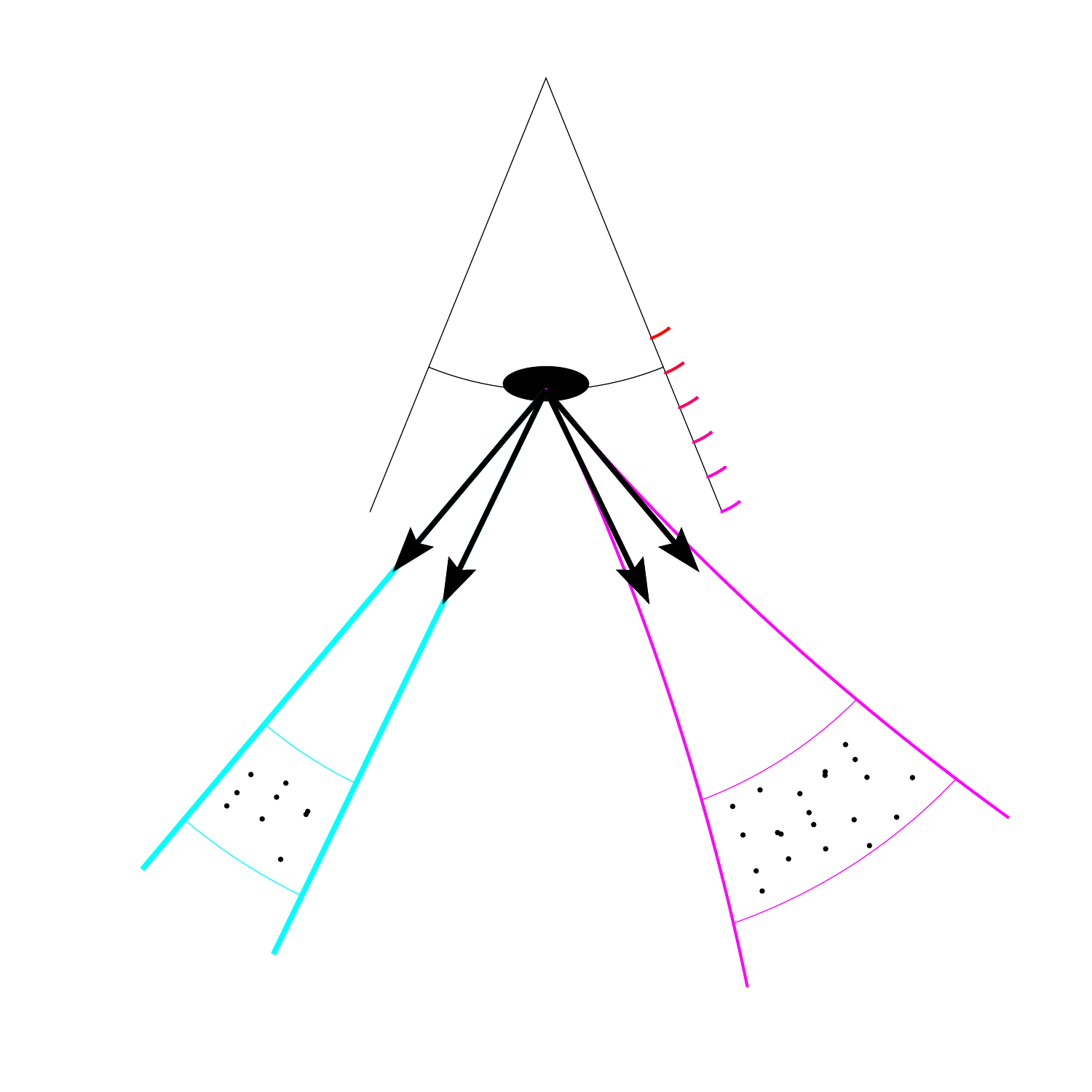} 
	\end{center}
	\caption[Volume perturbation]{\label{figframes:volumepert}
	\textbf{Volume perturbation}\\
	 Both bins appear to have the same angular size (the apparent angular aperture is the same). However, the bin on the right undergoes more lensing and corresponds to a bigger physical volume, and hence contains more galaxies (assuming the physical densities are equal). }
\end{figure}

In this Section, we present the mathematical expression of the \ac{GNC}. The detailed derivation can be found in \cite{Bonvin:2011bg} and we show only the important results here. The \ac{GNC} can be decomposed into two parts as
\boxemph{ \label{eqframes:NC}
\Delta(z, \boldsymbol n) = \delta_z (z, \boldsymbol n) + \frac{\delta V (z, \boldsymbol n)}{V(z)}\,.
}
The first term is the \emph{redshift density perturbation} defined as
\be
\delta_z(z, \boldsymbol n) \equiv \frac{\rho(z, \boldsymbol n)- \rho (z)}{ \rho (z)}\,,
\ee 
where $\rho(z)$ is the mean energy density at constant observed redshift $z$. The second term is the \emph{volume perturbation} and it quantifies the fact that the physical volume of a redshift bin may be different. It is defined as
\be 
V(z,\boldsymbol n) = V(z) +\delta V (z,\boldsymbol n)\,,
\ee 
where $V(z)$ is the average physical volume per bin, at fixed redshift $z$, see Fig~\ref{figframes:volumepert}.

\MYrem{Two different energy perturbations \label{remframes:twodiff}}{
The quantity $\delta_z$ is \emph{not} the usual relative energy perturbation of the energy-momentum tensor. To understand that, let us have another look at Fig.~\ref{figframes:setupredshift}. The photon was emitted on the time slice corresponding to $z_0$. Hence, the energy density at the emission position should be expanded around the background quantity at this specific redshift
\be
\rho(z, \boldsymbol n) = \rho_0 (z_0) + \delta \rho(z, \boldsymbol n)\,.
\ee 
This last term $\delta \rho(z, \boldsymbol n)$ is the perturbation to the energy density appearing in the expression of the energy-momentum tensor discussed in Section~\ref{eqcosmo:pertEMT}. On the other side, the redshift density perturbation $\delta_z(z,\bds n)$ is computed with respect to a given observed redshift which, depending on the chosen gauge, may not correspond to a constant time slice.
}
As usual, we can try to give a nice interpretation of Eq.~\eqref{eqframes:NC}. The first term is quite intuitive: In a given pixel, the \ac{GNC} is the same as the redshift density perturbation, quantifying the over- or under- energy density in this pixel. However, a redshift bin should be converted to a physical bin to really estimate the number of galaxies it contains. This is why it is also important to consider the volume perturbation. Indeed, because of lensing, the physical volume of a redshift bin may be slightly bigger or slightly smaller. If the physical volume is, say, slightly bigger than the average, we have $\delta V>0$, see Fig.~\ref{figframes:volumepert}. In this case, the bin we are observing contains more galaxy simply because it represents a bigger portion of space, which enhances the \ac{GNC}.

For completeness, we provide the full expression for the \ac{GNC}, taken from \cite{Bonvin:2011bg}. For a photon observed in direction $\bds n$ and emitted at comoving distance $r_\ms$, with comoving (gauge-invariant) velocity $\bds V$ the \ac{GNC} reads
\boxemph{
\Delta(z,\bds n) &=
\frac{D_\rho}{\bar \rho} - 3 (1+w) \Phi + \Phi + \Psi + \Hcal \left(
 \Phi^\prime - \partial_r (\bds V \cdot \bds n )\right) \\
&+ \left( \frac{ \Hcal^\prime }{\Hcal^2} + \frac{2}{r_\ms \Hcal} \right)
\left( \Psi - \bds V \cdot \bds n + \int_0^{r_\ms}\;  \Phi^\prime  +  \Psi^\prime \, \dd \lambda\right) \\
&+ \frac{1}{r_\ms} \int_0^{r_\ms}\; \left(2- \frac{r_\ms-r}{r} \Delta_\Omega\right) (\Phi + \Psi)\,,
}
where the integrals are performed over the unperturbed path and $w$ is the barotropic index of the background fluid.

\section{Invariance of the Number Counts}
\subsection{Jordan frame}
The expression given by Eq.~\eqref{eqframes:NC} holds in the Jordan frame, as again value of the reference length is constant, hence it is not necessary to \emph{convert} any quantities. The \ac{GNC} in the Jordan frames reads
\boxemph{ \label{eqframes:NC_J}
\tilde \Delta(z, \boldsymbol n) = \tilde  \delta_z (z, \boldsymbol n) + \frac{ \tilde{\delta V} (z, \boldsymbol n)}{\tilde V(z)}\,.
}

\subsection{Einstein frame}
\subsubsection{Expression for the Number Counts}
Things get a little bit more complicated in the Einstein frame, where the reference length does not have a constant value. We want to compute the redshift energy perturbation in this frame. To evaluate $\rho(z, \boldsymbol n)$, we need to think about the definition of the energy density. In theory, one should count the total mass of the galaxies and divide it by the physical volume. However, the mass having the dimension of an inverse length, their value has to be computed using the reference length as
\be
M(z, \boldsymbol n) = N(z, \boldsymbol n) \ell(z, \boldsymbol n)^{-1} m_0\,,
\ee
where we assumed that all the galaxies have constant mass $m_0$ is a constant mass in the Jordan frame. This assumption can be relaxed and does not change the final result. Then, in the Einstein frame, the redshift perturbation density reads
\begin{align}
\delta_z(z, \bds{n}) 
&= 
\frac{\rho(z, \bds{n}) - \rho_0(z)}{\rho_0(z)} 
=
\frac{
\frac{N(z, \bds{n})  \ell(z, \bds{n}) ^{-1}}{V(z, \bds{n}) } -
\frac{N_0(z) \ell_{0}(z)^{-1}}{V(z)} 
}
{	\frac{N_0(z) \ell_{0}(z)^{-1}}{V(z)}  } \,.
\label{eqframes:deltarhozE}
\end{align}
Now, as pointed out in Remark~\ref{remframes:twodiff} and shown in Fig.~\ref{figframes:setupredshift}, we can write the reference length as
\be 
\ell(z, \boldsymbol n) = \ell_0(z_0) + \delta \ell(z, \boldsymbol n)
= \ell_0(z) - \delta z \frac{\dd \ell_0}{\dd z_0}+ \delta \ell(z, \boldsymbol n)\,,
\ee 
where we used the definition of the redshift perturbation Eq.~\eqref{eqframes:diffz}. The next steps follow the same logic and the final result is
\boxemph{ \label{eqframes_NC_E}
\Delta(z, \boldsymbol n) =
\delta_z(z, \boldsymbol n) 
+ \frac{\delta V (z, \boldsymbol n)}{V(z)} 
-\frac{\delta z(z, \boldsymbol n)}{\ell_0(z)} \frac{\dd \ell_0}{\dd z_0}
+\frac{\delta \ell(z, \boldsymbol n)}{\ell_0(z)}
}


Now that we obtained the expression for the \ac{GNC} in the Einstein frame, what should be done is to show that it is indeed equal to the \ac{GNC} in the Jordan frame, by computing explicitly the relationship between the redshift energy perturbation and the volume perturbation in both frames. Before doing so, we state and proof a cute theorem which will prove valuable for these tasks.

\subsubsection{A little theorem}

\MYthm{Relative perturbations in the Einstein frame}{\label{thmframes:deltaf}
Let us consider a quantity $f(z, \boldsymbol n)$ whose length dimension is $n$. We define its redshift perturbation in the Einstein frame as
\be 
\frac{\delta f}{f} \equiv \frac{\delta f (z, \boldsymbol n)}{f_0(z)}
=
\frac{ f (z, \boldsymbol n)- f_0(z)}{f_0(z)}\,,
\ee
where $f_0(z)$ is the average value of the function at redshift $z$, and equivalently in the Jordan frame. 

Then, the quantities in both frames are related as
\be
\frac{\delta f}{f} = \frac{\delta \tilde f}{\tilde f} -n \frac{\delta z}{\ell_0} \frac{\dd \ell_0}{\dd z_0} +n \frac{\delta \ell}{\ell_0}\,.
\ee 
}
\begin{proof}
The proof follows the same lines as the computation of the redshift density perturbation $\delta_z(z, \bds n)$ above. The value of the function in both frames are related as
\begin{align}
f(z, \boldsymbol n) &= \tilde{f}(z,\boldsymbol n) \ell(z, \boldsymbol n)^n \,, \\
f_0(z)&=\tilde{f}_0(z_0) \ell_0(z)^n\,.
\end{align}
Using again that 
\be
\ell(z, \boldsymbol n)= \ell_0(z_0) + \delta \ell (z, \boldsymbol n)
\ee 
and 
\be 
\delta z = z -z_0\,,
\ee
the result can be shown directly.

\end{proof}

\subsubsection{Redshift density perturbation}
The redshift density perturbation corresponds to a quantity with length dimension $n=-4$. Using the Theorem \ref{thmframes:deltaf}, we obtain
\be  \label{eqframes:deltaz_EF}
\delta_z = \tilde{\delta}_z 
+4\frac{\delta z}{\ell_0} \frac{\dd \ell_0}{\dd z_0} 
-4\frac{\delta \ell}{\ell_0}\,.
\ee 
This relation can also be derived explicitly relating $\delta_z$ to $\delta \rho$ and using the dictionary for the perturbations given by Eq.~\eqref{eqframes:deltarho}, the definition of the redshift involving the reference lengths Eq.~\eqref{eqframes:redshiftEF} and the dictionary for the Hubble constant Eq.~\eqref{eqframes:Hubble}. This derivation does not contain any major difficulties, and more details can be found in the article.

\subsubsection{Volume perturbation}
The physical volume has length dimension $n=3$ and the Theorem \ref{thmframes:deltaf} reads in this case
\be  \label{eqframes:deltaV_EF}
\frac{\delta V}{V} =\frac{\delta \tilde{V}}{\tilde V}
-3\frac{\delta z}{\ell_0} \frac{\dd \ell_0}{\dd z_0} 
+3\frac{\delta \ell}{\ell_0}\,.
\ee 
Again, this relation can be derived in a more detailed fashion which explained in the article and in \cite{Bonvin:2011bg} (mostly in Section II B.). We again encourage the interested reader to have a look at the specific parts and we won't go into more details.

\subsection{Comparison}
We now have all the ingredients in our hands to compare the expressions of the \ac{GNC} in both frames. Combining the dictionary for the redshift perturbation Eq.~\eqref{eqframes:deltaz_EF} and for the volume perturbation Eq.~\eqref{eqframes:deltaV_EF}, we get
\be 
\delta_z + \frac{\delta V}{V}= 
\tilde{\delta}_z V
+\frac{\delta \tilde{V}}{V}
+\frac{\delta z}{\ell_0} \frac{\dd \ell_0}{\dd z_0} 
-\frac{\delta \ell}{\ell_0}
= \tilde{\Delta}(z, \boldsymbol n)
+\frac{\delta z}{\ell_0} \frac{\dd \ell_0}{\dd z_0} 
-\frac{\delta \ell}{\ell_0}\,,
\ee 
where we used the definition of the \ac{GNC} in the Jordan frame Eq.~\eqref{eqframes:NC}. Comparing this result with the expression of the \ac{GNC} in the Einstein frame Eq.~\eqref{eqframes_NC_E}, we see directly that 
\boxemph{
\tilde{\Delta}(z, \boldsymbol n)
=
{\Delta}(z, \boldsymbol n)\,,
}
or in other words, the value of the \ac{GNC} is frame-independent! To make a link with the definition of physical observables presented in Chapter~\ref{secintro:Obs}, here the \ac{GNC} is an observable, i.e. $\Delta = \mathcal{O}$, while a change of frame is one of the transformations $\mathcal{T}$ we introduced. The physical quantity such as $\delta V$ or $\rho_z$ are elements of $\bds{\mathcal{Q}}$ which do transform under $\mathcal{T}$.

\MYrem{Disformal transformations}{
Regarding disformal transformations (see Section~\ref{secframes:disformal}), we showed in \cite{Ghosh:2022} that the number count is also invariant in these cases. The proof goes along the same lines, the only difference is that time is also modified, and new terms should be considered. In the end, as we did here, all the new terms cancel out in the final expression.
}

\stopchap

\chapter{Conclusion\label{chapframes:conc}}
In this Part, we introduced the central notion of conformally related frames. These are metrics who differ only by a position-dependent global multiplication. We argue that a change of frame is equivalent to a local change of units, where the reference length of a given physical process depends on space and time, as was already done in \cite{Chiba:2013,Dicke:1962}.

We defined the two most important frames, the Einstein frames in which gravity is not coupled to the scalar field, but it non minimally coupled to the matter fields, and the Jordan frame, where the gravity and the scalar field are coupled, but where the matter is minimally coupled to gravity. We frames the discussion in the more general context of \ac{ST} theories, inspired for instance by \cite{Fujii:2003,Sotiriou:2015lxa}.

In a second part, we studied the Number Counts, first introduced in \cite{Bonvin:2011bg}. We presented this observable and provided some physical interpretation. The most important part of this work is the proof that the Number Counts does not depend on the frame. This was expected, but this work is a safety check to convince the community that the Einstein and the Jordan frames are indeed equivalent.

Additional works were made on the topic. For example, a new observable is defined in \cite{Matthewson:2022} where the Number Counts observable is defined with a redshift-dependent weight. In \cite{Galaverni:2021,Galaverni:20212}, they proceed to a Hamiltonian analysis in conformal frames in a Cosmological context. Finally, in \cite{Ghosh:2022}, we showed that the invariance of the Number Counts also holds if one considers the more general disformal transformations. Applying a disformal transformation to a Cosmological background modify not only the length scale, but also the time scale. The derivation is slightly different, but the logic behind is actually very fun.

In a future work, one should be able to show the invariance under conformal transformations of more physical observables. Even better would be a convincing proof that the frames are equivalent (or, why not, a counter example to show that they are not). Even if the debate does not seem to take a lot of attention, it is still worth noting that it is not closed yet.

\part{Rotation from Lensing}\label{Part:RotLensing}
\begin{figure}[h!]
	\centering
	\includegraphics[width = 0.9\textwidth]{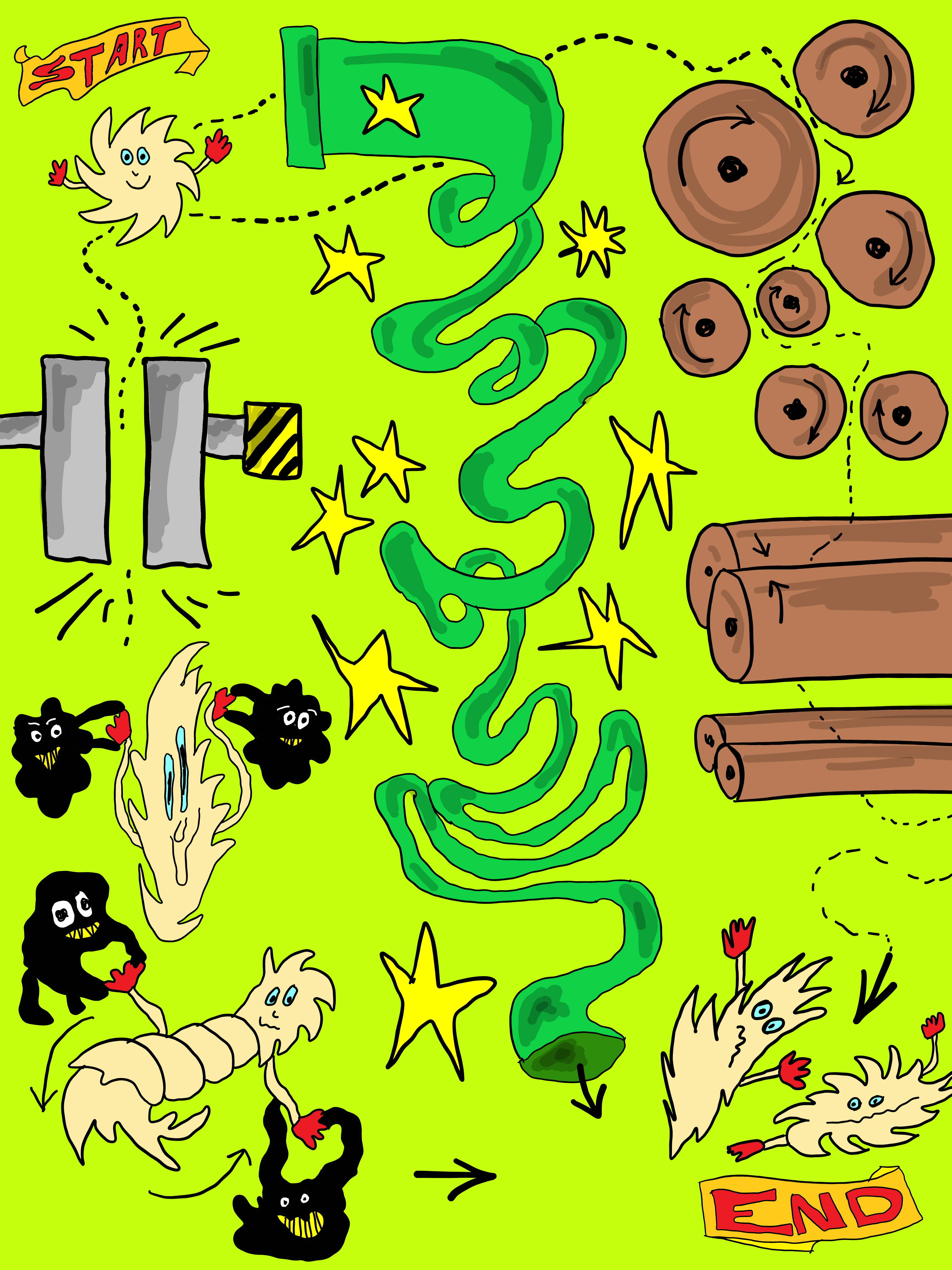} 
	\caption*{\textbf{Black Holes Tormenting Galaxies}, Juan Manuel Garc\'ia Arcos}
\end{figure}

\chapter{General Introduction\label{chaplensing:intro}}
In this Part, I present two articles that Giulia, Ruth and I wrote~\cite{Francfort:2021,Francfort:2022}. We started this project in August 2020, after the first lockdown. It somehow refueled my motivation after this weird semester.

This part is about gravitational weak lensing. Lensing is one of the main effects in General Relativity and Cosmology. It occurs as light travels through a curved spacetime, e.g. due to the presence of matter (luminous and dark). Being able to quantify the lensing undergone by the signals we measure on Earth is a very powerful tool to probe the content of the Universe, e.g. between us and a far away galaxy. In practice, lensing has a very small effect on individual sources. A statistical approach is then necessary to extract meaningful quantities, and reduce the noise. For instance, estimating angular correlation functions on various physical quantities is a powerful tool to estimate the values of Cosmological parameters, and in fine to constrain various Cosmological models, in particular \ac{LCDM}.

Lensing is mathematically described by the Jacobi map which itself depends on the second derivative of the lensing potential $\phi$, a redshift-dependent function defined on the celestial sphere. More explicitly, it is the integral of the Weyl potential along the line of sight between the observer and a given source of light. Its value is parametrised by $3$ quantities: The convergence, the net rotation and the shear. Lensing is a broad field of research and it would not be possible to give a thorough list of references. Nice reviews are for instance~\cite{Schneider:2006,Perlick:2004,Lewis:2009}. For more specific examples, see~\cite{Lepori:2020} for a numerical simulation (including vector perturbations), \cite{DiDio:2019} for an application to \ac{CMB} lensing, \cite{Yamauchi:2012} for an application to cosmic strings and~\cite{Adamek:2015,Cusin:2019} where lensing of gravitational waves is considered.

The convergence is related to the diagonal part of the Jacobi map. Theoretical predictions estimating the convergence from the Number Counts are presented in~\cite{Montanari:2015, Lepori:2021,Nistane:2022} and numerical simulations are found in~\cite{Scranton:2005, Liu:2021}. On the other side, the net rotation is the antisymmetric part of the Jacobi map. Typically, it is a second order effect and is neglected in the present work, see for example~\cite{Whittaker:2017,Fanizza:2022}.

Shear measurements are in general difficult. Their effect on the shape of the galaxy is small and these correlation functions are affected by intrinsic alignment of galaxies~\cite{Bartelmann:2010,Hirata:2004,Kirk:2010}. However, various large-scale projects, e.g. KiDs and DES, have already been done~\cite{Kids:2020,DES:2020,DES:2021,DES:2022}. These surveys will provide data, which motivate the necessity to study shear correlation functions from a theoretical point of view. Several authors have presented methods to compute shear correlation functions, see for example~\cite{Hildebrandt:2018,Kilbinger:2014,Bartelmann:2010,Schneider:2002}. In this work, the goal is to present a new method to build an estimator for shear correlation functions using galaxy orientation and light polarisation.

It is a well-known result that light polarisation is parallel transported, while galaxy shape is Lie transported~\cite{Perlick:2004}, which is correctly described by the Jacobi formalism. It has been shown that the emitted polarisation at the galaxy location is generally aligned with the intrinsic shape of the galaxy (in particular, as it will be explained, it is aligned with the semi-minor axis), see~\cite{Gardner:1966}. In our first paper~\cite{Francfort:2021}, we showed that if the main axes of the shear are not aligned with the main axes of the galaxy, lensing will induce a slight misalignment between the polarisation vector and the shape of the galaxy. This angle difference is a genuine signature of lensing. In particular, it is a probe of cosmic shear.

Polarisation has already been used in the past to probe gravitational lensing. For example, in~\cite{Kronberg:1991,Kronberg:1996,Burns:2004}, the angle between the polarisation vector and the main axes of a galaxy is used to estimate the mass of a foreground galaxy using light emitted by distant quasars. In \cite{Brown:2011}, they propose a method to use polarisation to build maps of Dark Matter distribution. In~\cite{Thomas:2017}, they use polarisation to estimate the net rotation part of the Jacobi map. In~\cite{Brown:2010,Camera:2016}, it is argued that polarisation measurements can mitigate the shot-noise, while in the series or articles~\cite{Whittaker:20152,Whittaker:2014, Whittaker:2015,Whittaker:2017,Thomas:2017}, a method to estimate cosmic shear using the intrinsic alignment of galaxy shapes is proposed. Finally, in~\cite{Harrison:2020}, the authors propose to use radio surveys to study weak lensing, but the polarisation is not really considered.

Our approach is similar to what has been done, yet it brings some novelties. We consider the rotation angle described above to build an estimator for the shear correlation functions, see our second article \cite{Francfort:2022}. This has not been done so far. Using polarisation information, together with observed galaxy orientation and morphology, and considering two couples of galaxies, we show how to build an estimator of the coordinate-independent shear correlation functions.

This Part is structured as follows. This introduction is Chapter~\ref{chaplensing:intro}. We present the main theoretical tools in Chapter~\ref{chaplensing:theory}, the most important being the derivation of the rotation angle. In Chapter~\ref{chaplensing:app} we discuss two applications of our formalism: A toy-model where lensing is only due to Schwarzschild lenses, and the Cosmological setup where we build an estimator for the shear correlation functions. We conclude the discussion in Chapter~\ref{chaplensing:conc}.

\chapter{Theoretical tools\label{chaplensing:theory}}
In this Chapter we present the theoretical framework to study lensing. In the first Section, we discuss two ways to transport vectors from one spacetime point to another: The Parallel Transport and the Lie Transport. These concepts are closely related to the Jacobi formalism presented in Section~\ref{seccosmo:jacobi}. Much more can be found on the topic, for example in the reviews of Bartelmann~\cite{Bartelmann:2010} or Perlick~\cite{Perlick:2004} as well as the thesis of Pierre Fleury~\cite{fleury_2015} (especially Chapter~2), a work of excellent overall quality!

In the second Section, we present the physical assumptions of our model regarding the properties of the galaxy, especially their shape and the orientation of the polarisation of the light they emit.

In the third Section, we define a genuine observable to quantify the rotation of the shape of a galaxy due to lensing (and in particular to shear).

\startchap

\section{How to Transport Vectors?}
\subsection{The Screen}
We discuss here in more details the parallel transport of vectors, whose formal definition is given by Eq.~\eqref{eqgr:partra}. The parallel transport is of great relevance in General Relativity as the $4$-velocity of any particle (could it be massive or massless) is parallel transported along itself. In other words, in this theory, trajectories of particles follow geodesics. In the context of lensing, the vector we consider is the photon's $4$-velocity $\bds k$. The parallel transport condition implies 
\be 
\bds \nabla_{\bds k} \bds k =0\,.
\ee 

We want however to be able to describe lensing as modification of the shape of a bundle of light (corresponding for example to a galaxy). In order to do so, we need a way to \emph{print} this shape on a \emph{screen} seen by an \emph{observer}. We would like to define those concepts precisely, but before doing so it is insightful to give a simple example.

Let us consider an observer at rest in Minkowski spacetime in Cartesian coordinates. Their $4$-velocity is given by 
\be
\bds{u}_\mo = \bpt_t\,.
\ee 
They observe a galaxy, whose image is made of several photons. Among them, they chose a reference one (say at the center of the galaxy). This photon comes from the $-\bds z$ direction and its trajectory and $4$-velocity are
\begin{align}
x^\mu(\lambda) &= \omega(\lambda, 0, 0, \lambda)\,, \\
\bds k &= \omega (\bpt_t + \bpt_z)\,,
\end{align}
where $\omega$ is the frequency of the photon as measured by the observer and $\lambda$ an affine parameter such that $x (\lambda_\mo)$ is the position of the observer. Along this trajectory, we define a set of fictitious observers with $4$-velocity
\be 
\bds u (\lambda) = \bpt_t\,.
\ee 
Note that, at the observer position, the $4$-velocity $\bds u_\mo$ and the fictitious velocity are equal, i.e.
\be 
\bds u (\lambda_\mo) = \bpt_t = \bds{u}_\mo\,.
\ee
At each point, we would like to define a screen on which the set of fictitious observers could print the shape of the galaxy centered on the reference photon. Which properties should such screens have? Relevant elements are the following.
\begin{itemize}
    \item The screen should be a $2$-dimensional spatial surface.
    \item It should be in the $(\bds{x}, \bds {y})$-plane to be orthogonal to the photon's trajectory.
    \item It should be spanned by basis vectors whose norm does not change along the trajectory.
\end{itemize}
The first and second conditions agree with the intuitive definition of a screen (How is the screen placed when you go to the movie theater?). The third definition can be intuitively understood. Indeed, if we want to make comparisons between the shapes on the different screens, it makes sense to ensure that the basis vectors do not change size. From these conditions, the only possibility is to assume that the screen is spanned by the two basis vectors (up to a global rotation)
\be 
(\bds{e}_1 , \bds{e}_2 ) = (\bpt_x, \bpt_y)\,.
\ee 
At each event of the photon's trajectory $x^\mu(\lambda)$, a point on the screen (for example the position of a given photon) is given by its coordinates $(\alpha^1, \alpha^2)$ in the $(\bds{e}_1 , \bds{e}_2 )$ basis. Using this specific example, we are now able to give a more general definition of a screen.
\MYdef{Screen}{ \label{deflensing:screen}
Given a photon's trajectory $x(\lambda)$ with its $4$-velocity $\bds k$ and an observer at $x^\mu(\lambda_\mo)$ with $4$-velocity $\bds u_\mo$, a \emph{screen} is a set of fictitious observers with $4$-velocity $\bds u(\lambda)$ along the photon trajectory such that $\bds u(\lambda_\mo) = \bds u_\mo$ together with a set of spatial orthonormal basis  $(\bds{e}_1(\lambda) , \bds{e}_2 (\lambda) )$. Moreover, each orthonormal basis must be defined such that
\begin{align}
\bds{e}_a  \cdot \bds{u}&=0\,, \\
\bds{e}_a  \cdot \bds{k} &=0\,, \\
\bds{\nabla}_{\bds k}\bds{e}_a  &\propto \bds{k}  \,,
\end{align}
where it is understood that these three conditions should hold at each point of the trajectory.

The first condition ensures that the screen is a spatial surface (as the vector $\bds u$ somehow represents the time coordinate of the observers). The third condition guarantees that the norm of the basis vectors do not change (the proof is direct using the second relation). 
}
\MYrem{Quasi-parallel transport}{
In Definition~\ref{deflensing:screen}, the last condition defines the \emph{quasi parallel transport}. It is not possible, in general, for the vectors $\bds{e}_a$ to be purely parallel transported along $\bds k$. However, such a strong condition is not necessary and we present the most general definition. Note also that the $4$-velocities of the fictitious observes $\bds{u}(\lambda)$ must not be parallel transported. More on these details are found in \cite{fleury_2015}.
}
\MYrem{Sachs basis}{
The orthonormal basis $(\bds{e}_1(\lambda) , \bds{e}_2 (\lambda) )$ is called the \emph{Sachs basis} and the screen it spans is called the \emph{Sachs screen}.
}

\begin{figure}[h!t]
	\begin{center}
	\includegraphics[width =0.6\textwidth]{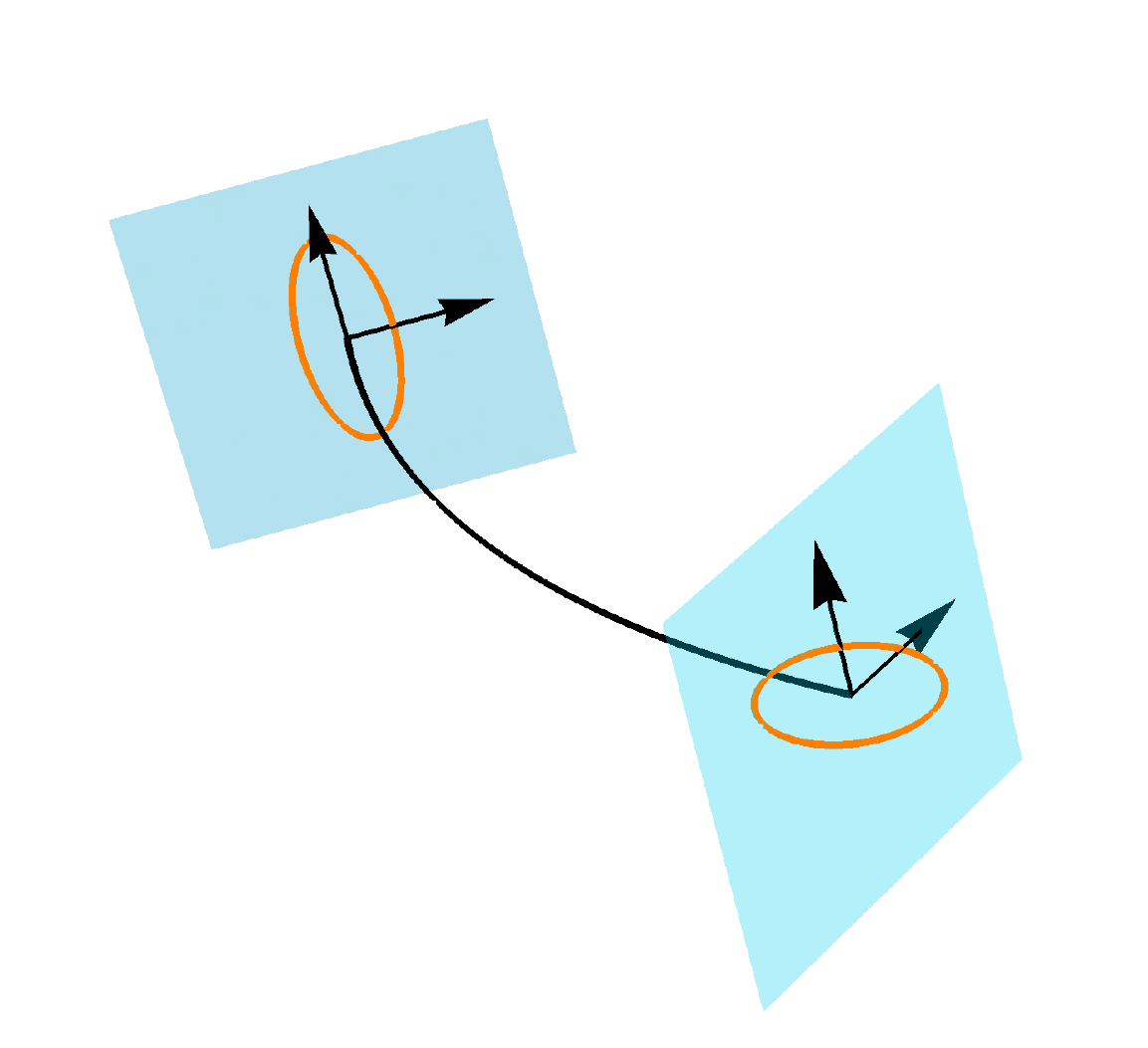} 
	\end{center}
	\caption[The Sachs screen]{\label{figlensing:screen}
	\textbf{The Sachs screen}\\
	The trajectory of the photon is shown in black. At two positions, the screen is represented, together with the orthonormal Sachs basis, represented by the arrows. On the screen, one can \emph{print} the shape of the galaxy as it would be seen by a fictitious observer. The images of, for example, a galaxy are \emph{printed} in orange on the screens, see next Section for more details.}
\end{figure}

The condition on the basis vector to be quasi-parallel transported is very useful. Indeed, recall that the polarisation vector of light is parallel transported along the trajectory, see for example \cite{DiDio:2019}. This implies that, when studying lensing of galaxy shape, using the polarisation vectors as basis vectors is a natural choice as their transport properties are known. We represent schematically in Fig.~\ref{figlensing:screen} the trajectory of a photon and the Sachs formalism.

\subsection{Lie Transport}
We discussed the Sachs formalism which allows us to define a screen, together with a parallel transported basis. We also defined fictitious observers at each point on the trajectory, together with such a screen. Let us assume that the light was emitted by a galaxy. At each point, those observers could project the shape of the galaxy on the screen. We would like to understand how this shape changes along the photon's trajectory. 

The light emitted by the galaxy is made from a bundle of photons. Each of these photons follows a null geodesic. However, because all of these photons are travelling on different spacetime points, we expect the deflection of these trajectories to be different. We can describe this qualitatively using the theory of \emph{geodesic deviation}. 

\begin{figure}[h!]
	\begin{center}
	\includegraphics[width =0.55\textwidth]{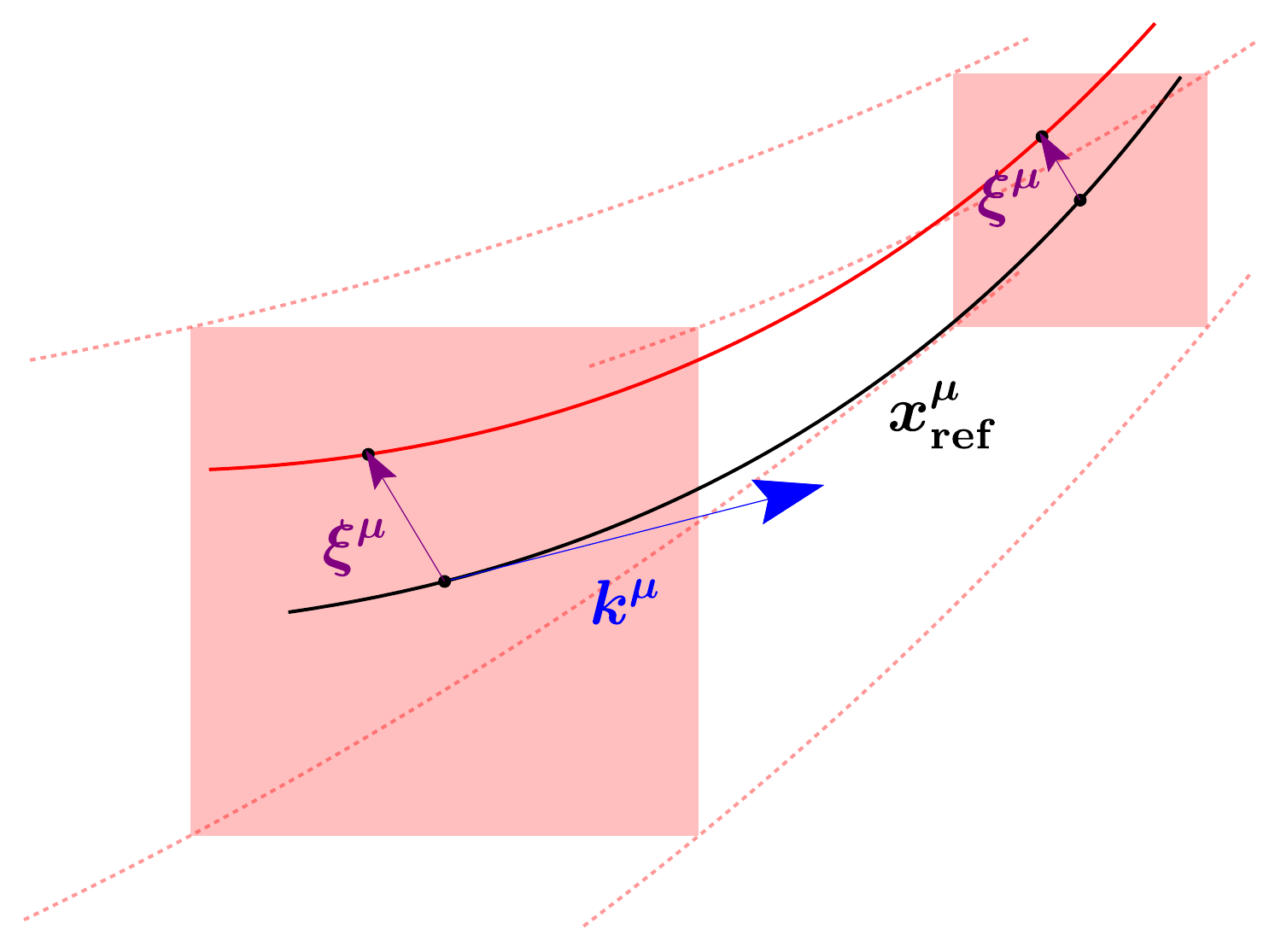}
	\end{center}
	\caption[Deviation vector]{\label{figlensing:deviation}
	\textbf{Deviation vector}\\
	The reference ray is shown in black and its $4$-velocity in blue, while an arbitrary ray is shown in red. The Sachs screen are shown in pink. The deviation vector for two different affine parameters is shown in purple. Image inspired from \cite{fleury_2015}.} 
\end{figure}

Let us consider a bundle of travelling neighbouring photons. One of them is arbitrarily chosen to be the reference one, with trajectory $x_{\mathrm{ref}}^\mu(\lambda)$ and $4$-velocity  $\bds{k}_{\mathrm{ref}}$. For another fixed ray with trajectory $x^\mu(\lambda )$ and trajectory  $\bds{k}$, we define the deviation vector as
\be
\xi^\mu = x^\mu - x_{\mathrm{ref}}^\mu\,.
\ee
Note that $\bds \xi$ can indeed be seen as a $4$-vector on the tangent space at $x_{\mathrm{ref}}^\mu(\lambda)$ as the two rays are considered very close to each other, see Fig~\ref{figlensing:deviation}. We can determine the acceleration of the vector $\bds \xi$, see \cite{Francfort:2021} for more details. The final result is
\boxemph{ \label{eqlensing:geo_devi}
\ddot\xi^\alpha = - \tensor{R}{^\alpha_{\mu\beta\nu}} k^\mu \xi^\beta k^\nu\,,
}
where a dot indicates here a derivative with respect to the affine parameter $\lambda$. This relation is the geodesics deviation equation and it provides a vivid interpretation of the Riemann tensor. We can also show directly that it implies that the Lie Derivative (see Section~\ref{seccosmo:Lie}) of the deviation vector vanishes along the geodesics, i.e.
\boxemph{ \label{eqlensing:LieDer}
\pounds_{\boldsymbol k} {\bds \xi} =0\,. 
}
In other words, the deviation vector is Lie transported along the geodesics. This relation is very important for us as it describes how the shape of a light bundle is deformed when it travels through a curved spacetime. 

\subsection{Comparison with the Jacobi Formalism\label{seclensing:jacobi}}
In the two previous Sections, we have seen two important concepts. First, we defined a screen along the trajectory of light. Second, we stated the geodesics deviation equation describing how the shape of a bundle of light is deformed as it travels through spacetime. We would like to relate those two concepts together and relate them with the Jacobi formalism presented in~Section~\ref{seccosmo:jacobi}.

Recall the Jacobi map $\bds A$ was defined through
\boxemph{
\label{eqlensing:Jacobi_map}
\bds{\alpha}_\me &= \bds{A} \bds{\alpha}_\mo\,, \\
 \label{eqlensing:Jacobi_matrix}
\bds{A} &=\begin{pmatrix}
1-\kappa - \gamma_1 & - \gamma_2 - \omega \\
-\gamma_2  + \omega &   1-\kappa + \gamma_1 
\end{pmatrix}\,.
}
Note that this is the more general expression of the Jacobi matrix (at linear order), where a global (counterclockwise) rotation by an angle $\omega$ is also included. In the relation given by Eq.~\eqref{eqlensing:Jacobi_map}, $\bds{\alpha}_\me $ (resp. $\bds{\alpha}_\mo$) quantifies the position of a given ray at the emission (resp. observer) position. Those positions are defined with respect to the Screen basis. 

For example, in the Cosmological setup of Section~\ref{seccosmo:jacobi}, we have
\begin{align} \label{eqlensing:e1}
\bds{e}_1 &= \bpt_\theta\,, \\
\label{eqlensing:e2}
\bds{e}_2 &= \sin^{-1} \bar \theta \bpt_\varphi\,, 
\end{align}
where $(\bar \theta, \bar \varphi)$ is a reference position, corresponding to the trajectory of the reference light ray. If we consider an observer at rest, it is easy to check that all the conditions in the Definition~\ref{deflensing:screen} are satisfied. This is nothing else as saying that the $2$-dimensional space spanned by $(\bds e_1, \bds e_2)$ is a Sachs screen.

However, the choice given by Eq.~\eqref{eqlensing:e1} and Eq.~\eqref{eqlensing:e2} is just one possibility in this specific Cosmological context. In general, the expression for the Sachs basis can be anything as long as it satisfies the conditions of Definition~\ref{deflensing:screen}.

In other words, the lensing map given by its most general linear expression Eq.~\eqref{eqlensing:Jacobi_matrix} together with the relation given by Eq.~\eqref{eqlensing:Jacobi_map} describe how the shape of images is deformed with respect to a parallel transported basis. This transformation can also, as we described in the previous Section, be quantified by the Lie transport of the images. We will explain how the Jacobi map can be useful in Section~\ref{Seclensing:General_idea}, but first we would like to give a physical interpretation to the vectors $\bds \alpha_\mo$ and $\bds \alpha_\me$.

\subsection{Mapping Angles to Distances}
In the definition of the Jacobi map Eq.~\eqref{eqlensing:Jacobi_map}, the vectors $\bds \alpha_\mo$ and $\bds \alpha_\me$ do not have a clear physical interpretation. We want to discuss here how to give them a better signification.

The vector $\bds \alpha_\mo$ lies on the plane at the observer's position. A sensible way to define it goes as follows. Again, we should choose a reference light ray. For any other ray, the norm of the vector $\bds \alpha_\mo$ is the angle it forms with the reference ray, and its orientation is simply given by the relative position of the ray with respect to the reference ray. This defines the vectors $\bds \alpha_\mo$ without the ambiguity of a global rescaling.

Regarding the vector at the emission $\bds \alpha_\me$, we define $\bds \xi_\me$ to be the physical distance between the ray and the reference ray as
\be \label{eqlensing:def_xi} 
\bds \xi_\me= (1+\kappa) \DA \bds \alpha_\me\,,
\ee 
where $\DA$ is a quantity with units of length we would like to determine. Combining the definition of Jacobi map Eq.~\eqref{eqlensing:Jacobi_map} and the definition of $\bds{\xi}_\me$, we get
\be \label{eqlensing:xi_alpha}
\bds \xi_\me= (1+\kappa) \DA \bds A \bds \alpha_\me\,.
\ee 
We assume that, at the observer, the bundle of vectors $\bds{\alpha}_\mo$ form a solid angle $\Omega_\mo$ and that at the emission, the rays form a physical surface of area $\mathcal{A}_\me$. Using the relation between $\bds \xi_\me$ and $\bds \alpha_\me$, we obtain at linear order
\be 
\mathcal{A}_\me = (1+2\kappa) \DA^2 \vert \bds A \vert \Omega_\mo\,.
\ee 
Using the definition of the Jacobi map given by Eq.~\eqref{eqlensing:Jacobi_map}, we have  $\vert \bds A \vert  = 1-2\kappa$, which implies 
\be 
\mathcal{A}_\me =  \DA^2  \Omega_\mo\,.
\ee 
This is exactly the definition for $\DA$ of the angular diameter distance presented in Section~\ref{secintro:distances}. This gives a physical interpretation to the length $\DA$. Combining these results yields
\boxemph{
\bds \xi_\me = \bds D \bds \alpha_\mo\,,
}
where we defined the dimensional Jacobi matrix
\be  \label{eqlensing:jacobi_dimensional}
\bds D = (1+\kappa)\DA \bds{A} \equiv \bar{D}_{\mathrm A} \bds A\,,
\ee 
and where $\bar{D}_{\mathrm A}$ is the non-perturbed angular diameter distance of the source.
\MYrem{Vocabulary}{
In general, the matrix $\bds D$ is what people called the \emph{Jacobi matrix}. However, here, we will only need its dimensionless counterpart $\bds A$, which we call the \emph{Jacobi matrix} with a slight vocabulary abuse.
}
The evolution of the dimensional Jacobi equation (more details can be found in \cite{fleury_2015,Francfort:2021, Perlick:2004} for example) is given by
\boxemph{
\label{eqlensing:jacDdotdot}
&\ddot{\bds D} =  \bds R \jacD \,,  \\
\label{eqlensing:jacD0}
&\bds D_\mo = 0 \,, \\
\label{eqlensing:jacDp0}
&\dot{\bds D}_\mo =- \mathbbm{1}\,,
}
where the two last conditions are evaluated at the observer position and $\mathbbm{1}$ is the identity matrix. A dot denotes derivation with respect to the future-oriented affine parameter $\lambda$. Recall that $\lambda_\ms <0$ at the source position (and $\lambda_\mo =0$ at the observer), hence the negative sign, as we intuitively want $\bds \alpha_\mo$ and $\bds \xi_\me$ to be in the same direction. The matrix $\bds R$ is given by
\begin{align} \label{eq:RPhiPsi}
\bds R&=\left(\begin{array}{cc}
\Phi_{00}&0\\
0&\Phi_{00}
\end{array}
\right)+
\left(
\begin{array}{cc}
-\operatorname{Re}(\Psi_0)&\mathrm{Im}(\Psi_0)\\
\mathrm{Im}(\Psi_0)&\mathrm{Re}(\Psi_0)
\end{array}
\right)\,,
\end{align}
with
\begin{align}
\label{eqlensing:Psi0}
\Psi_0&=-C_{\mu\nu\rho\sigma}k^{\nu}k^{\rho}\overline{m}^{\mu}\overline{m}^{\sigma}\,, \\
\label{eqlensing:Phi00}
\Phi_{00}&=-\frac{1}{2}R_{\mu\nu}k^{\mu}k^{\nu}\,, \\
\boldsymbol{m} &= \frac{\bolde_1 + \ii \bolde_2}{\sqrt{2}}\,.
\end{align}
where an overbar indicates complex conjugation, $\bds k$ is the photon $4$-momentum,  $\tensor{R}{_\mu_\nu}$ is the Ricci tensor and $C_{\mu\nu\rho\sigma}$ the Weyl tensor. The relations Eq.~\eqref{eqlensing:jacDdotdot}, Eq.~\eqref{eqlensing:jacD0} and Eq.~\eqref{eqlensing:jacDp0} allow in principle to determine completely the Jacobi matrix (if the trajectory of the photon and the Sachs basis are known).

\section{Physical Assumptions}
\subsection{Shape of Galaxies}
We would like to use the shape of galaxies to study lensing (particularly the shear). To this aim, we need to make a few assumptions about the shape of the galaxies.

Our main assumption is that galaxies have an elliptical shape in $3$ dimensions. When we observe a galaxy, we actually see its $2$-dimensional projection on the Sachs screen at our position. However, it it intuitively understood that the projection of a $3$-dimensional ellipse on a $2$-dimensional plane is still an ellipse. If the semi-major, respectively semi-minor, axis has size $R$, respectively $r$, the eccentricity of the ellipse is 
\be 
\varepsilon = \sqrt{1- \frac{1}{a^2}}\,,
\ee 
where
\be
a = \frac{R}{r}
\ee 
is the aspect ratio of the ellipse.
\MYrem{Eccentricity}{Note that the eccentricity of a given ellipse is invariant under a scaling of the ellipse given by $(r,R)=\lambda(\tilde r, \tilde R)$ with $\lambda>0$. This will prove useful as we can simply use the observed shape on a screen whose size is much smaller than the size of a galaxy! Moreover, note that $\varepsilon=0$ corresponds to a circle while for $\varepsilon\rightarrow  \infty $, the ellipse becomes a straight line.}

\subsection{Polarisation of Light}
We also would like to use the direction of the polarisation of light to extract information about the true shape of the galaxy. This information has already been proposed or used to be an interesting observable, see for example \cite{Brown:2010, Burns:2004, Mahatma:2020, Sereno:2004,Sereno:2004b,Whittaker:2014}. More precisely, it has been already been mentioned  that light polarisation is aligned with the principal axes of the galaxy, see~\cite{Huff:2013,Brown:2010,Brown:2011}. The explanation is that the dominant source of polarisation is the synchrotron polarisation emitted by the electrons moving because of the magnetic field of the galaxy. The dominant component of this field is in the galactic plane and is in general aligned with the galaxy semi-major axis, see \cite{Gardner:1966}. The polarisation of the light emitted by the rotating electrons is orthogonal to the magnetic field and to the line of sight, hence it is aligned with the semi-minor axis. Note that we can only consider galaxies  where the emitted frequency is around $\SI{50}{\Ghz}$, as lower frequency light is depolarised due to Faraday rotation \cite{Kronberg:1993}. In \cite{Stil:2008}, the observable correlation is explicitly shown, with a typical difference of $\SI{5}{\degree}$ between the two vectors.
\begin{figure}[h!t]
	\begin{center}
	\includegraphics[width =0.55\textwidth]{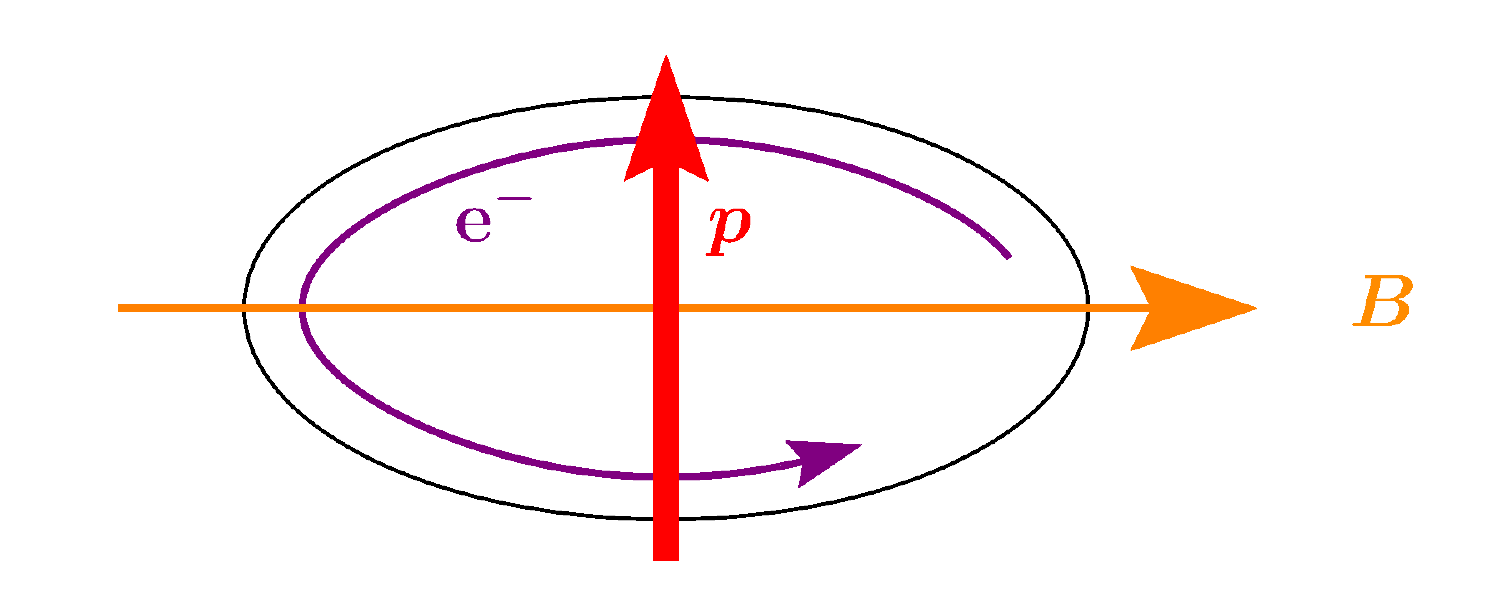}
	\end{center}
	\caption[Polarisation of galaxies]{\label{figlensing:polarisation}
	\textbf{Polarisation of galaxies}\\
	The black ellipse represents the galaxy. The magnetic field $\bds B$ is aligned with the semi-major axis, while the polarisation vector $\bds p$ of the emitted light is aligned with the semi-minor axis.} 
\end{figure}

From now on, we take this observation as granted, and also assume that it holds at the observed level, i.e. that the projected polarisation on the screen is aligned with the projected principal axes of the galaxy shape. Moreover, note that, strictly speaking, the polarisation is defined to be the direction of the electric field. This direction is actually aligned with the semi-minor axis, but this is not relevant as it can be defined up to a rotation of $\SI{90}{\degree}$.

\section{Rotation of Ellipses}
\subsection{General idea\label{Seclensing:General_idea}}
We now have everything to explain the main idea of this Chapter. Let us summarise what we learned. First, the shape that we observe in the sky can be mathematically described as a set of points on a $2$-dimensional screen. Second, the axes of this screen have to be parallel transported. Third, the deformation of the shape on this screen with respect to this basis is described by the Jacobi formalism, presented in Section~\ref{seccosmo:jacobi}. Fourth, the polarisation of the light is parallel transported and is aligned with the main axes of the galaxy at the emission position.

\begin{figure}[h!t]
	\begin{center}
		\includegraphics[width =0.4\textwidth,height=0.4\textheight]{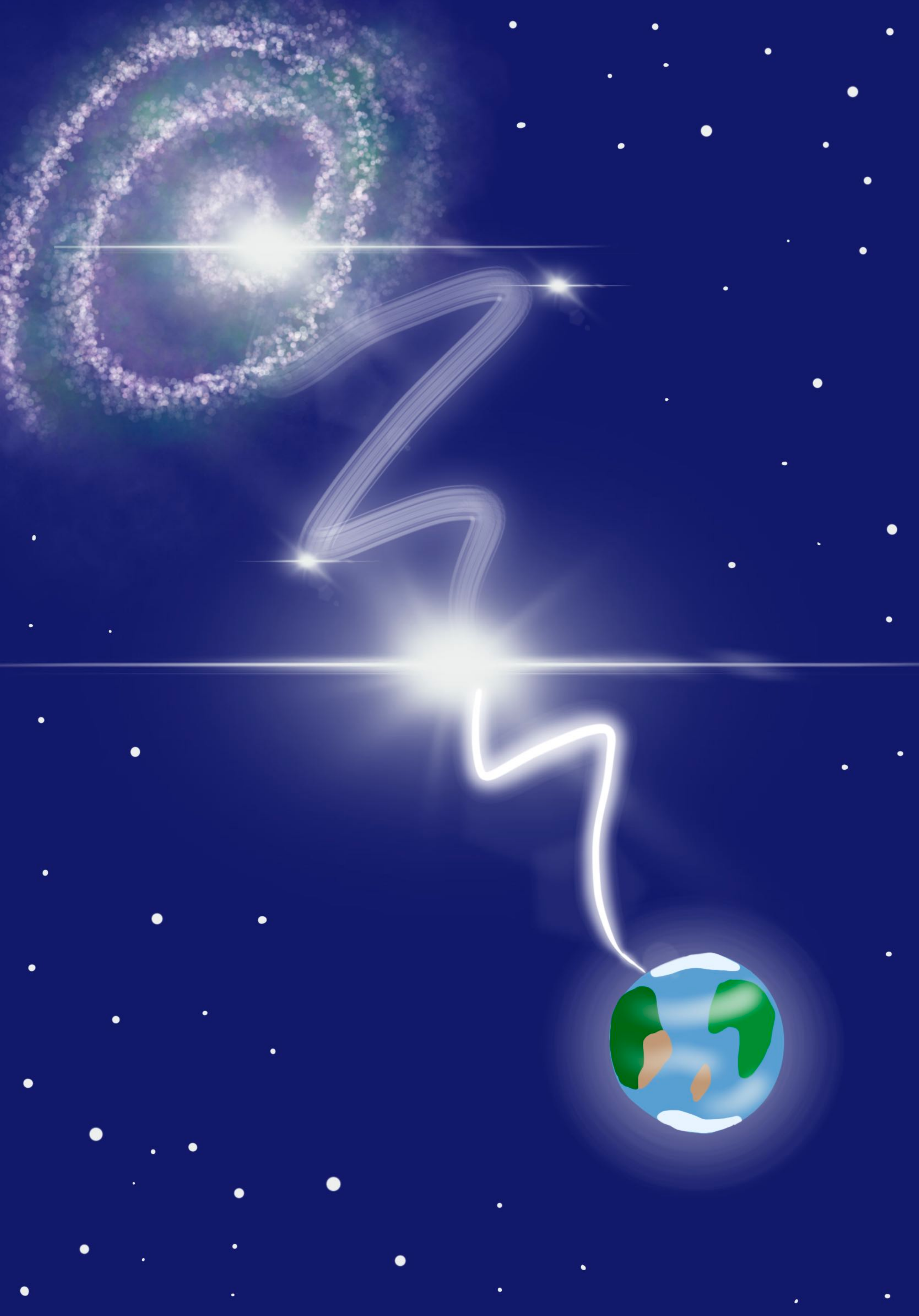} 
	\end{center}
	\caption*{\label{figlensing:jess}
		\textbf{Lumi\`ere propag\'ee et d\'evi\'ee}, Jessica Mascellino
	} 
\end{figure}

In other words, the polarisation vectors (e.g. the electric and magnetic field) are parallel transported along the light trajectory and can be considered as basis vectors for the screen. At the emission position, these vectors are aligned with the main axes of the galaxy. If there were no lensing, say there is only you and this far away galaxy, the shape of the galaxy will not be modified and will stay the same with respects to the Sachs basis. Hence, at the observed position, the polarisation vectors would still be aligned with the main axes of the galaxy. In a more realistic situation, the Universe is filled with matter between this far away galaxy and the Earth. The shape of the galaxy is Lie transported, which is different from the parallel transported undergone by the polarisation vectors. Hence, at the observed position, the polarisation vector does not coincide anymore with the main axes! Such a discrepancy is a true signature of lensing and this is what we would like to quantitatively study in the next Section. 

\subsection{Rotation of the Principal Axis}
We would like to study the rotation of the main axes of an ellipse under the Jacobi transformation given by 
\begin{align} \label{eqlensing:alphae}
\bds{\alpha}_\me &= \bds A \bds{\alpha}_\mo\,, \\
\bds{\xi}_\me &= (1+\kappa) \DA \bds A \bds{\alpha}_\mo \quad \mathrm{with} \\
\bds{A} &= \begin{pmatrix}
1-\kappa - \gamma_1 & - \gamma_2 - \omega  \\
-\gamma_2 + \omega &   1-\kappa + \gamma_1 
\end{pmatrix}\,, \quad \mathrm{and} \\
\bds \xi_\me &= (1+\kappa) \DA \bds \alpha_\me\,.
\end{align}
where $\bds{\alpha}_\me$ (resp. $\bds{\alpha}_\mo$) represent the position of one light on the screen at the emission (resp. observer) position, and $\bds{\xi}_\me$ is the physical distance vector at the emission position.

Let us take a step back and describe qualitatively what we want. The parameters at hand are the convergence $\kappa$, and the shear $\gamma$ together with the angle $\chi$ it forms with the Sachs basis, or equivalently its components defined as 
\begin{align}
   \gamma_1 &= \gamma \cos 2 \chi \,, \\
   \gamma_2 &= \gamma \sin 2 \chi\,.
\end{align}

When we started the project, with Giulia and Ruth, we thought that the relevant quantities was the net rotation $\omega$. However, we soon realised that it vanishes in the Schwarzschild case, and that it is of second order in the Cosmological setup. We were a bit puzzled. Indeed, if there is no net rotation in the Jacobi matrix, does this mean that the image of the galaxy does not rotate neither? If this were true, there would be no hope to observe a rotation of the principal axes with respect to the polarisation direction.

\begin{figure}[h!t]
	\begin{center}
	\includegraphics[width =0.55\textwidth]{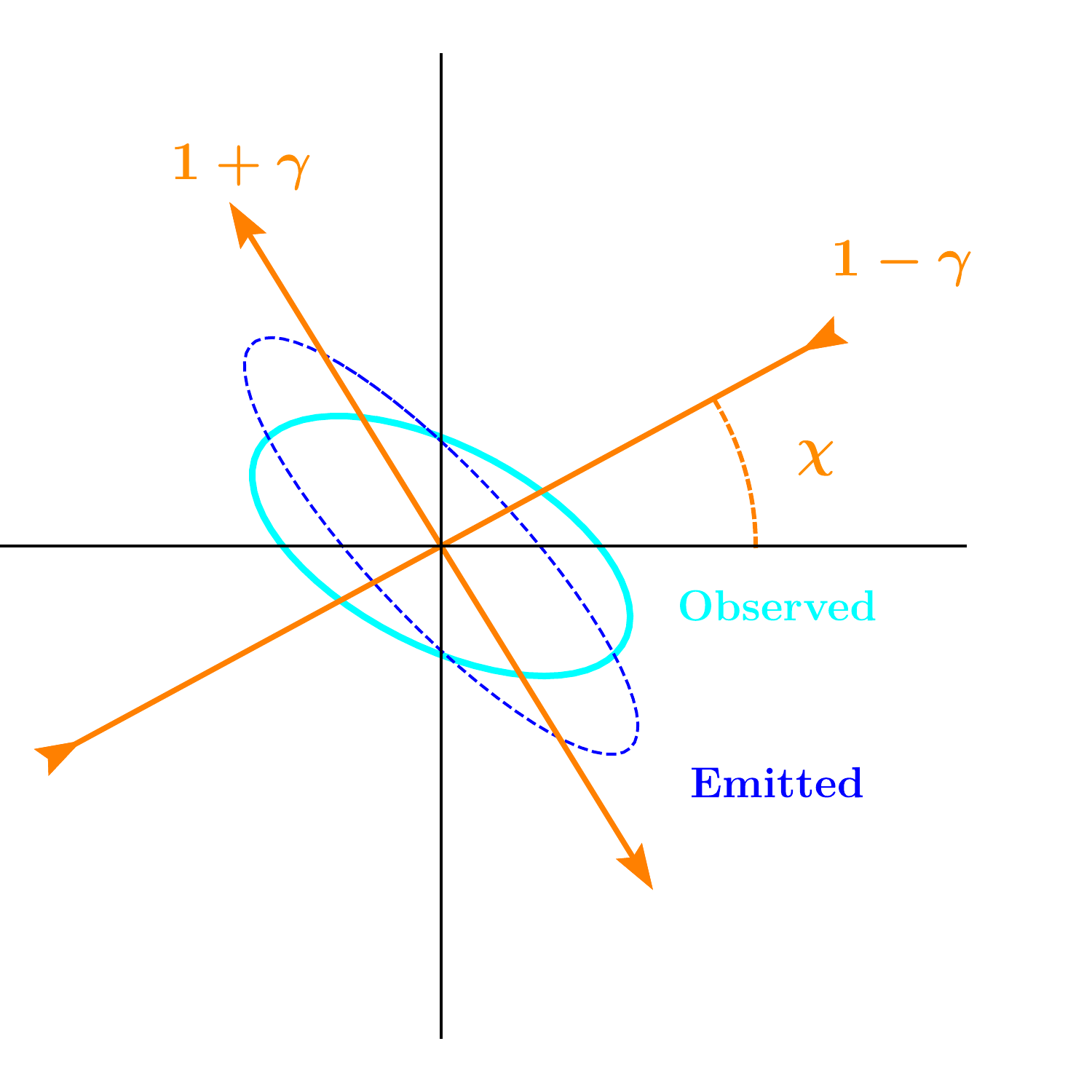}
	\end{center}
	\caption[Rotation from shear]{\label{figlensing:shear_effect}
	\textbf{Rotation from shear}\\
	The principal axes of the shear do not coincide with the main directions of the observed ellipse (in cyan, full line). The source ellipse (in blue, dashed) is obtained applying the Jacobi map. The semi-major axis is rotated under this process!} 
\end{figure}

However, we realised soon after that an ellipse can undergo a rotation because of the shear, provided its direction do not coincide with the axes of the ellipse. Indeed, in this situation, the ellipse would be twisted and the position of its axes would rotate! This is shown schematically in Fig.~\ref{figlensing:shear_effect}.
We want to describe this quantitatively. We assume that the galaxy shape forms an ellipse at the observer. In other words, the vector $\bds{\alpha}_\mo$ satisfies 
\be  \label{eqlensing:condalphao}
\bds{\alpha}_\mo^{\mathrm T} \bds{M}_\mo \bds{\alpha}_\mo=1\,,
\ee 
with 
\be  \label{eqlensing:matrixo}
\bds{M}_\mo =  \begin{pmatrix}
r_\mo^{-2}&0 \\
0& R_\mo^{-2}
\end{pmatrix}\,.
\ee 
It is direct to show that the condition Eq.~\eqref{eqlensing:condalphao} together with Eq.~\eqref{eqlensing:matrixo} represent an ellipse with semi-minor (dimensionless) axis $r_\mo$ in the $\bds e_1$ direction and semi-major (dimensionless) axis $R_\mo$ in the $\bds e_2$ direction.

To study the shape of the galaxy at the emission position, we use Eq.~\eqref{eqlensing:alphae} and plug it into Eq.~\eqref{eqlensing:condalphao} to get
\be 
\bds{\alpha}_\me^{\mathrm T} \bds{M}_\me \bds{\alpha}_\me=1\,,
\ee 
with
\be 
\bds{M}_\me = (\bds{A}^{-1})^{\mathrm T}  \bds{M}_\mo \bds{A}^{-1} \,.
\ee 
The eigensystem of the matrix $\bds{M}_\me$ contains all the information about the shape and the orientation of the ellipse at the emission. Generalising the case of a diagonal matrix given by $\bds{M}_\mo$, we infer that the eigenvectors of $\bds{M}_\me$ correspond to the direction of the main axes while the eigenvalues $\lambda_{\pm}$ are related to their length via $\lambda_+=R_\me^{-2}$ and $\lambda_-=r_\me^{-2}$. A more formal proof can be given considering the usual diagonal decomposition of a symmetric matrix under an orthogonal change of coordinates.

At linear order in $\kappa$, $\gamma_1$ and $\gamma_2$, the computations can be done using the standard techniques to determine eigenvalues and eigenvectors. We present only the results here. At the emission, we get
\boxemph{
R_{\me} &= \DA R_{\mo} \left( 1+ \gamma_1 \right)\,, \\
r_{\me} &= \DA r_{\mo} \left( 1- \gamma_1 \right)\,, \\
\delta \alpha &= \gamma_2 \left( \frac{2-\varepsilon_\mo}{\varepsilon_\mo^2}\right)\,,
}
where $R_{\me}$ and $r_{\me}$ are the main axes of the ellipse tilted with an angle $\delta \alpha$ with respect to the Sachs basis, and 
\be 
\varepsilon_\mo = \sqrt{1-\frac{r_\mo^2}{R_\mo^2}}
\ee 
is the eccentricity at the observer. Moreover, the eccentricity at the emission is 
\be 
\varepsilon_\me = \varepsilon_\mo + 2 \gamma_1 \frac{1- \varepsilon_\mo^2}{ \varepsilon_\mo}\,.
\ee 

Here, we considered an ellipse at the observer whose main axes coincide with the Sachs basis. We can be even more general and assume that the semi-minor axis forms an angle $\beta_\mo$ with the horizontal axes. The formulas can be easily adapted. In this general case, the formulas are
\boxemph{
\label{eqlensing:Re_fin}
R_\me &= \DA R_\mo (1+\gamma \cos (2\chi-2\beta_\mo)  )\,, \\
\label{eqlensing:re_fin}
r_\me &=\DA  r_\mo (1-\gamma \cos(2\chi-2\beta_\mo) )\,, \\
\label{eqlensing:delta_alpha_fin}
\delta \alpha &= \gamma  \sin(2\chi-2\beta_\mo) \frac{2-\varepsilon_\mo}{\varepsilon_\mo^2}\,, \\
\label{eqlensing:varepsi_fin}
\varepsilon_\me &= \varepsilon_\mo  + 2 \gamma \cos(2\chi-2\beta_\mo) \frac{1- \varepsilon_\mo^2}{ \varepsilon_\mo}\,. 
}

\subsection{Physical Interpretation}
We can give a physical interpretation to the formulas given by Eq.~\eqref{eqlensing:Re_fin}, Eq.~\eqref{eqlensing:re_fin}, Eq.~\eqref{eqlensing:delta_alpha_fin} and Eq.~\eqref{eqlensing:varepsi_fin}. 
The first thing to note is that all the formulas only depend on the difference $\chi-\beta_\mo$. Recall that $\chi$ parametrises the orientation of the shear, while $\beta_\mo$ quantifies the orientation of the semi-minor axis with respect to $\bds e_1$ on the Screen space. When one set of Sachs screen has been chosen along the light trajectory, it is always possible to rotate all the screens by a constant angle. By doing so, the orientation of the shear $\chi$ and the position of the observed galaxy $\beta_\mo$ will both be modified, however their difference will be kept constant. Hence, we expect all the observables, such as the eccentricity and the rotation of the main axes, to be invariant under such a change.

Second, we note that if 
\be
\chi - \beta_\mo =  n\frac\pi 2\,,
\ee
with $n\in \mathbb Z$, then $\delta \alpha=0$. This condition simply means that the axes of the shear are aligned with the main axes of the galaxy. In such a situation, the galaxy is only distorted, but its direction is not affected, as discussed previously.

The opposite case happens if
\be
\chi - \beta_\mo =\frac \pi 4+  n\frac\pi 2\,.
\ee
In this case, the eccentricity is not affected: The ellipse undergoes a pure rotation and its size is not modified. 

One last comment can be made about the area of the ellipse. At the emission, it is given by
\be
\mathcal{A}_\me = \pi R_\me r_\me = \DA^2 \mathcal{A}_\mo\,,
\ee 
with $\mathcal{A}_\mo= \pi R_\mo r_\mo $, as expected from the definition of the angular diameter distance.

To conclude this Section, we want to stress that among all the formulas, we will only be interested about the rotation of the main axis of the ellipse $\delta \alpha$. Indeed, as explained above, the polarisation vector is aligned with the shape of the galaxy at the emission. Hence, the difference angle $\delta \alpha$ is indeed a physical observable at the observer, provided the polarisation direction can be measured. However, the rotation angle has a non-trivial dependence on the eccentricity which is not well-defined around $\varepsilon_\mo=0$ (corresponding to a circle). It will be more convenient to change slightly its definition.

\MYdef{Scaled rotation}{
We define the \emph{scaled rotation} as
\be 
\Theta \equiv \delta \alpha \frac{\varepsilon_\mo^2}{2-\varepsilon_\mo^2}\,.
\ee 
}
It can be shown that its error is well-defined for an ellipse with $\varepsilon\approx 0$, see \cite{Francfort:2021} for more details. Moreover, note that the scaled rotation is a true observable: $\delta \alpha$ can be measured as explained above, and the eccentricity $\varepsilon_\mo$ can be determined if the ratio of the axes is known, which is certainly possible if one sees the shape of the galaxy. Using some trigonometric identities, and the definition of the shear components, we obtain the final formula for the scaled rotation
\boxemph{ 
\label{eqlensing:delta_alpha_good}
\Theta = \gamma_2 \cos 2 \beta_\mo - \gamma_1 \sin 2 \beta_\mo\,.
}
The expression for the rotation of an ellipse given by Eq.~\eqref{eqlensing:delta_alpha_good} is the main relation of the paper. This relation will become very important in Section~\ref{seclensing:cosmo} when we use it in a Cosmological context to study the statistical properties.

\stopchap

\chapter{Applications\label{chaplensing:app}}
In this Chapter we present two applications of the results derived previously, especially the formula for the scaled rotation. In the first Section, we apply the formalism to the Schwarzschild setup, where we can compute the shear at linear order in the mass of the lens. We present various toy models in which we estimate the statistics of the scaled rotation for different redshifts of the source, with a foreground constituted of a distribution of several Schwarzschild lenses.

In the second Section, we turn our attention to a cosmological setup to refine this model. The scaled rotation is a physical observable allowing us to probe the cosmic shear between the observer and the source. We build an observable to compute the shear correlation function and discuss the number of galaxies needed to obtain a final result with an acceptable signal-to-noise ratio.

\startchap

\section{Schwarzschild Lens\label{seclensing:SCH}}
\subsection{Setup}
We consider a Schwarzschild lens, whose metric is given in the usual Spherical Coordinates by
\be 
\bds g = 
- \left(1 - \frac{2m}{r} \right) \bds\dd  t^2
+\left(1 - \frac{2m}{r} \right)^{-1} \bds \dd r^2 
+ r^2 \bds \dd \Omega^2\,.
\ee 
We defined $m=\GN M$, $M$ being the mass of the lens. The Schwarzschild radius is given by $2m$. The lens could be for example a supermassive Black Hole, or a whole galaxy.

\begin{figure}[h!t]
\begin{center}
\includegraphics[width =0.49\textwidth]{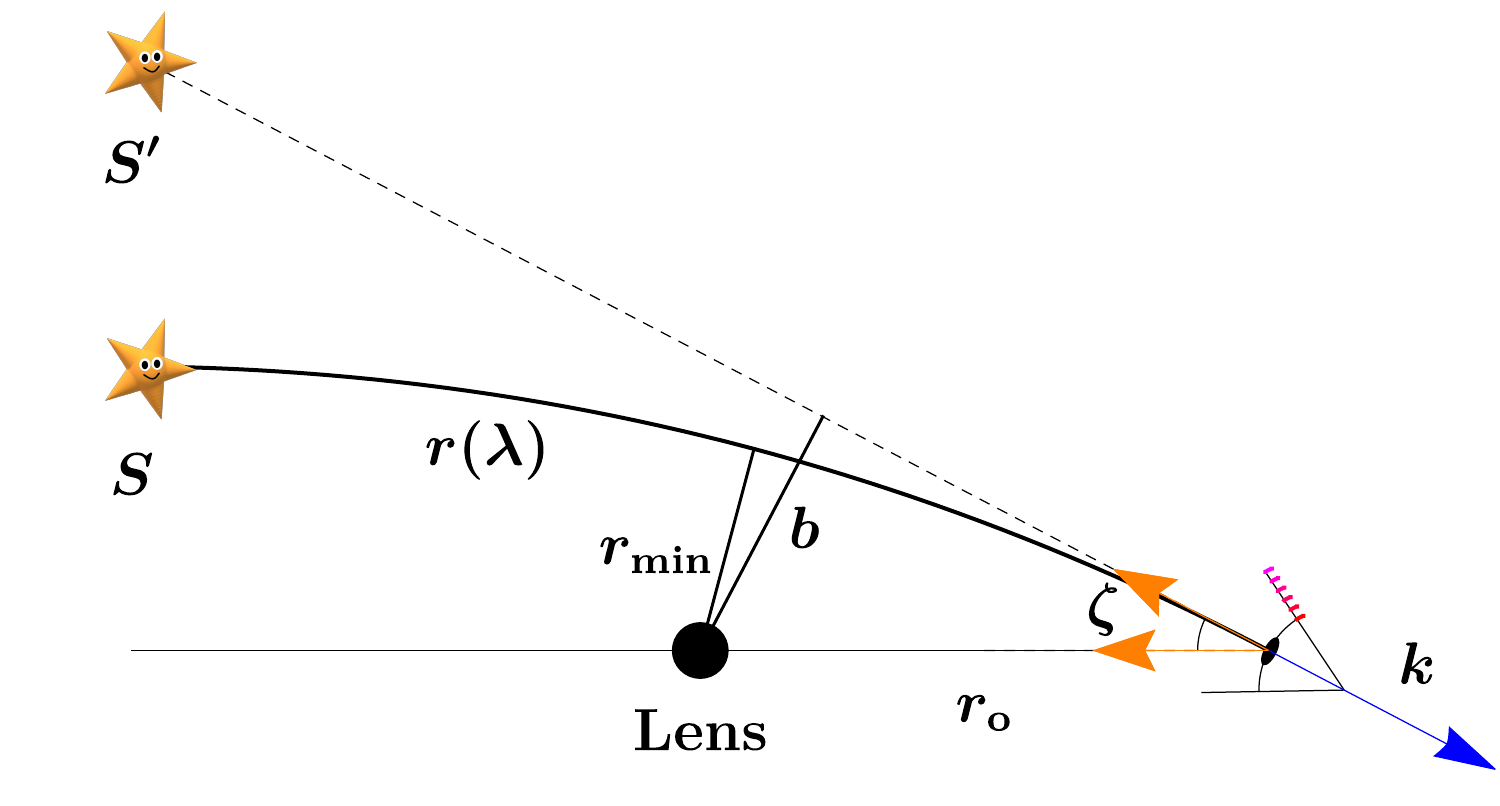} 
\includegraphics[width =0.49\textwidth]{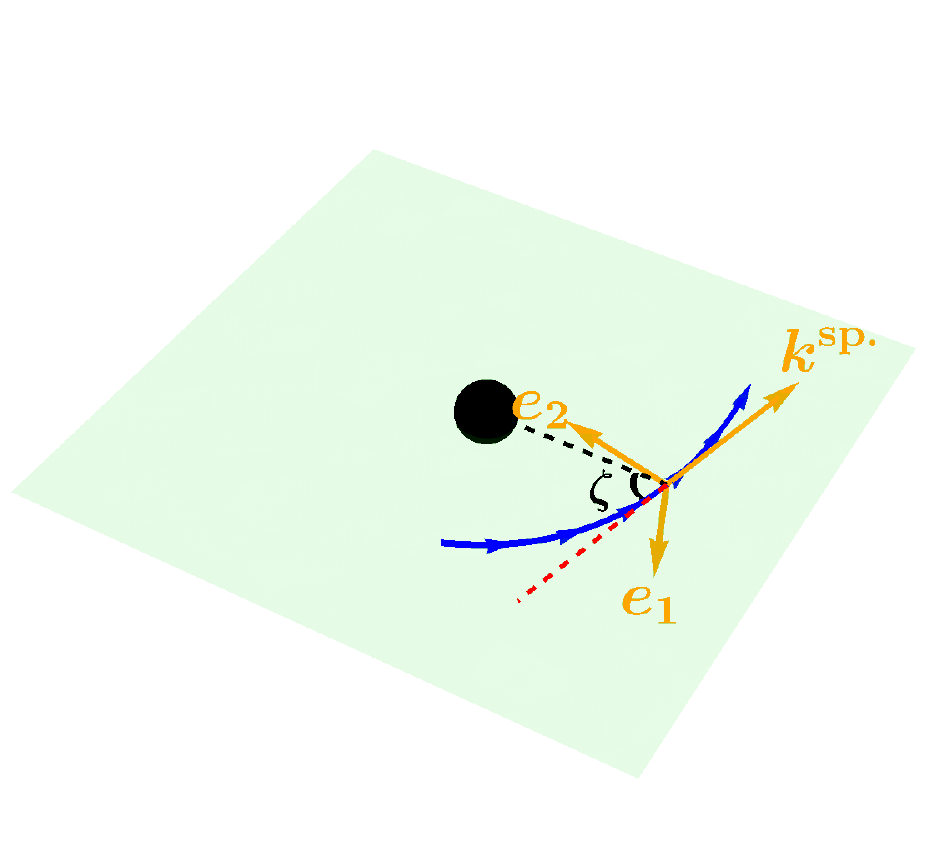} 
\end{center}
	\caption[Lensing around a Schwarzschild Lens]{\label{figlensing:SCH_setup}
	\textbf{Lensing around a Schwarzschild Lens}\\
	\textbf{Left:} The position of the star is $S$ and its apparent position is $S^\prime$. \\
	\textbf{Right:} The Sachs basis is $(\bolde_1, \bolde_2)$ and $\bds{k}^{\mathrm{sp.}}$ is the spatial part of the $4$-momentum.\\
	Figures taken from \cite{Francfort:2021}.
	} 
\end{figure}

We consider the setup shown in Fig.~\ref{figlensing:SCH_setup}, taken from \cite{Francfort:2021}. The observer sees a galaxy (or a star) at $S^\prime$, while the \emph{true} position is at $S$. The $4$-momentum of the photon is
\begin{align} \label{eqlensing:kmu}
\bds k 
&=
\frac{E}{1-\frac{2m}{r}}\bpt_t +\dot r\bpt_r +\frac{L}{r^2}\bpt_\varphi\,, \\
\dot{r}^2 &= E^2 - \frac{L^2}{r^2} \left( 1-2 \frac mr\right)\,,
\end{align}
where $E$ is the dimensionless energy and $L$ its angular momentum, both of which are conserved. Here, a dot denotes a derivative with respect to an affine parameter $\lambda$. As usual, we set $\theta = \frac \pi 2$. The distance between the observer and the lens is $r_{\mo}$, and $\zeta$ is the angle between the lens direction and the incoming direction, as seen by the observer. The impact parameter of the unperturbed trajectory with the same angle $\zeta$ is $b$, while the \emph{true} impact parameter is $r_{\mathrm{min}}$. The angle $\zeta$ is given by
\be  \label{eqlensing:anglezeta}
\sin \zeta = \frac{b}{r_\mo} \sqrt{1- 2 \frac{m}{r_\mo}}
\approx \frac{b}{r_\mo}  \,.
\ee
We consider a family of static observers along the trajectory whose $4$-velocities are
\be 
\boldsymbol{u} =\frac{1}{\sqrt{ 1- \frac{2m}{r}}} \bpt_t\,.
\ee 
The impact parameter is given by (see for example \cite{Renzini:2017} for more details)
\be
r_\textrm{min.} = b \left( 1 - \frac{m}{b} + \mathcal O\left[\left(\frac{m}{b}\right)^2\right] \right).
\ee

\subsection{The Jacobi map}
We would like now to compute the Jacobi map at linear order in the parameter $m/b$. In other words, we assume that the light is travelling sufficiently far from the lens. For example, if we assume that the Schwarzschild lens is a foreground galaxy, we require that the light emitted by a background galaxy does not pass close to the foreground one. This is certainly a good assumption: If this would not be the case, the background galaxy could not even be observed.

The Sachs basis is given by 
\begin{alignat}{2} \label{eqlensing:e1SCH}
    \bolde_1& =r^{-1}\bpt_\theta\,, &&\\
    \label{eqlensing:e2SCH}
    \bolde_2 &= \rho_r \bpt_r + \rho_\varphi \bpt_\varphi\,, && \mathrm{with} \\
    \rho_r &= \frac{b (2 m-r)}{r^2}\,, && \mathrm{and} \\
    \rho_\varphi &= \pm \frac{1}{r} \sqrt{1- \frac{b^2}{r^2} + \frac{2mb^2}{r^3}} \,.&&
\end{alignat}
It is straightforward to check that these choices satisfy the conditions given in Definition~\ref{deflensing:screen}. The first Sachs vector is quite easy to guess as the trajectory lies in the Equatorial plane. The Ansatz for the second vector is taken to ensure $\bolde_1 \cdot \bolde_2 =0$ and the specific values for $\rho_t$ and $\rho_\varphi$ are found using $\bolde_2 \cdot \bds k =0$ and $\bolde_2 \cdot \bolde_2=1$. The Sachs basis is shown in Fig.~\ref{figlensing:SCH_setup}.


We now determine the dimensional Jacobi matrix At zeroth order in the parameter $m/b$, it is direct to shot that the solution of Eq.~\eqref{eqlensing:jacDdotdot}, Eq.~\eqref{eqlensing:jacD0} and Eq.~\eqref{eqlensing:jacDp0} is
\be 
\bds D = -
\begin{pmatrix}
\lambda &0 \\ 0 & \lambda 
\end{pmatrix}\,,
\ee 
as at this order the spacetime is flat and $\Psi_0=\Phi_{00}=0$. At linear order, the matrix $\bds R$ can be explicitly computed. Note that, as $\Psi_0$ and $\Phi_{00}$ are already first order quantities, it is sufficient to take the background values for $\bds k$ and $\bds e_a$ in Eq.~\eqref{eqlensing:Psi0} ans Eq.~\eqref{eqlensing:Phi00}, i.e. setting $m=0$ in Eq.~\eqref{eqlensing:kmu} and Eq.~\eqref{eqlensing:e2SCH}. The final result is
\begin{align}
\bds R &= \Psi_0
\begin{pmatrix}
-1&0 \\ 0 & 1
\end{pmatrix}\,, \quad \mathrm{with} \\
\Psi_0 &= \frac{3mb^2}{r^5}\,.
\end{align}
Due to the symmetry of the matrix $\bds R$, we pose the following Ansatz for the dimensional Jacobi matrix
\be  \label{eqlensing:Jacobi_ansatz}
\bds D = -
\begin{pmatrix}
\lambda + m f^{(1)}&0 \\ 0 & \lambda-m f^{(1)}
\end{pmatrix}\,.
\ee 
The function $f^{(1)}$ satisfies the following differential equation
\begin{align} \label{eqlensing:f1}
    m \ddot{f}^{(1)} &= - \lambda \Psi_0\,, \\
    f^{(1)}_0&=0\,, \\
    \dot{f}^{(1)}_0&=0\,.
\end{align}
As the function $f^{(1)}$ is a first-order perturbation, it is sufficient to use the background solution for $r$ in the expression for $\Psi_{0}$. It is given by 
\be \label{eq:rla}
r(\lambda ) = \sqrt{\lambda^2+r_\mo^2 +2\lambda \sqrt{r_\mo^2-b^2}} \,, 
\ee 
which is nothing else as the equation of a straight line starting at $r_\mo$ and with impact parameter $b$. The equation for $f^{(1)}$ Eq.~\eqref{eqlensing:f1} can be solved analytically but the full solution does not give any physical insight. We will assume that the impact parameter is much small than the distances at hand
\begin{align} 
\label{eqlensing:brs}
    b &\ll r_\ms\,, \\
    \label{eqlensing:bro}
    b &\ll r_\mo\,,
\end{align}
where $r_\ms$ is the distance between the lens and the source. In this regime, the solutions reads
\be
m f^{(1)}(\lambda)= - m \frac{4r_\mo}{b^2} (r_\mo + \lambda) + \mathcal{O}(b^2)\,.
\ee 
Recalling the decomposition of the dimensional Jacobi matrix Eq.~\eqref{eqlensing:jacobi_dimensional}, we get
\be 
\bds D = -\lambda
\begin{pmatrix}
(1-\gamma)&0 \\ 0 &  (1+\gamma)
\end{pmatrix}\,,
\ee 
where $-\lambda_\ms >0$ is the angular diameter distance to the source (recall $\lambda<0$ between the source and the observer). Comparing this expression with the Ansatz Eq.~\eqref{eqlensing:Jacobi_ansatz} yields to the final expression for the shear
\boxemph{ \label{eqlensing:gamma_final}
\gamma = - m \frac{f^{(1)}(\lambda_\ms)}{\lambda_\ms} \approx 
\frac{4mr_\mo}{b^2} \left( \frac{r_\ms}{r_\mo+r_\ms} \right)\,,
}
where we used that, at the background level, $\lambda_\ms = - (r_\mo + r_\ms)$\,. Note that all the other quantities vanish: $\kappa=\chi=\omega=0$. The fact that the convergence vanishes is a consequence of $\bds{R}=0$, see for example \cite{fleury_2015}. The net rotation $\omega$ vanishes by symmetry: If you're looking at the image with the Black Hole, say, at your right, there is no reason why the galaxy shape would rotate clockwise or counterclockwise. Last, the angle $\chi$ vanishes simply because we chose \emph{wisely} our Sachs basis.
\MYrem{Behaviour of the solution for $b=0$}{
The final solution for the shear Eq.~\eqref{eqlensing:gamma_final} seems to be divergent in the regime $b\approx 0$. However, the weak lensing regimes imposes that we observe rays coming outside the Einstein radius of the lens
\be 
\zeta \gg \theta_{\mathrm E}\,,
\ee 
where
\be
\theta_{\mathrm E} = \sqrt{\frac{4m r_\ms}{r_\mo(r_\mo + r_\ms)}}
\ee 
is the Einstein angle. Using the expression for $\zeta$ Eq.~\eqref{eqlensing:anglezeta} and for the shear Eq.~\eqref{eqlensing:gamma_final}, we get
\be 
\gamma = \left( \frac{\theta_{\mathrm E} }{\zeta}\right)^2 \ll 1
\ee 
under our assumption.
}

\subsection{Statistical Analysis}
\subsubsection{General Setup}
In this Section, we present a cute toy-model to show how the result derived above can be useful in a more realistic situation. Recall that the scaled rotation (in absolute value) is by\footnote{Here we adopt the subscript for \emph{source} instead of \emph{emitted} as it is more adapted to the context.}
\be
\Theta = \vert \delta \alpha \vert \frac{\varepsilon_\mo^2}{2-\varepsilon_\mo^2} 
= \vert \delta \alpha \vert\frac{\varepsilon_\ms^2}{2-\varepsilon_\ms^2} \,,
\ee
where the last equality holds as $\delta \alpha$ is already a first order quantity, and we have at this order $\varepsilon_\mo  = \varepsilon_\ms$. The rotation angle $\delta \alpha$ is given by Eq.~\eqref{eqlensing:delta_alpha_fin}. Combining these relations with the result for the shear Eq.~\eqref{eqlensing:gamma_final} yields (using $\gamma=\gamma_1$ as $\chi=0$)
\be 
\Theta = \frac{4m}{b^2} \frac{\DL \DLS}{\DS} \vert \sin 2 \beta_\ms \vert\,,
\ee 
where $r_\mo=\DL$ is the distance between the lens and the observer, $r_\ms=\DLS$ is the distance betwee n the lens and the source and $r_\mo + r_\ms=\DS$ is the distance between the observer and the source. Recall that $\beta_\ms$ is the angle between the semi-minor axis and the first Sachs vector $\bolde_1$ at the source position. Note that, as before, there is no need to distinguish between $\beta_\mo$ and $\beta_\ms$ as the scaled rotation $\Theta$ is already a first order quantity.

We would like now to define some statistical quantities that can be measured. For example, for a given collection of sources at redshift $z_\ms$ (up to a the width of the bin), what is the probability to have $\Theta> \bar \Theta$ where $\bar \Theta$ is some arbitrary fixed value? Or what is the PDF of the statistical quantity $\Theta$? We can answer these questions but we need some statistical tools. We give here the main ingredients, and more details can be found in the Appendix~\ref{app:optical_depth}.

\subsubsection{Statistical Ingredients\label{seclensing:stat_ingred}}
Let us consider the lens at distance $\DL$ and redshift $z_\ell$, and a source at distance $\DS$ and redshift $z_\ms$. We define the cross section $\sigma(\Theta, z_\ell, z_\ms, m)$ as the area of the region (projected on the plane perpendicular to the line of sight and passing through the lens) in which the scaled rotation would be bigger than $\Theta$. Its is simply a circle with radius $b$ whose area is given by
\be  \label{eqlensing:sigmaSCH1}
\sigma(\Theta, z_\ell, z_\ms, m) = \pi b^2 
=
\frac{4\pi m}{\Theta} \frac{\DL \DLS}{\DS} \vert \sin 2 \beta_\ms \vert \,.
\ee
The optical depth for a source at a fixed redshift $z_\ms$ and a scaled rotation of fixed value $\Theta$ is
\be  \label{eqlensing:tauSCH1}
\tau(\Theta, z_\ms)= \mathcal{A} \int_{0}^{z_\ms} n^{\mathrm{phys}}(z,M) \frac{\dd r}{\dd z} \sigma(\Theta, z, z_\ms, M)\, \dd M \, \dd z\,,
\ee
see Appendix~\ref{app:optical_depth} for more details. We used $m=\GN M$ to recover the lens of the mass. The integral spans all the potential lens between the observer ($z=0$) and the source ($z=z_\ms$). The physical density of galaxies of mass $M$ at redshift $z$ is given by $n^{\mathrm{phys}}(z,M)$. We also integrated over the orientation of the galaxies using a flat distribution, which gives the constant prefactor
\be 
\mathcal{A} = \frac{2}{\pi} \int_{0}^{\frac \pi 2}\; \sin(2 \beta_\ms)\, \dd \beta_\ms = \frac{2}{\pi}\,.
\ee
Using the relation between the physical distance $r$ and the redshift $z$ given by
\be 
\frac{\dd r}{\dd z}  = \frac{1}{(1+z)H(z)}\,,
\ee 
and the expression for the cross section Eq.~\eqref{eqlensing:sigmaSCH1}, we get
\be  \label{eqlensing:tauSCH2}
\tau(\Theta, z_\ms)
=
\mathcal{A} \int_{0}^{z_\ms} \frac{1}{(1+z)H(z)} 
\left(  \frac{4\pi \GN M}{\Theta } \frac{\DL \DLS}{\DS} \right)
n^{\mathrm{phys}}(z,M) \, \dd M \, \dd z\,.
\ee
We can relate the mass $M$ of a lens inside its Einstein radius to its velocity dispersion $\sigma_v$ (see \cite{Oguri:2019} for example)
\be 
M = \frac{4\pi^2}{G} \sigma_v^4 \frac{\DL \DLS}{\DS}\,.
\ee 
The density of galaxies using this variable satisfies
\be 
n^{\mathrm{phys}}(z,M) \, \dd M = (1+z)^3 n(z,\sigma_v)\,  \dd \sigma_v\,,
\ee 
where $n(z,\sigma_v)$ is the comoving number of galaxies (using today's value as a reference). The factor $(1+z)^3$ encodes the dilution of galaxies (matter) as the Universe is expanding. Combining these different relations, we can obtain the optical depth, together with the probability density function $p(\Theta, z_\ms)$ of the scaled rotation, namely
\boxemph{
\tau(\Theta, z_\ms) &= \frac{\mathcal A}{\Theta} \mathcal{J}(z_\ms)\,, \\
P(>\Theta, z_\ms) &= 1- \eee^{-\tau(\Theta, z_\ms)}\,, \\
p(\Theta, z_\ms) &= - \frac{\partial P(>\Theta, z_\ms) }{\partial \Theta}
= \frac{\mathcal A}{\Theta^2}\mathcal{J}(z_\ms) \exp\left( -\frac{\mathcal A}{\Theta} \mathcal{J}(z_\ms)\right)\,, \quad \mathrm{with} \\
\label{eqlensing:Jzs}
\mathcal{J}(z_\ms)&=
\int_{0}^{z_\ms} \frac{16\pi^3(1+z)^2}{H(z)} 
\left( \frac{\DL \DLS}{\DS} \right)^2 \sigma_v^4 
n(z,\sigma_v) \, \dd z \, \dd \sigma_v \,.
}

\subsubsection{Models for the Galaxy Number Density\label{seclensing:models}}
To get some physical insight, we can first take the fitted values for $n(z,\sigma_v)=n(0,\sigma_v)$ coming from the SDSS observations, see Bernardi \cite{Bernardi_2010}, in which the comoving density of galaxies does not take into account a potential evolution of galaxies and only depends on $\sigma_v$, i.e.
\be 
n(z,\sigma_v)=n(\sigma_v)\,. 
\ee
The expression is given by
\begin{align}
\label{eqlensing:Bernardi}
n(\sigma_v) &= 
\phi^\star 
\left( \frac{\sigma_v}{\Sigma_\star}\right)^{\alpha} 
\exp\left(- \left(\frac{\sigma_v}{\Sigma_\star}\right)^{\beta}\right) \frac{1}{\Gamma\left(\frac{\alpha}{\beta}\right)} \frac{\beta}{\sigma_v}\,, \quad \mathrm{with}\\
\Sigma^\star &= 113.78\mathrm{km\times s^{-1}}\approx 0.0038\,, \\
\phi^\star &= 1.65\times 10^9 H_0^3\,, \quad \textrm{and} \\
(\alpha,\beta) &=( 0.94,1.85)\,. 
\end{align}
With the galaxy number density Eq.~\eqref{eqlensing:Bernardi}, the integral over $\sigma_v$ in Eq.~\eqref{eqlensing:Jzs} reads
\be 
\int_0^\infty\; \sigma_v^4 n(\sigma_v)\, \dd \sigma_v 
=
(\Sigma^\star)^4 \phi^\star
\frac{\Gamma\left( \frac{4+\alpha}{\beta} \right)}{\Gamma\left( \frac{\alpha}{\beta} \right)}\,.
\ee 
The only integral left is
\be
\mathcal{I}(z_\ms) =
\int_0^{z_\ms} \;
\frac{(1+z)^2}{H(z)} 
\left( \frac{\DL \DLS}{\DS} \right)^2
\, \dd z
=
\int_0^{z_\ms} \;
\frac{1}{H(z)} 
\left( \frac{\chi(0,z)\chi(z,z_\ms)}{\chi(0,z_\ms)} \right)^2
\, \dd z\,,
\ee 
where $\chi(z_1,z_2)$ is the comoving distance (see Section~\ref{secintro:distances}) from $z_1$ to $z_2$ given by
\be 
\label{eqlensing:confdist}
\chi(z_1,z_2) = \frac{1}{a_0}\int_{z_1}^{z_2}\; \frac{1}{H(z)}\, \dd z\,,
\ee 
and
\begin{align}
    \DL &= \frac{\chi(0,z)}{1+z}\,, &
    \DLS &= \frac{\chi(z,z_\ms)}{1+z_\ms}\,, &
    \DS &= \frac{\chi(z,z_\ms)}{1+z_\ms}\,.
\end{align}
The final expression for the function $\mathcal{J}(z_\ms)$ reads
\be
\mathcal{J}(z_\ms) = \frac{16\pi^3 (\Sigma_\star)^4 \phi^\star}{H_0^3}
\frac{\Gamma\left( \frac{4+\alpha}{\beta} \right)}{\Gamma\left( \frac{\alpha}{\beta} \right)}
H_0^3\mathcal{I}(z_\ms)\,.
\ee 
The factors of $H_0^3$ were explicitly added to show that the various terms are dimensionless. We can also use another parametrisation for the galaxy number density $n(z,\sigma_v)$ given by the Illustris model, see \cite{Torrey:2015,Cusin:2019}, and Appendix~\ref{app:Illustris} for more details. This models can take into account the redshift evolution of galaxy density.

\subsubsection{Results}
\begin{figure}[ht!]
	\begin{center}
	\includegraphics[width = 0.49\textwidth]{./Pics/Lensing/PDF_O.pdf} 
	\includegraphics[width = 0.49\textwidth]{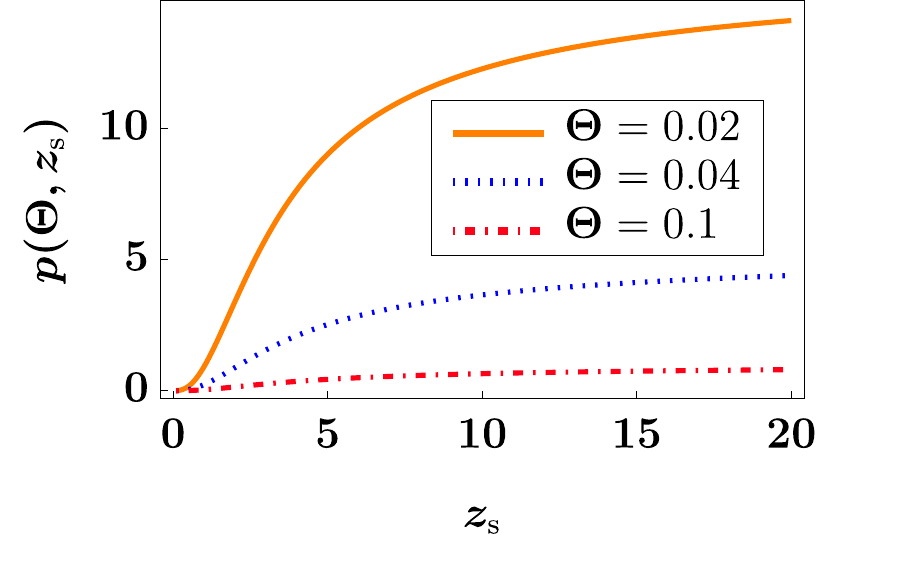} 
	\end{center}
	\caption[Probability density function of the scaled rotation]{\label{figlensing:PDF}
		\textbf{Probability density function}\\
		Function of the scaled rotation and the source redshift, computed with the Illustris model taking into account the evolution of the galaxy density.\\
		\textbf{Left:} PDF for various redshifts. When $z_\ms$ increases, the PDF for large values of $\Theta$ also increases as the path of the light takes more time, hence the probability of getting rotated increases.\\
	 \textbf{Right:} PDF for various values of $\Theta$. A low value for $\Theta$ corresponds to a higher probability as the cross section for such a rotation is bigger. Note that here too, the PDF is always increasing with $z_\ms$.}
\end{figure}

\begin{figure}[ht!]
    \begin{center}
    \includegraphics[width = 0.75\textwidth]{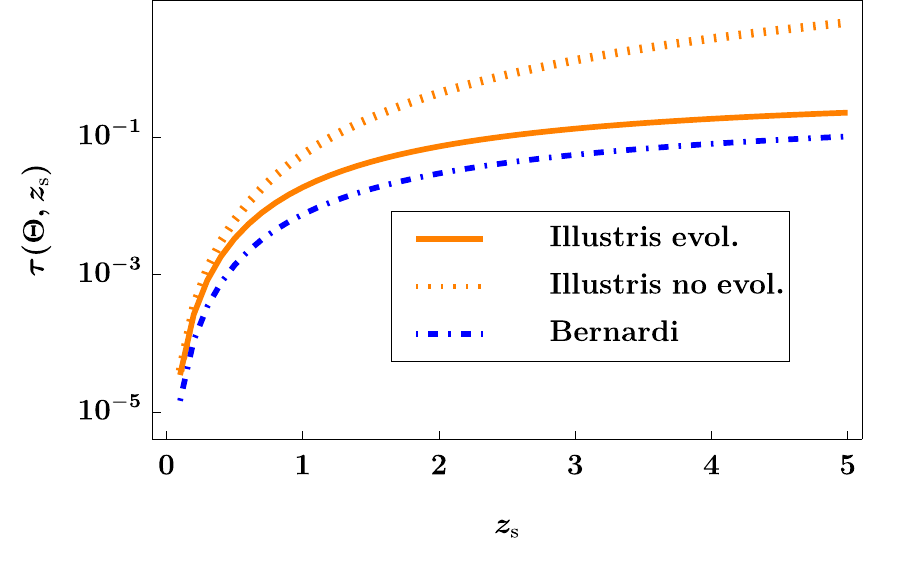}
    \end{center}
    \caption[Optical depth]{\label{figlensing:tau}
    \textbf{Optical depth}\\
    Comparison of the three models for the galaxy number density (with $\Theta=0.02$ fixed)}
\end{figure}

In Fig.~\ref{figlensing:PDF} we show the PDF $p(\Theta, z_\ms)$ for various values of the scaled rotation $\Theta$ and source redshift $z_\ms$. In both cases, it is clear that the overall probability increases with $z_\ms$, the further the source, the bigger the \emph{opportunity} for the light to undergo a bigger deflection. In Fig.~\ref{figlensing:tau} we show the optical depth for the three models (the simple toy model, and the Illustris simulation with and without the redshift evolution). We see that for a small source redshift (up to $z_\ms\approx 1$), all the models agree at $\sim10\%$. For higher redshifts, the Illustris without evolution is different from the other two models. It is clear that the final result strongly depends on the model chosen for $n(z,\sigma_v)$.

\section{Cosmic shear\label{seclensing:cosmo}} 
\subsection{General Idea}
In this Section, we want to apply our result in a cosmological context to estimate the cosmic shear. The cosmic shear is expressed in terms of the derivatives of the lensing potential $\phi(z,\boldsymbol n)$ which is itself a stochastic quantity. This means that its exact value in a specific direction and at a specific redshift cannot be predicted from our cosmological theory. However, statistical properties of the cosmic shear can be predicted from a given model. For example, its expectation value $\langle  \phi(z,\boldsymbol n) \rangle$ vanishes in most theories. More interesting, the two-point correlation functions $\langle \phi(z_1,\boldsymbol n_1) \phi(z_2, \boldsymbol n_2) \rangle$ (or equivalently its power spectrum $C_\ell(z_1,z_2)$) has a non-trivial expression which depends on the cosmological model or on cosmological parameters. We want to build an estimator for such a quantity, in order to compare it with theoretical predictions.

\begin{figure}[ht!]
	\begin{center}
	\includegraphics[width = 0.55\textwidth]{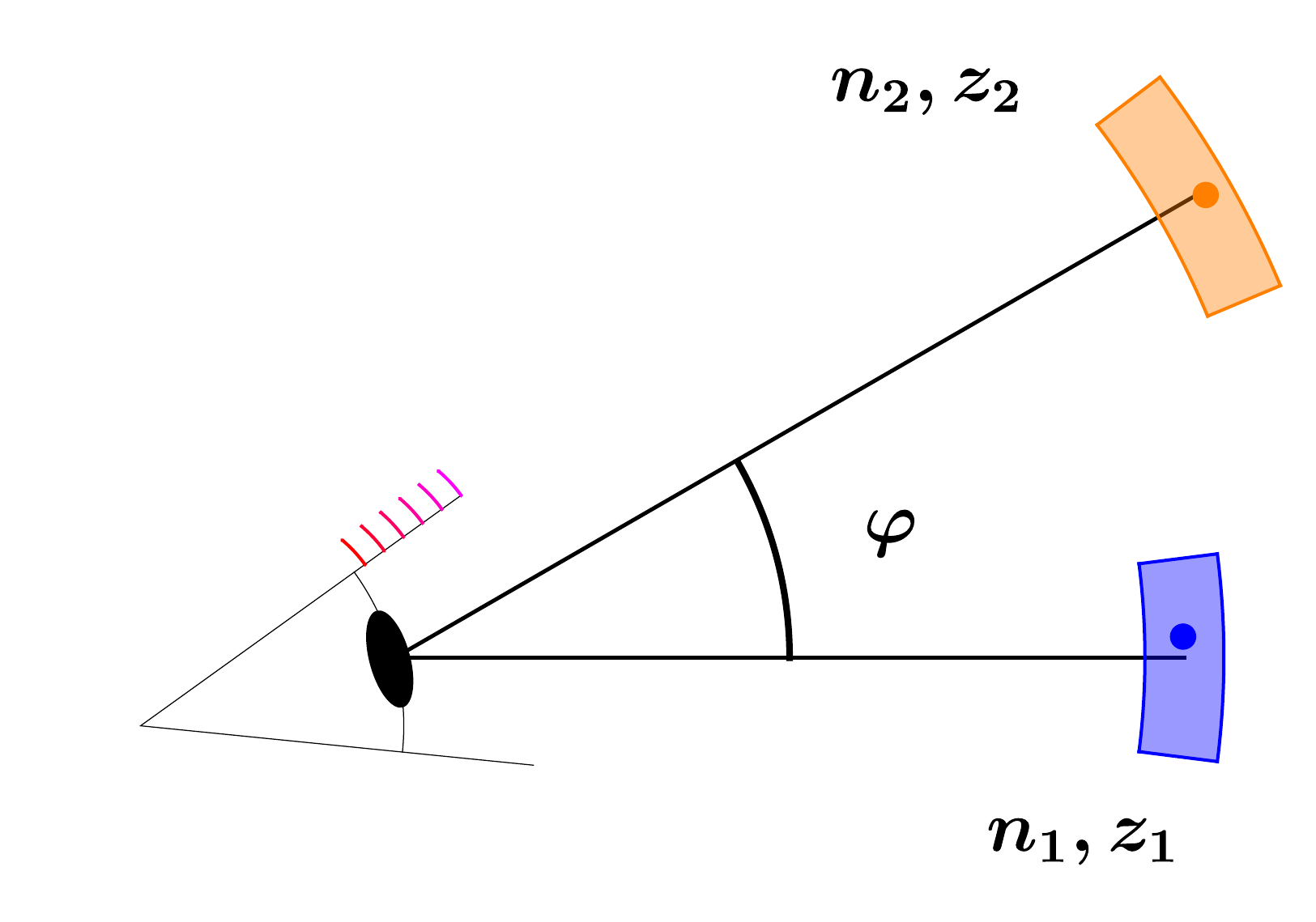}
	\end{center}
	\caption[Estimation of the correlation function]{\label{figlensing:setup_bins}
		\textbf{Estimation of the correlation function}\\
		A couple of galaxies in two bins separated by an angle $\varphi$ allows the computation of $\langle \Theta^2 \rangle$. To such couples are needed to estimate the two correlation functions $\zeta_+$ and $\zeta_-$.}
\end{figure}

The main idea is the following. The shear has two independent components, for example $\gamma_1$ and $\gamma_2$, but other examples are possible. From this quantity, we can build two independent correlation functions: $\zeta_+(z_1,z_2,\varphi)$ and $\zeta_-(z_1,z_2,\varphi)$ which we will define precisely below. They can only depend on the angle between the two directions because the cosmological background is isotropic. Each single galaxy measurement provides one value of the scaled rotation $\Theta$. The (two-point) correlation function of the scaled rotation can be schematically related with the shear correlation function as
\begin{align} 
\Theta &\sim   \gamma_1 +\gamma_2 \,,\\
\langle \Theta^2 \rangle  &\sim \Lambda_+ \zeta_+ + \Lambda_- \zeta_-\,,
\end{align}
where $\Theta^2$ simply means here that we need two galaxies, and the coefficients $\Lambda_{\pm}$ are arbitrary functions. To determine completely the independent functions $\zeta_{\pm}$, we need two such observables. In other words, we need two pairs of galaxies, each pairs being separated by a constant angle $\varphi$ to determine the two independent correlation functions $\zeta_\pm$. The observation of such a pair is shown in Fig.~\ref{figlensing:setup_bins}. Note that, in practice, the galaxies are located in bins of a given angular and redshift size.

\subsection{Theoretical Results\label{seclensing:cicf}}
\subsubsection{Coordinate Independent Shear}
We want to define here the relevant correlation functions. The first guess would be
\be 
\zeta_{ab}(z_1,z_2,\mu) \equiv \langle \gamma_a( z_1, \boldsymbol n_1) \gamma_b (z_2, \boldsymbol n_2)\rangle \,,
\ee 
with $a,b=1,2$. We use here 
\boxemph{
\mu \equiv \cos \varphi= \boldsymbol n_1 \cdot \boldsymbol n_2\,,
}
where $\varphi$ is the angle between the two pixels. We will use a slight notation abuse and use $\varphi$ or $\mu$ indifferently in what follows, as one notation or the other may be more convenient.

The components of the shear are given by (see Section~\ref{seccosmo:lens_map} and Section~\ref{seccosmo:jacobi})
\begin{align}
\label{eqlensing:gamma1expr}
\gamma_1(z,\boldsymbol n) &= - \frac 12 (\nabla_1 \nabla_1 - \nabla_2 \nabla_2) \phi\,, \\
\label{eqlensing:gamma2expr}
\gamma_2(z, \boldsymbol n) &= - \nabla_1 \nabla_2 \phi\,,
\end{align}
where $\phi$ is the lensing potential. It is clear from this definition that the components $\gamma_1$ and $\gamma_2$ depend on the coordinate system, as the covariant derivatives are defined with respect to $(\theta, \varphi)$.

\begin{figure}[ht!]
	\begin{center}
	\includegraphics[width = 0.55\textwidth]{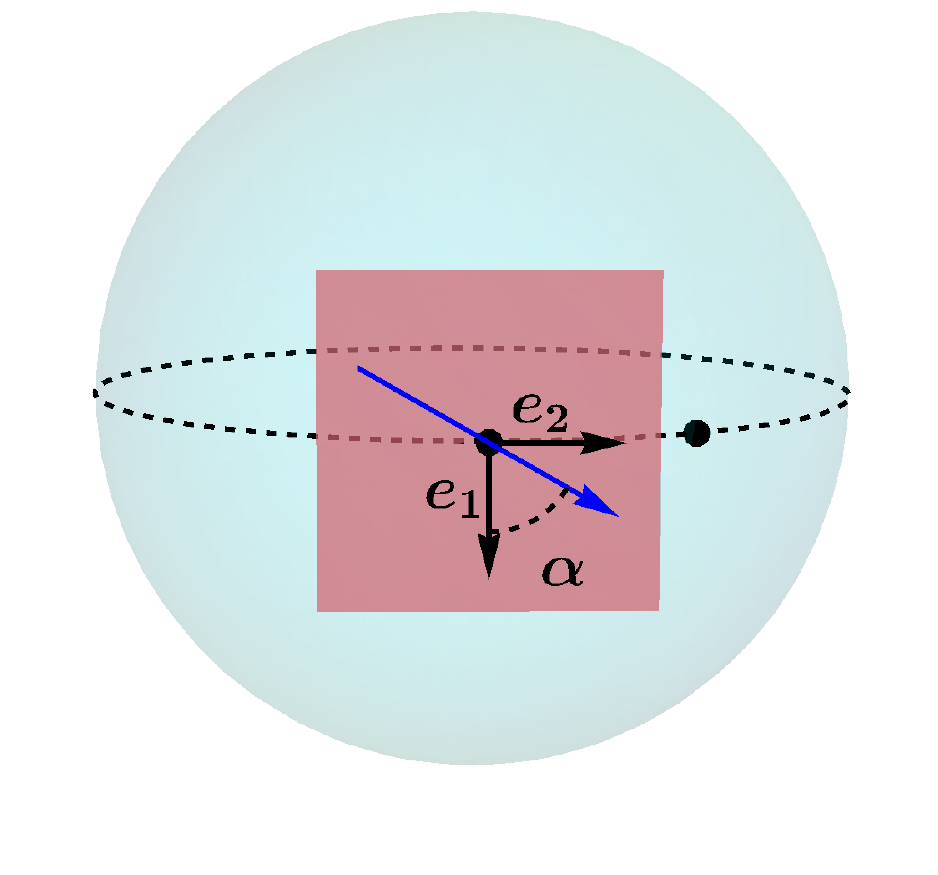}
	\end{center}
	\caption[Coordinate independent shear]{\label{figlensing:shear_alpha}
		\textbf{Coordinate independent shear}\\
		The black dot represents the galaxies and the black line is the Equator connecting them. This defines the basis $(\bolde_1, \bolde_2)$ of the tangent space (purple square). The shear $\gamma_\alpha$ is computed in the direction of the green arrow, and does not depend on the system of coordinates. }
\end{figure}

It is possible to define a coordinate independent shear, see for example \cite{Ghosh:2018}. The logic goes as follows: The shear can be seen as a rank-$2$ tensor given by
\be
\bds{\Gamma} = 
\begin{pmatrix}
- \gamma_1 & -\gamma_2 \\
- \gamma_2 & \gamma_1\,
\end{pmatrix}\,,
\ee 
whose components depend on the coordinate system. For a direction $\boldsymbol n$, one defines an orthonormal basis of the tangent space given by $(\bolde_1, \bolde_2)$ associated with the coordinates $(\theta, \varphi)$. We define the vector in direction $\alpha$ as
\be 
\bolde_{\alpha} \equiv \cos \alpha \, \bolde_1 + \sin \alpha\,  \bolde_2 \,,
\ee 
With this notation, the coordinate independent shear is defined as
\boxemph{ \label{eqlensing:gamma_alpha}
\gamma_{\bolde_\alpha}=\gamma_{\alpha} \equiv \bolde_{\alpha} \bds{\Gamma} \bolde_{\alpha}^{\mathrm{T}}
= - \gamma_1 \cos 2 \alpha - \gamma_2 \sin 2 \alpha\,.
}
This shear is independent of the coordinate system, provided the direction $\bolde_\alpha$ is given. The independent correlation function can be defined as follows. Let two galaxies be located at $(z_1,\boldsymbol n_1)$ and $(z_2,\boldsymbol n_2)$ with $\boldsymbol n_1 \cdot \boldsymbol n_2=\cos \varphi = \mu$. The great circle connecting them defines an \emph{Equator}, from which the usual Spherical coordinates $(\theta, \varphi)$ are easily defined. An illustration is shown in Fig.~\ref{figlensing:shear_alpha}.

\MYrem{Ambiguity in the definition}{Note that there is are a ambiguity, as one can define new coordinates $(\tilde \theta , \tilde \varphi ) = (\pi - \theta, \varphi + \varphi_0)$ where $\varphi_0$ is an arbitrary constant. For $\varphi$ this is not problematic as only consider the angle between galaxies is relevant. For $\theta$, the only effect is to define $\tilde \alpha = \alpha + \pi$ which leaves invariant $\gamma_\alpha$ defined by Eq.~\eqref{eqlensing:gamma_alpha}. 
}

\subsubsection{Coordinate Independent Correlation Functions}
When this coordinate system has been defined unambiguously, the invariant correlation functions are defined as
\boxemph{
\zp(z_1,z_2, \mu) &= \langle \gamma_0(z_1, \boldsymbol n_1)\gamma_\pi(z_2, \boldsymbol n_2) \rangle \,, \\
\zc(z_1,z_2, \mu) &= \langle \gamma_{\pi/4}(z_1, \boldsymbol n_1)\gamma_{5\pi/4}(z_2, \boldsymbol n_2) \rangle \,.
}
\begin{figure}[ht!]
	\begin{center}
	\includegraphics[width = 0.49\textwidth]{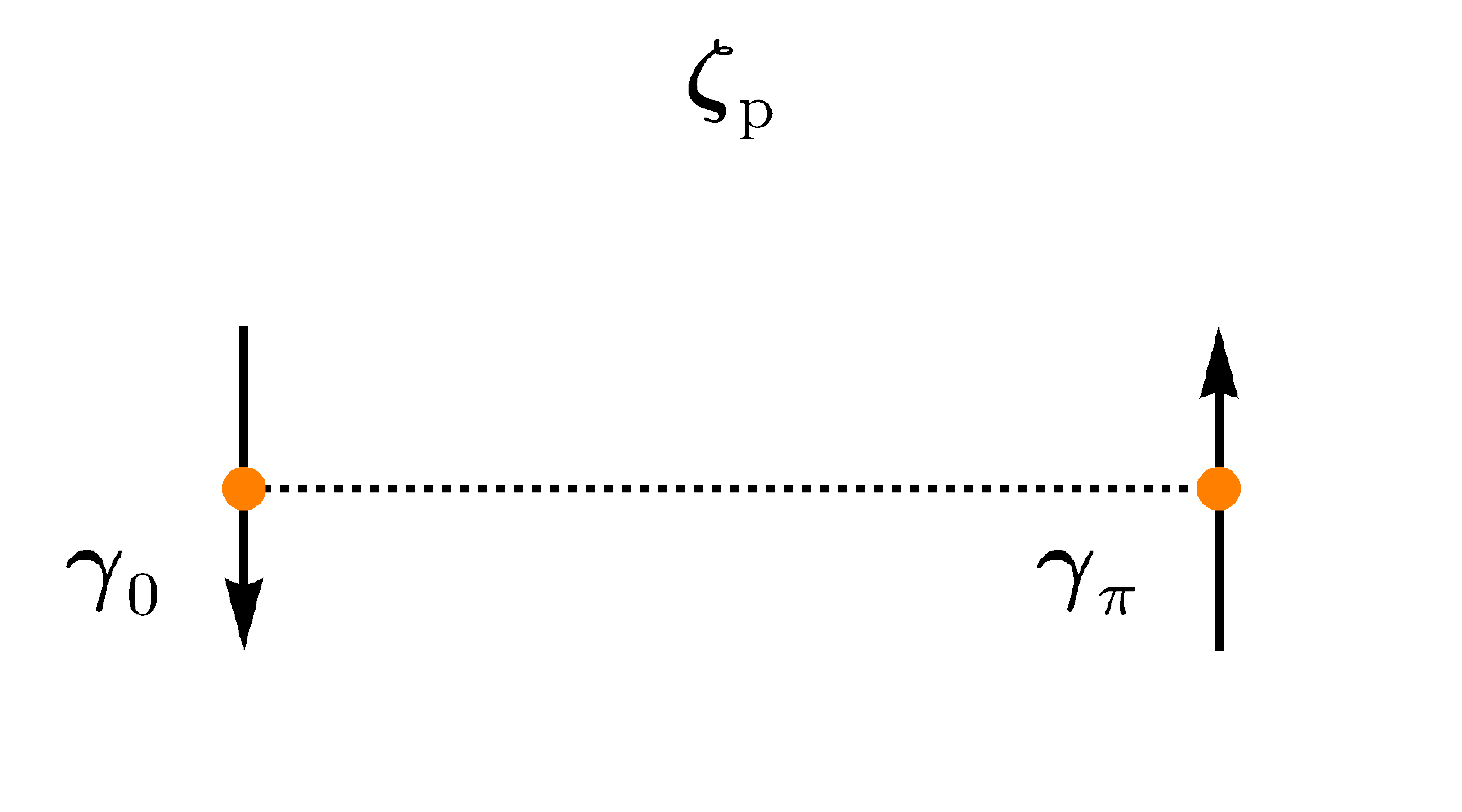} 
	\includegraphics[width = 0.49\textwidth]{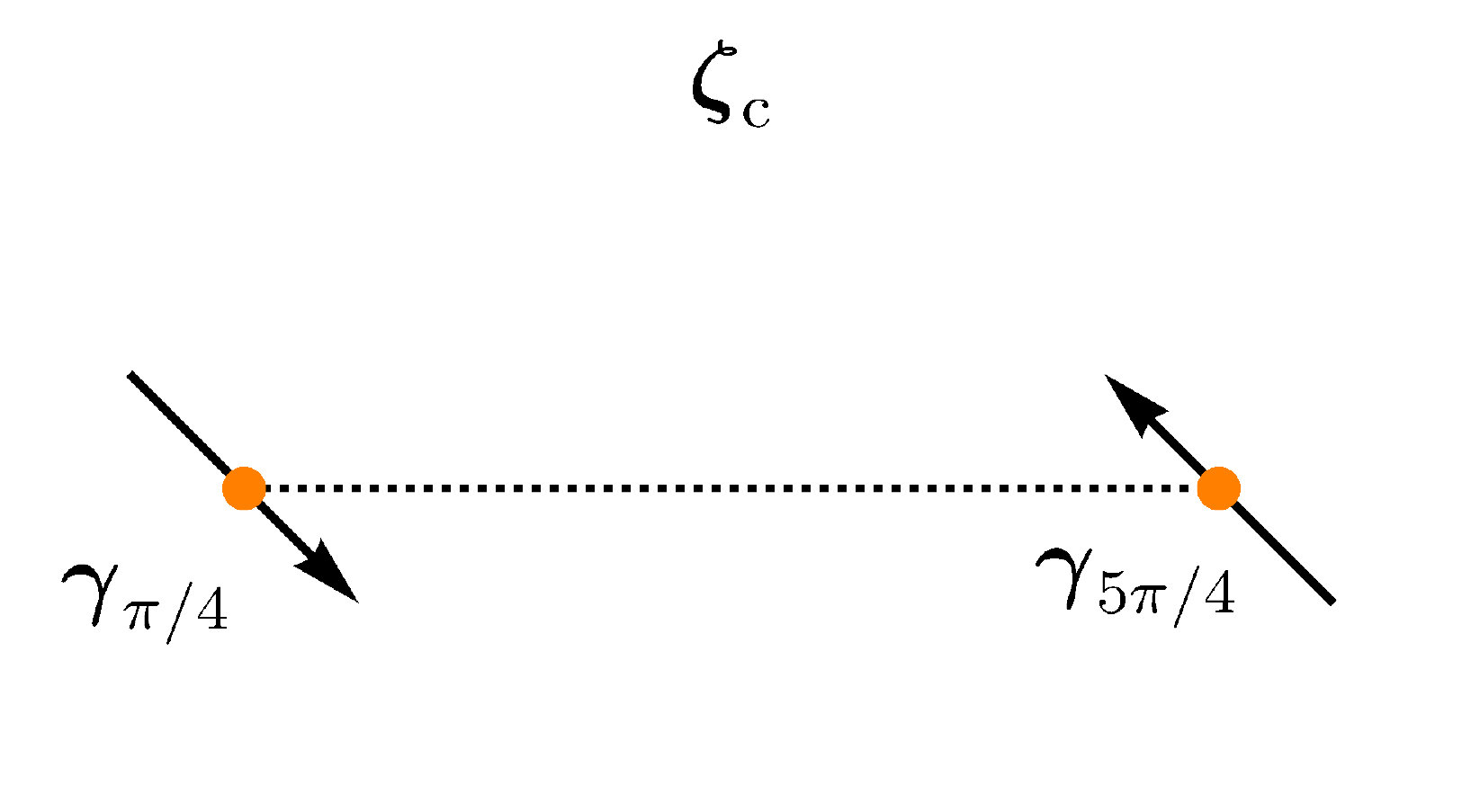} 
	\end{center}
	\caption[Invariant correlation functions]{\label{figlensing:zetapc}
		\textbf{Invariant correlation functions}\\
		The dots represent the galaxies and the dashed line represents the Equator. The angle are computed with respect to $\bolde_1  = \bpt_\theta$ (towards the South).}
\end{figure}
In this specific coordinate system, one has
\begin{alignat*}{2}
\gamma_0 & = \gamma_\pi       && = -\gamma_1\,, \quad \mathrm{and}\\
\gamma_{\pi/4} & =  \gamma_{5\pi/4} && =  - \gamma_2\,.
\end{alignat*}
Hence, the invariant correlation functions are given in this system by
\begin{align}
\label{eqlensing:zetapequa}
\zp(z_1,z_2,\mu ) &= \langle \gamma_1(z_1, \boldsymbol n_1)\gamma_1(z_2, \boldsymbol n_2) \rangle \,, \\
\label{eqlensing:zetacequa}
\zc(z_1,z_2,\mu ) &= \langle \gamma_2(z_1, \boldsymbol n_1)\gamma_2(z_2, \boldsymbol n_2) \rangle \,, \\
\boldsymbol n_1 \cdot \boldsymbol n_2 &= \mu\,.
\end{align}
We stress that, even if these relations seem to depend on the coordinate systems, the correlation functions do not. The coordinate system defined above and represented in Fig.~\ref{figlensing:shear_alpha}, and these relations hold only in this system. 

\subsubsection{Explicit Computations\label{seclensing:explicit}}
We want now to express these correlation functions in terms of the power spectrum $\Clphi{\ell}(z_1,z_2)$. The computations are easily performed in the $(+,-)$ basis. The main relations are (see Section~\ref{seccosmo:stats} for more details)
\begin{align} 
\label{eqlensing:gammapm}
\gamma^{\pm} &= \gamma_1 \pm \ii \gamma_2\,, \\
\label{eqlensing:gammap}
\gamma^+(z,\boldsymbol n) &= -\frac 12 \sum_{\ell=2,m} \phi_{\ell,m}(z) \rll  \;  \YLMS{\ell}{m}{+2}(\boldsymbol n)\,, \\
\label{eqlensing:gammam}
\gamma^-(z,\boldsymbol n) &= -\frac 12 \sum_{\ell=2,m} \phi_{\ell,m}(z) \rll\;  \YLMS{\ell}{m}{-2}(\boldsymbol n)\,, \\
\rll &= \sqrt{\frac{(\ell+2)!}{(\ell-2)!}}\,, \\
\phi(z, \bds{n})& =\sum_{\ell,m} \almphi{\ell}{m}(z) \YLM{\ell}{m}(\bds{n})\,.
\end{align}
Moreover, we define the functions $\tilde{P}_{\ell}(\mu)$ and $\tilde{Q}_{\ell}(\mu)$ as
\begin{alignat}{3}
  \frac{2\ell+1}{16\pi} \tilde{P}_{\ell}(\mu)   &~\equiv~&    - \sqrt{\frac{2\ell+1}{4\pi}}\; _{+2}\tensor{Y}{_{\ell,-2}}(\theta,\pi/2) ~=~
        - \sqrt{\frac{2\ell+1}{4\pi}}\;
        _{+2}\tensor{Y}{_{\ell,-2}}(\theta, \pi/2)  \,, \\
  \frac{2\ell+1}{16\pi} \tilde{Q}_{\ell}(\mu) &~\equiv~&     - \sqrt{\frac{2\ell+1}{4\pi}}\; _{+2}\tensor{Y}{_{\ell,+2}}(\theta,\pi/2) ~=~
        - \sqrt{\frac{2\ell+1}{4\pi}}\;
        {_{-2}\tensor{Y}{_{\ell,-2}}(\theta,\pi/2)}  \,,
\end{alignat}
more details are given in Appendix.~\ref{app:PQ}.

We compute first the correlation function $\langle \gamma^+(z_1, \boldsymbol n_1)\gamma^-(z_2, \boldsymbol n_2) \rangle$. Recall that $\boldsymbol n_1$ and $\boldsymbol n_2$ are given by
\begin{alignat}{2}
    \theta_1 &= \frac \pi 2 \quad  & \varphi_1 &= 0\,,   \\
    \theta_2 &= \frac \pi 2 & \varphi_2 &= \varphi\,.
\end{alignat}
The correlation function is given by
\begin{align} \label{eqlensing:gpmi1}
    \langle \gamma^+(z_1, \boldsymbol n_1)
    \gamma^-(z_2, \boldsymbol n_2) \rangle &=
    \frac 14 \sum_{\ell_1,\ell_2,m_1,m_2} \langle \almphi{\ell_1}{m_1}(z_1) \almphi{\ell_2}{m_2}(z_2)\rangle
    \rlla{\ell_1} \rlla{\ell_2} \, 
    \YLMS{\ell_1}{m_1}{+2}(\bds n_1) \, 
    \YLMS{\ell_2}{m_2}{-2}(\bds n_2) \\
    \label{eqlensing:gpmi}
    &= 
    \frac 14 \sum_{\ell} 
    \Clphi{\ell}(z_1,z_2) \rll^2
    \left(
    \sum_m 
    \YLMS{\ell}{m}{+2}(\bds n_1) \, 
    \YLMSStar{\ell}{m}{+2}(\bds n_2) 
    \right)\,, 
\end{align}
where we used the conjugation properties Eq.~\eqref{eqcosmo:ylmconj} and Eq.~\eqref{eqcosmo:alpphistar} together with the definition of the power spectrum Eq.~\eqref{eqcosmo:Clphi}. The sum can be done explicitly using the addition theorem for the Spin Weighted Spherical Harmonics (see Appendix~\ref{sec:appADD} with $s_1=+2$ and $s_2=-2$) and gives
\begin{align}
\label{eqlensing:sumi}
\sum_m 
\YLMS{\ell}{m}{+2}(\bds n_1) \, 
\YLMSStar{\ell}{m}{+2}(\bds n_2) 
&=
- \sqrt{\frac{2\ell+1}{4\pi}} \YLMS{\ell}{-2}{+2}\left(\varphi, \frac \pi 2\right) \\
&=
\frac{2\ell+1}{16\pi} \tilde{P}_\ell(\mu)\,,
\end{align}
Combining the expressions Eq.~\eqref{eqlensing:gpmi} and Eq.~\eqref{eqlensing:sumi}, we get the final expression for the correlation function
\boxemph{
\langle \gamma^+(z_1, \boldsymbol n_1)
\gamma^-(z_2, \boldsymbol n_2) \rangle &=
 \sum_{\ell}  \frac{2\ell+1}{64\pi}
\Clphi{\ell}(z_1,z_2) \rll^2 \tilde{P}_\ell(\mu)\,.
}

The other correlation functions can be computed following the same steps. We use the indices $a,b=1,2$ and $\alpha, \beta = +,-$. The final result is
\begin{align}
\langle \gamma^\alpha(z_1, \boldsymbol n_1)
\gamma^\beta(z_2, \boldsymbol n_2) \rangle &=
 \sum_{\ell}  \frac{2\ell+1}{64\pi}
\Clphi{\ell}(z_1,z_2) \rll^2\, \Pi^{\alpha \beta}_\ell (\mu) \,, \\
\tensor{ \Pi}{^{\alpha \beta }_\ell}(\mu)&=
\begin{pmatrix}
\tilde{Q}_\ell(\mu) & \tilde{P}_\ell(\mu) \\
\tilde{P}_\ell(\mu) & \tilde{Q}_\ell(\mu) 
\end{pmatrix}\,.
\end{align}
The correlation function in the $(\bolde_1, \bolde_2)$ basis can be found inverting the relation Eq.~\ref{eqlensing:gammapm} to express $\gamma_1$ and $\gamma_2$ as
\begin{align}
\gamma_a &= \tensor{M}{_{a \alpha}} \gamma^{\alpha} \,, \\
\bds M &= 
\frac 12 
\begin{pmatrix}
1 & 1\\
-\ii & \ii
\end{pmatrix}\,.
\end{align}
With these notations, we get
\boxemph{
\label{eqlensing:gammaab}
\langle \gamma_a(z_1, \boldsymbol n_1)
\gamma_b(z_2, \boldsymbol n_2) \rangle &=
 \sum_{\ell}  \frac{2\ell+1}{64\pi}
\Clphi{\ell}(z_1,z_2) \rll^2\, \Pi{_{ab,\ell}}(\mu) \,, \\
\label{eqlensing:Piab}
\tensor{\Pi}{_{ab,}_\ell}(\mu)
=
\tensor{M}{_{a \alpha}}
\tensor{M}{_{b \beta}}
\tensor{\Pi}{^{\alpha \beta }_\ell}(\mu)
&= \frac 12 
\begin{pmatrix}
 \tilde{P}_\ell(\mu)+\tilde{Q}_\ell(\mu) &0 \\
0 &  \tilde{P}_\ell(\mu)-\tilde{Q}_\ell(\mu)
\end{pmatrix}\,.
}

\subsubsection{Final Results}
Recall that these computations were done in the preferred coordinate system with both galaxies at the Equator. In this system, the invariant correlation functions are given by Eq.~\eqref{eqlensing:zetapequa} and Eq.~\eqref{eqlensing:zetacequa}. The final expression can be obtained using Eq.~\eqref{eqlensing:gammaab}, which yields
\boxemph{
\zp(z_1,z_2,\mu) &= \sum_{\ell} 
\frac{2\ell+1}{128\pi}
\Clphi{\ell}(z_1,z_2) \rll^2  \left(
\tilde{P}_\ell(\mu)+\tilde{Q}_\ell(\mu) 
\right)\,, \\
\zc(z_1,z_2,\mu) &= \sum_{\ell} 
\frac{2\ell+1}{128\pi}
\Clphi{\ell}(z_1,z_2) \rll^2  \left(
\tilde{P}_\ell(\mu)-\tilde{Q}_\ell(\mu) 
\right)\,.
}
It is tempting to define
\be  \label{eqlensing:zetapmdef}
\zeta_\pm \equiv\frac 12( \zp \pm \zc)\,,
\ee 
to get
\boxemph{
\label{eqlensing:zetapfinal}
\zeta_+(z_1,z_2,\mu) &= \sum_{\ell} 
\frac{2\ell+1}{128\pi}
\Clphi{\ell}(z_1,z_2) \rll^2  \tilde{P}_\ell(\mu)\,, \\
\label{eqlensing:zetamfinal}
\zeta_-(z_1,z_2,\mu) &= \sum_{\ell} 
\frac{2\ell+1}{128\pi}
\Clphi{\ell}(z_1,z_2) \rll^2 \tilde{Q}_\ell(\mu)\,.
}
Finally, using the orthogonality relation for the functions $\tilde{P}_\ell(\mu)$ and $\tilde{Q}_\ell(\mu)$ given in Appendix~\ref{app:PQ}, we get
\boxemph{
\label{eqlensing:zetapint}
\int_{-1}^{+1} \zeta_+(z_1,z_2,\mu) \tilde{P}_\ell(\mu)\, \mathrm{d}\mu &= \frac{1}{4\pi} \Clphi{\ell}(z_1,z_2) \rll^2 \,, \\
\label{eqlensing:zetamint}
\int_{-1}^{+1} \zeta_-(z_1,z_2, \mu) \tilde{Q}_\ell(\mu)\, \mathrm{d}\mu &= \frac{1}{4\pi} \Clphi{\ell}(z_1,z_2) \rll^2 \,.
}
The relations Eq.~\eqref{eqlensing:zetapfinal}, Eq.~\eqref{eqlensing:zetamfinal},  Eq.~\eqref{eqlensing:zetapint} and Eq.~\eqref{eqlensing:zetamint} are the main theoretical results of this Section. They relate the invariant correlation functions, which are observable, to the power spectrum of the lensing potential, which is a theoretical quantity predicted by the specific cosmological model at hand.

\subsection{Estimator for the Correlation Functions\label{seclensing:estcorrfun}}
We want now to define an estimator, using observable quantities, to quantify the invariant correlation functions $\zeta_\pm$. The logic goes as follows. For two fixed redshifts, we consider a couple of galaxies: One located at $(z_1,\bds n_1)$ and the other located at $(z_2, \bds n_2)$, with $\bds n_1 \cdot \bds n_2 =\mu$. With these galaxies, we build the preferred coordinate system as explain in the previous Section. We then measure the scaled rotation of each galaxy and define 
\be 
\Xi\equiv \Theta(z_1, \bds n_1, \beta_1)   \Theta(z_2, \bds n_2, \beta_2)\,,
\ee
where $\beta_i$ is the angle between the semi-minor axis of galaxy $i$ and the first basis vector $\bolde_1$.

However, we have two independent correlation functions to estimate, and hence we need two different observables: Another couple of galaxies located at $(z_1, \bds n_1^\prime)$ and $(z_2, \bds n_2^\prime)$ separated by the same angular difference $\bds n_1^\prime \cdot \bds n_2^\prime =\mu$ and located at the same redshifts is needed. From this couple, we define a second observable as
\be 
\Xi^\prime\equiv \Theta(z_1, \bds n_1^\prime, \beta_1^\prime)   \Theta(z_2, \bds n_2^\prime, \beta_2^\prime)\,.
\ee
Using the expression for the scaled rotation
\be 
\Theta(z,\bds n, \beta) = \gamma_2(z, \bds n) \cos 2 \beta - \gamma_1(z,\bds n) \sin 2 \beta\,,
\ee
its two-point correlation function (in the preferred coordinate system) is
\be 
\langle \Theta(z_1,\bds n_1, \beta_1) \Theta(z_2, \bds n_2, \beta_2) \rangle
=
\zeta_+ \cos\left((2(\beta_1 - \beta_2)\right )
+ \zeta_-\cos\left(2(\beta_1 + \beta_2)\right)\,,
\ee 
where we used the expression for $\langle \gamma_1 \gamma_1 \rangle$ and $\langle \gamma_2 \gamma_2 \rangle$ given by Eq.~\eqref{eqlensing:zetapequa} and Eq.~\eqref{eqlensing:zetacequa}, and the definitions of $\zeta_\pm$ given by Eq.~\eqref{eqlensing:zetapmdef} (together with Eq.\eqref{eqlensing:Piab} to see that the cross-correlation vanishes)

We will now assume that $\Xi$ and $\Xi^\prime$ are estimators of this two-point correlation functions. In other words, we have
\boxemph{
\label{eqlensing:Xizeta}
\Xi &= \hat{\zeta}_+(z_1,z_2,\mu) \cos\left((2(\beta_1 - \beta_2)\right )
+  \hat{\zeta}_-(z_1,z_2,\mu)  \cos\left((2(\beta_1 + \beta_2)\right )\,, \\
\label{eqlensing:Xipzeta}
\Xi^\prime &= \hat{\zeta}_+(z_1,z_2,\mu)  \cos\left((2(\beta^\prime_1 - \beta^\prime_2)\right )
+  \hat{\zeta}_-(z_1,z_2,\mu)  \cos\left((2(\beta^\prime_1 + \beta^\prime_2)\right )\,.
}
It is possible to solve Eq.~\eqref{eqlensing:Xizeta} and Eq.~\eqref{eqlensing:Xipzeta} to express the estimators for the correlation functions as
\begin{align} 
        \label{eqlensing:zetapest}
        \hat{\zeta}_+(z_1,z_2,\mu) &= \Xi\, F_1(\beta_1^\prime, \beta_2^\prime, \beta_1, \beta_2) 
        + \Xi^\prime\, F_1(\beta_1, \beta_2,\beta_1^\prime, \beta_2^\prime) \,, \\
        \label{eqlensing:zetamest}
        \hat{\zeta}_-(z_1,z_2,\mu) &= \Xi\, F_2(\beta_1^\prime, \beta_2^\prime, \beta_1, \beta_2) 
        + \Xi^\prime\, F_2(\beta_1, \beta_2,\beta_1^\prime, \beta_2^\prime) \,,
\end{align}
and the explicit expression for the functions $F_1$ and $F_2$ are found in Appendix~\ref{app:estimators}. Note that in the expressions for the estimators Eq.~\eqref{eqlensing:zetapest} and Eq.~\eqref{eqlensing:zetamest} there is no dependence on the various angles $\beta$. It is a slight abuse of notation, but the idea is that for each galaxy quadruplet, we obtain one estimator for $\zeta_+$ and one for $\zeta_-$. Even if each of these individual quantities depends on the angles $\beta$, their final average should be independent on the galaxies orientation and only depend on the angle $\varphi$ between the pixel and the redshifts. Finally, we see that by construction, these estimators satisfy
\be
 \langle\hat{\zeta}_{\pm}(z_1,z_2,\mu)\rangle = \zeta_{\pm}(z_1,z_2,\mu)\,,
\ee
in other words the estimators Eq.~\eqref{eqlensing:zetapest} and Eq.~\eqref{eqlensing:zetamest} are unbiased.

\subsection{Error Estimation}
\subsubsection{General Idea}
Let us recap now what should be done in practise to estimate the correlation functions $\zeta_\pm(z_1,z_2,\mu)$, together with their errors. For each galaxy, one measures
\begin{itemize}
\item its position in the sky $\bds n$;
\item its redshift $z$;
\item the rotation angle $\delta \alpha$ between its major-axis and the direction of the polarisation of light;
\item the position of the semi-minor axis of the galaxy $\beta$;
\item its eccentricity.
\end{itemize}
Note that the position of the semi-minor axis $\beta$ can only be measured when a couple of galaxies has been chosen, which defines the preferred coordinate system.

Each of these quantities is measured with a given error. An estimator is thus given by\footnote{We drop the plus/minus index for clarity, the logic being the same for both estimators.} $\hat \zeta_{i}$ with an associated error  $\Delta \hat \zeta_{i}$, where $i$ represents one estimator (associated with a given quadruplets of galaxies). Even is the individual errors are high (e.g. of the same order or even higher than the signal itself), it is still possible to obtain a precise enough estimator, see \cite{Planck:VIII}. To achieve this goal, we need to construct the best estimator from several individual estimators.

\subsubsection{Best Estimator}
With a given catalogue, we build a set of estimators $\hat \zeta_{i}$, the number of which depends on the particular catalogue. From this set, we define the best estimator with weights $w_i$ as
\be 
\hat \zeta^{\mathrm{BE}} = \sum_i w_i \hat \zeta_i\,,
\ee 
and its error is 
\be 
(\Delta \hat \zeta^{\mathrm{BE}})^2  = \sum_i w_i^2 \Delta \hat \zeta_{i}^2\,.
\ee
The weights satisfy $\sum_i w_i =1$ and their values is given imposing the condition that the \ac{SNR}, defined as
\be 
\mathrm{SNR} = \frac{\hat \zeta^{\mathrm{BE}}}{\Delta \hat \zeta^{\mathrm{BE}}}\,,
\ee 
is maximal. It is straightforward to find the solution which is\footnote{We do not use Einstein summation convention here.} (see \cite{Francfort:2022} for more details)
\begin{align}
w_i &= \frac 1Z \frac{1}{\hat \zeta_i \tau_i^2}\quad \mathrm{with} \\
Z&= \sum_{i} \frac{1}{\hat \zeta_i \tau_i^2 } \quad \mathrm{and} \\
\tau_i &= \frac{\Delta \hat \zeta_i}{ \hat \zeta_i}\,.
\end{align}
In this case, the \ac{SNR} is
\be \label{eqlensing:SNR}
\mathrm{SNR} = \sqrt{\sum_i\frac{1}{ \tau_i^2}}\,.
\ee 

\MYrem{Divergent estimators}{
Note that, from the expressions Eq.~\eqref{eqlensing:Xizeta} and Eq.~\eqref{eqlensing:Xipzeta}, it is clear that some values for the angles may lead to divergent values for the estimators, for example if $\beta_1 - \beta_2 \sim \beta_1^\prime - \beta_2^\prime \sim \pi/4$, the value of $\hat{\zeta}_+$ will be arbitrarily high. However, the error on this estimator will be also high. When we compute the best estimator, the weights $w_i$ associated with such measurements will be close to zero, and in the end, those \emph{ill-defined} pairs will not contribute to the final result.
}
Note that, in practise, this is not an easy task. For example, with a catalogue with $\Ng\sim 10^6$ galaxies, it is possible to extract $\Ne\sim10^{24}$ estimators (from the number of quadruplets). Assuming a computing power of $10^9$ (!) operations per second, computing $\zeta^{\mathrm{BE}}$ would take roughly $\SI[parse-numbers=false]{10^{15}}{\second}\sim \SI[parse-numbers=false]{10^{8}}{\year}$, which is not very doable.

Without a precise catalogue and real data at hand, we cannot go much further. To extract some physical insights, we will make the following assumptions: All the relative errors $\tau_i$ satisfy
\be 
\tau_i  \lesssim \tau_0\,,
\ee
where $\tau_0$ is a constant quantifying the maximum relative error. In this case, the \ac{SNR} given by Eq.~\eqref{eqlensing:SNR} is
\be
\mathrm{SNR}  \gtrsim \frac{\sqrt{\Ne}}{\tau_0}\,,
\ee 
where $\Ne$ is the number of estimators at hand, which we wish now to estimate.

\subsubsection{Number of Estimators}
We want to count the number of estimators we can extract to estimate $\zeta(z_1,z_2,\mu)$. First, we estimate the number of couples of galaxies subtending an angle $\varphi$ with $\mu = \cos \varphi$. We consider angular pixels of aperture $\delta \theta$. The solid angle spanned by these pixels is given by (at lowest order in $\delta \theta$)
\be
\delta \Omega = \delta \theta^2 \pi\,.
\ee 
The total number of pixels is then
\be 
\Np = \fsky \frac{4\pi}{\delta \Omega } = \frac{4\fsky}{\delta \theta^2}\,,
\ee 
where we assumed that only a fraction of the sky $\fsky$ is covered by the survey.

Once such a pixel has been chosen as \emph{first pixel}, we need to count the number of pixels subtended by an angle $\varphi$ around it. Assuming that the first pixel is fixed at the North Pole, the solid volume forming an angle $\varphi\pm \delta \theta/2$ is given by 
\be 
\delta \Omega_{\varphi} = 2\pi \int_{\varphi-\delta \theta/2}^{\varphi + \delta \theta/2}\; \sin \theta \, \mathrm{d} \theta =
2 \pi \delta \theta \sin \varphi\,.
\ee 
In this volume, the number of pixels is given by
\be
N(\varphi) =  \frac{\delta \Omega_{\varphi}}{\delta \Omega} 
= \frac{2 \sin(\varphi)}{\delta \theta}\,.
\ee
Finally, the number of couples of pixels separated by an angle $\varphi$ is given by 
\be
\Nc(\varphi) = \frac{1}{1+\delta_{z_1,z_2}} \Np \times N(\varphi) = \frac{1}{1+\delta_{z_1,z_2}}\frac{8 \fsky \sin \varphi}{\delta \theta^3}\,.
\ee
The fraction with the Kronecker delta prevents over-counting if we consider auto-correlation ($z_1=z_2$).

In a given pixel at redshift $z$, the number of observed galaxies is $\Ng(z)$. Hence, the total number of pairs of galaxies is 
\be 
\Np(z_1,z_2,\varphi) = \Ng(z_1) \Ng(z_2) \Nc(\varphi) \,.
\ee 

Finally, an estimator is built with two pairs of galaxies. The number of estimators is then given by
\boxemph{
\Ne(\varphi,z_1,z_2)
\simeq \frac{1}{1+3\delta_{z_1,z_2}} \frac{1}{2}\left(\Ng(z_1) \Ng(z_2) \frac{8 \fsky \sin \varphi}{\delta \theta^3}\right)^2\,,
}
where we assumed that the number of galaxies is sufficiently high to assume $\Np-1\approx \Np$, and we used that $\delta_{z_1,z_2}^2 =\delta_{z_1,z_2} $.

As explained in the previous section, if $\Ne$ is the number of estimators, the \ac{SNR} would scale as $\sqrt{\Ne}$ (see Eq.~\eqref{eqlensing:SNR}). However, the underlying assumption is that all the estimators (and their errors) are independent from each other. This is certainly not always the case: If galaxies are observed in the same pixels, the errors should be somehow related. For example, if there are strong magnetic fields in a given direction, the polarisation direction measured from all these galaxies will carry the same error. For this reason, a more pessimistic assumption would be that the number of estimators is simply the numbers of pairs of pixels
\boxemph{
\Ne(z_1,z_2,\varphi) \approx  \frac{1}{1+\delta_{z_1,z_2}} \frac{8 \fsky \sin \varphi}{\delta \theta^3}\,,
}
where we again assumed that the number in the brackets is sufficiently large.

\subsection{Results\label{seclensing:result_cosmo}}
In principle, we could stop here, as an explicit catalogue of galaxies is needed to go further. However, we provide here a very simple simulated example for the correlation functions $\zeta_\pm(z_1,z_2,\mu)$. We use CLASS \cite{Blas:2011, Lesgourgues:2011} to compute the power spectrum $\Clphi{z_1,z_2}$, with $\ell_{\mathrm{max}}=20'000$ and the parameters taken from Planck \cite{Planck:VI} are
\begin{align}
h&=0.6781\,, & h^2\Omega_{\mathrm{cdm}}&= 0.0552278\,, & h^2\Omega_{\mathrm b}&=  0.0102921\,,  \\
\log(10^{9}A_{\mathrm s})&= 0.742199\,, & n_{\mathrm s} &= 0.9660499\,, &
T_{\mathrm{CMB}} &= \SI{2.7255}{\kelvin}\,.
\end{align}
In principle, one should compute the sums given by Eq.~\eqref{eqlensing:zetapfinal} and Eq.~\eqref{eqlensing:zetamfinal} to compute the correlation functions. However, the functions $\tilde{P}_\ell$ and $\tilde Q_\ell$ oscillate violently (especially for large $\ell$) which makes numerical computations difficult. Instead, we can use the flat sky approximation which is valid for small angles, see \cite{Bartelmann:2010, durrer_2021, Kilbinger:2014,Hu:2000,Bunn:2006,White:2017,White:1997,Bernardeau:2010} or Appendix~\ref{applensing:FS} for more details. The correlation functions are then given by
\boxemph{
\label{eqlensing:zetapFS}
\zeta_+(z_1,z_2,\varphi) &= \frac{1}{2\pi} \int_0^{\infty}\; \ell^5 \, J_0(\ell \varphi)\,  \frac{1}{4} 
\Clphi{\ell}(z_1,z_2)\, \mathrm{d}\ell\,,\\
\label{eqlensing:zetamFS}
\zeta_-(z_1,z_2,\varphi) &= \frac{1}{2\pi} \int_0^{\infty}\; \ell^5\,  J_4(\ell \varphi) \, \frac{1}{4} 
\Clphi{\ell}(z_1,z_2)\, \mathrm{d}\ell\,,
}
where $J_n(x)$ is the Bessel function of order $n$. Moreover, we take
\begin{align}
    \fsky &\approx 0.7\,, \\
    \delta \theta &= 5^\prime \approx 1.4\times 10^{-3}\,, \\
    \tau_0 & \approx 10^3\,,
\end{align}
inspire from the values of SKA2 \cite{Bull:2016}. This yields
\be 
\mathrm{SNR} \approx \frac{40}{\sqrt{1+\delta_{z_1,z_2}}} \sqrt{\sin \varphi}\,.
\ee 

\begin{figure}[ht]
\begin{adjustbox}{max width=1.35\linewidth,center}
\centering
\subfloat{{\includegraphics[width=0.42\textwidth]{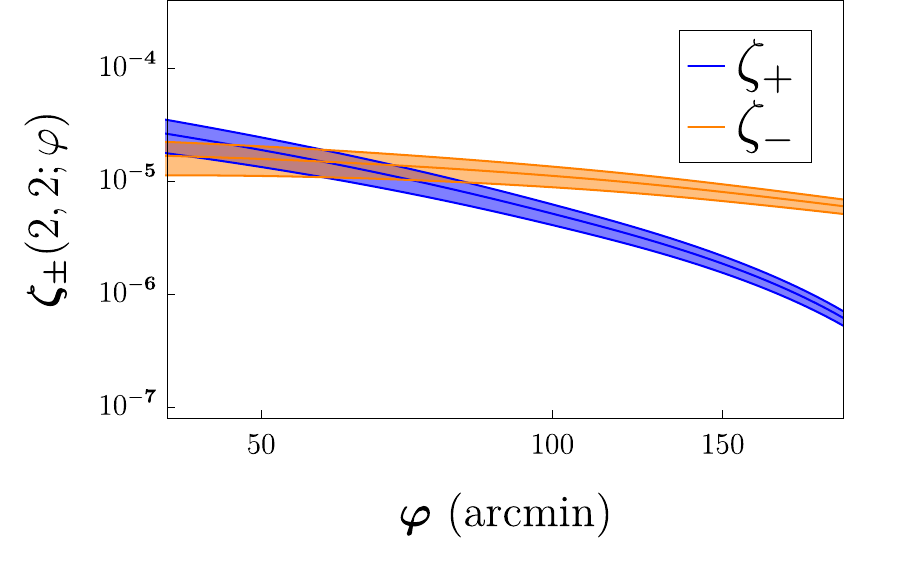} }}
\Fantome{0.42}
\Fantome{0.42}
\end{adjustbox}
\begin{adjustbox}{max width=1.35\linewidth,center}
\centering
\subfloat{{\includegraphics[width=0.42\textwidth]{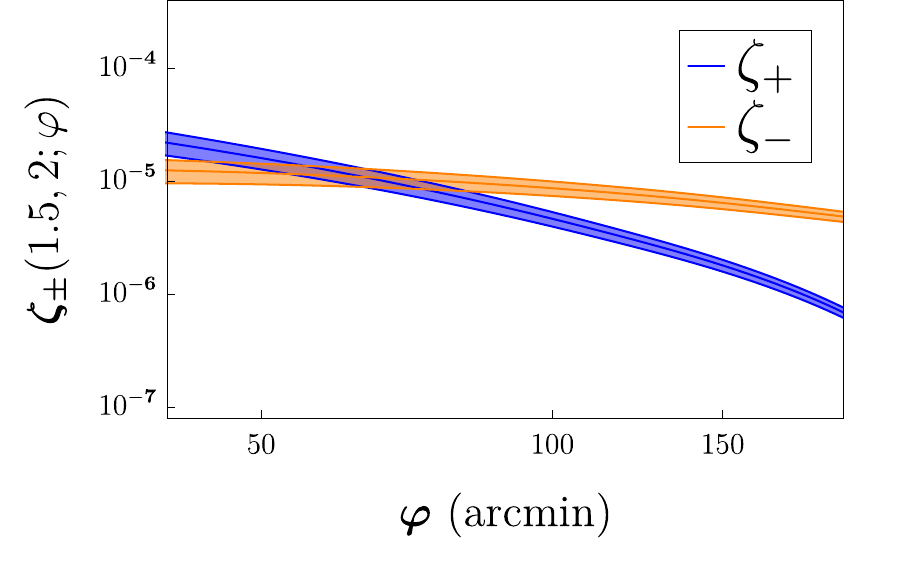} }}
\subfloat{{\includegraphics[width=0.42\textwidth]{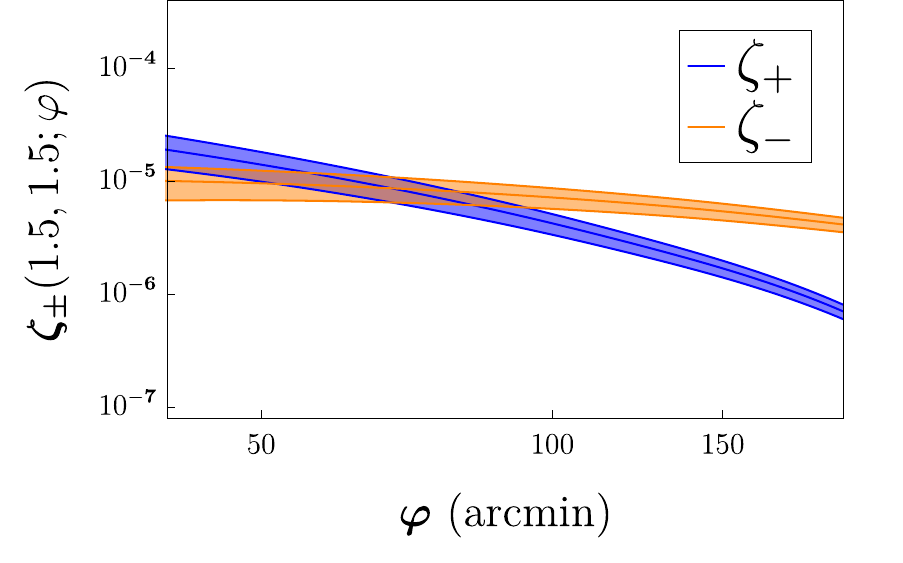} }}
\Fantome{0.42}
\end{adjustbox}
\begin{adjustbox}{max width=1.35\linewidth,center}
\centering
\subfloat{{\includegraphics[width=0.42\textwidth]{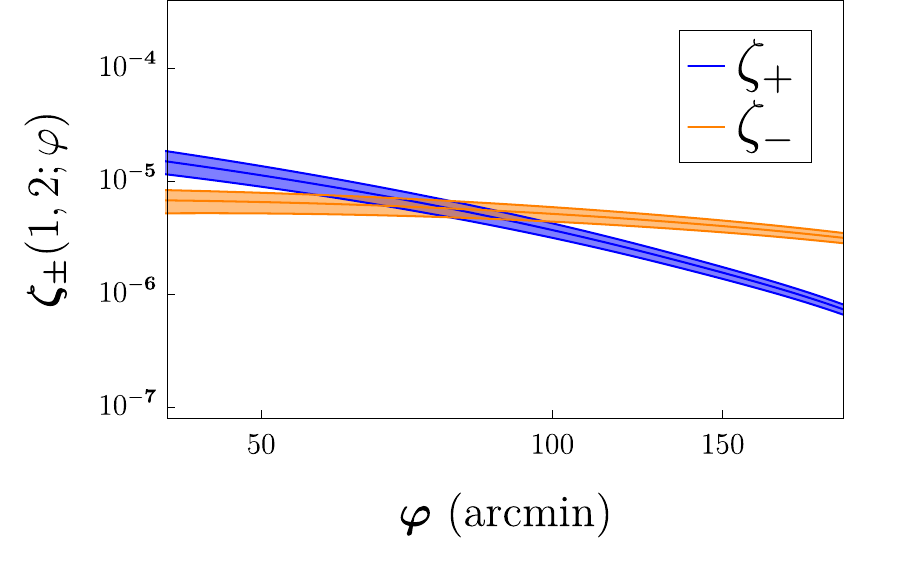} }}
\subfloat{{\includegraphics[width=0.42\textwidth]{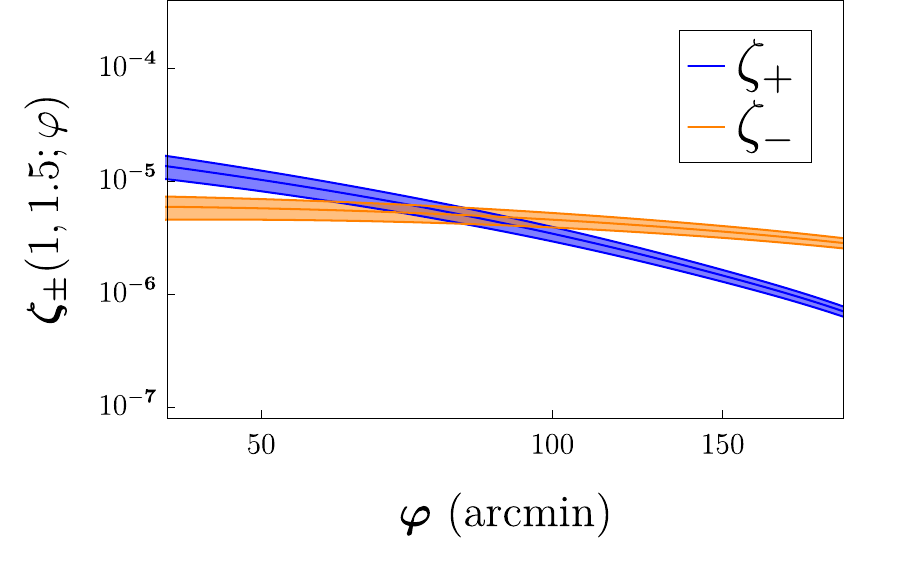} }}%
\subfloat{{\includegraphics[width=0.42\textwidth]{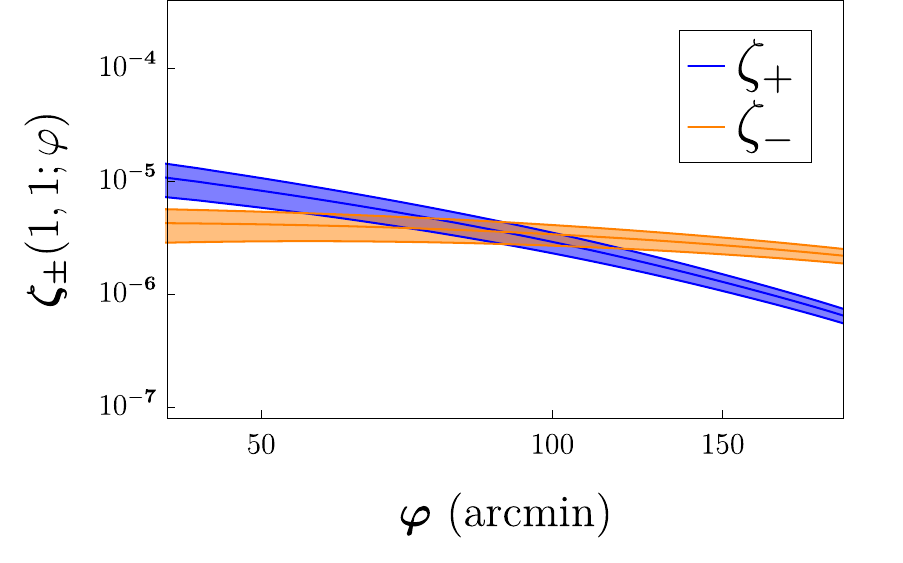} }}
\end{adjustbox}
\caption[Correlation functions]{
	\textbf{Correlation functions $\zeta_\pm(z_1,z_2;\varphi)$ for $z=1,\, 1.5,\, 2$}\\
	The values for $\zeta_+$ and $\zeta_-$ is computed with the flat sky approximation and the power spectrum $\Clphi{\ell}(z_1,z_2)$ is obtained with {\textsc CLASS}. Using a relative error of the order $10^3$ for each single measurement and an angular aperture of $5^\prime$, we can reach a relatively good SNR (the error being of the order of $10\%$).}%
\label{figlensing:corr_funs}%
\end{figure}

The results for all the correlations with $z=1,\, 1.5,\, 2$ are shown in Fig.\ref{figlensing:corr_funs}. This correspond to light emitted respectively $\SI{8}{\Gyear}$, $\SI{9.5}{\Gyear}$ and $\SI{10.5}{\Gyear}$ ago, i.e. when the rate of stars formation was high. The results agree with \cite{Kilbinger:2014}. We see that the correlation functions is not very sensitive to the redshift, i.e. the result are similar, with less than one order of magnitude of difference.

\stopchap

\chapter{Conclusion\label{chaplensing:conc}}
In this Part, we studied weak lensing and built a new estimator for shear correlation functions. We first reviewed the theory of weak lensing together with its main tools.

We presented two different transports of vectors in General Relativity: Parallel Transport and Lie Transport. While the polarisation vector is parallel transported, the galaxy shape is Lie Transported. We argued that, while the polarisation vector is aligned with the galaxy shape at the source position, this is not the case at the observer position. This important result was formalised with the computation of the rotation angle, quantifying the discrepancy between the polarisation and the observed galaxy shape. We showed that this quantity is a signature of cosmic shear (provided the axes of the shear are not aligned with the axes of the galaxy).

We then presented a toy-model application of our result, where we applied it to the lensing of light by Schwarzschild galaxies, using the various models for galaxy density, see \cite{Torrey:2015,Cusin:2019,Bernardi_2010}.

In the second applications, we estimated cosmic shear correlation functions using the rotation angle as an observable. This quantity was already used in the past, for example to study the weak lensing and the intrinsic alignment of galaxies \cite{Brown:2010,Brown:2011,Camera:2016,Thomas:2017}. However, our method is new in the sense that we use this rotation angle to estimate cosmic shear. We described how to obtain one estimator for the correlation functions using $4$ galaxies (two pairs of galaxies). An important advantage of our method is that it is not affected by intrinsic alignment and we can be agnostic about its effects.

We also showed that, even if the error on a single galaxy measurement can be of the same order, or higher, than the signal itself, this problem can be circumvented by considering a high number of galaxies. For example, we showed that we can reach a \ac{SNR} of order $50$ even if the relative error on the shape and polarisation are of order $10^{3}$.

Future survey are expected to include polarisation measurements, for example SKA2, \cite{Bull:2016}. In this context, it will provide very useful to have different independent ways to estimate cosmic shear, in order to constraint the various Cosmological models. Hence, this method can be useful as it is different from what has already been done.

\part{Gravitational Waves and Effective Theory of Gravity}\label{Part:BH}
\begin{figure}[ht!]
	\centering
	\includegraphics[width = 0.9\textwidth]{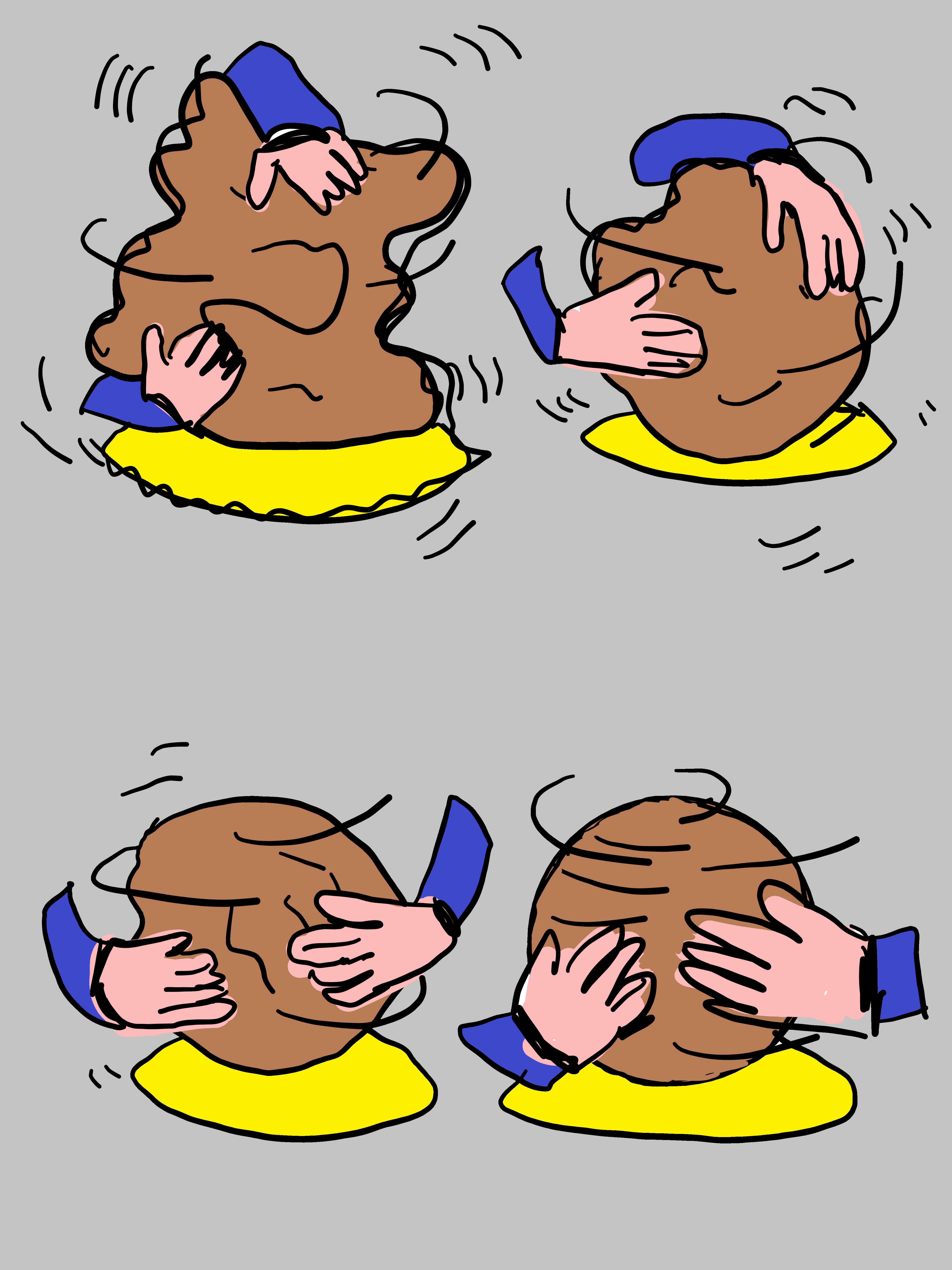} 
	\caption*{\textbf{Clay}, Juan Manuel Garc\'ia Arcos}
\end{figure}

\chapter{General Introduction\label{chapR3:intro}}
In this Part, I present the work we did together with Claudia and Jun  during my $3$ months visit in London in Fall 2019. This was a lovely time for me, the visit was a total success. Also, 2019 was a nice period, confinement was a concept only relevant when talking about strong interactions, and I did not know the meaning of a PCR. Little that I knew... Anyway, let me tell you what I did in London, which resulted in~\cite{Francfort:2020}.

Gravitational waves emitted by colliding Black Holes were first observed by LIGO in 2016 \cite{Abbott:2017,Abbott:20172,Goldstein:2017}. Among other results, these observations constrained the speed of gravitational waves which is very close to the speed of light, the relative error being smaller than $10^{-10}$. In this new era of Gravitational waves observations, it seems necessary to study as extensively as possible their behaviour in various theories of gravity and in various background, see for example \cite{Baker:2022,Ezquiaga:2022,Hohmann:2022,Martinovic:2019} for such predictions in different contexts. 

We already know from Electrodynamics that the speed of photons in a curved background can differ from unity \cite{Lafrance:1994,Drummond:1980}. This speed can even be superluminal. This is actually not in contradiction with causality as explained in \cite{Hollowood:2007, Hollowood:2008, Hollowood:2009,Hollowood:2010,Hollowood:2011,Hollowood:2012, Hollowood:2015,Shore:1995,Shore:2000}. The same analysis has been done for gravitational waves. For example, in \cite{deRham:2018}, the authors examine how the speed of the gravitational waves can depend on their frequency and argue why they should be exactly luminal at high energies probed by LIGO for example. In \cite{deRham:20202} the authors examine the propagation of gravitational waves around a FRLW background and show that the speed can differ from unity. 

The general idea is that variation from unity of the speed of gravitational waves can be caused by interactions with very heavy fields or particles. To study this in details, we would need for example a theory of of Quantum Gravity, or at least of Gravity at higher energies... which we do not have! A good framework to circumvent this problem is given by Effective Field Theories of gravity. The main idea is to consider a theory valid at low-energies (for example the Einstein-Hilbert action), and to add \emph{by hand} various higher order terms, staying agnostic about their origin and the specific value of any free parameters, even though various arguments can provide constraints on several terms, see for example \cite{Endlich:2017,Gruzinov:2006,deRham:2021}  where they use arguments about the causality of a theory to constraint higher curvature terms. For a general introduction to Effective Theories, see for example \cite{Donoghue:1994,Donoghue:1995, Donoghue:2012,Burgess:2003}. 

In our work, we consider a particular class of Effective Theories where the higher order terms are Riemann-squared and Riemann-cubed. Such theories were already considered in the framework of string theories, see for example \cite{Gross:1986,Metsaev:1986}. More recently, the speed of gravitational waves and light in curved spacetimes in this framework was discussed in \cite{deRham:2020}, while in \cite{Baumann:2019} they apply the formalism to probe the mechanism of inflation at high energy. In general, the effective terms are treated non-perturbatively. For an example where this is not the case, see \cite{Cayuso:2020}, where a full treatment can lead to interesting results, different from General Relativity.

Effective Theories can also be applied to study Gravity around Black Holes, see for example \cite{Cardoso:2018} for an Effective Theory with Riemann-cubed terms and the Schwarzschild metric or \cite{Cano:2019,Cano:2020,Cano:2021} for examples around a rotating Kerr Black Hole. Studying Effective Theories in a Black Hole context is particularly interesting as merging Black Holes are the origin of a good portion of the Gravitational waves we observe on Earth. In particular, the frequencies of the ringdown signal emitted by the final Black Hole after its formation are quantified. These are called the \ac{QNM}, and the specific values they can take is a genuine signature both of the Black Hole parameters, and of the Theory of Gravity. The theory of \ac{QNM} has been studied a lot since the seminal results of Regge and Zerilli \cite{Regge:1957, Zerilli:1970} about the gravitational perturbations around Schwarzschild Black Holes. Since then, a lot of work has been done, see for example \cite{Berti:2009, Maggiore:2008,Leung:1992,Leung:1999,Silva:2019} and \cite{Pani:2013, Nollert:1999} for more complete reviews.

The structure of this Part goes as follows. Chapter~\ref{chapR3:intro} is the present introduction. In Chapter~\ref{chapR3:SCH}, we review some well-known concepts of the Schwarzschild metric. In particular, we describe the parametrisation of Gravitational perturbations around this specific background and derive the so-called master equation. Then we discuss in details the concept \ac{QNM}. Finally we present a quantitative way to estimate the speed of the gravitational waves around \ac{SSSS}. In Chapter~\ref{chapR3:EFT}, we present the Effective Theory we want to consider. We derive the perturbed Black Hole solution and the perturbed master equation for the Gravitational perturbations. We compute their speed, and the correction to the \ac{QNM}, based on a method presented in \cite{Cardoso:2019,McManus:2019,Kimura:2020}. Finally, we conclude in Chapter~\ref{chapR3:concl}.


\chapter{Schwarzschild Black Hole}
\label{chapR3:SCH}
In this Chapter, we present the Schwarzschild Black Holes and the formalism to describe \ac{GW} therearound. We describe these using the formalism presented in Section~\ref{seccosmo:ssss}. In particular, we detail the decomposition between odd and even modes and define the master equation, a Schr\"odinger-like equation governing the evolution of the perturbations for each mode. We explain the concept of \ac{QNM} and briefly present some well-known results about their values in General Relativity. Finally, we argue how the speed of \ac{GW} can be computed using the effective metric formalism. Most of what is presented in the Chapter can be found in standard textbooks, e.g. \cite{Chandrasekhar:1984,Nollert:1999,Pani:2013}.

\startchap

\section{The Schwarzschild Background}
\subsection{The Metric}
In this Section, we would like to discuss in more details the Schwarzschild metric which we briefly presented in Section~\ref{seclensing:SCH}. The Schwarzschild metric is defined, in the usual spherical coordinates, as
\boxemph{ \label{eqR3:SCH}
\gS &= 
- f(r) \bds \dd t^2
+ f(r)^{-1} \bds \dd r^2 
+ r^2 \, \bds \dd  \Omega^2\,, \\
f(r) &= 1- \frac{2m}{r}\,.
}
The Schwarzschild metric is a solution of the vacuum Einstein's Equations Eq.~\eqref{eqgr:EFE}, i.e. they satisfy $\bds{R}= \bds{G}=0$. The constant $m$ is the only parameter of this metric and is given by $m=\GN M_{\mathrm{BH}}$, where $M_{\mathrm{BH}}$ is the physical mass (the Komar mass) of the Black Hole. 

This metric seems to be ill-defined at $r=\rg\equiv 2m =2\GN M$. This radius is the Schwarzschild radius of the Black Hole and corresponds to the event horizon (which gives the metric Eq.~\eqref{eqR3:SCH} the interpretation of a Black Hole!). It is tempting to think that there is some physical singularity on this spatial sphere. This is actually not the case. To check this, it is possible to compute some invariant scalar and show that they behave well at $r=\rg$. As the Ricci and the Einstein tensors vanish, the only tensor at hand is the Riemann tensor. A typical scalar is the Kretschmann scalar given by
\be 
K  \equiv \tensor{R}{_{\mu \nu \rho \sigma }}\tensor{R}{^{\mu \nu \rho \sigma }} = 12 \frac{\rg^2}{r^6}\,,
\ee 
which is well-defined at $r=\rg$. The metric seems to be singular at the event horizon, but it is possible to perform some changes of coordinates to write it in a form which also holds for $r<\rg$.

The only physical singularity is located at $r=0$, which is in some sense the \emph{position} of the Black Hole. As of today, nobody really knows what happens here...

\subsection{Geodesics}
We quickly present the geodesics equation in this metric. We take a slightly more general (stationary and isotropic) metric given by 
\boxemph{ \label{eqR3:metricAB}
\bds g =  
-A(r) \, \bds \dd t^2
+B(r)^{-1}\,  \bds \dd r^2 
+ r^2 \, \bds \dd  \Omega^2\,,
}
where $A(r)$ and $B(r)$ are arbitrary radial functions. Consider a particle moving in this gravitational field, with trajectory $x^\mu(\lambda)$.. As usually, we can define the energy $E$ and the angular momentum $L$ by 
\begin{align}
E &= A(r) \dot t^2\,, \\
L &= r^2 \dot \varphi\,, 
\end{align}
where a dot represents here a derivative with respect to the affine parameter $\lambda$. The radial equation of motion of the particle is implicitly given by
\begin{align}
    \dot r^2 = B(r) \left(\gamma + \frac{E^2}{A(r)} - \frac{L^2}{r^2} \right) \,,
\end{align}
with $\gamma=-1$ for a massive particle and $\gamma =0$ for a massless particle.

\section{Gravitational Waves \label{eqR3:G_waves_def}}
\subsection{Perturbed Einstein Equations}
We use the results from Section~\ref{seccosmo:ssss}, in which we defined the perturbations around a \ac{SSSS}. We will work in the Regge-Wheeler gauge with $2$ odd perturbations ($h_0$ and $h_1$) and $4$ even perturbations ($H_0$, $H_1$, $H_2$ and $K_0$). Moreover, as we consider a static background, we can also specify the time evolution of these variables. Hence, the only time dependence comes from the factor $\eee^{-\ii \omega t}$, and all these functions depend only on $r$. We will omit the $\ell$, $n$ and $\omega$ dependence and write $F(r)$, for $F=h_0, \dots$. Explicitly, the perturbations are given by
\boxemph{ 
\label{eqR3:oddRW}
(h_{\mo})_{\mu \nu}&=
\eee^{-\ii \omega t}
\begin{pmatrix}
0 &0  &  - h_0 \sin^{-1}\theta\, \partial_\varphi &  h_0 \sin\theta \partial_\theta  \\
0 & 0 & - h_1 \sin^{-1}\theta\, \partial_\varphi &  h_1 \sin\theta\, \partial_\theta \\
- h_0 \sin^{-1}\theta\, \partial_\varphi &  - h_1 \sin^{-1}\theta\, \partial_\varphi &0&0\\
h_0 \sin\theta\, \partial_\theta &  h_1 \sin\theta\, \partial_\theta & 0&0 
\end{pmatrix} Y_{\ell,m}\,, \\
& \nonumber \\
\label{eqR3:evenRW}
(h_{\me})_{\mu \nu}&=
\eee^{-\ii \omega t}
\begin{pmatrix}
-  A(r)\,  H_0\, &H_1\,    & 0 & 0\\
H_1\,     &  
\frac{H_2}{B(r)} & 0 &0 \\ 
0  &0  &  r^2 K_0\,   &  0\\ 
0 &  0  &   0&   r^2 K_0 \sin^2 \theta\,  \\
\end{pmatrix} Y_{\ell,m} \,.
}
Note that, for the Schwarzschild metric, one has
\be 
A(r)=B(r)=1- \frac{2m}{r} = f(r)\,.
\ee 
The full metric is then given by
\be 
\bds g = \gS + \varepsilon_2\,  \bds h\,.
\ee
Here, $\varepsilon_2$ is a bookkeeping parameter without any physical meaning. It only represents the gravitational perturbations around the background metric.

The Einstein's Equations read (in vacuum)
\be 
\bds{R}=0\,.
\ee 
They are trivially satisfied by the background Schwarzschild metric. At the perturbed level, they can be written as
\be  \label{eqR3:dRmn}
\delta \bds{R} =0\,,
\ee 
where it is understood that this quantity is at order $\varepsilon_2$, coming from the perturbation $\bds h$. They are the $10$ equations of motion we need to solve to characterise the behaviour of the perturbations. The explicit expressions can be found in \cite{Nollert:1999}.

As explained in Section~\ref{seccosmo:ssss}, the odd and the even perturbations do not interact with each other. As a matter of fact, we are left with $3$ odd and $7$ even equations, as expected from the original number of individual degrees of freedom. If we write $\mathcal{E}_{\mu \nu}\equiv \delta R_{\mu \nu}$, the schematic form of the perturbed equations of motion is
\be \label{eqR3:Emn}
\mathcal{E}_{\mu \nu}
=
\begin{pmatrix}
\mathrm{E} &\mathrm{E}  & \mathrm{E} & \mathrm{O} \\
\mathrm{E} & \mathrm{E} & \mathrm{E} &  \mathrm{O}\\
\mathrm{E} & \mathrm{E} & \mathrm{E} & \mathrm{O} \\
 \mathrm{O}& \mathrm{O} &\mathrm{O}  & \mathrm{E} \\
\end{pmatrix}\,,
\ee 
where O and E stand for \emph{Odd} and \emph{Even} respectively.
\MYrem{Simplifications}{
Using the spherical symmetry of the background metric, it is sufficient to set $m=0$ and $\theta = \pi/2$ in the equations of motion. The $\ell$-dependence only appears through the usual variable $J$ given by
\boxemph{
J^2 \equiv  \ell (\ell+1)\,.
}
}

\subsection{Master Equations}
\subsubsection{General idea}
The equation of motion given by Eq.~\eqref{eqR3:dRmn} are not very useful if considered in this form. Indeed, they are coupled second order differential equations. Hence, it is necessary to give them a very good massage to extract any relevant physical information. Let us explain qualitatively how to proceed.

The general philosophy goes as follows. We know that \ac{GW} only have two degrees of freedom. This means that out of the $7$ functions, only two are truly independent: $\psi_{\mo}$ and $\psi_{\me}$, where the first one corresponds to odd modes and the second ones to even modes. Note that $\psi_{\me/\mo}$ does not necessarily correspond to the metric perturbations. The variables $\psi_{\me/\mo}$ are called \emph{Master Variables} and the equation of motion they satisfy is the \emph{Master Equation}.

In the framework of gravitational perturbations around Black Holes, the master equation for both modes takes the form of a Schr\"odinger equation
\boxemph{ \label{eqR3:Mastergen}
f(r) \big[f(r) \psi_{\mo/\me}^\prime(r) \big]^\prime +
\big[V_\omega(r)\omega^2 - f(r)V_{\mo/\me}(r) \big] \psi_{\mo/\me}(r) =0 \,.
} 
In this equation, $V_{\mo/\me}(r)$ represents an effective radial potential, which differs between the two types of perturbations and which depends on $J$ and $m$ and $V_\omega(r)$ is a radial function, linked with the speed of the waves (see Section~\ref{secR3:speed}). Here, the function $f$ is defined above and corresponds to the Schwarzschild case, but we will see below in Remark~\ref{RemR3:Jerémie} that other choices are possible.

\MYrem{Tortoise coordinates}{
Defining the \emph{Tortoise coordinates} $x$ via 
\be 
\dd x = \frac{1}{f(r)} \dd r\,,
\ee 
the master equation Eq.~\eqref{eqR3:Mastergen} becomes
\be
\frac{\dd \psi_{\mo/\me}}{\dd x^2} + [V_\omega\,  \omega^2 - f\, V] \psi_{\mo/\me} =0\,,
\ee 
which is exactly similar (up to a constant multiplication and some redefinition) to the time-independent Schr\"odinger equation. For the Schwarzschild metric, the explicit coordinate transformation is given by
\be  \label{eqR3:Tortoise}
r + \log \left( \frac{r}{\rg}-1\right) = x\,,
\ee 
which is shown in Fig.~\ref{figR3:Tortoise}.
\begin{figure}[h!t]
	\begin{center}
	\includegraphics[width =0.6\textwidth]{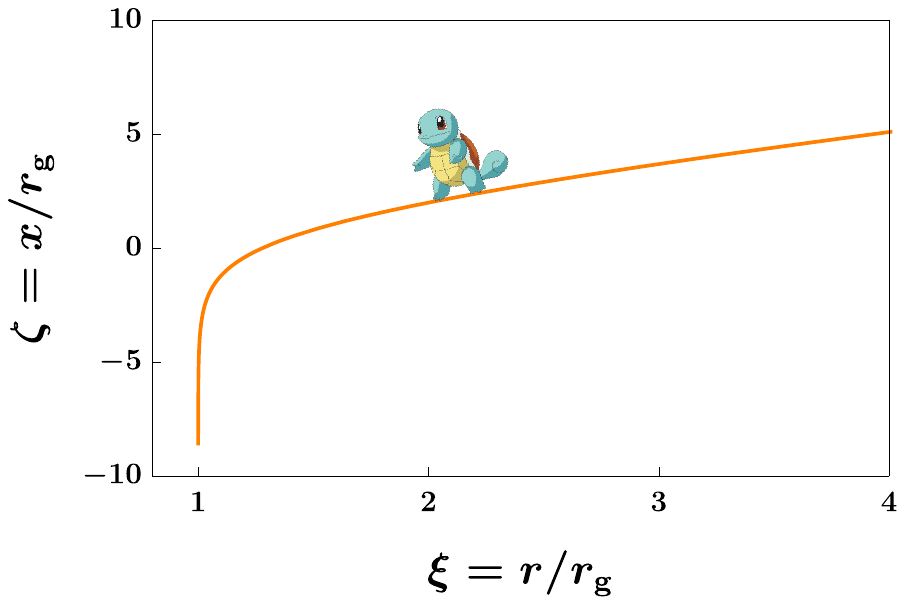} 
	\end{center}
	\caption[Tortoise coordinate]{\label{figR3:Tortoise}
		\textbf{Tortoise coordinates}\\
	Tortoise coordinates in dimensionless units: The turtle can walk as long as they fancy, they will never reach $\xi=1$, $\zeta=-\infty$.} 
\end{figure}
}

\MYrem{J\'er\'emie's trick}{ \label{RemR3:Jerémie}
When we work with the metric perturbations, the master equation does not take directly the form given by Eq.~\eqref{eqR3:Mastergen}, but is rather a general second order differential equation
\be  \label{eqR3:generic}
\phi_1^{\prime \prime} + \alpha(r) \phi_1^\prime +  \big[\alpha_\omega(r) \omega^2 - \alpha_V(r) \big]  \phi_1=0\,,
\ee
with arbitrary functions $\alpha(r)$, $\alpha_\omega(r)$ and $\alpha_V(r)$. This expression is equivalent to
\be  \label{eqR3:master_gen1}
F_1 \big[F_1 \phi_1^\prime\big]^\prime + \big[V_{\omega,1}(r) \omega^2 - F_1\, V_{r,1}(r)\big] \phi_1 =0\,,
\ee 
with 
\begin{align}
\label{eqR3:rulealpha}
\frac{F_1^\prime}{F_1} &= \alpha\,,\\
V_{\omega,1} &= F_1^2 \alpha_\omega\,, \\
V_{r,1} &= F_1 \alpha_V\,.
\end{align}
Hence, any generic second order differential equation of the form of Eq.~\eqref{eqR3:generic} can be cast into a master equation with a particular function $F$. Moreover, with the field redefinition
\be 
\phi_1 = \frac{1}{\sqrt{\chi}} \phi_2\,,
\ee 
the master equation Eq.~\eqref{eqR3:master_gen1} becomes
\be
F_2 \big[ F_2 \phi_2^\prime\big]^\prime + \big[V_{\omega,2}(r) \omega^2 - F_2 V_{r,2}(r)\big] \phi_2 =0\,,
\ee 
with
\boxemph{
    \label{eqR3:ruleF}
    F_2 &= \frac{F_1}{\chi}\,, \\
    \label{eqR3:ruleVo}
    V_{\omega,2} &= \frac{V_{\omega,1}}{\chi^2}\,, \\
    \label{eqR3:ruleVr}
    V_{r,2} &= \frac{V_{r,1}}{\chi} - \frac 34 F_1  \frac{(\chi^\prime)^2}{\chi^3} + \frac{1}{2\chi^2} (F_1\, \chi^\prime)^\prime\,.
}
The bottom line here is that the master equation is not unique: It is always possible to perform a field redefinition, provided the function $F$ and the potentials $V_\omega$ and $V_r$ are modified according to Eq.~\eqref{eqR3:ruleF}, Eq.~\eqref{eqR3:ruleVo} and Eq.~\eqref{eqR3:ruleVr}. Note that if $F_1(\rg)=0$, then $F_2(\rg)=0$, which limits the potential range of the possible functions.

In practice, the equation of motion is of the general form of Eq.~\eqref{eqR3:generic}. It can be cast into a master equation solving Eq.~\eqref{eqR3:rulealpha}, and the function $F$ can be modified using Eq.~\eqref{eqR3:ruleF}. We call this process \emph{J\'er\'emie's trick}, and we will see its relevance in Chapter~\ref{chapR3:EFT}.
}

\subsubsection{Odd perturbations \label{secR3:odd_BG_master}}
We discuss here the odd perturbations. As indicated in\footnote{Is it really an equation?} Eq.~\eqref{eqR3:Emn}, we have three equations:~$\Ecal_{t,\varphi}$, $\Ecal_{r,\varphi}$, and $\Ecal_{\theta,\varphi}$ and two variables:~$h_0$ and $h_1$. Recall that at this stage, the equation only depend on $r$ (and contains remanent terms proportional to $\omega$ and $J$ coming from the time and angular derivatives).

\begin{figure}[h!t]
	\begin{center}
	\includegraphics[width =0.6\textwidth]{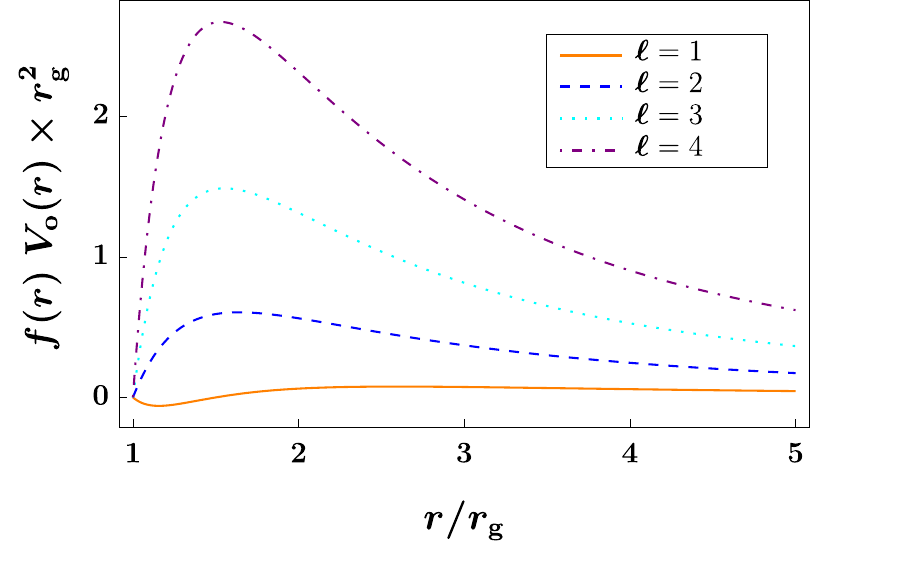} 
	\end{center}
	\caption[Regge potential]{\label{figR3:Odd_pot}
		\textbf{Regge potential}\\
	The Regge potential (multiplied by the function $f(r)$) for the first $4$ values of $\ell$ in dimensionless units} 
\end{figure}

These three equations are not independent, and an algebraic relation can be found between them and their derivatives, which is expected as we only have two variables. As a matter of fact, the functions $h_0$ can be expressed in terms of $h_1$ and $h_1^\prime$ using $\Ecal_{\theta,\varphi}$, see Appendix~\ref{secAppBH:bgodd} for the explicit expression. Once this has been done, $\Ecal_{r,\varphi}$ reads
\be 
\frac{\rg-r}{4r}h_1^{\prime \prime }
+ \frac{2r-5\rg}{4r^2} h_1^\prime
+ \frac{5\rg^2+r\rg (J^2-6)-r^2(J^2-2) + r^4 \omega^2}{4r^3 (\rg-r)} h_1=0\,.
\ee 
Let us use J\'er\'emie's trick to massage this equation. Using the notation of Eq.~\eqref{eqR3:generic}, we have
\be 
\alpha(r) = - \frac 5r + \frac{3}{r-\rg}\,,
\ee 
and hence it is a master equation with (up to a constant prefactor)
\be 
F_1(r) = \frac{(r-\rg)^3}{r^5}\,,
\ee 
where we used the relation Eq.~\eqref{eqR3:rulealpha}. We want to get a master equation with 
\be 
f(r)= 1- \frac{\rg}{r}\,,
\ee 
which is possible with the field redefinition
\begin{align}
h_1(r) &= \frac{1}{\sqrt{\chi}}\psi_\mo(r)\,, \\
\chi(r) &= \frac{F_1(r)}{f(r)} = \frac{(r-\rg)^2}{r^4}\,,
\end{align}
where we used the transformation rule Eq.~\eqref{eqR3:ruleF}. The master equation is then
\boxemph{ \label{eqR3:Master_Odd}
&f (f \psi_\mo^\prime)^\prime + (\omega^2 - f\, V_\mo(r)) \psi_\mo =0\,,\\
\label{eqR3:VRegge}
&V_{\mo}(r)=\frac{J^2}{r^2} - 3 \frac{\rg}{r^3}\,.
}
This is the famous result found by Regge \cite{Regge:1957}, the function $V_\mo(r)$ being the \emph{Regge-Wheeler potential}\footnote{Some authors include the function $f(r)$ in the definition of the potential}. The functions $f(r) V_\mo(r)$ for the first $4$ values of $\ell$ are shown in Fig.~\ref{figR3:Odd_pot}

\subsubsection{Even perturbations \label{secR3:even_BG_master}}
We would like to do the same for even perturbations. Now, \emph{Rien ne va plus~!}... The equations are more numerous, there are more variables and overall the process is more tedious. Let us try to massage them well enough to decipher their secrets.

We have now $7$ equations:~$\Ecal_{t,t}$, $\Ecal_{r,r}$, $\Ecal_{\theta,\theta}$, $\Ecal_{\varphi,\varphi}$, $\Ecal_{t,r}$, $\Ecal_{t,\theta}$ and $\Ecal_{r,\theta}$, and $4$ variables:~$H_0$, $H_1$, $H_2$ and $K_0$. Using $\Ecal_{\theta, \theta}$ and $\Ecal_{\varphi, \varphi}$, it is direct to show that $H_0+H_2=0$, which allows us to eliminate one of them. The second one of these variables can then be eliminated using $\Ecal_{t,r}$. We are then left with $H_1$ and $K_0$ only. It is possible to go even a bit further: Using $\Ecal_{t,\theta}$ allows us to express $K_0^\prime$ as a function of $H_1$ and $K_0$. Then, $K_0^\prime$ and its derivatives can be substituted in the other equations. Finally, using $\Ecal_{r,r}$ and $\Ecal_{r,\theta}$ makes it possible to express $H_1^\prime$ as a function of $H_1$ and $K_0$. The full expressions are given in Appendix~\ref{secAppBH:bgeven}. The goal now is to find the master equation. We know that we have only one physical degree of freedom for the even perturbation $\psi_\me$. The procedure is slightly more subtle than for the odd perturbations. 

First, we assume that the master variable $\psi_\me$ is a linear combination of the metric perturbations, i.e.
\be  \label{eqR3:Ansatzphie}
\psi_\me (r) = \frac{1}{\omega}f_H(r) H_1(r) + f_K(r) K_0(r)\,.
\ee 
Recall that all the other variables can be expressed as functions of $H_1$ and $K_0$, which justifies this choice (e.g. there could not be any term proportional to $H_0$ or $K_0^\prime$). We describe here how to determine the functions $f_H$ and $f_K$ from scratch.

We impose the master equation 
\be  \label{eqR3:Master_Even_Ansatz}
f \big[f \psi_\me^\prime\big]^\prime + \big[V_\omega(r) \omega^2 - f V_\me(r) \big] \psi_\me\,.
\ee 
Inserting the Ansatz Eq.~\eqref{eqR3:Ansatzphie} into the master equation Eq.~\eqref{eqR3:Master_Even_Ansatz}, and using the expressions for $H_1^\prime$ and $K_0^\prime$ (given in Appendix~\ref{secAppBH:bgeven}), we are left with an expression containing only $H_1$ and $K_0$. From this expression, we can isolate the quantity
\be 
V_{\omega,r}(r) \equiv V_\omega(r) \omega^2 - f V_\me(r)\,,
\ee 
and extract the radial part given by
\be 
V_r(r) = V_{\omega,r}(r) - \omega^2 \frac 12 \left(\partial_\omega^2 V_{\omega,r} \right)\vert_{\omega=0}\,.
\ee 

The key point is that here, with generic functions $f_H$ and $f_K$, the potential $V_r(r)$ would still contain terms proportional to $H_1$ and $K_0$, which we do not want. Hence we impose the conditions
\begin{align}
    \label{eqR3:c1}
    E_H &\equiv \frac{\delta V_r}{\delta H_1} =0\,, \\
    \label{eqR3:c2}
    E_K &\equiv \frac{\delta V_r}{\delta K_0} =0\,, 
\end{align}
This is not straightforward, and one needs to play a little bit before getting the final solution, which we provide here as a \emph{Deux Ex Machina} (the use of \emph{Mathematica} is recommanded!). As a matter of fact, combining $E_H$ and $\partial_\omega^2 E_K$ allows us to eliminate $f_H^{\prime \prime}$. The remaining equation is algebraic for $f_K$ and yields
\be  \label{eqR3:condfK}
f_K(r) =\ii  \frac{ r^2 f_H(r)}{r-\rg}\,.
\ee 
Substituting this solution for $f_K(r)$ in either condition Eq.~\eqref{eqR3:c1} or Eq.~\eqref{eqR3:c2} leads to a differential equation for $f_H(r)$ only, which is directly solved by
\be 
f_H(r) = \frac{r-\rg}{r(J^2-2)+3\rg}\,,
\ee 
which implies, using Eq.~\eqref{eqR3:condfK},
\be 
f_K(r)= \ii \frac{r^2}{r(J^2-2)+3\rg} \,.
\ee 

\begin{figure}[h!t]
	\begin{center}
	\includegraphics[width =0.6\textwidth]{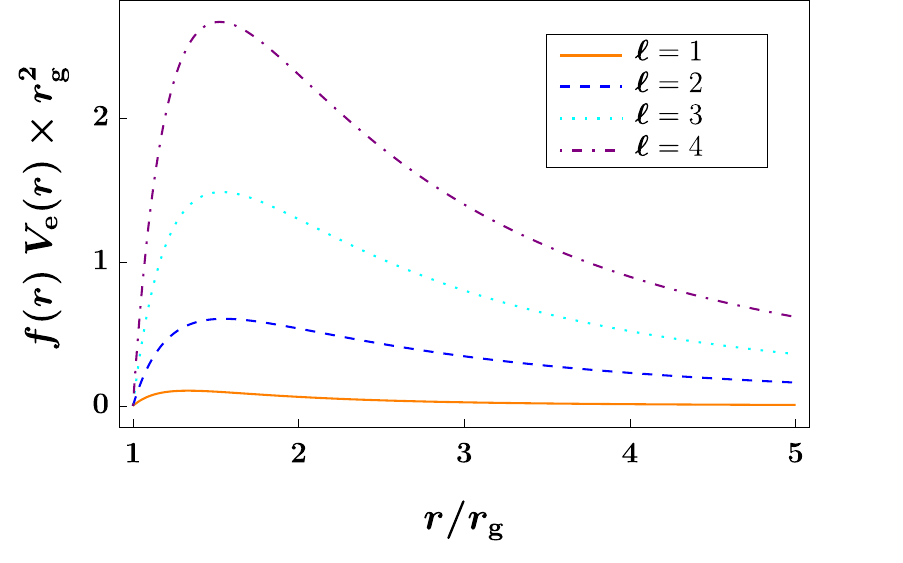} 
	\end{center}
	\caption[Zerilli potential]{\label{figR3:Even_pot}
	\textbf{Zerilli potential}\\
	The Zerilli potential (multiplied by the function $f(r)$) for the first $4$ values of $\ell$ in dimensionless units} 
\end{figure}

We are now able to define $\psi_\me$ using Eq.~\eqref{eqR3:Ansatzphie}, and the master equation reads
\boxemph{ \label{eqR3:Master_Even}
&f \big[f \psi_\me^\prime\big]^\prime + \big[\omega^2 - f\, V_\me(r)\big] \psi_\me =0\,.\\
\label{eqR3:VZerilli}
&V_\me(r)=\frac{J^2-2}{3r^2} +  \frac{\rg}{r^3} + \frac 23 \frac{4-3J^4+J^6}{((J^2-2)r+3\rg)^2}\,.
}
This equation is also famous. It was first derived by Zerilli \cite{Zerilli:1970}, using the seminal work of Regge. The potential $V_\me(r)$ is the Zerilli potential, and the product $f(r)V_\me(r)$ is shown in Fig.~\ref{figR3:Even_pot} for the first $4$ values for $\ell$.

\MYrem{Similarity of the two potentials}{
From Fig.~\ref{figR3:Odd_pot} and Fig.~\ref{figR3:Even_pot}, it seems that the potentials are roughly equal for $\ell\geq 2$. Using the explicit expressions Eq.~\eqref{eqR3:VRegge} and Eq.~\eqref{eqR3:VZerilli}, it is direct to show that
\be
V_\mo(r) - V_\me(r) = \mathcal{O}(\ell^{-2})\,,
\ee
which implies that these potentials are equal for large values of $\ell$.
}
\section{Quasinormal Modes}
\subsection{General Idea}
In this Section, we would like to discuss and compute the \ac{QNM} of a Schwarzschild Black Hole. Before going deep into the technical details, let us present the topic from another point of view.

Remember when you were a kid (or even as a grown-up!), you probably played with this soap bubble bottles. What happened is that the bubbles you had just produced were not perfectly spherical. Some of them were \emph{Siamese bubbles} which merged after some point, or two different bubbles could collide and merge to form one bigger bubble.

In all these situations, the rest of the story is the same: The final bubble is not spherical and undergoes some oscillations which are damped over time. This physical process has been discuss quantitatively, see \cite{Kornek:2010}. 

\begin{figure}[h!t]
	\begin{center}
	\includegraphics[width =0.7\textwidth]{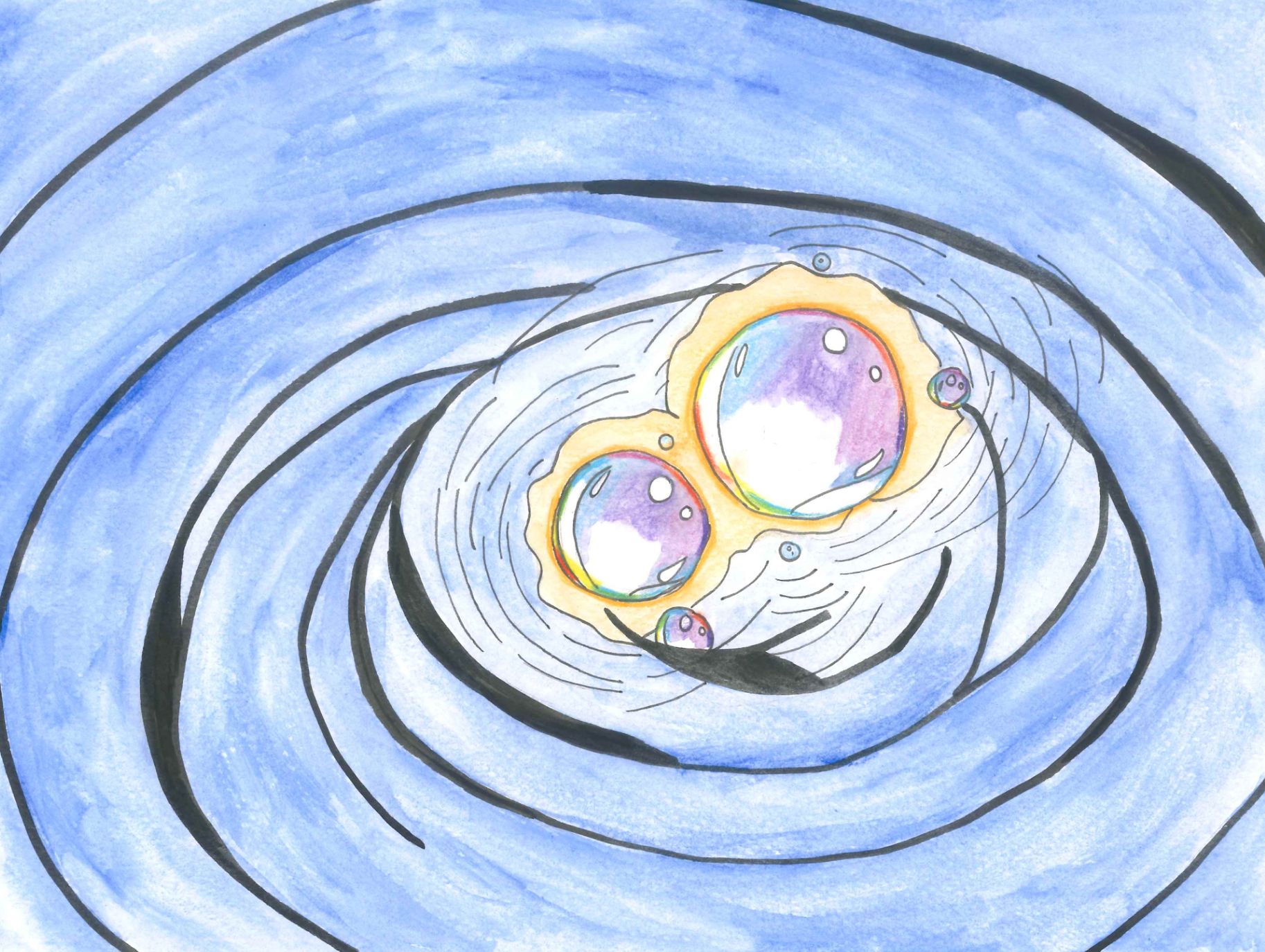} 
	\end{center}
	\caption*{\label{figR3:Bulles}
	\textbf{Bulles noires}, Beatriz Alvarez
	} 
\end{figure}

The same process happens when two Black Holes collide to produce one final, bigger, Black Hole. The latter is not completely spherical at the beginning but presents some \emph{wrinkles} and oscillates. These oscillations emit \ac{GW} that we can, eventually, detect. More important, and we will explain this later on, the frequencies of these oscillations are quantified. Their precise values depend on the physical properties of the Black Hole (e.g. its mass or angular momentum) and on the physical theory of gravity (e.g. \ac{GR}). These \ac{GW} travel through spacetime, and hence their amplitude decreases over time. They form the signal that we measure on Earth, e.g. with LIGO, and this signal is called the \emph{ringdown} of the Black Hole. For a quantitative, yet excellent, video on the topic, see \cite{PBS:1} by \emph{PBS Space Time}.

Being able to compute and measure these \ac{QNM} is very relevant as it allows us to build precise predictions to test our understanding of \ac{GW} propagation, spacetime and Black Holes. Moreover, from a theoretical and mathematical point of view, the theory of \ac{QNM} is very interesting in several aspects.

\subsection{A Mechanical Example}
To give a precise definition of \ac{QNM}, let me first use a mechanical analogy. Classical Mechanics is a theory we meet in everyday life, and using examples therefrom can in general be very fruitful before diving into a more advanced and less intuitive topic. This discussion has already been presented in more details, see for example \cite{Maggiore:2008} and we discuss here only the general ideas.

Consider a damped harmonic oscillator of mass $m$, spring constant $k$ and friction term $\gamma$. The differential equation describing the evolution of its position $x$ is 
\be 
m \ddot x = - k x - \gamma \dot x\,.
\ee 
Using the Ansatz
\be  \label{eqR3:XAnsatz}
x \sim \eee^{-\ii \omega  t}\,,
\ee 
we find
\be 
- m \omega^2 = -k + \ii \gamma \omega \,,
\ee 
which leads to the solution for the frequency
\be 
\omega = \pm \frac{\sqrt{4km-\gamma^2}}{2m} - \ii \frac{\gamma}{2m}\,.
\ee 
Plugging this into the solution Eq.~\eqref{eqR3:XAnsatz} yields to an oscillatory solution\footnote{We assume that the damping is sufficiently small, i.e. $4km>\gamma^2$.} with frequency
\be 
\omega_{\mathrm{R}} =  \pm \frac{\sqrt{4km-\gamma^2}}{2m}\,,
\ee 
and with exponential damping term given by $\eee^{ \omega_{\mathrm I} t}$ with
\be 
\omega_{\mathrm I} = -\frac{\gamma}{2m} <0\,.
\ee 
The bottom line is that the final solution is indeed an exponential solution, as usual, but with an imaginary frequency corresponding to the damping term. This motivates the following definition.
\MYdef{Quasinormal modes}{
A Quasinormal Mode (\ac{QNM}) is an oscillatory mode of the form $\eee^{-\ii \omega t}$ where $\omega$ is a complex number. The real part of $\omega$ is the oscillatory part, while the imaginary part of $\omega$ is a damping term (provided it is negative).
}
With this definition of \ac{QNM}, we understand why they are ubiquitous in Physics: Any real oscillatory system possesses \ac{QNM}. Put a banana on a table and make it oscillate slightly, or observe a swing which has just been left by a kid if you're not convinced.

The emission of \ac{GW} by a Black Hole is by essence a dissipative system: The waves either fall back into the horizon, or leave away towards infinity. In both cases, the amplitude of the wave is expected to decrease over time!

\subsection{Results}
We have just discussed the nature of \ac{QNM}. The next natural question is: How to compute those frequencies explicitly, for example in our setup with Perturbation around a \ac{SSSS} corresponding to a Black Hole. This is a field of investigation in Gravitational Physics per se, and we briefly present here the method introduced by Leaver \cite{Leaver:1985, Leaver:1986}.

To compute the \ac{QNM}, one needs an additional assumption, which we briefly mentionned. Recall that the master equation Eq.~\eqref{eqR3:Mastergen} is weirdly similar to the Schr\"odinger equation. In Quantum Mechanics, the energies are quantified. The mathematical origin of this comes, in principle, from the boundary conditions of the system. For example, in the usual \emph{Particle in a Box} problem, the condition $\psi(0)=\psi(L)=0$ imposes some conditions on the momentum $k$ of the wave, henceforth on its energy. In our situation, we have to boundary conditions:
\begin{itemize}
    \item no wave is coming from infinity ;
    \item no wave is coming from inside the horizon.
\end{itemize}

These two conditions lead to the quantification of the \ac{QNM}. The details can be found in \cite{Leaver:1985} (see Section 2). We present here the main ideas. It is possible to rewrite the master equation (for the odd case) as
\be  \label{eqR3:Leaver_master}
r(r-\rg) \Psi^{\prime \prime} + \Psi^{\prime} + \left( \frac{\omega^2 r^3}{r-\rg}-J^2 + \frac{3\rg}{r}\right) \Psi=0\,,
\ee 
where $\Psi$ is a master variable (different from the one we used before).

The first boundary condition, in the Tortoise coordinates, reads
\be 
\Psi(t,x) \sim \eee^{-\ii \omega t} \eee^{\ii \omega x}\,,
\ee 
corresponding to a wave escaping at $x=\infty$. Using the definition of the Tortoise coordinates Eq.~\eqref{eqR3:Tortoise}, we get
\be 
\Psi(r) \sim  r^{\ii \omega} \eee^{\ii \omega r}\,,
\ee 
where we used $x \approx r$ as $r\gg \rg$.

The second boundary condition reads
\be 
\Psi(t,x) \sim \eee^{-\ii \omega t} \eee^{-\ii \omega x}\,,
\ee 
corresponding to a wave falling into the Black Hole at $x=-\infty$. In radial coordinates, this yields
\be 
\Psi(r) \sim (r-\rg)^{-\ii \omega \rg}\,.
\ee 
A solution of the differential equation Eq.~\eqref{eqR3:Leaver_master} satisfying the boundary conditions is
\be \label{eqR3:Leaver_series}
\Psi(r)=(r-\rg)^{-\ii \omega \rg } r^{2\ii \omega \rg } \eee^{\ii \omega (r-\rg)}
\sum_{n=0}^{\infty} a_n \left(\frac{r-\rg}{r}\right)^n\,,
\ee 
Plugging this Ansatz in the master equation yields to the recursion formula
\begin{align}
    0&= \alpha_0 a_0 + \beta_0 a_0\,, \\
    0&=\alpha_n a_{n+1} +  \beta_n a_n + \gamma_n a_{n-1}\,, \\
    \alpha_n &=n^2 + \left(-2\ii\omega \rg+2 \right)n - 2\ii \omega \rg +1\,, \\
    \beta_n &= -(2n^2 + (-8\ii \omega \rg +2) - 8 \omega^2 \rg^2-4\ii \omega \rg+ J^2 -3)\,, \\
    \gamma_n &= n^2 -4\ii \omega \rg n -4\,.
\end{align}
The argument of Leaver goes as follows: The allowed frequencies $\omega$ are the ones for which the series in the Ansatz Eq.~\eqref{eqR3:Leaver_series} converges. This is the case if $\omega$ is a solution of the continuous fraction equation
\boxemph{
\beta_0 - \frac{\alpha_0 \gamma_1}{ \beta_1 - \frac{\alpha_1 \gamma_2}{\beta_2 - \dots}}=0\,.
}
This condition can be truncated at a given order setting $\alpha_j=0$ for $j\geq j_{\mathrm{max}}$. For a given values of $j_{\mathrm{max}}$, we get $2j_{\mathrm{max}}$ numerical solutions for the frequency $\omega$.

\begin{figure}[ht!]
	\begin{center}
	\includegraphics[width =0.6\textwidth]{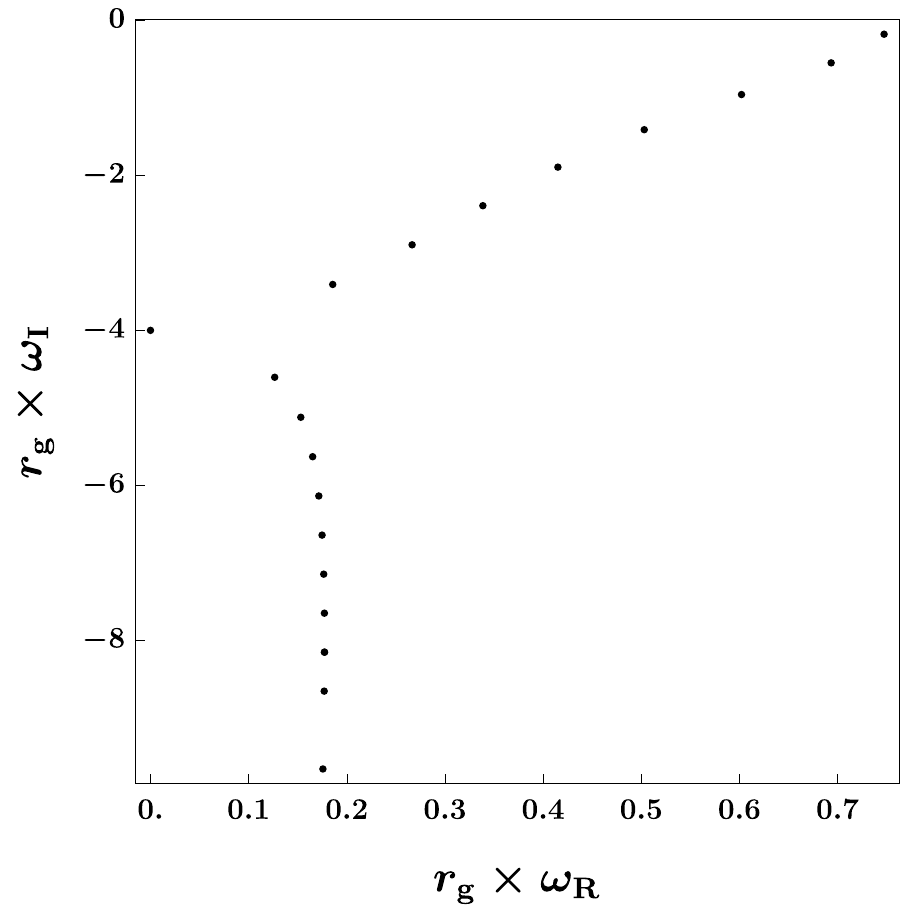}
	\end{center}
	\caption[Quasinormal modes - Schwarzschild]{\label{figR3:Freq_BG} 
	\textbf{Quasinormal modes}\\
	\ac{QNM} for the Schwarzschild Black Hole with $\rg=2m$ and for $\ell=2$. We show the first $20$ modes with decreasing $\omega_{\mathrm I}$. Note the mode with $n=8$ is exactly real, $\omega_8 \rg\approx -4$, see Section~5.1 of \cite{Berti:2009} for more explanations.} 
\end{figure}

We will not reproduce these computations, as they have been done several times. We only present them in Fig.~\ref{figR3:Freq_BG}. The values are taken from  \cite{Berti:2009}, and the same analysis can be done for even perturbations. An interesting fact: The Schwarzschild Black Hole is said to be \emph{isospectral}, i.e. the odd and the even modes have the same \ac{QNM} frequencies, see \cite{Chandrasekhar:1984, Moulin:2019} for more explanations on this topic. The \ac{QNM} frequencies are ordered with decreasing $\omega_{\mathrm I}$, the classification $n$ index being the \emph{Overtone Number}, starting at $n=0$. The first frequencies (for $\ell=2,3,4$) are given by (see e.g. \cite{Konoplya:2004}\footnote{The sign of the complex part of $\omega$ can differ depending on the sign convention of the exponential.})
\begin{alignat}{2}
\omega_0 = \frac{0.747343-0.177925\ii}{\rg}\,, &&\quad  \ell=2\,, \\
\omega_0 = \frac{1.199887-0.185406\ii}{\rg}\,, &&\quad \ell=3\,, \\
\omega_0 = \frac{1.618357-0.188328\ii}{\rg}\,, &&\quad \ell=4\,.\\
\end{alignat}

\section{Speed of Gravity\label{secR3:speed}}
\subsection{Effective Metric}
The second useful information we can extract from the master equation is the speed of the \ac{GW}. We present here the general formalism and derive the main formula. 

Recall that the master equation for a generic field $\psi(r)$ is
\be \label{eqR3:Master_speed}
F \big[F \psi^\prime\big]^\prime  + \big[V_\omega(r)\omega^2- F V_r(r)\big] \psi =0\,,
\ee 
where $F$ is an arbitrary function satisfying $F(\rg)=0$, while $V_\omega$ and $V_r$ are arbitrary radial (recall $V_r$ generally depends on $J$). 

We would like to rewrite the master equation Eq.~\eqref{eqR3:Master_speed} in the form of a generic Laplacian, namely
\be \label{eqR3:eff_eq}
Z^{\mu \nu} \mathcal{D}_\mu \mathcal{D}_\nu  \Psi  =0\,,
\ee 
where
\be  \label{eqR3:AnsatzPsi}
\Psi(t,r,\theta, \varphi) = \eee^{-\ii \omega t} \psi(r) \YLM{\ell}{m}(\theta, \varphi)\,,
\ee 
is a generic field depending on all the variables. In the relation given by Eq.~\eqref{eqR3:eff_eq}, 
\be
\bds{Z} = Z_{\mu \nu} \bdd{x}^\mu \bdd{x}^\nu
\ee
is an effective metric, and $\bds{\mathcal D}$ is its associated covariant derivative. The effective metric can be interpreted as the metric \emph{felt} by the \ac{GW}. We will assume that the effective metric has the same structure as the background metric, i.e.
\be \label{eqR3:eff_Z}
\bds Z = -Z_t(r) \bdd t^2 + Z_r(r)^{-1} \bdd r^2 + r^2 Z_\Omega(r) \bdd  \Omega^2\,.
\ee 
Plugging the Ansatz for $\Phi$ Eq.~\eqref{eqR3:AnsatzPsi} in the generic Laplacian Eq.~\eqref{eqR3:eff_eq}, and using the form of the effective metric Eq.~\eqref{eqR3:eff_Z} yields (setting $m=0$ and $\theta=\pi/2$ at the end)
\begin{align}
\label{eqR3:effective_details}
\psi^{\prime \prime}
+ & \left( \frac 2r +\frac{Z_\Omega^\prime}{Z_\Omega}+ \frac 12 \frac{Z_r^\prime}{Z_r} + \frac 12 \frac{Z_t^\prime}{Z_t}\right) \psi^{\prime}\\
+ &  \frac{\omega^2}{Z_t Z_r} \psi(r)
-  \frac{J^2}{r^2 Z_r Z_\Omega} \psi(r) =0\,,
\end{align}
which should be equal to the master equation Eq.~\eqref{eqR3:Master_speed}
\be  
\psi^{\prime \prime} + \frac{F^\prime}{F} \psi^{\prime} + \frac{V_\omega \omega^2}{F^2} \psi - \frac{V_r}{F} \psi\,.
\ee
Comparing the terms proportional to $\omega^2$ gives directly
\be 
Z_r = \frac{F^2}{Z_t V_\omega}\,.
\ee 
Plugging this solution into Eq.~\eqref{eqR3:effective_details} and comparing the terms proportional to $\psi^\prime$ leads to\footnote{We set the integration constants to get convenient expressions for $Z_\Omega$ and $Z_t$.}
\be
Z_\Omega(r) = \frac{J^2 \sqrt{V_\omega}}{r^2}\,.
\ee
Finally, comparing the last term proportional to $\psi$ gives
\be 
Z_t(r) = \frac{F(r) V_r(r)}{\sqrt{V_\omega(r)}}\,.
\ee 
Combining all these results, we get the expression for the effective metric
\boxemph{ \label{eqR3:Eff_met_final}
\bds Z = - \frac{F(r) V_r(r)}{\sqrt{V_\omega(r)}} \bdd t^2 
+ \frac{V_r(r) \sqrt{V_\omega(r)}}{F(r)} \bdd r^2
+ J^2 \sqrt{V_\omega(r)} \bdd \Omega^2\,.
}

\MYrem{Form of the effective metric}{
It is interesting to see that the effective metric only depends on the structure of the master equation, and not directly on the metric given by Eq.~\eqref{eqR3:metricAB}. Obviously, in fine, the potentials $V_\omega$ and $V_r$ do depend implicitly on the background metric and on the underlying theory of gravity, as we will see in the next Chapter.
}

\subsection{Speed of the Waves}
In this Section, we would like to compute the speed of a \ac{GW} satisfying the Laplace equation with the effective metric $\bds Z$ Eq.~\eqref{eqR3:eff_eq}. We present here a heuristic derivation of the formulas. We take an Ansatz of the form
\be 
\Psi \sim \eee^{-\ii \omega t} \eee^{\ii k_r r} =\eee^{\ii k_\mu x^\mu}\,,
\ee 
with $\bds k =- \omega \bdd t + k_r \bdd r$. Plugging this Ansatz into the Laplacian Eq.~\eqref{eqR3:eff_eq} gives
\be 
Z^{\mu \nu} k_\mu k_\nu =0\,,
\ee 
or
\be 
-Z^{tt} \omega^2 = Z^{rr} k_r^2\,.
\ee 
The phase velocity is given by the condition 
\be  \label{eqR3:phase}
 k_\mu x^\mu =0\,,
\ee 
or
\be  \label{eqR3:omegatkr}
\omega t = k_r r\,.
\ee 
Assume now that the waves travels during a small coordinate time $\dd t$ and a radial coordinate distance $\dd r$. The relation Eq.~\ref{eqR3:omegatkr} still holds at this infinitesimal level. The physical time the waves travels is given by 
\be 
\dd  t_{\mpp}^2 = -g_{tt} \dd t^2\,,
\ee 
and likewise, the physical distance is
\be 
\dd  r_{\mpp}^2 = g_{rr} \dd r^2 \,.
\ee 
Using the relation Eq.~\eqref{eqR3:omegatkr}, the velocity is given by
\be 
\cS^2 = \frac{\dd r_{\mpp}^2}{\dd t_{\mpp}^2}
= - \frac{g_{rr}}{g_{tt}}\left(\frac{\omega}{k_r}\right)^2\,.
\ee 
Finally, using the condition on $k$ and $\omega$ given by Eq.~\eqref{eqR3:phase}, the final formula for the speed of the \ac{GW} is (recall that $Z^{tt}<0$)
\boxemph{ \label{eqR3:cs_formule}
\cS^2 = \frac{Z^{rr}g_{rr}}{Z^{tt}g_{tt}} = \frac{F(r)^2}{A(r) B(r)V_\omega(r)}\,,
}
where we used the generic form for the metric Eq.~\eqref{eqR3:metricAB} and the expression of the effective metric Eq.~\eqref{eqR3:Eff_met_final}.

Regarding the perturbations around the Schwarzschild metric, we have for both modes (see the master equations Eq.~\eqref{eqR3:Master_Odd} and Eq.~\eqref{eqR3:Master_Even})
\be
A(r)=B(r)=F(r)=f(r)=1- \frac{\rg}{r}\,,
\ee 
and
\be 
V_\omega(r)=1\,,
\ee
which implies trivially 
\be 
\cS =1\,.
\ee 
In other words, around a Schwarzschild Black Hole, \ac{GW} travel at the speed of light. We already knew this, but the work we did in this Section is not useless, as we will see in the next Section that this result does not hold at the EFT level.

\MYrem{Invariance of the speed}{
The expression for the speed of the waves Eq.~\eqref{eqR3:cs_formule} seems problematic: The function $F$, as explained in Rem.~\ref{RemR3:Jerémie} is not well-defined. This is actually not a problem, as the potential $V_\omega$ is also modified when the function $F$ is modified. It can be checked directly using the transformation rules Eq.~\eqref{eqR3:ruleF} and Eq.~\eqref{eqR3:ruleVo} that the expression for the speed Eq.~\eqref{eqR3:cs_formule} is invariant under such a transformation (as the metric, and hence the functions $A$ and $B$, is not modified). We conclude that the formula we found is indeed a physical observable as it does not depend on the arbitrary choice for the function $F$.
}

\MYrem{Dependence on the coordinates}{
Note that, in principle Eq.~\eqref{eqR3:cs_formule} depends on the radial coordinate. This is irrelevant for the Schwarzschild metric as it turns out that the final expression does not, but this will turn out to be very important when we consider the EFT at the perturbed level. 
}
\MYrem{Which speed did we compute?}{
We computed the speed of gravity $\cS$, but with respect to which frame? In the derivation, we used that the distances and time were computed with the metric $\bds g$. If we want to compare this speed with other speed (say the speed of light), we need to assume which metric is \emph{felt} by light. We will then make the assumption that we are working in the Jordan frame. This implies that other fields (including light and matter) are minimally coupled with gravity, and follow geodesics of the metric $\bds g$, see Section~\ref{secframes:action} where we discussed this with more details. We can then use this metric to compute distances, times, and assume that the speed of light is $c_{\mathrm{l}}=1$ in this frame.  
}

\stopchap

\chapter{Effective Theory of Gravity\label{chapR3:EFT}}
In this Chapter, we present an effective theory of gravity parametrised by Riemann-squared and Riemann-cubed terms added to the Einstein-Hilbert Lagrangian. A similar theory is considered for example in \cite{Cardoso:2018,Metsaev:1986}. We build explicitly the Lagrangian and argue why most of the terms can effectively be neglected in our framework, as we work at first order around a Schwarzschild solution. We derived the \ac{SSSS} Schwarzschild-like Black Hole in this theory, and derive the master equations governing the \ac{GW} in this perturbed background. The main results of this chapter are the derivation of the speed of the \ac{GW}, and the corrections to the \ac{QNM} of the Black Hole following \cite{Cardoso:2019,McManus:2019,Kimura:2020}.

\startchap

\section{A Perturbed Black Hole}
\subsection{General Idea}
In this Section, we want to present an effective field theory of gravity. Effective field theories are a domain of research and of interest per se, and I am far from pretending that I master, not even understand, it. As usual, a lot of references exist on the topic, see for example \cite{Donoghue:1994,Donoghue:1995, Donoghue:2012,Burgess:2003} or the nice article written by Georgi \cite{Georgi:1993}, which was very useful to me at the beginning of PhD.

The idea goes as follows: When we probe a physical system, we may only be able do it at low energy. For example, our detectors may not be sensitive to photons with energy higher than a given cutoff. A more intuitive example is simply friction: When you try to move forward in a swimming pool, you are not sensitive to the individual, tiny, water molecules dragging you back. Rather, you feel at the human scale a friction force which you model as $F  = - \gamma v$, knowing that this is simply an approximation capturing out all the microscopical effects of the water. In this specific example, the coefficient $\gamma$ is not a fundamental physical parameter, but is simply an effective coefficient in the theory.

We will follow the same approach for gravity. The Einstein-Hilbert action given by Eq.~\eqref{greq:EH} describes our theory of gravity. But what if some very massive field are \emph{hidden} because we are not able to produce them? As explained above, we may be able to still describe their effects on low-energy processes by adding some effective terms in the Lagrangian, as we explain in the following section.

More precisely, in this Chapter, we will consider a Lagrangian of the form
\be 
\Lcal = \LEH + \varepsilon\,  \mathcal{L}_{\mathrm{EFT}}\,,
\ee 
where $\mathcal{L}_{\mathrm{EFT}}$ is the effective Lagrangian coming from the effects of high energy particles, and $\varepsilon \ll 1$ is a small perturbative parameter. More details are provided in the next Section. Our goal is to find a solution of this theory in the form
\boxemph{ \label{eqR3:g_decompos}
\bds g = \gS + \varepsilon\,  \delta \bds{g} + \varepsilon_2\, \bds h\,.
}
In this Ansatz, $\gS$ is the usual Schwarzschild background given by Eq.~\eqref{eqR3:SCH}. The second term $\delta \bds{g}$ represents a correction to the Black Hole solution due to the effective Lagrangian $\mathcal{L}_{\mathrm{EFT}}$, which is supposed to be small (hence proportional to $\varepsilon$). We will assume that this perturbed metric is still a \ac{SSSS}. The second term describes the gravitational perturbations around the Black Hole, similar to what we did in Section~\ref{eqR3:G_waves_def}. Here, $\varepsilon_2$ has no particular physical meaning, but is rather a bookkeeping parameter representing the gravitational perturbations.

\subsection{The Lagrangian}
Recall that the Einstein-Hilbert Lagrangian\footnote{We do not include in the Lagrangian the square root determinant.} is given by
\be 
\Lcal_{\mathrm{EH}} = \frac{\Mpl^2}{2} R\,,
\ee
where we used the definition of the Planck mass
\be 
\frac{1}{8\pi\GN} = \Mpl^2\,.
\ee 
We would like to estimate the order of magnitude of this Lagrangian, assuming that the metric is close to the Schwarzschild Black Hole. Obviously, the value of the Lagrangian per se vanishes, as this is a solution of the Einstein's Equations. However, we can simply think that $R$ represents a typical component of the Riemann tensor, which does not vanish. 

The curvature $R$ has the units of an inverse length squared. In the Schwarzschild metric, there is only one length: The Schwarzschild radius. Hence, we expect that
\be 
R\sim \frac{1}{\rg^2}\,,
\ee 
and 
\be 
\Lcal_{\mathrm{EH}} \sim \frac{\Mpl^2}{\rg^2}\,.
\ee 

We want to build the effective Lagangian. However, we know that General Relativity and the Schwarzschild solution describe the Physics of the Solar System very well: We do not want the perturbative Lagrangian to be \emph{big}, and hence we need to find a very small dimensionless quantity to multiply $\LEH$ therewith. Let $M$ be an arbitrary mass of the same order of magnitude as the Planck mass (or higher). Then, the quantity (with $\Lpl$ the Planck length)
\be
\frac{1}{M \rg} \sim \frac{\Lpl}{\rg}\,,
\ee 
is very small if we consider a typical astronomical Black Hole whose radius is much bigger than the Planck Length. Hence, we would like to find a Lagrangian of the form
\be 
\Lcal_{3} \sim \Lcal_{\mathrm{EH}} \times \frac{1}{M \rg}  \sim \frac{\Mpl^2}{M \rg^3}\,,
\ee 
where the number $3$ indicates the power of $\rg$ we need in the denominator. With the Riemann tensor only, we cannot get odd powers of $\rg$, so we also need to use the covariant derivative $\bds \nabla$ which has the units of an inverse length, or $\rg^{-1}$. Schematically, this reads
\begin{align} 
\label{eqR3:Runits}
    \bds{R} &\sim 2\,, \\
    \label{eqR3:nablaunits}
    \bds \nabla&\sim 1\,.
\end{align}
The only way to get $3$ is (schematically) $\bds{R}+\bds \nabla =3$, hence we need to construct a Lagrangian with one covariant derivatives and one Riemann component. The problem is the following: The Riemann components have an even number of indices, while the covariant derivative only has one index. It then impossible to contract all the indices with these two objects only, so we conclude that $\Lcal_{3}=0$.

We consider the next order given by 
\be 
\Lcal_{4} \sim \Lcal_{\mathrm{EH}} \times \left(\frac{1}{M \rg}\right)^2  \sim \frac{\Mpl^2}{M^2 \rg^4}\,,
\ee 
Now, things are more interesting and we have two possibilities to get a term proportional to $\rg^{-4}$ in the Lagrangian: $\bds{R}+\bds{R}$ or $\bds{R}+\bds{\nabla} + \bds \nabla$. With the first option, we have three possibility:
\begin{align}
\Lcal_{4,1} &= \tensor{R}{}_{\mu \nu \rho \sigma}\tensor{R}{^{\mu \nu \rho \sigma}}\,, \\
\Lcal_{4,2} &= \tensor{R}{}_{\mu \nu}\tensor{R}{^{\mu \nu}}\,, \\
\Lcal_{4,3} &= R^2\,. 
\end{align}
With the second option, the only possibility where all the indices are contracted is
\be 
\Lcal_{4,4}=\nabla_\mu \nabla_\nu \tensor{R}{^{\mu \nu}}\,.
\ee 
This term is a divergence, and hence its integral vanishes because of Stokes's theorem, hence it can be ignored. Finally, the fourth order Lagrangian is
\be  \label{eqR3:L4_first}
\Lcal_{4} = c_1 \Lcal_{4,1} +  c_2 \Lcal_{4,2} +  c_3 \Lcal_{4,3}\,,
\ee 
where $c_j$ are dimensionless coefficients that we keep unspecified for now.

This would be the end if we did not know the definition of the Gauss-Bonnet term
\be  \label{eqR3:GB}
\Lcal_{\mathrm{GB}}\equiv R_{\mathrm{GB}} =  \tensor{R}{}_{\mu \nu \rho \sigma}\tensor{R}{^{\mu \nu \rho \sigma}} - 4 \tensor{R}{}_{\mu \nu}\tensor{R}{^{\mu \nu}} + R^2 = 
\Lcal_{4,1}- 4 \Lcal_{4,2} + \Lcal_{4,3}.
\ee 
This term is interesting as its value in $4$ dimensions is fixed and only depends on the topology of the manifold. Such a term does not contribute to the action as its variation trivially vanishes. Using then its definition Eq.~\eqref{eqR3:GB}, we can re-express $\Lcal_{4,1}$ and write
\be 
\Lcal_{4} = c_1 \Lcal_{\mathrm{GB}} +  (c_2+4c_1) \Lcal_{4,2} +  (c_3-c_1) \Lcal_{4,3} = (c_2+4c_1) \Lcal_{4,2} +  (c_3-c_1) \Lcal_{4,3}\,,
\ee 
where we used that effectively $\Lcal_{\mathrm{GB}}=0$ as a Lagrangian in an action. The bottom line is that we only need to consider two terms in the Lagrangian and write
\boxemph{ \label{eqR3:L4final}
\Lcal_{4} = c_1 R^2 + c_2 \tensor{R}{}_{\mu \nu}\tensor{R}{^{\mu \nu}}\,.
}
\MYrem{The coefficients}{
In both Lagrangians Eq.~\eqref{eqR3:L4_first} and Eq.~\eqref{eqR3:L4final}, we did not specify anything about the coefficients $c_j$, which is also part of the philosophy of the effective theory formalism. We do not really know why those terms are here or where they come from. We only know that they are allowed to exist, so we consider them with generic coefficients. This is why we also replaced $c_2+4c_2$ and $c_3-c_1$ by $c_1$ and $c_2$, as these coefficients do not posses any meaning per se. Moreover, we should have multiplied them by $\Mpl^2/M^2$ to be consistent, but this quantity being a dimensionless number, we can just absorb it and redefine the coefficients. We will do this in what follows and we will not mention it systematically. We will present in Section~\ref{secR3:Gwaves_return} an example where these coefficients can be computed exactly. 
}
We continue our journey and consider the next order
\be 
\Lcal_5 \sim \LEH \times  \left(\frac{1}{M \rg}\right)^2 \sim \frac{\Mpl^2}{M^3r_g^5}\,.
\ee 
This case is similar to $\Lcal_3$: The odd powers of $\rg$ forces us to use an odd number of covariant derivatives, which makes it impossible to contract all the indices with the Riemann tensor. Hence, the term $\Lcal_5$ cannot appear.

The last term we will consider is
\be 
\Lcal_6 \sim \LEH \times  \left(\frac{1}{M \rg}\right)^4 \sim \frac{\Mpl^2}{M^4r_g^6}\,.
\ee 
This is more interesting: To obtain a power of $6$, we have three options : $\bds{R}+\bds{R}+\bds{R}$, $\bds{R}+\bds{R}+\bds{\nabla}+\bds{\nabla}$ and $\bds{R}+\bds{\nabla}+\bds{\nabla}+\bds{\nabla}+\bds{\nabla}$. The last option ought to be of the form $\bds{\nabla}\bds{\nabla}\bds{\nabla}\bds{\nabla}\bds{R}$, which is a total derivative corresponding to a boundary term, and we can ignore it.

\begin{enumerate}
\item Let us focus on the terms of the form $\bds{R}+\bds{R}+\bds{\nabla}+\bds{\nabla}$ first.  We have several options.
    \begin{enumerate}
            \item \textbf{Two Ricci scalars}\\
            This implies that both covariant derivatives must be contracted together, which leaves us with only two options : $\nabla_\mu R \nabla^\mu R$ and $ R \nabla_\mu \nabla^\mu R = R  \Box R$. These two options are equivalent at the action level as they can be obtained through an integration by part, i.e.
            \be 
            \nabla_\mu R \nabla^\mu R = \nabla_\mu (\nabla^\mu R) -  R  \Box R\,,
            \ee
            and the first term does not contribute to the action. Doing this may artificially change the coefficient in front of $  R  \Box R$, but as explained before, this is not a problem as the various coefficients do not have any physical interpretation per se in an effective context. The first term of $\Lcal_6$ is then
            \be 
            \Lcal_{6,1} = d_1 R \Box R\,. 
            \ee 
        \item \textbf{One Ricci scalar and one Ricci tensor}\\
        We cannot contract the Ricci tensor anymore, as this would lead to the previous case. Hence, the Ricci tensor should be contracted with the covariant derivatives, i.e. $R \nabla_\mu \nabla_\nu R^{\mu \nu}$.  
        However, the Bianchi identity of the Einstein tensor Eq.~\eqref{eqgr:bianchiG} implies
        \be 
        \nabla_\nu R^{\mu \nu} \propto \nabla^\mu R\,,
        \ee
        and hence
        \be 
        R \nabla_\mu \nabla_\nu R^{\mu \nu}\propto R \Box R\,,
        \ee 
        which has already been considered.
        \item  \textbf{Two Ricci tensors}\\
        Contracting a covariant derivative with a Ricci tensor would lead to a term with the Ricci scalar (using again the Bianchi identity). The first option is
        \be
        \Lcal_{6,2} = d_2 R_{\mu \nu} \Box R^{\mu \nu}\,,
        \ee 
        up to a boundary term as usual. The second option is $\nabla_\mu R^{\nu \alpha} \nabla_\nu \tensor{R}{^\mu_\alpha}$. An integration by part would lead to
        \be 
         \tensor{R}{^\nu_\alpha} \nabla _\mu \nabla_\nu \tensor{R}{^\mu^\alpha}
         =  \tensor{R}{^\nu_\alpha} \left(\nabla _\nu \nabla_\mu \tensor{R}{^\mu^\alpha} 
         + \tensor{R}{^\mu_{\rho \mu \nu}} R^{\rho \nu}
         +\tensor{R}{^\alpha_{\rho \mu \nu}} R^{\mu \rho}\right)\,,
        \ee 
        where we used the definition of the curvature. The first term simplifies because of the Bianchi identity, and the second and third terms are of the form $\bds{R}+\bds{R}+\bds{R}$, which we consider below.
        \item \textbf{One Ricci scalar and one Riemann tensor}\\
        Such a term would necessarily contains a Ricci tensor as the Riemann tensor has to be contracted at least once, hence the case has already been considered above.
        \item \textbf{One Ricci tensor and one Riemann tensor}\\
        The only possible term would be $R^{\mu \nu} \nabla_\alpha \nabla_\beta \tensor{R}{^{\alpha \beta}_{\mu \nu}}$. However, using the second Bianchi identity Eq.~\eqref{eqGR:riemid3}, the term with the Riemann tensor can be expressed as a combination of the covariant derivative of the Ricci tensor, which has already been considered.
        \item \textbf{Two Riemann tensor}\\
        It is possible to show that a term with two Riemann tensors and two covariant derivatives can be rewritten as a linear combination of the terms above and of Riemann-cubed terms, using the same properties and the fact that $\nabla_\mu R^{\mu \nu \rho \sigma}$ can be rewritten in terms of the Ricci tensor only.
    \end{enumerate}
\item We now turn our attention to the terms of the form $\bds{R}+\bds{R}+\bds{R}$. The idea is the same as above. One needs to carefully list all the possibilities, and uses all the tricks (integration by part, identity of the tensors, symmetries) to check which terms have already been considered. We do not give all the details here, but only the final result which reads
\begin{align}
\sum_{j=3}^{10} d_j \Lcal_{6,j} = 
&d_3 R^3
+d_4 R R_{\mu \nu}R^{\mu \nu}
+d_5 R R_{\mu \nu \rho \sigma}R^{\mu \nu \rho \sigma}
+ d_6 \tensor{R}{_\mu^\nu}\tensor{R}{_\nu^\rho}\tensor{R}{_\rho^\mu} \\ \nonumber
&+ d_7 R^{\mu \nu} R^{\rho \sigma} R_{\mu \nu \rho \sigma}
+ d_8 R^{\mu \nu} R_{\mu \rho \sigma \alpha} \tensor{R}{_\nu^{ \rho \sigma \alpha}} \\ \nonumber
&+ d_9 \tensor{R}{_{\mu \nu}^{\rho \sigma}}\tensor{R}{_{\rho \sigma}^{\alpha \beta}}\tensor{R}{_{\alpha \beta}^{\mu \nu}}
+ d_{10} \tensor{R}{_\mu^\rho_\nu^\sigma}\tensor{R}{_\rho^\alpha_\sigma^\beta}\tensor{R}{_\alpha^\mu_\beta^\nu}\,.
\end{align}
\end{enumerate} 
Finally, the Lagrangian $\Lcal_6$ is given by
\boxemph{ \label{eqR3:Lcal6_final}
\Lcal_6 =\frac{1}{M^2} \Bigl( &d_1  R \Box R + d_2  R_{\mu \nu} \Box R^{\mu \nu} +d_3 R^3 \\ \nonumber
 +&d_4 R R_{\mu \nu}R^{\mu \nu}
+d_5 R R_{\mu \nu \rho \sigma}R^{\mu \nu \rho \sigma}
+ d_6 \tensor{R}{_\mu^\nu}\tensor{R}{_\nu^\rho}\tensor{R}{_\rho^\mu} \\ \nonumber
+& d_7 R^{\mu \nu} R^{\rho \sigma} R_{\mu \nu \rho \sigma}
+ d_8 R^{\mu \nu} R_{\mu \rho \sigma \alpha} \tensor{R}{_\nu^{ \rho \sigma \alpha}} \\ \nonumber
+& d_9 \tensor{R}{_{\mu \nu}^{\rho \sigma}}\tensor{R}{_{\rho \sigma}^{\alpha \beta}}\tensor{R}{_{\alpha \beta}^{\mu \nu}}
+ d_{10} \tensor{R}{_\mu^\rho_\nu^\sigma}\tensor{R}{_\rho^\alpha_\sigma^\beta}\tensor{R}{_\alpha^\mu_\beta^\nu} \Bigr)
 \,.
}
The final Lagrangian and action read
\boxemph{ \label{eqR3:Final_Lag}
\Lcal &= \LEH + \Lcal_4 + \Lcal_6\,, \\
S[\bds{g},c_i,d_j] &= \int\; \sqrt{-g}\, \Lcal\, \dd^4 x\,,
}
where it is implied that the $\varepsilon$ factor is absorbed in the coefficients $c_j$ and $d_j$.

Before turning our attention to a quantitative analysis, let us estimate the value of the small parameter $\varepsilon$ introduced in Eq.~\eqref{eqR3:g_decompos}. As explained above, we have three different Lagrangians, the order of magnitude of which being given by
\begin{align}
    \LEH &\sim \frac{\Mpl^2}{\rg^2}\,, \\
    \Lcal_4 &\sim \frac{1}{\rg^4}\,, \\
    \Lcal_6 &\sim \frac{1}{M^2 \rg^6}\,,
\end{align}
We require the two last Lagrangians to be smaller than the Einstein-Hilbert one. 

For $\Lcal_4$ the condition reads
\be 
\varepsilon \equiv \frac{\Mpl^2}{\rg^2} = \left(\frac{\Lpl}{\rg}\right)^2 \ll 1\,.
\ee 
This condition is definitely satisfied for astronomical Black Hole. For a mass of the order of one solar mass, we have
\be 
\varepsilon \sim  10^{-80}\,,
\ee
which perfectly justifies the perturbative approach we are using for such Black Holes. Only the extreme case of a Primordial Black Hole whose mass is of the order of the Planck mass (see \cite{Carr:2020} for example) would lead to
\be 
\varepsilon \sim 0.5\,.
\ee 
At this length scale, the Lagrangian $\Lcal_4$ would not be perturbative anymore and more sophisticated approach would be necessary. Moreover, at this length scale, we can no longer assume that massive fields can be integrated out and the knowledge of a better theory is necessary.

The condition on $\Lcal_6$ is more interesting. It leads to
\be
\frac{1}{M^2 \Mpl^2 \rg^4}  \ll 1\,.
\ee 
Expressing this parameter in more intuitive variables yields
\be \varepsilon \sim 
10^{-160} 
\left( \frac{M}{\Mpl}\right)^{-2}
\left( \frac{\MBH}{\Msun}\right)^{-4}\ \ll 1\,.
\ee 
This condition is definitely valid for astrophysical Black Holes with $\MBH \sim 50\Msun$, even if the integrated fields satisfy $M\sim 10^{-60}\Mpl$. For these reasons, the perturbative dimensionless parameter is given by
\boxemph{
\varepsilon =
\frac{1}{M^2 \Mpl^2 \rg^2}\,.
}

\subsection{So many Terms... \label{secBH:somany}}
\subsubsection{A toy-model}
In this Section, we would like to describe how the final Lagrangian Eq.~\eqref{eqR3:Final_Lag} can be effectively simplified, provided we study perturbations around the Schwarzschild Black Hole. Before going into the technical details, we again present a simple mechanical toy-model. This model was proposed as an exercise about perturbation theory and Oscillations around an equilibrium position for second year students following the course \emph{M\'ecanique II} during the Fall Semester 2019.

Consider a non-relativistic particle of mass $m$ moving in a potential of the form
\be 
V(x) = V_0(x) + \varepsilon\, \delta V(x)\,,
\ee 
where $V_0(x)$ is a background potential and $\varepsilon\, \delta V(x)$ is a small perturbation. The equation of motion is
\be  \label{eqR3:eom_toy}
m \ddot x = - V^\prime(x)\,.
\ee 
Assume now that we know that $x_0$ is a stable equilibrium position of the background potential $V_0(x)$, i.e. $V_0^\prime(x_0)=0$ and $V_0^{\prime \prime} (x_0) >0$. In the full theory, the new equilibrium position is
\be 
\tilde x_0 = x_0 + \varepsilon\, \delta x\,,
\ee 
and $\delta x$ is found imposing $V'(\tilde x_0)=0$ at order $\varepsilon$. The computations are straightforward and yield
\be 
\delta x = - \frac{\delta V^{\prime}(x_0)}{V_0^{\prime \prime}(x_0)}\,,
\ee 
which is well-defined as $V_0^{\prime \prime}(x_0)\neq 0$.

Now, we consider oscillations around the equilibrium. At the background level, we have
\be 
x(t) = x_0 + \varepsilon_2\, y(t)\,,
\ee 
where $\varepsilon_2$ is the bookkeeping parameter describing the oscillations. Plugging this Ansatz into the equation of motion Eq.~\eqref{eqR3:eom_toy} yields to the very well-known result
\be 
m \ddot y = - y V^{\prime \prime}(x_0)\,,
\ee
from which we can extract the frequency 
\be 
\omega_0^2 = \frac{V^{\prime \prime}(x_0)}{ m}\,.
\ee 
At the perturbed level, the oscillatory solution reads
\be 
x(t) = x_0 + \varepsilon\, \delta x + \varepsilon_2\, y(t)\,,
\ee 
and the equation of motion is
\be 
m \ddot y =- y \Bigl(
V_0^{\prime \prime} + \varepsilon\, V_0^{\prime \prime \prime} \delta x + \varepsilon\, \delta V^{\prime \prime} 
\Bigr)\,,
\ee 
where all the derivatives are evaluated at $x=x_0$. We can extract the perturbed frequency as
\be
\tilde \omega_0 =  \frac{ V_0^{\prime \prime} + \varepsilon\, V_0^{\prime \prime \prime} \delta x + \varepsilon\, \delta V^{\prime \prime} }{m}\,.
\ee 

Let us stress three important points about this toy-model, which will illuminate the discussion below.
\begin{enumerate}
    \item Two terms appear in the correction of the frequency: The first term is a consequence of the displacement of the equilibrium due to $\delta x$, and the second one comes from the perturbation $\delta V$.
    \item If the first derivative $\delta V'(x_0)=0$, the equilibrium position is not modified, i.e. $\delta x =0$.
    \item If, in addition to $\delta V'(x_0)=0$, the second derivative $\delta V^{\prime \prime}(x_0)=0$, the frequency is not modified.
\end{enumerate}

\subsubsection{A useful trick}
In the final effective Lagrangian Eq.~\eqref{eqR3:Final_Lag}, there are $2+10=12$ independent parameters, which is very large. We can however simplify this action. To see this, let us consider a simpler version with the Lagrangian given by
\be \label{eqR3:Lcal_example}
\Lcal = \frac{\Mpl^2}{2} \left(R + \varepsilon\,   R_{\rho \sigma} \Pi^{\rho \sigma \alpha \beta } R_{\alpha \beta}\right)\,,
\ee 
where $\bds{\Pi}$ is an arbitrary tensor, which can potentially contains derivatives. The equation of motion are
\be  \label{eqR3:eom_pi}
G_{\mu \nu} + \varepsilon\left(
\delta( R_{\rho \sigma} \Pi^{\rho \sigma \alpha \beta } R_{\alpha \beta})_{\mu \nu}- \frac 12 g_{\mu \nu}R_{\rho \sigma} \Pi^{\rho \sigma \alpha \beta } R_{\alpha \beta}\right) =0\,,
\ee 
where the second term schematically represents the term containing $\delta g^{\mu \nu}$ and the last term comes from the variation of the square root determinant.

We first require the background solution to be close to the Schwarzschild solution, i.e.
\be 
\bar{\bds g} = \gS + \varepsilon\,  \delta \bds g\,.
\ee 
Here comes the trick: As we work at first order, in the term proportional to $\varepsilon$ in the equations of motion Eq.~\eqref{eqR3:eom_pi}, we can set $\bds g = \gS$. It is direct to show that every term is proportional to the Ricci tensor, which vanishes on the Schwarzschild solution. Hence, the equation is 
\be 
\bds{G}(\gS + \varepsilon \, \delta \bds g)=0\,.
\ee 
As we want the perturbation $\delta \bds g$ to be small, the trivial solution is $\delta g =0$. In other words, the term we considered does not contribute to the modified background.

We now turn our attention to gravitational perturbations of the form (using $\delta \bds g =0$) $\bds g = \gS + \varepsilon_2\, \bds h$. The equations of motion are still given by Eq.~\eqref{eqR3:eom_pi}, but we will expand them at linear order in $ \varepsilon_2$. At the \ac{GR} level ($\varepsilon=0$), the equations of motion are simply
\be  \label{eqR3:solh2bg}
(\delta G_{\mu \nu})_{\rho \sigma} h^{\rho \sigma}=
(\delta R_{\mu \nu})_{\rho \sigma} h^{\rho \sigma}=
(\delta R)_{\rho \sigma} h^{\rho \sigma}=
0\,,
\ee
where we used the relation between the Ricci and the Einstein tensor (or simply said, if one vanishes, so does the other). In the equations of motion Eq.~\eqref{eqR3:eom_pi}, all the terms quadratic in $\bds h$ can be neglected, and we are left with
\be 
G_{\mu \nu} + \varepsilon\left(
(\delta R_{\rho \sigma})_{\mu \nu} \Pi^{\rho \sigma \alpha \beta } R_{\alpha \beta} 
+ R_{\rho \sigma}\Pi^{\rho \sigma \alpha \beta } 
(\delta R_{\alpha \beta})_{\mu \nu}\,.
\right)=0\,.
\ee 
As before, in the second term, we can use the \ac{GR} solution and use that $\bds{R}(\gS + \bds h) =0$ to cancel both terms. We are left with
\be 
G_{\mu \nu} =0\,,
\ee 
i.e. even at the gravitational perturbations level, the second term in the Lagrangian Eq.~\eqref{eqR3:Lcal_example} does contribute.

\MYrem{Connection with the mechanical toy-model}{
Using the notation we introduced in the previous section, we have first  $x_0\sim \gS$, $\delta V\sim  R_{\mu \nu} R^{\mu \nu}$, and we want to compute $\delta x \sim \delta \bds g$. Schematically, we have (neglecting the indices)
\be 
V_0 \sim R\,, \quad \delta V^\prime \sim R \delta R =0\,.
\ee
The linear term does not contribute when evaluated at $\bds g = \gS$ (or $x=x_0$). Hence, the displacement vanishes and we have indeed $\delta x \sim \delta \bds g =0$. At the oscillations level
\be 
\delta V^{\prime \prime} \sim R \delta \delta R + \delta R \delta R =0,
\ee 
which also vanishes when the zeroth order equations of motion are used.
}

\subsubsection{Many spurious terms}
With this trick in mind, the Lagrangian becomes much simpler. Basically, any term containing two Ricci's can be eliminated, so effectively speaking:
\be 
c_1=c_2=d_1=d_2=d_3=d_4=d_6=d_7=0\,,
\ee
and the Lagrangian becomes
\begin{align} 
\Lcal = &\frac{\Mpl^2}{2} R + \frac{\varepsilon}{M^2}  (d_5 \Lcal_{6,5} + d_8 \Lcal_{6,8} + d_9 \Lcal_{6,9} + d_{10} \Lcal_{6,10})\,.
\end{align}
We can actually go even further. The Lagrangians $\Lcal_{6,10}$ and $\Lcal_{6,5}$ can be expressed as
\begin{align}
\label{eqR3:Lca10_expr}
\Lcal_{6,5} &= 4\Lcal_{6,8} + \sum_{j\in \mathcal{I}} f_{j,5} \Lcal_{6,j}\,, \\
\Lcal_{6,10} &= \frac 12\Lcal_{6,9} - 3 \Lcal_{6,8} + \frac 38 \Lcal_{6,5} + \sum_{j\in \mathcal{I}} f_{j,6} \Lcal_{6,j}\,, \\
\end{align}
where $\mathcal{I} = \{3,4,6,7 \}$, see \cite{Cano:2019} for the explicit formula. Replacing $\Lcal_{6,5}$ and $\Lcal_{6,10}$ using these relations yields
\be
d_5 \Lcal_{6,5} + d_8 \Lcal_{6,8} + d_9 \Lcal_{6,9} + d_{10} \Lcal_{6,10} =
\left(d_8 + 4d_5 - \frac 32 d_{10} \right) \Lcal_{6,8}
+ \left(d_9 + \frac 12 d_{10}\right) \Lcal_{6,9}\,,
\ee
where we neglected the terms corresponding to $j\in \mathcal{I}$.

To summarise, it is sufficient to make the computations with $(d_8, d_9)$ only. The complete expressions for the observables, with the coefficients $d_5$ and $d_{10}$, are given by substituting
\be  \label{eqR3:replace_d89}
(d_9, d_9) \rightarrow 
\left(  d_8 + 4  d_5 - \frac{3}{2}d_{10} , d_9 + \frac 12 d_{10} \right)\,.
\ee 

\subsection{The Equations of Motion}
We present now briefly the equations of motion, assuming that only the terms $\Lcal_{6,8}$ and $\Lcal_{6,9}$ are present. Moreover, if their respective equations of motion contains any term with a Ricci tensor or scalar, they can be neglected (the idea follows the same lines as the proof above). Hence, the equations of motion are

\boxemph{ \label{eqR3:EOM_FINAL}
\Ecal_{\mu \nu} &= \mathcal{E}^{\mathrm{EH}}_{\mu \nu} + d_8 \mathcal{E}^{8}_{\mu \nu} + d_9 \mathcal{E}^{9}_{\mu \nu}  \\
\mathcal{E}^{\mathrm{EH}}_{\mu \nu}& = \frac{\Mpl^2}{2}\left( R_{\mu \nu} - \frac 12  R g_{\mu \nu} \right)\,, \\
\mathcal{E}^{8}_{\mu \nu} &= \frac{ 1}{2M^2} \left(
-\nabla^\rho \nabla_\nu \tensor{C}{_\mu_\rho}
-\nabla^\rho \nabla_\mu \tensor{C}{_\nu_\rho} 
+g_{\mu \nu} \nabla_\rho \nabla_\sigma  C^{\rho \sigma } 
+\Box C_{\mu \nu}
\right)\,,  \\
\mathcal{E}^{9}_{\mu \nu} &= \frac{1}{2 M^2} 
\left(6 \tensor{R}{_\mu^{\rho \sigma \alpha}}  \tensor{C}{_{\nu\rho \sigma \alpha}}
- g_{\mu \nu} \tensor{R}{_{\rho \sigma}^{\alpha\beta}} \tensor{C}{^{\rho \sigma}_{\alpha\beta}} 
+12 \nabla_\rho\nabla_\sigma \tensor{C}{_\mu^\rho_\nu^\sigma}\right)\,, \\
\tensor{C}{_{\mu \nu}^{\gamma \sigma}} &= \tensor{R}{_{\mu\nu}^{\alpha \beta}} \tensor{R}{_{\alpha \beta}^{\gamma \sigma}}\,, \\ \tensor{C}{_\mu^\nu} &= \tensor{C}{_\alpha_\mu^\alpha^\nu}\,.
}

\subsection{A Spherically Symmetric Static Black Hole}
In this section, we finally have all the ingredients in our hand to construct the perturbed Schwarzschild background. We assume that the metric is 
\be 
\bds g \equiv \gS + \varepsilon \, \delta \bds g 
= - A(r) \bdd t^2 + \frac{1}{B(r)} \bdd r^2 + r^2 \bdd \Omega^2\,,
\ee 
and we suppose that the functions $A$ and $B$ are of the form
\begin{align}
    A(r)&= f(r) + \varepsilon\,  \delta A(r)\,, \\
    B(r)&= f(r) + \varepsilon\,  \delta B(r)\,,
\end{align} 
with
\be 
f(r) = 1 - \frac{\rg}{r}\,.
\ee 
The Ansatz can be plugged into the equations of motion Eq.~\eqref{eqR3:EOM_FINAL}, recalling that in the perturbed term, the $\varepsilon$ correction can be ignored. The resulting equation are not straightforward to solve, but we postulation a solution of the form
\begin{align}
    \delta A(r) = \alpha_1 \left( \frac{\rg}{r}\right)^{n_1} + 
    \alpha_2 \left( \frac{\rg}{r}\right)^{n_2}\,, \\
        \delta B(r) = \beta_1 \left( \frac{\rg}{r}\right)^{n_1} + 
    \beta_2 \left( \frac{\rg}{r}\right)^{n_2}\,.
\end{align}
The resulting equations are schematically of the form
\be 
f_0(r) r^0 + f_1(r) r^1 = g_1(r) r^{7-n_1} + g_2(r) r^{7-n_2}\,,
\ee 
which is solved if $n_1=6$ and $n_2=7$ (or vice-versa). Plugging these values, it is straight forward to find the coefficients $\alpha_i$ and $\beta_j$ and the solution reads
\begin{align} \label{eqR3:Apert}
    A(r) &= 1 - \frac{\rg}{r} + \varepsilon \left( \alpha_6  \left( \frac{\rg}{r}\right)^{6} +\alpha_7  \left( \frac{\rg}{r}\right)^{7} \right)\,, \\
    \label{eqR3:Bpert}
      B(r) &= 1 - \frac{\rg}{r} + \varepsilon \left( \beta_6  \left( \frac{\rg}{r}\right)^{6} +\beta_7  \left( \frac{\rg}{r}\right)^{7} \right)\,, 
\end{align}
with
\begin{alignat}{2}
\alpha_6 &= -6d_8\,, \quad &  \alpha_7 &=9 d_8 + 10 d_9\,, \\
\beta_6 &= 36(d_8 + 3d_9) \quad & \beta_7 &= -33 d_8 - 98 d_9 \,.
\end{alignat}

The Black Hole horizon is displaced, as $A(\rg), B(\rg)\neq 0$. The position of the horizon is defined (at first order in $\varepsilon$) by
\be 
A(\rg + \varepsilon \, \delta \rg ) = B(\rg + \varepsilon \, \delta \rg) =0\,.
\ee 
The solution is the same for both functions and is given by
\boxemph{ \label{eqR3:rH}
\rH = \rg + \varepsilon\,  \delta \rg = \rg \big(1- \varepsilon (3d_8 + 10d_9) \big)\,.
}
In this effective theory, we have two different radii. The first unperturbed radius is related to the Black Hole mass as $\rg = 2 \GN \MBH$, while the physical horizon $\rH$ is given by Eq.~\eqref{eqR3:rH}, with $\rH\neq 2 \GN \MBH$.

\section{Gravitational Waves, the Return \label{secR3:Gwaves_return}}
\subsection{The Magic of Perturbation Theory}
We determined the \ac{SSSS} solution given by
\be 
\bds g = \gS + \varepsilon \, \delta \bds g\,.
\ee
We want now to determine the master equations of the \ac{GW} around this perturbed solution, in our effective theory. The parametrisation for the perturbations is the same as for the unperturbed case, see Eq.~\eqref{eqR3:oddRW} and Eq.~\eqref{eqR3:evenRW}. Note that the functions $A(r)$ and $B(r)$ can be taken at zeroth or first order without modifying the final results.

We present here the general logic to get the master equation at the perturbed level. At the background level, we have for each modes a collection of fields $\bds \Phi$, which includes all the perturbations and their derivatives. For example, in our two cases, they read
\begin{align}
\bds{\Phi}_{\mo} &= \{h_0, h_1, h_0^\prime, h_1^\prime, \dots \}\,, \\
\bds{\Phi}_{\me} &= \{H_0, H_1, H_2, K_0,  H_0^\prime, H_1^\prime, H_2^\prime, K_0^\prime, \dots \}\,. 
\end{align}
When we solved the background solution, we actually found a finite number of fields with respect to which all the others fields can be expressed. Mathematically speaking, there is a finite set $\bds{\Psi}_\mf$ of \emph{free}\footnote{Here, the term \emph{free} has nothing to do with the usual free fields in QFT.} fields and a infinite subset $\bds \Psi_\mc$ of constrained fields such that 
\begin{align}
    \bds{\Psi}_\mf &\cap \bds{\Psi}_\mc = \varnothing  \,, \\
    \bds \Psi_\mf &\cup \bds \Psi_\mc = \bds \Psi  \,. 
\end{align}
The constrained fields can be expressed as linear functions of the free fields as 
\be \label{eqR3:sol_Bg_genform}
 \bds \Psi_\mc = \bds \Xi \bds \Psi_\mf\,,
\ee 
where $\bds \Xi$ is a (infinite) linear operator. In our cases, we have
\begin{alignat}{2}
\bds{\Psi}_{\mo, \mf} &= \{h_1, h_1^\prime \} \quad & \bds{\Psi}_{\mo, \mc} &= \{h_0, h_0^\prime, h_0^{\prime\prime}, h_1^{\prime \prime}, \dots \} \,, \\
\bds{\Psi}_{\me, \mf} &= \{H_1,K_0 \} \quad & \bds{\Psi}_{\me, \mc} &= \{H_0, H_2, H_0^\prime, H_1^\prime , H_2^\prime, K_0^\prime,\dots \} \,.
\end{alignat}

We now turn our attention to the first order equations. They are of the form
\be 
\Ecal = \Ecal_{\mathrm{GR}}(\bds \Psi) + \varepsilon\,  \delta \Ecal(\bds \Psi)\,,
\ee 
in which we can use the magic of perturbation theory: The second term being first order, we can plug therein the background solutions Eq.~\eqref{eqR3:sol_Bg_genform} to eliminate all the fields $\bds \Psi_\mc$. The equations then read
\be 
\Ecal(\bds \Psi) = \Ecal_{\mathrm{GR}}(\bds \Psi) + \varepsilon\,  \delta \Ecal(\bds \Psi_\mf)\,.
\ee 
Moreover, at the background level, we can massage the equations in a certain way to express all the constraints fields $\bds \Psi_\mc$ as functions of the free fields $\Psi_\mf$. We can apply the \emph{same} massage to the full equations of motion to express them at first order. At each step of the process, we can reuse the magic of perturbation theory and replace $\bds \Psi_\mc$ by $\bds{\Psi}_\mf$ in the terms proportional to $\varepsilon$. The final result is an expression of the form
\be \label{eqR3:sol_pert_genform}
 \bds \Psi_\mc = (\bds \Xi+ \varepsilon\,  \delta \bds \Xi) \bds \Psi_\mf\,,
\ee 
from which we can fully eliminate the constrained fields and then extract the master equation, as we did for the background case.

\subsection{Master Equations}
\subsubsection{Odd perturbations\label{secR3:odd_pert}}
The massage needed for odd perturbations is detailed in~Section~\ref{secR3:odd_BG_master}. The final equation reads
\begin{align}  \label{eqR3:eqh1_pert_fin}
\frac{\rg-r}{4r}h_1^{\prime \prime }
&+ \frac{2r-5\rg}{4r^2} h_1^\prime
+ \frac{5\rg^2+r\rg (J^2-6)-r^2(J^2-2) + r^4 \omega^2}{4r^3 (\rg-r)} h_1 \\ \nonumber
&+ \varepsilon \big(\alpha_{h_1}(r) h_1 + \alpha_{h_1^{\prime}}(r) h_1^\prime \big) =0\,,
\end{align}
and the explicit formulas are given in Appendix~\ref{secAppBH:perodd}.
\MYrem{Schwarzschild radius}{
Recall that in the zeroth order term $\rg=2\GN M$ is \textbf{not} the Schwarzschild radius of the Black Hole, while at first order all these values are equal, i.e. $\rg=\rH=2\GN M$.
}
We want to find the master equation
\be 
\label{eqR3:Master_eq_EFT}
F(r) \big[F(r) \psi_{\mo}^\prime \big]^\prime + 
\big[ V_\omega(r)\omega^2 - F(r)V_{\mo,1}(r) \big] \psi_{\mo}(r) =0 \,.
\ee 
Recall J\'er\'emie's trick (Remark~\ref{RemR3:Jerémie}): We are free to choose the function $F(r)$ (as long as it vanishes on the Black Hole horizon). Here, for reasons that will be clear later, we take
\be 
F(r)=1 - \frac{\rH}{r}\,,
\ee 
where $\rH$ is the Schwarzschild radius at the perturbed level.

These manipulations lead to the final master equation with
\boxemph{ \label{eqR3:Vomega_pert}
V_{\omega}(r) &= 1  + \varepsilon \big( d_8\,  V_{\omega,8}(r)  d_9\,  V_{\omega,9}(r) \big)\,, \\
 V_{\omega,8}(r) &= \frac{6\rg}{r^6} 
(r^5+ r^4 \rg + r^3 \rg^2+ r^2 \rg^3+r \rg^4-4 \rg^5)\,,   \\
 V_{\omega,9}(r) &= \frac{4\rg}{r^6}  
(5 r^5  +5 r^4 \rg+5 r^3 \rg^2+ 5 r^2 \rg^3 +77 r \rg^4-94 \rg^5)\,, \\
\label{eqR3:deltaVodd}
V_{\mo,1}(r) &=  V_\mo(r) + \varepsilon\, \delta V_{\mo}(r)\,,
}
where $V_\mo(r)$ is the Regge potential given by Eq.~\eqref{eqR3:VRegge}. The expression for $ \delta V _{\mo}(r)$ is given in Appendix~\ref{secAppBH:perodd}. As we will see in the next Section, the potential $V_\omega(r)$ is the same for both modes, hence we omit the subscript odd/even. Note that the Regge potential $V_\mo(r)$ is expressed here in terms of the Black Hole mass $\MBH$. Inverting the relation between $\rg$ and $\rH$ yields
\be 
2\GN \MBH = \rg = \rH \big(1+ \varepsilon (3d_8 + 10d_9) \big)\,,
\ee 
which will bring extra first order corrections. This will reveal important when we compute the corrections to the \ac{QNM}.

\subsubsection{Even perturbations\label{secR3:even_pert}}
Dealing with the even perturbations is slightly more complicated than with the odd ones. The expressions and the computations are longer as the number of variables and of equations is higher. As for the odd modes, the massage is described in Section~\ref{secR3:even_BG_master}. Using the background solution to simplify the first order part of the equation, we are left with $H_0^\prime$ and $K_1^\prime$ expressed as functions of $H_0$ and $K_1$. As before, we assume that the master variable is given by
\be \label{eqR3:mastervariable_even}
\psi(r) = \frac{\ii}{(J^2-2)r + 3 \rg} \left(
r^2 K_0(r) \big[1+\varepsilon\, \delta f_K(r) \big] 
- \ii \frac{\sqrt{AB} r}{\omega} H_1(r) \big[1+\varepsilon\, \delta f_H(r) \big]
\right)\,,
\ee
with $A(r)$ and $B(r)$ given by Eq.~\eqref{eqR3:Apert} and Eq.~\eqref{eqR3:Bpert}. One last field redefinition given by 
\be  \label{eqR3:field_redef_even}
\psi(r) =  \big[1+\varepsilon \chi(r) \big] \psi_\me(r)
\ee 
is necessary to bring the master equation in the same form as for the odd perturbations
\be 
\label{eqR3:Master_eq_even}
F(r) \big[F(r) \psi_{\me}^\prime \big]^\prime + \big[V_\omega(r)\omega^2 - F(r)V_{\me,1}(r) \big] \psi_{\me}(r) =0 \,,
\ee 
with
\be 
F(r)=1 - \frac{\rH}{r}\,,
\ee 
and 
\be \label{eqR3:perturbed_Ve}
V_{\me,1}(r) =  V_\me(r) + \varepsilon\,  \delta V _{\me}(r)\,,
\ee 
$V_\me(r)$ being the Zerilli potential given by Eq.~\eqref{eqR3:VZerilli}. Moreover, $V_\omega(r)$ takes the same form as in the odd case (Eq.~\eqref{eqR3:Vomega_pert}). The explicit expressions for the other quantities are given in Appendix~\ref{secAppBH:pereven}.

\MYrem{Defining the correct field redefinition}{
Note that it is not straightforward at all to define the functions $\delta f_K$, $\delta f_H$ and $\chi$. The general idea is to define $\psi_\me$ in terms of $K_0$ and $H_1$. We can isolate $\delta V_{\me}(r)$ by imposing the master equation Eq.~\eqref{eqR3:Master_eq_even}, and we impose that the final expression does not depend on $K_0$ nor on $H_1$, as we did in the background case. These conditions, with an Ansatz about the form of $\delta f_K$ and $\delta f_H$ yield the final results.
}

\subsection{Discussion}
\begin{figure}[ht]
	\begin{center}
	\includegraphics[width =0.65\textwidth]{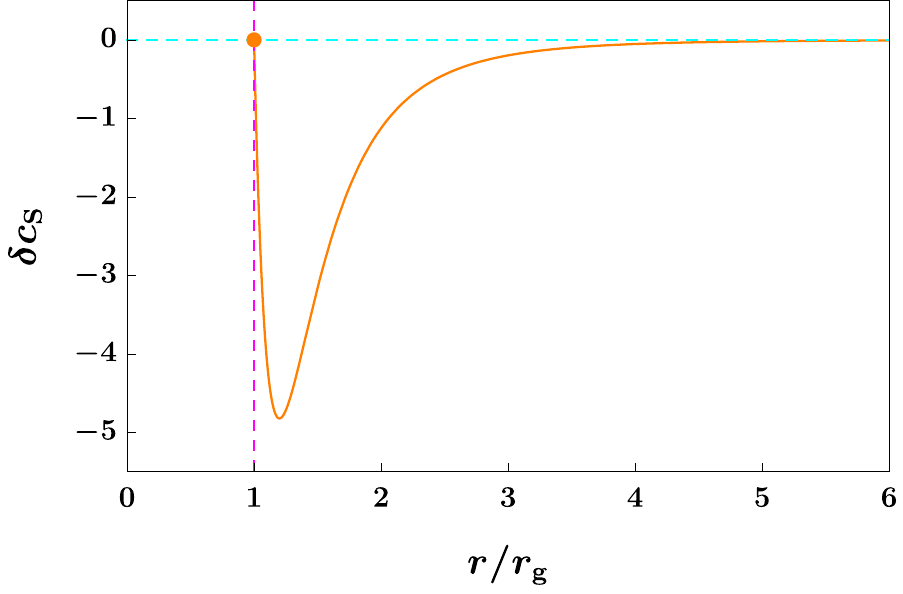} 
	\end{center}
	\caption[Speed difference]{\label{figR3:deltaC}
	\textbf{Speed difference}\\
	 Speed difference as a function of the dimensionless radius $\xi = r/\rg$. The minimum is located at $(\xi, \delta \cS)\sim(6/5, -5)$.} 
\end{figure}

With the expression for the potential $V_\omega(r)$ given by Eq.~\eqref{eqR3:Vomega_pert}, we can directly compute the speed of the \ac{GW} using Eq.~\eqref{eqR3:cs_formule}. This yields
\boxemph{ \label{eqR3:csvalue}
\cS = 1-\varepsilon\, 72 (2d_9+d_{10} ) \frac{\rg^5(r-\rg)}{r^6}\,,
}
where we restored the coefficient $d_{10}$ using Eq.~\eqref{eqR3:replace_d89}. This is the main result of this Section. We define the speed difference as
\be
\delta \cS \equiv \frac{\cS-1}{\varepsilon (2d_9+d_{10} )}\,,
\ee 
which we show in Fig.~\ref{figR3:deltaC}. It clear both from the analytical expression and from the figure that the sign of the speed difference is always negative $\delta \cS<0$. 
As a specific example, we consider three cases where the lightest massive particle which has been integrated out is either a scalar ($s=0$), a fermion ($s=1/2$) or a boson ($s=1$). The values of the coefficients $d_9$ and $d_{10}$ are taken from \cite{deRham:20202,Avramidi:1986,Avramidi:1990}. From the relation Eq.~\eqref{eqR3:csvalue}, the \ac{GW} are \emph{superluminal} if $2d_{9}+d_{10}<0$ and \emph{subluminal} if $2d_{9} + d_{10}>0$. The results are shown in Tab.~\ref{tabR3:d910}. We see that in the fermion case, the waves are superluminal, while they are subluminal in the scalar and boson case.
\begin{center}
\begin{table}
\centering
\begin{tabular}{c| c  | c   }
$s$ & $2d_9 + d_{10}$ & Speed \\ \hline \hline 
& &\\
$0$ & $+\frac{1}{1260}$ & Subluminal \\
& \\
$1/2$ & $-\frac{1}{630}$  & Superluminal \\
& \\
$1$ & $+\frac{1}{420}$ & Subluminal \\
\end{tabular}
\caption{\label{tabR3:d910} Velocity of the gravitational in three examples}
\end{table}
\end{center}
It can be surprising to see that \ac{GW} can be superluminal in some cases. In general, subluminality is a necessary condition to ensure that the theory does not violate causality. Naively, one would expect that an effective theory with $\cS>1$ should be eliminated, and for example conclude that the lightest particle integrated out could not be a fermion. However, some authors pointed out that superluminality is not a signature of causality violation, see for example the work of Hollowood \cite{Hollowood:2007, Hollowood:2008, Hollowood:2009,Hollowood:2010,Hollowood:2011,Hollowood:2012, Hollowood:2015} or \cite{deRham:2020,deRham:2021,Chen:2021}.

Another important point is that see from the expression Eq.~\eqref{eqR3:csvalue} that the speed of \ac{GW} exactly vanishes at $r=\rg=\rH$ (recall that these two last quantities are equal at first order). Moreover, using the expression for the effective metric Eq.~\eqref{eqR3:Eff_met_final}, it is direct to show that
\be
Z_t(\rH)=Z_r(\rH)=0\,.
\ee 
This means that the horizons seen by the \ac{GW} and by the photons are actually the same. This result is actually more general and is related to the \emph{Horizon Theorem}, see for example \cite{Hawking:1973, Shore:1995, Shore:2000, Hollowood:2009}. Even if we did the computations at first order, this result should also holds at all orders.

The fact that both species see the horizon at the same location can be understood as follows. We have two metrics at hand: The usual metric $\bds g$, and the effective metric $\bds Z$. We can compute the Riemann tensors of each metric, respectively $\bds{\mathcal{R}}_g$ and $\bds{\mathcal{R}}_Z$. Let us assume that both horizons are different, i.e. $\rH(\bds{g}) \neq \rH(\bds{Z})$. We can compute a scalar invariant, for example the hybrid Kretschmann
\be 
\Upsilon \equiv R_{g,\mu \nu \rho \sigma}  R^{Z,\mu \nu \rho \sigma}\,.
\ee 
In our effective theory, the terms we added are finite when evaluated at $\rH$, hence we do not expect such a scalar to be divergent when we turn on the small perturbation. However, computing $\Upsilon$ explicitly leads to a divergent part of the form
\be 
\Upsilon \propto \frac{1}{[r-\rH(\bds{g})][r-\rH(\bds{Z})]}\,,
\ee 
which is not defined on either horizon. Hence, we need $\rH(\bds{g}) = \rH(\bds{Z})$ to get a well-defined behaviour for such quantities at the horizon.

\section{Quasinormal modes}
\subsection{General Method}
In this Section, we compute the corrections to the \ac{QNM} frequencies. We follow the method developed by Cardoso et al. \cite{Cardoso:2019,McManus:2019,Kimura:2020}. They develop a systematic approach to compute corrections to \ac{QNM} around Black Holes, let us summarise their method.

Let $\rH$ be the Black Hole radius and $\psi$ a master variable. We do not specify if we study odd or even variables as the logic is the same. The generic master equation reads
\be  \label{eqR3:Cardoso}
F  \big[F \psi^\prime\big]^\prime + \big[\Omega^2 - F V(r)\big] \psi =0\,.
\ee 
with
\be 
F(r) = 1- \frac{\rH}{r}\,,
\ee 
and $\Omega$ is a constant frequency. It can be a priori different from the frequency we measure $\omega$, we will discuss this below. The potential $V$ can be decomposed as
\be 
V = V_{\mathrm{GR}} + \varepsilon\,  \delta W\,,
\ee 
where the first term is the \ac{GR} contribution, while the second term is a perturbation of any kind. 

An important comment should be made here: The \ac{GR} potential $V_{\mathrm{GR}}$ should be expressed in terms of $\rH$, which brings a correction to $\delta W$ as $\rg = \rH + \mathcal{O}(\varepsilon)$. In our context, it corresponds to the higher terms in the Lagrangian, but the potential can also come from considering slowly rotating Black Holes.

The perturbed potential should then be decomposed as an infinite series
\be 
\delta W(r) = \rH^2 \sum_j \alpha_j \left( \frac{\rH}{r} \right)^j\,.
\ee 
In the article, the authors computed a set of complex numbers $e_j$ from which the perturbed frequency can be computed as
\begin{align}
\Omega &= \Omega_0 + \varepsilon\, \delta \omega\,, \\
\delta \omega &=  \sum_j e_j \alpha_j\,,
\end{align}
where $\Omega_0$ is the \ac{QNM} in \ac{GR}. In principle, the coefficients $e_j$ will depend on the modes we consider (e.g. odd or even) as well as on the angular number $\ell$.

To apply this method to our case, there is one subtlety. Our master equations is of the form
\be  \label{eqR3:Master_us}
F  \big[F \psi^\prime\big]^\prime + \big[V_\omega \omega^2 - F V(r)\big] \psi =0\,.
\ee 
This is not exactly equivalent to Eq.~\eqref{eqR3:Cardoso} as there is a function in front of the frequency. This is discussed in Appendix B of \cite{Cardoso:2019}. We present here the key ideas.

First, we can use the fact that in our cases
\be 
V_\omega(r) = 1 + \varepsilon \delta V_\omega(r)
\ee 
both for the odd and for the even perturbations. The master equation Eq.~\eqref{eqR3:Master_us} can be identically rewritten as
\be  \label{eqR3:master_shift}
F(r)  \big[F(r) \psi^\prime\big]^\prime +  \left[
\big[1+  \varepsilon\delta V_\omega(\rH) \big] \omega^2  - F(r) \left(V(r) -  \varepsilon \omega^2 \frac{\delta V_\omega(r)- \delta V_\omega(\rH)}{F(r)} \right)
\right] \psi =0\,.
\ee
Hence, assuming that the potential $V$ is decomposed as
\be 
V(r) = V_{\mathrm{GR}}(r) + \varepsilon \, \delta V(r)\,,
\ee
the effective perturbation $\delta W$ appearing in Eq.~\eqref{eqR3:Cardoso} is then
\be 
\delta W =  \delta V- \omega^2 \frac{\delta V_\omega(r)- \delta V_\omega(\rH)}{F(r)}\,.
\ee 
Moreover, in the final equation Eq.~\eqref{eqR3:master_shift}, the frequency we compute is actually a rescaled frequency $\Omega$ defined as
\be 
\Omega^2 = \big[1+ \varepsilon \, \delta V_\omega(\rH) \big] \omega^2 \,.
\ee 
This relation can be checked by directly comparison between the generic master equation Eq.~\eqref{eqR3:Cardoso} and the master equation we obtain in our particular case Eq.~\eqref{eqR3:master_shift}. Hence, the final frequency is given by 
\be 
\omega =  \left(1- \frac 12  \varepsilon\,  \delta V_\omega(\rH) \right)  \Omega 
= \left(1- \frac 12   \varepsilon \, \delta V_\omega(\rH) \right)  (\Omega_0 + \varepsilon \delta \omega ) = 
\Omega_0 + \varepsilon \left( \delta \omega - \frac 12 \Omega_0 \delta V_\omega(\rH)\right)\,.
\ee 
This is the correction to the \ac{QNM} for a Black Hole with Schwarzschild radius $\rH$. We want to compare this frequency to a Black Hole with the same mass. In order to do so we express the background frequency as
\be 
\Omega_0 = \frac{\xi_0}{\rH}\,,
\ee
where $\xi_0$ is dimensionless complex number. Using the generic relation between the radius of the Black Hole and its mass (see Eq.~\eqref{eqR3:rH})
\be 
\rH = \rg + \varepsilon \delta \rg\,,
\ee 
we get
\boxemph{
\omega = \omega_0 +\varepsilon \left( 
\delta \omega - \frac 12 \omega_0 \delta V_\omega(\rH) - \frac{\delta \rg}{\rg}\right)\,,
}
where we used that $\Omega_0=\omega_0$ at first order and $\omega_0 = \xi_0/\rg$. As $\rg=2\GN \MBH$, this relation makes it possible to compute the \ac{QNM} for a given Black Hole mass.

\subsection{Results}
\begin{figure}[h!t]
	\begin{center}
	\includegraphics[width =0.6\textwidth]{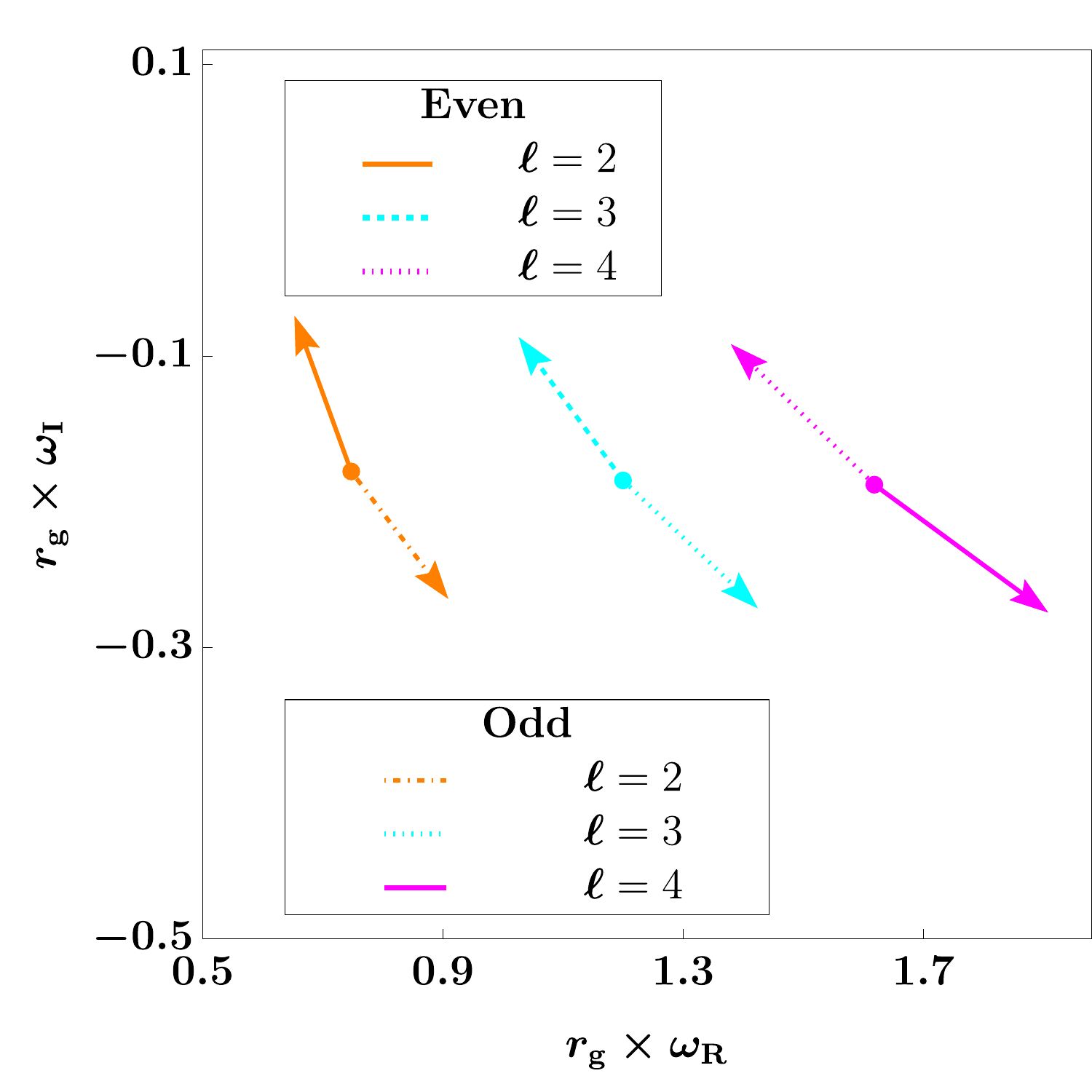} 
	\end{center}
	\caption[Displacement of the Quasinormal Modes]{\label{figR3:QNM_pert}
	\textbf{Displacement of the Quasinormal Modes}\\
	Direction of the displacement of the \ac{QNM} due to the $d_9$ term, assuming $d_9>0$. Obviously, the lengths do not correspond to anything. The shift in the imaginary part is negative for even modes and positive for odd modes.} 
\end{figure}

As an example, we compute the numerical corrections to the \ac{QNM} for the first overtone number ($n=0$), i.e. the background frequencies are $\xi_0=0.747343-0.177925 \ii$, $1.199887-\ii 0.185406 \ii$ and $1.618357-\ii 0.188328$, respectively for  $\ell=2,3,4$. We define the relative perturbation as
\be 
\delta \equiv \left( \frac{\mathrm{Re}(\omega-\omega_0)}{\varepsilon\, \mathrm{Re}(\omega_0)},\frac{\mathrm{Im}(\omega-\omega_0)}{\varepsilon\, \mathrm{Im}(\omega_0)}  \right)
= d_8 \delta_8 + d_9 \delta_9\,.
\ee 
We show in Tab.~\ref{tabR3:QNM} the numerical corrections we obtain for the first frequency. , and in Fig.~\ref{figR3:QNM_pert}, we show the direction of the displacement in the $(\omega_{\mathrm{R}}, \omega_{\mathrm I})$ plane for the $\delta_9$ contribution only, assuming $d_9>0$. Interestingly, the imaginary part increases for odd modes and decrease for even modes. Recall that the imaginary part is proportional to the damping (or inversely proportional to the lifetime of the wave).
\begin{center}
\begin{table}[h!]
\centering
\begin{tabular}{c | c c c }
& $\ell=2$               & $\ell=3$               & $\ell=4$  \\ \hline 
$\delta_8^{\mo}$ & $\mathcal{O}(10^{-8})$ & $\mathcal{O}(10^{-7})$ & $\mathcal{O}(10^{-6})$  \\
$\delta_9^{\mo}$ & $(21.07,47.53)$ & $(18.29,46.27)$ & $(17.55,45.70)$ \\
\hline \hline 
$\delta_8^{\me}$ & $\mathcal{O}(10^{-4})$  & $\mathcal{O}(10^{-6})$  &  $\mathcal{O}(10^{-6})$   \\
$\delta_9^{\me}$ & $(-12.33,-58.12)$ & $(-14.15,-51.82)$ & $(-14.46,-50.16)$
\end{tabular}
\caption{\label{tabR3:QNM} Quasinormal Modes Corrections}
\end{table}
\end{center}

It seems from the numerical results that the corrections coming from $d_8$  are very small. As a matter of fact, these should be exactly $0$. Indeed, if we perform a metric redefinition given by
\be  \label{eqR3:g_tilde_def}
\bds g = \tilde{\bds g} - \frac{2\varepsilon}{\Mpl^2} \delta \bds g,
\ee 
the Einstein Hilbert action 
\be 
\mathcal{S}_{\mathrm{EH}} =  \int\; \sqrt{-g} \frac{\Mpl^2}{2} R\, \bdd x^4
\ee 
becomes
\begin{align}
\mathcal{\tilde{S}}_{\mathrm{EH}} &= \mathcal{S}_{\mathrm{EH}} + \varepsilon\,  \delta \mathcal{S}_{\mathrm{EH}}\,, \\
\delta \mathcal{S}_{\mathrm{EH}} &= \int\; \sqrt{-\tilde{g}}
\left( \tilde{R}^{\mu \nu} - \frac{\tilde R}{2} \tilde{g}^{\mu \nu}\right) \delta g_{\mu \nu}
\, \bdd x^4 \,.
\end{align}
In particular, choosing 
\be \label{eqR3:deltag_fieldredef}
\delta g_{\mu \nu} = d_8 \left(
\frac 12 g_{\mu \nu}\tensor{R}{_{\mu \nu \rho \sigma}}\tensor{R}{^{\mu \nu \rho \sigma}} - \tensor{R}{_\mu_{\nu \rho \sigma}}\tensor{R}{_\mu^{\nu \rho \sigma}}
\right)
\ee
yields
\be 
\delta \mathcal{S}_{\mathrm{EH}} = -d_8  \int\; \sqrt{-\tilde{g}} R^{\mu \nu} \tensor{R}{_\mu^{\nu \rho \sigma}} \tensor{R}{_\nu_{\nu \rho \sigma}}\, \bdd x^4\,.
\ee 
Considering the effective Lagrangian Eq.~\eqref{eqR3:Lcal6_final}, this correction exactly cancels out the term $\Lcal_{6,8}$. 

Let us assume now that only the term $\Lcal_{6,8}$ is present in the theory. After performing the field redefinition Eq.~\eqref{eqR3:deltag_fieldredef}, we are left with \ac{GR}, where the \ac{QNM} are known, i.e.
\be 
\tilde{\omega} = \omega_{\mathrm{GR}}\,,
\ee
where $\tilde{\omega}$ corresponds to the frequency in the frame associated with $\tilde{\bds g}$. It is direct to show that the correction given by Eq.~\eqref{eqR3:deltag_fieldredef} has the the usual time dependence $\sim \eee^{-\ii \tilde{\omega} t}$. Hence, after performing the inverse field redefinition Eq.~\eqref{eqR3:g_tilde_def}, the time dependence stays the same, which means that $\tilde \omega$ is also the frequency in the original frame associated with the metric $\bds g$. Because of these considerations, the term proportional to $d_8$ does not contribute to the \ac{QNM}.
\MYrem{What about the other terms?}{
Do you remember all the terms we eliminated in Section~\ref{secBH:somany}? We were able to do it only because we are considerations perturbation around the Schwarzschild metric. However, this comment is independent on the metric background. Moreover, it can be shown that all the terms in the effective Lagrangian except $d_9$ and $d_{10}$ can be eliminated with an appropriate field redefinition. This means that, in this effective theory of gravity, and at linear order, the corrections to the \ac{QNM} will only depend on the combination $2d_9 + d_{10}$ (recall $d_{10}$ can also be eliminated, see Eq.~\eqref{eqR3:replace_d89}). Roughly speaking, only the terms that are genuinely Riemann-cubed contribute to the \ac{QNM} corrections.
}
\MYrem{What about the speed?}{
Could we also conclude that the speed of the \ac{GW} also depends only on $2d_9 + d_{10}$ only, as it is actually the case in our specific example? This is actually more subtle. Performing the field redefinition will add some non-minimal interactions between the matter fields and the metric. Hence, in the new frame, the speed of light is not $1$ anymore. However, we still expect the ratio of the velocities to be frame-independent, as we discussed in Part~\ref{Part:Frames}. \\
It can be formally shown that the ratio of the speed of \ac{GW} and of light is frame-independent, see \cite{deRham:20202}.
}

\stopchap

\chapter{Conclusion\label{chapR3:concl}}
In this Part, we first reviewed some well-known facts about the Schwarzschild Black Hole. We presented the formalism to study gravitational perturbations around a \ac{SSSS} Black Hole, and its decomposition into odd and even modes.

We discussed the concept of \ac{QNM}, which is central in Gravitational Physics and Astrophysics. Gravitational Waves have been detected in 2015, see \cite{Abbott:2017, Abbott:20172} after the collision of two Black Hole. The ringdown of this signal can be decomposed in a sum of different modes, the frequencies of which are called \ac{QNM}. The frequencies of the \ac{QNM} are quantified, and the list of this frequency is the identity card both of the Black Hole and of the theory of Gravity governing the production and the propagation of these \ac{GW}. Being able to predict these frequencies in various models and under different assumptions is necessary if we want to extract useful information from these gravitational waves signals.

We then quantitatively explained how the speed of a gravitational wave can be defined, provided a master equation. The definition we provided is observer invariant as it does not depend on the master equation one decides to use to perform the analysis. We developed the formalism of the effective metric, which can also be used to determine the position of the Black Hole horizon as seen by the gravitational waves.

Once these basic, yet important, concepts were defined, we turned our attention to a specific Effective Theory of Gravity. Effective Theories are very useful as they allow us to capture the effects of heavy fields, without knowing exactly the identities of those, or any high-energy Physical theory, see \cite{Cardoso:2018,Cano:2019,Cano:2020,Garcia-Saenz:2022,Gross:1986,Metsaev:1986,Franciolini:2018, Tattersall:2017, Tattersall:2019} for various examples on the topic of \ac{QNM}. The philosophy of an Effective Theory is to stay agnostic about this unknown theory, and only consider its effects on low-energy processes by adding an arbitrary linear combination of every possible terms in the Lagrangian up to a given (mass) dimension. In our case, we considered terms of the forms Riemann-squared and Riemann-cubed. We argued that, even if one should technically consider $12$ different terms, $10$ of them can be actually removed if we work perturbatively around a Schwarzschild Black Hole. We were then left with only two new terms in the action, namely $\Lcal_{6,8}$ and $\Lcal_{6,9}$, parametrised by two dimensionless coefficients $d_8$ and $d_{9}$ and one mass $M$ corresponding, roughly, to the mass of the lightest fields which was integrated out.

The first step of our analysis in this Effective Theory was to determine the \ac{SSSS} correction to the Schwarzschild Black Hole. In the usual Spherical coordinates, the terms $g_{tt}$ and $g_{rr}$ receives corrections proportional to $r^{-6}$ and $r^{-7}$. We also computed the position of the new horizon $\rH$ which is different from the usual Schwarzschild radius $\rg$, hence the usual relation does not hold anymore in our theory $\rH\neq 2\GN \MBH$.

Second, we derived a master equation for the odd and the even perturbations. We used the method discussed in \cite{Cardoso:2019,McManus:2019,Kimura:2020} to compute numerically the corrections to the Quasinormal frequencies, based on a series expansion of the perturbed potential. We computed these frequencies for three different modes and showed that the contribution proportional to $d_8$ vanishes (up to some numerical errors, especially for even modes). We explained why this was expected, as the $\Lcal_{6,8}$ term can be removed by a field redefinition, under which the frequency is not modified. 

Lat but not least, we derived and expression for the speed of the Gravitational Waves around the Black Hole, which formed the main result of this work. This topic has received a lot of attention in the past, see \cite{deRham:2018,deRham:2020, deRham:20202}. We found that the sign of $\cS-1$ only depends on the coefficients of the Effective Theory (and not on the radial distance), and we consider examples where the gravitational waves can be either subluminal or superluminal. Even if the second could be surprising at first sight, one can actually not rule out such a theory, as was argued in \cite{Hollowood:2007, Hollowood:2008, Hollowood:2009,Hollowood:2010,Hollowood:2011,Hollowood:2012, Hollowood:2015} and \cite{deRham:2020}.

Finally, we derived the expressions for the effective metric seen by the Gravitational waves, and showed that they feel the same horizon as the other species (especially light). This is result is actually more general and is linked to the Horizon Theorem. Our analysis only holds at linear order in the perturbations, but this result should hold at any order.

One last comment: We showed qualitatively that the order of magnitude of the perturbations is extremely tiny around typical Black Holes. For example, if the Black Hole mass is around $50$ Solar Mass, and if the parameter in the Lagrangian is of order $M\sim 10^{-60}\Mpl$, the amplitude of the corrections would be of order $\epsilon\sim 10^{-40}$. This implies that no observations could distinguish between General Relativity and the theory we considered (at least considering astronomical Gravitational waves). Our work is therefore mostly theoretical and conceptual as we investigated the properties of the gravitational waves in a given Effective Theory of Gravity.

Future works could include, for example, a similar study around rotating Kerr Black Holes, as it was already done in \cite{Cano:2021,Eichhorn:2021}. Some other works considered the same kind of gravitational perturbations, but in a specific theory of Modified Gravity, for example in Horndesky's theory \cite{Kase:2021} or in theories of non-linear electrodynamics~\cite{Nomura:2021}.
\part{Conclusion and Outlook}\label{Part:Conc}
\begin{figure}[h!]
	\centering
	\includegraphics[width = 0.9\textwidth]{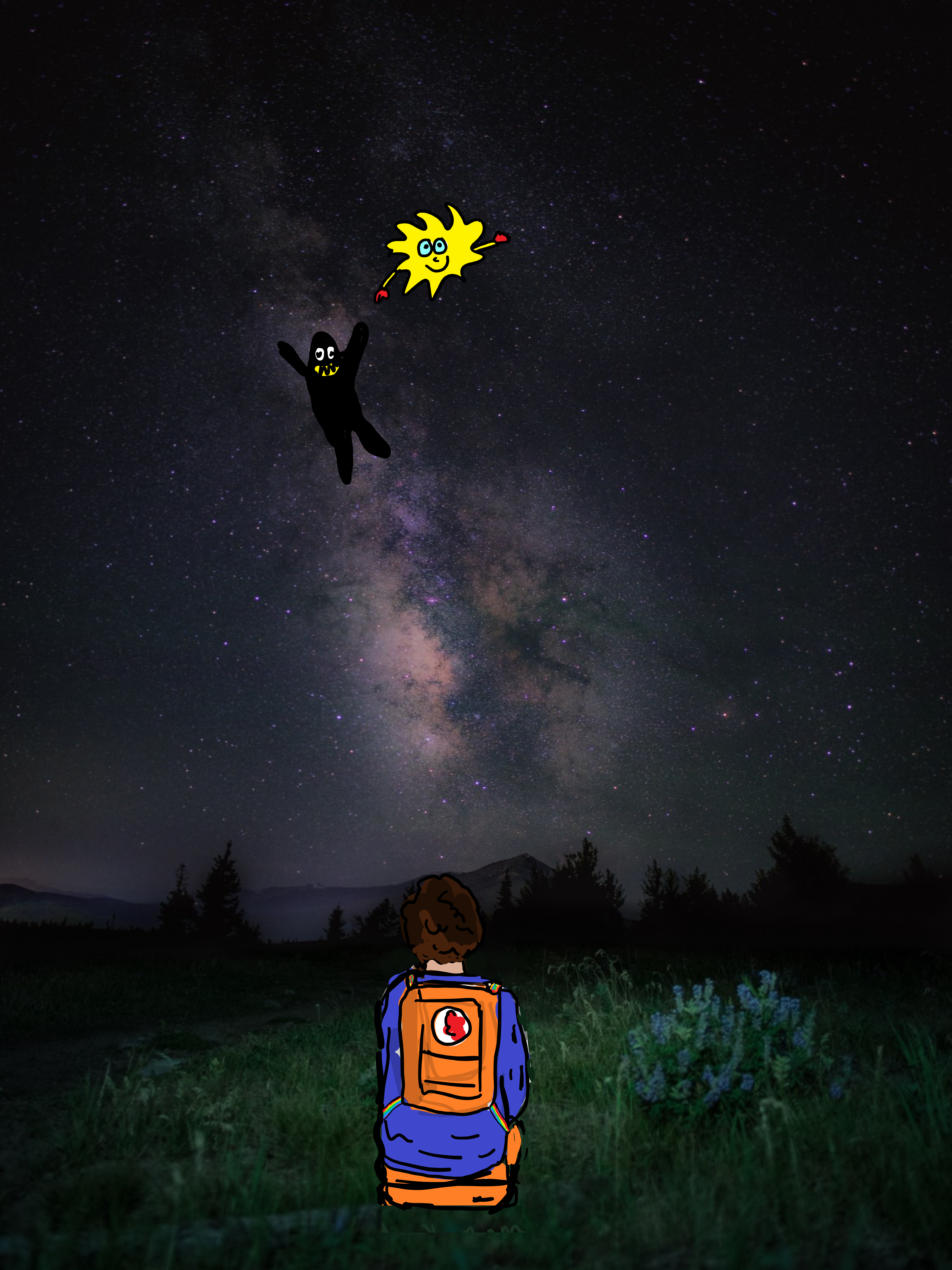} 
	\caption*{\textbf{Contemplation}, Juan Manuel Garc\'ia Arcos}
\end{figure}

\chapter{General Conclusion}
This is the end of our journey. For you it may have been a few hours, but for me it felt more like 4 years, probably a relativistic effect. Let us take a step back and summarise what we learned in this thesis.

The objective of this thesis was to study several topics with one idea in mind: How can we define clear observables, with the objective of testing our Cosmological models and parameters? As Cosmology becomes more and more data driven, it is very important to be able to have in our hands several different, and independent, ways to perform measurements and test our theories. Let us review the various observables presented in this thesis.

First, in Part~\ref{Part:Frames}, I studied conformally related frames, where both metrics are related to each other via a spacetime dependent global factor. I considered two specific frames: The Einstein and the Jordan frames. In the Einstein frame, the Einstein-Hilbert action takes the usual form, while matter fields are non-minimally coupled to gravity. The opposite is true in the Jordan frame: The Einstein-Hilbert action is not standard as there is a coupling to the scalar field, but the matter is minimally coupled to gravity. I presented the physical interpretation of working with two different frames: This is nothing but a spacetime-dependent (local) change of units. I argued that this should, in fine, not change the predictions regarding physical observables. Then, I introduced the Galaxy Number Counts as a physical observable quantifying the angular fluctuations of the number of observed galaxies. I showed that this quantity does not depend on the frame, as expected. This shows that the Galaxy Number Counts is a genuine observable, or the other way around, that both frames are physical. This also means that computations can be performed in either frames, as long as sufficient care is taken in the interpretation of the results. The Galaxy Number Counts can itself help us to test Cosmological models, for example by inferring the convergence correlation function, whose power spectrum is directly linked to the lensing power spectrum.

Then, in Part~\ref{Part:RotLensing}, I turned my attention to the weak lensing formalism. After briefly summarising the main concepts of lensing, I showed that images of galaxies whose main axes are not aligned with the principal direction of the shear undergo a rotation. I argued that this is a good candidate for a physical observable if the polarisation direction is also measured. Indeed, this vector tends to be aligned with the axes of the ellipse. As the polarisation is parallel transported along the path, its position defines a Sachs basis with respect to which the images of galaxies rotate. A non-zero angle between the polarisation of light and the main axes of the ellipse is a signature of shear. Its observation can be used to probe the shear along the line of sight between the observer and the source. With this idea in mind, I proposed a new way to probe cosmic shear by evaluating this angle for several galaxies in a catalogue. I explicitly showed how to estimate coordinate invariant shear correlation functions using this observable. Finally, I argued that this method can be competitive, i.e. yield low signal-to-noise ratio, if the number of observed galaxies is sufficiently high. Correlation functions, or the power spectrum, their equivalent in Fourier space, are again a good test of our Cosmological models, which motivates the will to define several independent ways to estimate them.

Finally, in Part~\ref{Part:BH}, I considered a Schwarzschild Black Hole in a Modified Gravity theory. I first recalled some well-known facts about the Schwarzschild Black Hole: The decomposition of gravitational perturbations into odd and even modes, the Master Equation and the Regge and Zerrili potentials. With these tools, I argued that we can define the speed of gravity in a non-ambiguous way, and I presented the central concept of Quasinormal Modes of Black Holes. Second, I defined an Effective Field Theory of Gravity, in which we assume that high energy fields exist but are not observable at low-energy scales. These fields still have an effect on low-energy Physics through their interaction with, for example, the metric. This interaction can be explicitly determined by integrating out these fields from the high-energy theory. I thus constructed the most general effective Lagrangian, up to the fourth power of the Riemann tensor (matter fields were not considered in this specific model). I argued that most terms in the Lagrangian are actually negligible, due to the specific Schwarzschild case I considered. Then, I proceeded with the same analysis as for the General Relativity case. I derived the background perturbed Black Hole and showed that the position of the horizon is slightly displaced. I computed the corrected speed of the Gravitational Waves and the correction to the Quasinormal Modes. Two main results were that the horizon is the same for both gravitational waves and other matter particles and that the speed of the gravitational waves can be either subluminal or superluminal, depending on the nature of the heavy particles integrated out. I argued that, in fine, this actually does not lead to any causality issues. Finally I also showed that the amplitude of the various corrections in this model are small, and that no experimental tests can be extracted therefrom. Hence, the main goal of this project was more to show some conceptual aspects of Effective Theories, rather than pretend to build realistic observables.

\chapter{Outlook}
\emph{What's next?}
\vspace{0.3 cm}

As mentioned in Chapter~\ref{chap:historical_intro}, Cosmology has become a strongly observational science. To give only one example, the first Gravitational Waves have been observed in 2015, and the mischief was repeated several times since then! The \ac{LCDM} model describes particularly well the observations we have made so far: Observations are in perfect agreement with the \ac{FRLW} model and the theoretical \ac{CMB} power spectrum.

But remember, this is not the end of the story. The Hubble tension, dark matter and dark energy are three ghosts wandering above our heads. To solve these main problems, and all the others I kept silent, we need to face two challenges. First, it is necessary to come up with new models that differ from \ac{GR} enough so as to predict these problematic observations, without spoiling all the old results, e.g. regarding the Solar System or in accordance with the \ac{FRLW} model. Second, we also need to find ways to perform extremely precise observations to confront these models or to be able to detect a potential candidate for the dark components.

To make a connection with the present work, let us remember that in Part~\ref{Part:BH}, we estimated that the effective theory we considered would add correction of magnitude $\varepsilon \sim 10^{-100}$, or smaller. Comparing this with the sensitivity of the LIGO experiment $\sim 10^{-30}$, this is a double failure: This theory does not make any useful prediction departing from \ac{GR}, and we are far from having the technology to probe such perturbations. However, this is not the end of the story. Black Hole Physics is an intensely explored field nowadays and there are high hopes that we will find a theory in which departure from \ac{GR} is detectable.

Regarding lensing, as I briefly pointed out, new galaxy surveys are planned. These will include new observational information, for example light polarisation. Be it thanks to the method I presented, or any other one, further research will doubtless sharpen our knowledge and lead to more precise estimation of the Cosmological parameters. 

This is the situation that Cosmology faces now. There are, however, reasons for optimism. Until today, humankind has shown a wonderful ability to come up with new ideas and new technologies to face never ending challenges (with the little cost that came along -- the destruction of our planet, LOL). This can only sound promising for the future of the field!

\appendix
\part{Appendices}\label{Part:App}
\begin{figure}[ht!]
	\centering
	\includegraphics[width = 0.9\textwidth]{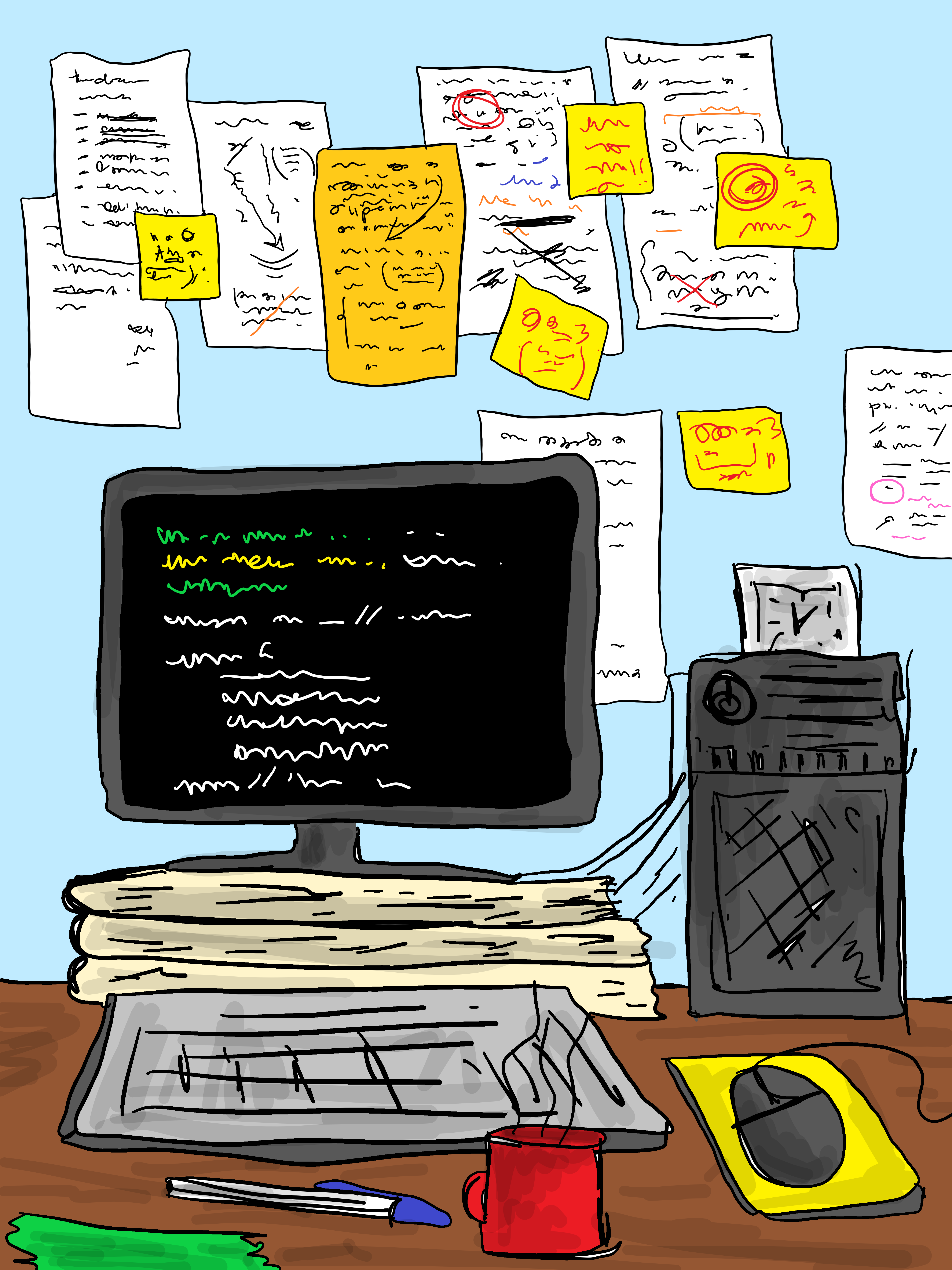} 
	\caption*{\textbf{The Desk}, Juan Manuel Garc\'ia Arcos}
\end{figure}

\chapter{Rotation from Lensing}
\startchap
\section{Optical Depth\label{app:optical_depth}}
\begin{figure}[ht!]
    \centering
	\includegraphics[width = 0.6\textwidth]{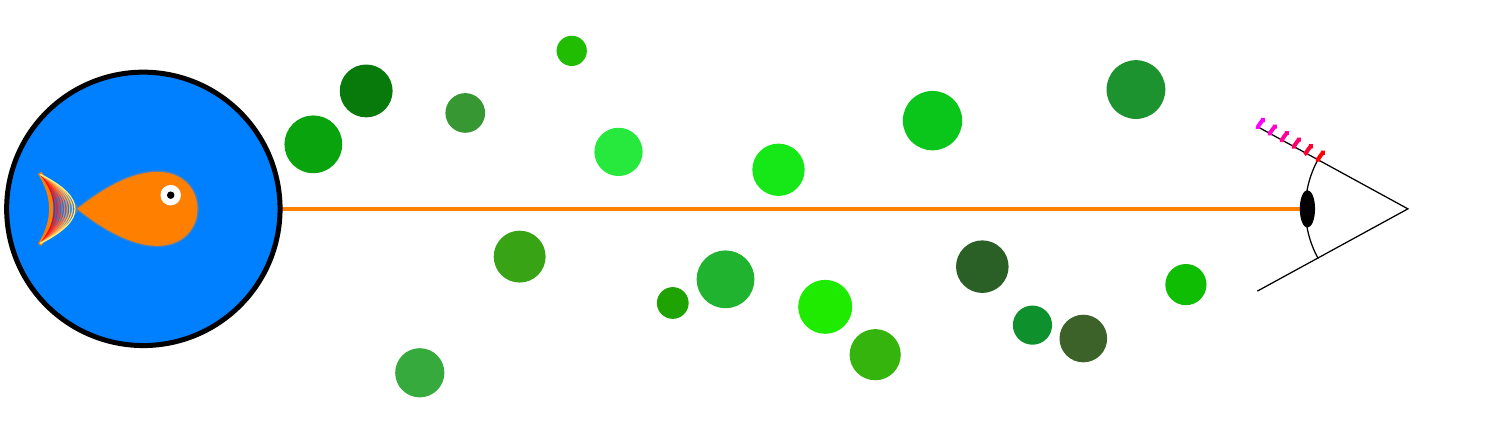} 
	\caption[Optical depth]{\label{figapp:optical}
	\textbf{Optical depth - Hide and seek with the fish}\\
		The observer is looking inside a forest filled with trees of various sized $\sigma$ and randomly distributed with density $n$. What is the probability that the observer can see the blue pond on the left, at a distance $L$?}
\end{figure}

We derive here the results presented in Section~\ref{seclensing:stat_ingred}. We consider the following situation. An observer is located at $x=0$ while a light source is located at $x=L$, where $x$ represents a distance. The space between the observer and the source is filled with randomly distributed targets. We assume that if light passes close enough to a target, it undergoes some \emph{modification}. We do not precise the modification we have in mind, we only assume that its intensity $\Theta$ can be measured. In the example of Section~\ref{seclensing:stat_ingred}, this is the scaled rotation and the targets are Schwarzschild lenses. We assume that, to obtain at least a given intensity $\Theta_0$, the light must pass in a surface of size $\sigma(\Theta_0,x)$ around a target located at distance $x$ from the observer. This quantity represents the cross-section of the targets, and we assume that it depends on the distance and on the intensity of the rotation. The dependence on $\Theta_0$ satisfies the reasonable assumption $\sigma(0,x)=\infty$. The situation is illustrated in Fig~\ref{figapp:optical}.

We want to compute $P(>\Theta_0,L)$, the probability of observing the source with a rotation bigger than $\Theta_0$. To compute this quantity, we compute first $P(\leq\Theta_0,x)$, the probability for the light ray to undergo a rotation less or equal than $\Theta_0$. We will go backwards and start at $x=0$ to finally reach $x=L$. The initial condition is 
\be  \label{eqapplensing:ini}
P(\leq \Theta_0,0)=1\,,
\ee
for all values of $\Theta_0$.

First, let us assume that $P(\leq\Theta_0,x)$ is known. We want to evaluate $P(\leq\Theta_0,x+\dd x)$. To do so, we need to consider the conditional probability
\be  \label{eqapplensing:conditional}
P(\leq \Theta_0 , x+\dd x \vert \leq \Theta_0, x) = 1-P(>\Theta_0, x \rightarrow  x+ \dd x)\,,
\ee
where $P(>\Theta_0, x \rightarrow  x+ \dd x)$ is the probability of undergoing a rotation bigger than $\Theta_0$ between $x$ and $x+\dd x$. This is intuitively understood as follows: If you have reached $x$ without problem, the probability for you to reach $x+\dd x$ without problem is almost $1$, up to the probability you get in trouble between $x$ and $x+\dd x$. 

We evaluate now this last term. We assume that the light beam travels in a tube of sectional area $\dd A$ and length $\dd x$. The number of targets is 
\be 
\dd N= n  \, \dd x \, \dd A\,,
\ee 
where $n$ is the density of targets, which can depend on $x$ and $\Theta_0$, but we do not write it explicitly to not clutter the notations. The probability of hitting a target is the ratio of the targets area $\dd A_\sigma = \sigma \, \dd N$ with the total area $\dd A$. Here, we assumed that the length $\dd x$ is sufficiently small to avoid overlapping targets (recall the surface of an individual target is $\sigma$). This probability is then
\be  \label{eqapplensing:nsx}
P(> \Theta_0 , x \rightarrow \dd x) =  \frac{\dd A_\sigma}{\dd A}=\frac{\sigma \, \dd N}{\dd A} = n \sigma\,\dd x\,.
\ee 
Using Bayes' Theorem, we can express the conditional probability in Eq.~\eqref{eqapplensing:conditional} as 
\begin{align} \label{eqapplensing:bayes}
P(\leq \Theta_0 , x+\dd x \vert \leq \Theta_0, x) &=
\frac{P(\leq \Theta_0, x \vert \leq \Theta_0 , x+\dd x ) P(\leq \Theta_0 , x+\dd x)}{P( \leq \Theta_0, x)} \\ \label{eqapplensing:bayes2}
&=
\frac{P(\leq \Theta_0 , x+\dd x)}{P( \leq \Theta_0, x)}
\,,
\end{align}
where we used the obvious fact that 
\be 
P(\leq \Theta_0, x \vert \leq \Theta_0 , x+\dd x )=1\,,
\ee 
i.e. if you reach your endpoint without problem, we can be sure you did not get into trouble anywhere before. Combining the various relations Eq.~\eqref{eqapplensing:conditional}, Eq.~\eqref{eqapplensing:nsx} and Eq.~\eqref{eqapplensing:bayes2} yields
\be 
P(\leq \Theta_0, x + \dd x) = P(\leq \Theta_0,x) (1- n \sigma \, \dd x)\,.
\ee 
The solution of this differential equation, together with the initial condition Eq.~\eqref{eqapplensing:ini}, is
\begin{align}
P(\leq \Theta_0, x ) &= \exp\left( - \tau(\Theta_0,x)\right)\,, \\
\tau(\Theta_0,x ) &= \int_0^{y}\; n(\Theta_0,y) \sigma(\Theta_0,y)\, \dd y\,,
\end{align}
where we explicitly wrote the dependence on $\Theta$ and on the position. Furthermore, we can assume that the targets have also internal parameters $\chi$ (e.g. the mass in our example) over which we integrate, and in a cosmological setup, the redshift $z$ is a better parameter than the distance. This gives the final relation
\boxemph{
P(\leq \Theta_0, z_\ms) &= \exp(-\tau(\Theta_0, z_\ms))\,, \\
\tau(\Theta_0, z_\ms) &= \int_0^{z_\ms}\; n(\Theta_0, z,\chi) \sigma(\Theta_0,z,\chi) \frac{\dd r}{\dd z}\, \dd \chi \, \dd z\,.
}
Note that here the density $n$ should be the physical density of targets. Moreover, the inverse probability is given by
\be 
P(>\Theta_0, z_\ms) = 1- P(\leq\Theta_0, z_\ms) 
\ee 
and its PDF $p(\Theta, z_\ms)$ satisfies
\boxemph{
    P(>\Theta_0, z_\ms) &= \int_{\Theta_0}^{\infty}\; p(\Theta, z_\ms)\, \dd \Theta\,, \\
    p(\Theta, z_\ms) &= -  \frac{\partial  P(>\Theta, z_\ms)}{\partial \Theta}\,.
}

\section{Illustris' Simulation\label{app:Illustris}}
\begin{figure}[ht!]
\begin{adjustbox}{max width=1.35\linewidth,center}
    \centering
	\includegraphics[width = 0.555\textwidth]{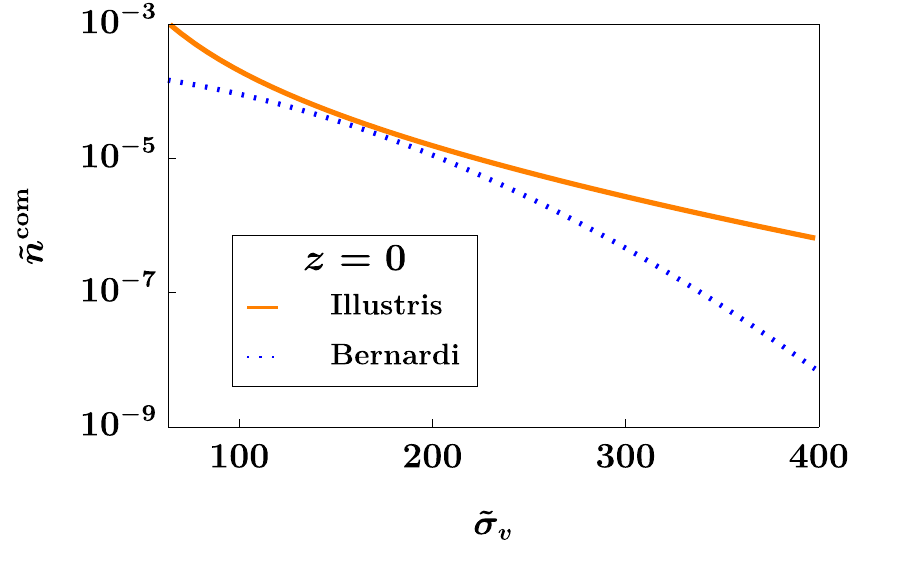} 
	\includegraphics[width = 0.55\textwidth]{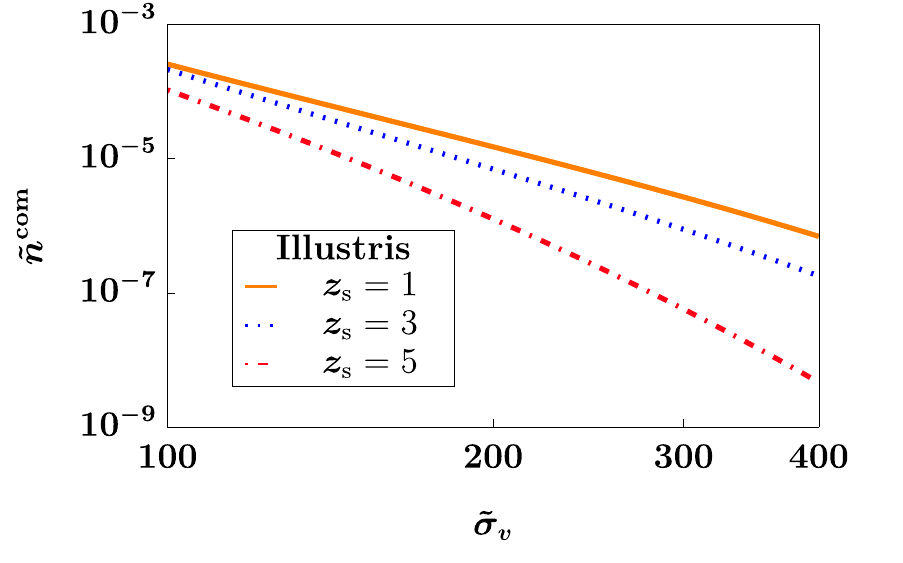} 
\end{adjustbox}
	\caption[Galaxy number density]{\label{figapp:models}
		\textbf{Galaxy number density, comparison between two models}\\
		\textbf{Left:} Comparison of the galaxy number density between Illustris' and Bernardi's models, evaluated at $z=0$\\
		\textbf{Right:} Illustris' model, evaluated for three different sources redshifts}
\end{figure}

We present here the parametrisation for the galaxy number density $n(z,\sigma_v)$ discussed in Section~\ref{seclensing:models}, based on the Illustris' hydronamical simulation, see \cite{Torrey:2015,Cusin:2019}. The cumulative galaxy number is
\begin{align}\label{eqapplensing:fit2}
\log_{10} \tilde{N}(>{\tilde{\sigma}}_v, z)=&
A(z)+\alpha(z) \left(\log_{10}\tilde{\sigma}_v-\gamma(z) \right) \\ \nonumber
&+\beta(z)\left(\log_{10}\tilde{\sigma}_v-\gamma( z)\right)^2 \\\nonumber &-\left(\tilde{\sigma}_v\times10^{-\gamma(z)}\right)^{1/\ln(10)}\,,
\end{align}
and the comoving galaxy number density is given by
\be 
\tilde{n}^{\mathrm{com}}(\tilde{\sigma}_v,z) = - \frac{\dd \tilde{N}(>\tilde{\sigma}_v,z)}{\dd \tilde{\sigma}_v}\,,
\ee
where we defined the dimensionless quantities
\begin{align}
    \tilde{N} &= N \times \SI{1}{\Megapc^3}\,,\\
    \tilde{\sigma}_v &=  \frac{\sigma_v}{\SI{1}{\kilom \times \sec^{-1}}}\,, \\
    \tilde{n}^{\mathrm{com}} & = n \times \SI{1}{\Megapc^3} \times {\SI{1}{\kilom\times  \sec^{-1}}}\,.
\end{align}
In Eq.~\eqref{eqapplensing:fit2} the functions $A$, $\alpha$, $\beta$ and $\gamma$ take into account a redshift evolution in the model. Their expressions are
\begin{align}
A(z)&=a_0+a_1 z+a_2 z^2\,,\\
\alpha(z)&=\alpha_0+\alpha_1 z+\alpha_2 z^2\,,\\
\beta(z)&=\beta_0+\beta_1 z+\beta_2 z^2\,,\\
\gamma(z)&=\gamma_0+\gamma_1 z+\gamma_2 z^2\,.
\end{align}
The values for the different parameters are given in Tab.~\ref{tabapp:coeffIllustris}.
\begin{table}[ht!]
\centering
\begin{tabular}{c|ccc}
$i$& $i=0$ &  $i=1$&  $i=2$  \\
 \hline
 $A_i$& 7.391498 &5.729400  &  -1.120552  \\
 $\alpha_i$ &-6.863393  &  -5.273271&  1.104114  \\
$\beta_i$ &2.852083  &1.255696  &  -0.286638  \\
 $\gamma_i$ &  0.067032& -0.048683 &  0.007648\\
 \hline
\end{tabular}
\caption{\label{tabapp:coeffIllustris} Coefficients for the redshift evolution parametrisation of the Illustris' simulation.}
\end{table}
We show in Fig.~\ref{figapp:models} a comparison between Illustris' and the Bernardi's model.
It is clear that the Illustris' simulation contains more galaxies. Using one or the other model can lead to very different results. Moreover, we see that the Illustris' simulation contains more galaxies at high redshift.

\section{Estimators for the Correlation Functions \label{app:estimators}}
The estimators for the correlation functions defined in Section~\ref{seclensing:estcorrfun} are
\begin{align} 
        \label{eqapp:zetapest}
        \hat{\zeta}_+(\mu,z_1,z_2) &= \Xi\, F_1(\beta_1^\prime, \beta_2^\prime, \beta_1, \beta_2) 
        + \Xi^\prime\, F_1(\beta_1, \beta_2,\beta_1^\prime, \beta_2^\prime) \,, \\
        \label{eqapp:zetamest}
        \hat{\zeta}_-(\mu,z_1,z_2) &= \Xi\, F_2(\beta_1^\prime, \beta_2^\prime, \beta_1, \beta_2) 
        + \Xi^\prime\, F_2(\beta_1, \beta_2,\beta_1^\prime, \beta_2^\prime) \,, \\
\end{align}
where the functions $F_1$ and $F_2$ are
\be\label{eqapp:F1}
F_1(\beta_1, \beta_2, \beta_1^\prime, \beta_2^\prime) =
\frac{\cos (2 (\beta_1 + \beta_2))}{ \cos(2(\beta_1^\prime-\beta_2^\prime)) \cos(2(\beta_1+\beta_2)) - \cos(2(\beta_1^\prime+ \beta_2^\prime)) \cos(2(\beta_1-\beta_2)) } \,,
\ee
and
\be \label{eqapp:F2}
F_2(\beta_1, \beta_2, \beta_1^\prime, \beta_2^\prime) =
\frac{\cos (2 (\beta_1 - \beta_2))}{ \cos(2(\beta_1^\prime+\beta_2^\prime)) \cos(2(\beta_1-\beta_2)) - \cos(2(\beta_1^\prime- \beta_2^\prime)) \cos(2(\beta_1+\beta_2)) } \,.
\ee

\section{Flat Sky Approximation\label{applensing:FS}}
\subsection{General Description}
We briefly present here the flat sky approximation. Some details can be found in \cite{Hu:2000,Bunn:2006,White:2017,White:1997,Bernardeau:2010}. Surprisingly enough, I did not find any reference covering the topic in a systematic and rigorous way.

\begin{figure}[ht!]
\begin{adjustbox}{max width=1.35\linewidth,center}
    \centering
	\includegraphics[width = 0.5\textwidth]{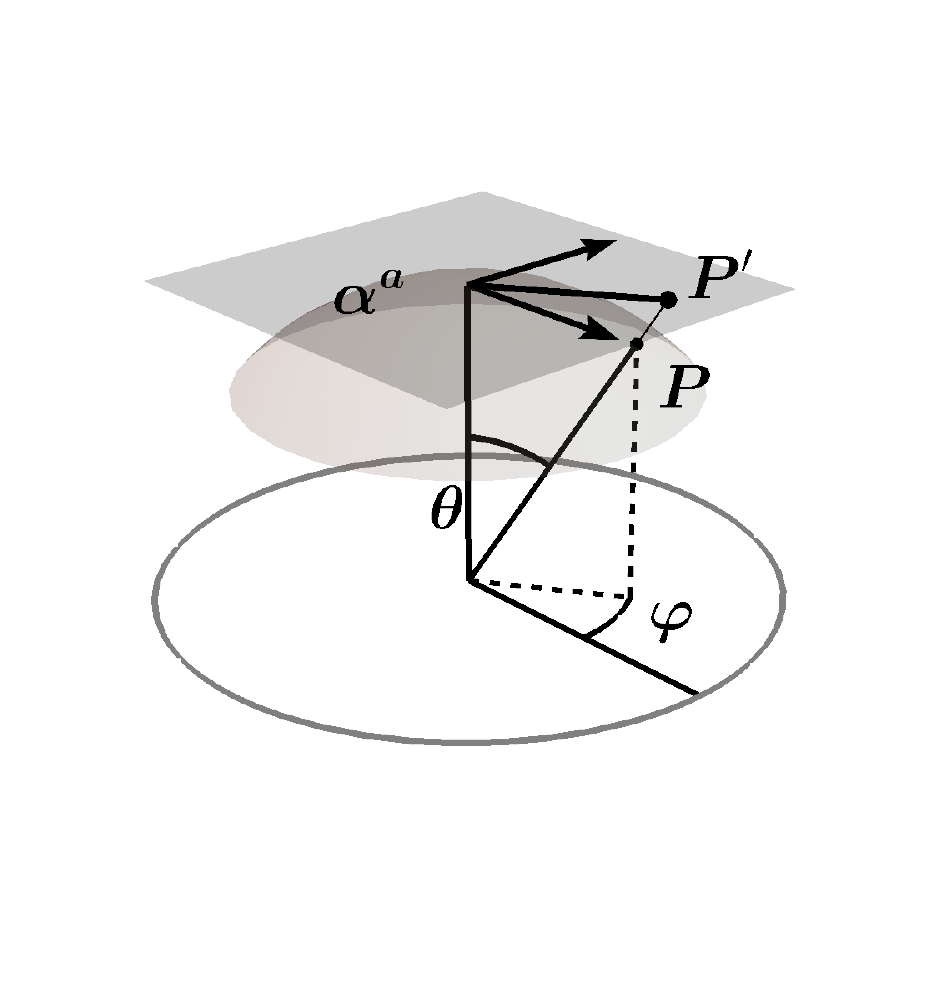} 
	\includegraphics[width = 0.5\textwidth]{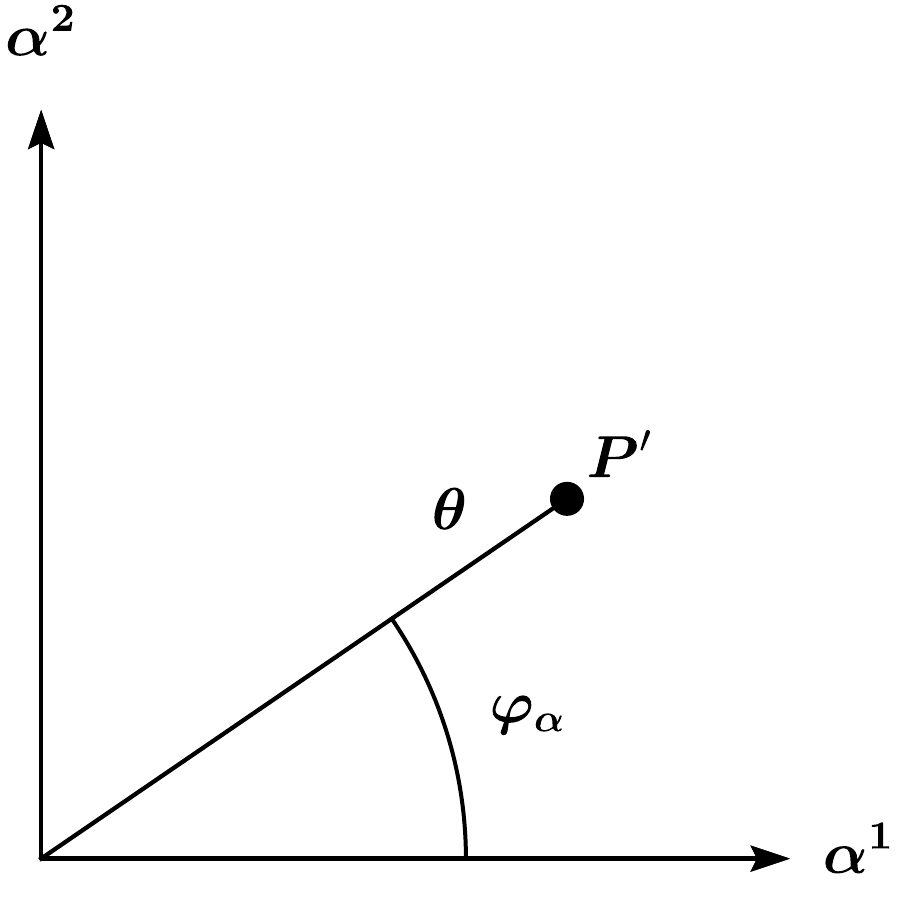}
\end{adjustbox}
	\caption[Flat sky approximation]{\label{figapp:Flat_sky_schema}
		 Representation of the flat sky approximation\\
		 \textbf{Left:} A point $P$ with coordinates $(\theta, \varphi)$ on the sphere is mapped onto a point $P^\prime$ on the flat sky plane.\\
		\textbf{Right:} The flat sky plane with coordinates $ (\alpha^1, \alpha^2 )$ seen from above. The point $P^\prime$ is located at $(\alpha^1, \alpha^2 )=\theta( \cos \varphi_\alpha, \sin \varphi_\alpha)$, with the simple relation $\varphi_\alpha = \varphi$.}
\end{figure}

The main idea if the following: For large values of $\ell$ (roughly for $\ell  > 100$), if we observe a small patch of the sky around a fixed direction $\bds n$, we can assume that it is a plane. To simplify the expressions, we assume that we observe the North Pole, but the final results are general (especially they hold even if we perform the computations on the Equator, as we did in Section~\ref{seclensing:result_cosmo}). Around this point, we cannot use the usual spherical coordinates, and we use the Cartesian coordinates. More precisely, a point $P$ is given by
\be 
P=(x,y, \sqrt{1-x^2-y^2})\,.
\ee 
We will work at linear order in $x$ and $y$, as we consider a very small patch around the North Pole. We project the point $P$ on the plane $z=1$ and obtain
\be
P^\prime = \left(\frac{x}{\sqrt{1-x^2-y^2}} ,\frac{y}{\sqrt{1-x^2-y^2}} ,1\right) \approx (x,y,1)\,.
\ee 
This is the definition of the flat sky coordinates of a point, in order words, we define the vector $\bds \alpha$ on the plane by
\be
\bds \alpha(P) = (\alpha^1 , \alpha^2 ) = (x,y)\,.
\ee
Moreover, it can be shown directly that the norm of the vector $\bds \alpha$ is given by the usual angle $\theta$ (which is also true at the Pole), and that the polar angle $\varphi_\alpha$ in the $\bds \alpha$ coordinates is also the usual polar angle $\varphi$. This setup is shown in Fig.~\ref{figapp:Flat_sky_schema}. 

Let us consider a real field $X$ on the sphere with the usual decomposition
\begin{align}  \label{eqlensingapp:Xn}
X (\bds n) &= \sum_{\ell,m} X_{\ell,m} \YLM{\ell}{m}(\bds n)\,, \\
\label{eqlensingapp:Xlm}
X_{\ell,m} &= \int\; X(\bds n) \YLMStar{\ell}{m}(\bds n)\, \dd \Omega^2\,.
\end{align}
Using the coefficients $X_{\ell,m}$, we want to define a flat sky field $\Xcal(\bds \alpha)$ on the $2D$-plane. First, we define its Fourier coefficients. Let $\bds \ell$ be a $2D$-vector of norm $\ell\in \mathbb{N}$ defined as
\be 
\bds{\ell} = \ell(\cos \varphi_\ell, \sin \varphi_\ell)\,.
\ee 
The Fourier coefficients in the plane are defined by
\be \label{eqapplensing:Xcal}
\Xcal(\bds \ell) \equiv \sqrt{\frac{1}{2\ell+1}} \sum_m \ii^{-m} X_{\ell,m} \eee^{\ii m \varphi_\ell}\,.
\ee 
We can extend analytically this field to all values of $\bds \ell$. In this definition, $m$ goes from $-\ell$ to $+\ell$ and $\varphi_\ell$ is the polar angle of $\bds \ell$ in the $2D$-Fourier plane. The inverse transformation is given by
\be 
X_{\ell,m} = \sqrt{2\ell+1} \, \ii^{m} \int \frac{\dd \varphi_\ell}{2\pi} \; \eee^{-\ii m \varphi_\ell } \Xcal(\bds \ell)\,,
\ee 
where $\bds \ell= \ell (\cos \varphi_\ell , \sin \varphi_\ell)$ has norm $\ell$. We define then the flat sky field as
\be \label{eqapp:flat_sky_field}
\Xcal(\bds \alpha) = \int\; \frac{\bdd \bds \ell^2}{2\pi} \Xcal(\bds \ell)\, \eee^{\ii \bds \ell \cdot \bds \alpha}\,.
\ee 
It is direct to show that the power spectrum agree, i.e.
\begin{align}
\langle  X_{\ell_1,m_1} X^\star_{\ell_2,m_2}\rangle &= C_{\ell_1}^X  \delta_{\ell_1,\ell_2}\delta_{m_1,m_2}\,, \\ \label{eqapp:pow_spec_flat}
\langle  \Xcal_{\bds \ell_1} \Xcal^\star_{\bds \ell_2}\rangle &= C_{\ell_1}^X  \delta (\bds \ell_1 - \bds \ell_2)\,. 
\end{align}
Moreover, it can be shown (but it is neither direct nor easy) that the field $\Xcal(\bds \alpha)$ is locally the original field, i.e.
\be 
 \Xcal(\alpha^1 , \alpha^2) = X(\theta, \varphi)\,,
\ee 
where $\theta$ and $\varphi$ depend on $\bds \alpha$ as explained above.

\subsection{Dictionary}
Technically, to use the flat sky approximation, we should use Eq.~\eqref{eqlensingapp:Xlm}, Eq.~\eqref{eqapplensing:Xcal} and Eq.~\eqref{eqapp:flat_sky_field} to replace $X(\bds n)$ by $\Xcal(\bds \alpha)$ at the beginning of any computation, and then use Eq.~\eqref{eqapp:pow_spec_flat} to compute correlations. However, we can use Eq.~\eqref{eqlensingapp:Xlm} and Eq.~\eqref{eqapp:flat_sky_field} and impose
\begin{align}
X(\bds n) &\rightarrow \Xcal(\bds \alpha)\,, \\
X_{\ell,m} &\rightarrow \Xcal(\bds \ell)\,.
\end{align}
to get the golden rule
\boxemph{
\YLM{\ell}{m}(\theta, \varphi) \rightarrow \frac{1}{2\pi} \eee^{\ii \bds{\ell}\cdot \bds{\alpha}}\,,
}
which is what we do in practice. In the computations, sums and integrals are replaced by
\boxemph{
\int\; \dots \, \bdd \Omega^2 & \rightarrow \int\; \dots \, \dd \bds \alpha\,, \\
\sum_{\ell,m} &\rightarrow \int\; \dots \, \dd \bds \ell\,,
}
where we stress that both $\bds \alpha$ and $\bds \ell$ are $2D$ vectors. In this approximation, the slashed derivatives become
\boxemph{
\sd & \rightarrow - \frac{\partial}{\partial \alpha^1} - \ii \frac{\partial}{\partial \alpha^2} \,, \\
\sds & \rightarrow - \frac{\partial}{\partial \alpha^1}  + \ii \frac{\partial}{\partial \alpha^2} \,, \\
(\sd)^2 \YLM{\ell}{m}(\theta, \varphi) &\rightarrow - \frac{\ell^2}{2\pi} \eee^{2\ii \varphi_\ell} \eee^{\ii \bds{\ell}\cdot \bds{\alpha}}\,, \\
(\sds)^2 \YLM{\ell}{m}(\theta, \varphi) & \rightarrow -\frac{1}{2\pi} \ell^2 \eee^{-2\ii \varphi_\ell}\eee^{\ii \bds{\ell}\cdot \bds{\alpha}}\,, \\
\rll\,   \YLMS{\ell}{m}{\pm 2} &\rightarrow - \ell^2 \frac{1}{2\pi} \eee^{\pm 2 \ii \varphi_\ell} \eee^{\ii \bds \ell \cdot \bds \alpha}\,.
}

\subsection{Explicit Computations}
We use this \emph{dictionary} to compute the correlation functions of Section~\ref{seclensing:explicit}. We consider two directions $\bds n_1$ and $\bds n_2$ with $\bds n_1 \cdot \bds n_2 = \cos \theta$ (Note that in Section~\ref{seclensing:cosmo}, the angle between the two direction is denoted $\varphi$ as computations are made in the equatorial plane). For example, the first correlation function (Eq.~\eqref{eqlensing:gpmi1}) is given by
\begin{align}
    \langle \gamma^+(z_1, \boldsymbol n_1)
    \gamma^-(z_2, \boldsymbol n_2) \rangle 
    &=
    \frac 14 \sum_{\ell,m} 
    \Clphi{\ell}(z_1,z_2) \rll^2
    \YLMS{\ell}{m}{+2}(\bds n_1) \, 
    \YLMSStar{\ell}{m}{+2}(\bds n_2) \\
    & \rightarrow \frac{1}{16\pi^2} \int\; \Clphi{\ell}(z_1,z_2)
    \ell^4 \eee^{\ii \bds \ell \cdot \bds \alpha_1} \eee^{-\ii \bds \ell \cdot \bds \alpha_2}\, \dd \bds{\ell}\,.
\end{align}
Without loss of generality, we can assume that $\bds \alpha_2 =0$. Using the results of the previous section, we get that the norm of $\bds \alpha_1$ is $\theta$.  We rotate the coordinates such that $\bds{\alpha}_1$ is in the first direction. The integral over the angle $\varphi_l$ can be computed first and reads
\be
    \int\; \eee^{\ii \bds \ell \cdot \bds \alpha_1}\, \dd \varphi_\ell     =
    \int\; \eee^{\ii \ell \theta \cos \varphi_\ell}\, \dd \varphi_\ell 
    =2\pi J_0(\ell \theta)\,,
\ee
where we used the definition of the Bessel function of order $n$
\be 
\int_0^{2\pi}\; \eee^{\pm\ii n \varphi_\ell} \eee^{\ii y \cos \varphi_\ell}\, \dd \varphi_\ell = 2\pi \ii^{n} J_n(y)\,.
\ee 
Using this, the sum in the correlation function Eq.~\eqref{eqlensing:gpmi} becomes
\begin{align}
    \langle \gamma^+(z_1, \boldsymbol n_1)
    \gamma^-(z_2, \boldsymbol n_2) \rangle 
    &=
    \frac{1}{8\pi}\int\; \Clphi{\ell}(z_1,z_2) \ell^5 J_0(\ell \theta)\, \dd \ell\,.
\end{align}
The correlations functions $\langle \gamma^+(z_1, \boldsymbol n_1)\gamma^+(z_2, \boldsymbol n_2) \rangle $ and $    \langle \gamma^-(z_1, \boldsymbol n_1)
\gamma^-(z_2, \boldsymbol n_2) \rangle $ are obtained following the same steps. The only differences is that the phases do not cancel, and we are left with a phase $\pm 4 \ii \ell \varphi_\ell$, leading to
\begin{align}
\langle \gamma^+(z_1, \boldsymbol n_1)
\gamma^+(z_2, \boldsymbol n_2) \rangle 
&=
\frac{1}{8\pi}\int\; \Clphi{\ell}(z_1,z_2)\ell^5 J_4(\ell \theta)\, \dd \ell\,, \\
\langle \gamma^-(z_1, \boldsymbol n_1)
\gamma^-(z_2, \boldsymbol n_2) \rangle 
&=
\frac{1}{8\pi}\int\;\Clphi{\ell}(z_1,z_2) \ell^5 J_4(\ell \theta)\, \dd \ell\,.
\end{align}
Recall the relations
\begin{align}
        \gamma^{\pm} &= \gamma_1 \pm \ii \gamma_2 \,, \\
        \zp(z_1,z_2, \theta)  &= \langle \gamma_1(z_1, \bds n_1) \gamma_1(z_2, \bds n_2) \rangle \,, \\
        \zc(z_1,z_2, \theta)  &= \langle \gamma_2(z_1, \bds n_1) \gamma_2(z_2, \bds n_2) \rangle \,, \\
        \zeta_\pm(z_1,z_2, \theta) &= \frac 12 \left(\zp(z_1,z_2; \mu) + \zc(z_1,z_2, \theta) \right)\,,
\end{align}
where the second and third relations hold in the preferred coordinates system described in Section~\ref{seclensing:cicf} (see also Eq.~\eqref{eqlensing:zetapequa} and Eq.~\eqref{eqlensing:zetacequa}). Combining all of this yields
\boxemph{
\zeta_+(z_1,z_2, \theta) &= \frac{1}{2\pi} \int_0^{\infty}\; \ell^5 \, J_0(\ell \theta)\,  \frac{1}{4} 
\Clphi{\ell}(z_1,z_2)\, \mathrm{d}\ell\,\\
\zeta_-(z_1,z_2, \theta) &= \frac{1}{2\pi} \int_0^{\infty}\; \ell^5\,  J_4(\ell  \theta) \, \frac{1}{4} 
\Clphi{\ell}(z_1,z_2)\, \mathrm{d}\ell\,,
}

As an example, we compute the correlation function for the lensing potential using the full sky formula given by
\be 
\zeta^\phi(z_1,z_2, \ \theta) =
\sum_{\ell} \frac{2\ell+1}{4\pi} \Clphi{\ell}(z_1,z_2) P_\ell(\cos \theta)\,,
\ee
and the flat sky approximation
\be
\zeta^\phi(z_1,z_2 , \ \theta) = \frac{1}{2\pi} \int\; \ell\, \Clphi{\ell}(z_1,z_2) J_0(\ell  \theta)\, \dd \ell\,.
\ee 
\begin{figure}[ht!]
	\centering
	\includegraphics[width = 0.65\textwidth]{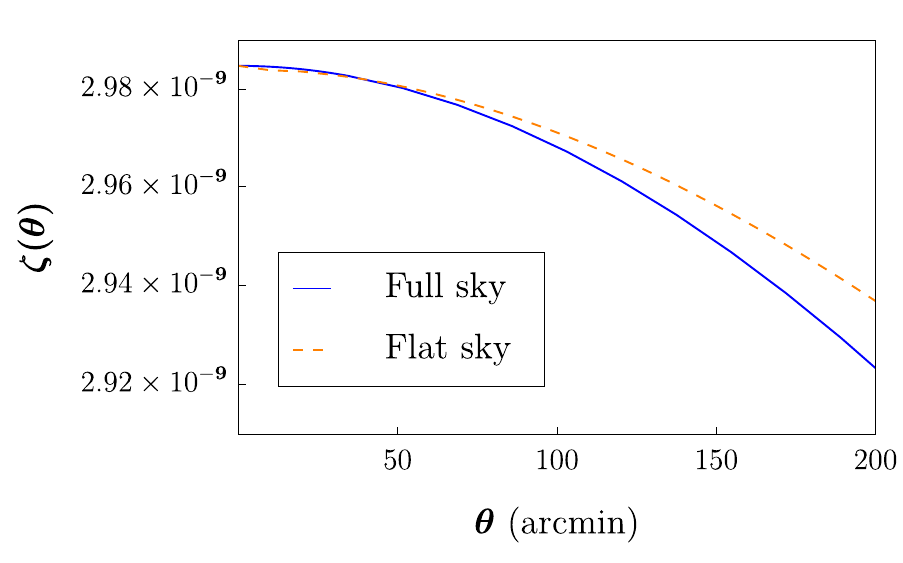} 
	\caption[Comparison between full sky and flat sky]{\label{figapp:FSFS}
		Correlation function for the lensing potential, comparison between the full sky formula and the flat sky approximation (with $z_1=z_2=1$)}
\end{figure}
The full sky formula is given by Eq.~\eqref{eqintro:corr_fun_full_sky}, and the flat sky approximation can be obtained following the same steps as above (see also \cite{durrer_2021}). We show in Fig.~\ref{figapp:FSFS} the correlation function of the lensing potential (with $z_1=z_2=1$) using the full sky formula and the flat sky approximation. We see that the relative error is of order of the percent for $\varphi<200\, \mathrm{arcmin}$. Both formulas were truncated at $\ell=25'000$.

\section{Spherical Harmonics\label{sec:appSH}}
\subsection{Explicit Formulas \label{sec:appYLM}}
We give the explicit expressions for the Spherical Harmonics for up to $s=2$ and $\ell=2$ in Tab.~\ref{tabcosmo:l1} and  Tab.~\ref{tabcosmo:l2}. The missing expressions can be deduced from the conjugation relation Eq.~\eqref{eqcosmo:ylmconj}.

\begin{center}
\begin{table}[h!]
\centering
\begin{tabular}{c|cc} 
\hline  &&\\
$m$    & $\tensor{Y}{_{1,m}}(\theta, \varphi)$ &  $_1\tensor{Y}{_{1,m}}(\theta, \varphi)$ \\ && \\ \hline \hline && \\
$-1$   & $\frac 12 \sqrt{\frac{3}{2\pi}} \eee^{-\ii \varphi} \sin \theta$ & $- \frac 14 \sqrt{\frac{3}{\pi}} \eee^{-\ii \varphi}  (1+\cos \theta)$   \\ && \\
$0$  & $\frac 12 \sqrt{\frac{3}{\pi}} \cos \theta$ & $\frac 12 \sqrt{\frac{3}{2\pi}} \sin \theta $ \\ &&  \\
$1$    & $- \frac 12 \sqrt{\frac{3}{2\pi}}\eee^{\ii \varphi} \sin \theta $ & $\frac 14 \sqrt{\frac{3}{\pi}}  \eee^{\ii \varphi}(-1+ \cos \theta )$  \\ && \\
\hline
\end{tabular}
\caption{\label{tabcosmo:l1} Spherical Harmonics of Spin Weight $s=0,1$ and $\ell=1$}
\end{table}
\end{center}

\begin{center}
\begin{table}
\centering
\begin{tabular}{ c|ccc}
\hline  &&&\\
$m$    & $\tensor{Y}{_{2,m}}(\theta, \varphi)$ &  $_1\tensor{Y}{_{2,m}}(\theta, \varphi)$ &  $_2\tensor{Y}{_{2,m}}(\theta, \varphi)$ \\ &&& \\ \hline \hline  &&&\\
$-2$   & $\frac 14 \sqrt{\frac{15}{2\pi}} \eee^{-2\ii \varphi} \sin^2 \theta$ & $-\frac 14 \sqrt{\frac{5}{\pi}} \eee^{-2\ii \varphi} (1+\cos \theta)\sin \theta$ & $\frac{1}{8} \sqrt{\frac{5}{\pi}} \eee^{-2\ii \varphi} (1+\cos \theta)^2$\\ 
$-1$   & $\frac 12 \sqrt{\frac{15}{2\pi}}\eee^{-\ii \varphi} \sin \theta \cos \theta $ & $- \frac 14 \sqrt{\frac{5}{\pi}} \eee^{-\ii \varphi} (2 \cos^2 \theta + \cos \theta -1)$ &
$- \frac 14 \sqrt{\frac{5}{\pi}} \eee^{-\ii \varphi} \sin \theta   (1+\cos \theta)$  \\&&& \\
$0$    & $\frac 18 \sqrt{\frac{5}{\pi}} (1+3 \cos(2\theta))$ & $\frac 12 \sqrt{\frac{15}{2\pi}} \sin \theta \cos \theta$ & $\frac 14 \sqrt{\frac{15}{2\pi}} \sin^2 \theta $ \\ &&&\\
$1$    & $- \frac 12 \sqrt{\frac{15}{2\pi}}\eee^{\ii \varphi}  \sin \theta \cos \theta$ & $\frac 14 \sqrt{\frac{5}{\pi}}  \eee^{\ii \varphi}(2 \cos^2 \theta - \cos \theta -1)$ & $\frac 14 \sqrt{\frac{5}{\pi}} \eee^{\ii \varphi}\sin \theta (-1+\cos(\theta))$ \\&&& \\
$2$    & $\frac 14 \sqrt{\frac{15}{2\pi}} \eee^{2\ii \varphi} \sin^2 \theta$ & $\frac 24 \sqrt{\frac{5}{\pi}}\eee^{2\ii \varphi}  \sin \theta (-1+\cos \theta)$ & $\frac 18 \sqrt{\frac{5}{\pi}} \eee^{2\ii \varphi}(1-\cos \theta)^2$
\\ &&& \\ \hline
\end{tabular}
\caption{\label{tabcosmo:l2}Spherical Harmonics of Spin Weight $s=0,1,2$ and $\ell=2$}
\end{table}
\end{center}

\newpage

\subsection{Addition Theorem \label{sec:appADD}}
The addition theorem of Spin Weighted Spherical Harmonics is useful to compute correlation functions (for example in Section~\ref{seclensing:cicf}). For generic spins, it reads
\be 
\sqrt{\frac{4\pi}{2\ell+1}} \sum_{m^\prime} \YLMS{\ell}{m^\prime}{s_1}(\theta_2, \varphi_2)  \YLMSStar{\ell}{m^\prime}{-s_2}(\theta_1, \varphi_1) 
=
\eee^{-\ii s_1 \gamma} \YLMS{\ell}{s_2}{s_1}(\beta, \alpha)\,.
\ee 
In this formula, the angles $(\alpha, \beta,\gamma)$ are defined via
\be
R_{\mathrm{E}}(\alpha, \beta, \gamma)=
R_{\mathrm{E}}(\varphi_1, \theta_1, 0)^{-1} 
R_{\mathrm{E}}(\varphi_2, \theta_2, 0)\,,
\ee 
and $R_\mathrm{E}(\alpha,\beta,\gamma)$ is the rotation matrix with the Euler angles $\alpha$, $\beta$ and $\gamma$ given by
\begin{align}
&R_{\mathrm{E}}(\alpha, \beta, \gamma) = \\
\nonumber
&\begin{pmatrix}
\cos \alpha \cos \beta \cos \gamma - \sin \alpha \sin \gamma &\ & - \cos \gamma  \sin \alpha -\cos \alpha  \cos \beta \sin \gamma  &\ & \cos \alpha \sin \beta  \\ 
 \cos \beta \cos \gamma \sin \alpha + \cos \alpha \sin \gamma&\ & \cos \alpha \cos \gamma - \cos \beta \sin \alpha \sin \gamma   &\ & \sin \alpha \sin \beta \\ 
- \cos \gamma \sin \beta  &\ & \sin \beta \sin \gamma  &\ & \cos \beta  
\end{pmatrix}\,.
\end{align}
The only relevant example is
\begin{alignat}{3}
    \theta_1 &= \frac \pi 2 \quad  & \varphi_1 &= 0\,, \quad && \\
    \theta_2 &= \frac \pi 2 & \varphi_2 &= \varphi\,, &&\\
    \alpha   &= \frac \pi 2 & \beta     &= \varphi & \gamma &= - \frac \pi 2\,.
\end{alignat}

\subsection{The Functions \texorpdfstring{$\tilde Q$}{TEXT} and \texorpdfstring{$\tilde P$}{TEXT}\label{app:PQ}}
In order to compute the shear correlation functions, it is useful to define the functions $\tilde{P}_{\ell}(\mu)$ and $\tilde{Q}_{\ell}(\mu)$ as
\begin{alignat}{3}
  \frac{2\ell+1}{16\pi} \tilde{Q}_{\ell}(\mu) &\equiv&     - \sqrt{\frac{2\ell+1}{4\pi}}\; _{+2}\tensor{Y}{_{\ell,+2}}(\theta,\pi/2) ~=~
        - \sqrt{\frac{2\ell+1}{4\pi}}\;
        {_{-2}\tensor{Y}{_{\ell,-2}}(\theta,\pi/2)}  \,, \\
  \frac{2\ell+1}{16\pi} \tilde{P}_{\ell}(\mu)   &\equiv&    - \sqrt{\frac{2\ell+1}{4\pi}}\; _{+2}\tensor{Y}{_{\ell,-2}}(\theta,\pi/2) ~=~
        - \sqrt{\frac{2\ell+1}{4\pi}}\;
        _{+2}\tensor{Y}{_{\ell,-2}}(\theta, \pi/2)  \,. 
\end{alignat}
From the orthonormality condition of the Spin Weighted Spherical Harmonics Eq.~\eqref{eqcosmo:yortho}, it follows that
\be \label{eq:intPQ}
\int_{-1}^{+1}\; \tilde{P}_{\ell_1}(\mu)
\tilde{P}_{\ell_2}(\mu)\, \mathrm{d}\mu =
\int_{-1}^{+1}\; \tilde{Q}_{\ell_1}(\mu)
\tilde{Q}_{\ell_2}(\mu)\, \mathrm{d}\mu =
\frac{32}{2\ell_1 +1} \delta_{\ell_1, \ell_2}\,.
\ee 
The explicit expressions for these polynomials for $\ell=2,\dots,5$ are given in Tab.~\ref{tabapp:PQ}.
\begin{center}
\begin{table}[h!]
\centering
\begin{tabular}{c|cc}
\hline  &&\\
$\ell$ & $\tilde{P}_\ell(\mu)$ & $\tilde{Q}_\ell(\mu)$ \\ &&\\ \hline \hline && \\ 
$2$ & $(\mu+1)^2$ & $(\mu-1)^2$ \\ && \\
 $3$  & $(\mu+1)^2(3\mu-2)$ & $(\mu-1)^2(3\mu+2)$ \\&& \\
$4$ & $(\mu+1)^2 (7\mu^2 -7\mu+1)$ & $(\mu-1)^2 (7\mu^2 +7\mu+1)$ \\ &&\\
$5$ & $(\mu+1)^2 (15\mu^3-18\mu^2+3\mu+1)$~~ &  ~~ $(\mu-1)^2 (15\mu^3+18\mu^2+3\mu-1)$\\ &&\\
\hline
\end{tabular}
\caption{The polynomials $\tilde{P}_\ell(\mu)$ and $\tilde{Q}_\ell(\mu)$. \label{tabapp:PQ}} \vspace{0.2cm}
\end{table}
\end{center}

\stopchap

\chapter{Gravitational Waves and Effective Theory of Gravity}
\startchap
\section{General Relativity}
\subsection{Odd Perturbations\label{secAppBH:bgodd}}
The relation between the metric perturbations for the odd modes defined in Section~\ref{secR3:odd_BG_master} is
\be 
h_0 = - \frac{\ii (\rg-r)}{r^3 \omega} \left( \rg h_1 + r (r-\rg) h_1^\prime \right)\,.
\ee 

\subsection{Even Perturbations\label{secAppBH:bgeven}}
The relation between the metric perturbations for the even modes defined in Section~\ref{secR3:even_BG_master} is

\be 
\begin{pmatrix}
    H_0  \\ H_1^\prime  \\ K_0^\prime
\end{pmatrix} = 
\bds{\Pi} 
\begin{pmatrix}
H_1 \\ K_0
\end{pmatrix}  \,,
\ee 
with
\be 
\bds{\Pi} = 
\begin{pmatrix}
    \frac{\ii \left(J^2 \rg -4 r^3 \omega ^2\right)}{2 r \omega  \left(\left(J^2-2\right) r+3 \rg \right)}
    &
    \frac{-2 \left(J^2-2\right) r^2+2 \left(J^2-3\right) r \rg+4 r^4 \omega ^2+3 \rg^2}{2 (r-\rg) \left(\left(J^2-2\right) r+3
    \rg\right)}  \\ & \\
    \frac{\left(4-3 J^2\right) r \rg+4 r^4 \omega ^2-6 \rg^2}{2 r (r-\rg) \left(\left(J^2-2\right) r+3 \rg\right)}
    &
    \frac{2 \ii r \omega  \left(\left(2-J^2\right) r^2+\left(J^2-4\right) r \rg+r^4 \omega ^2+\frac{9 \rg^2}{4}\right)}{(r-\rg)^2
\left(\left(J^2-2\right) r+3 \rg\right)} \\ & \\
    \frac{\ii \left(2 J^2 (\rg-r)+J^4 r+4 r^3 \omega ^2\right)}{2 r^2 \omega  \left(\left(J^2-2\right) r+3 \rg\right)}
    &
    \frac{\left(J^2-6\right) r \rg-4 r^4 \omega ^2+6 \rg^2}{2 r (r-\rg) \left(\left(J^2-2\right) r+3 \rg\right)}
\end{pmatrix}\,,
\ee
and
\be 
H_2 = - H_0\,.
\ee

\section{Effecticve Theory}
\subsection{Odd Perturbations\label{secAppBH:perodd}}
At the perturbed level, the functions defined in Section~\ref{secR3:odd_pert} (especially Eq.~\eqref{eqR3:eqh1_pert_fin} and Eq.~\eqref{eqR3:deltaVodd}) for the odd modes are
\begin{alignat}{1}
\alpha_{h_1}(r) =
&d_8 \frac{3 \rg^6}{{4 r^9 (r-\rg)^2}}
\Bigl(
10 r^5 \omega^2
-8 r^4 \rg \omega^2 
-6 \left(2 J^2+59\right) r^3 
+\left(23 J^2+1162\right) r^2 \rg\\ \nonumber
&-\left(11 J^2+1250\right) r \rg^2 
+444 \rg^3
\Bigr) 
\\ \nonumber
&+ d_9  \frac{\rg^5}{2 r^9 (r-\rg)^2}
\Bigl[
-144 r^6 \omega^2
+342 r^5 \rg \omega ^2
+4 r^4 \Bigl(180 \left(J^2-2\right)-47 \rg^2 \omega^2 \Bigr) \\ \nonumber
&
+18 \bigl(281-129 J^2\bigr) r^3 \rg 
+ \bigl(2479 J^2-6334\bigr) r^2 \rg^2\\ \nonumber
&
+\left(3314-877 J^2\right) r \rg^3
-588 \rg^4
\Bigr]\,, \\ \nonumber 
& \\
\alpha_{h_1^\prime}(r) =
&d_8 \frac{9 \rg^6}{{4 r^8 (r-\rg)}}
\Bigl(
34 r^2-61 r \rg+28 \rg^2
\Bigr)
\\ \nonumber 
&+d_9 \frac{3 \rg^5}{{2 r^8 (r-\rg)}}
\Bigl(
-240 r^3+834 r^2 \rg-913 r \rg^2+324 \rg^3
\Bigr)\,,
\end{alignat}
and
\begin{alignat}{1}
\delta V _{\mo}(r) = 
&d_8 \frac{3 \rg}{2r^9}
\Bigl(
2 \left( J^2-4\right) r^6
+\left(2  J^2-9\right) r^5 \rg
+2 \left( J^2-5\right) r^4 \rg^2\\ \nonumber
& +\left(2 J^2-11\right) r^3 \rg^3
+2 \left( J^2-6\right) r^2 \rg^4
+\left(197-22  J^2\right) r \rg^5-120 \rg^6
\Bigr) 
\\ \nonumber
&+ d_9  \frac{\rg}{r^9}
\Bigl(
10 \left(J^2-4\right) r^6
+5 \left(2 J^2-9\right) r^5 \rg 
+10 \left(J^2-5\right) r^4 \rg^2\\ \nonumber
& +5 \left(2 J^2-11\right) r^3 \rg ^3
+1450 \left(J^2-6\right) r^2 \rg ^4\\ \nonumber
&+\left(21769-1754 J^2\right) r \rg ^5-12840 \rg ^6
\Bigr)\,.
\end{alignat}

\poubelle{
\begin{alignat}{2}
\alpha_{h_1}(r) =
&d_8 \frac{3 \rg^6}{{4 r^9 (r-\rg)^2}}
\Bigl(&&
10 r^5 \omega^2
-8 r^4 \rg \omega^2 
-6 \left(2 J^2+59\right) r^3 
+\left(23 J^2+1162\right) r^2 \rg\\ \nonumber
&&&-\left(11 J^2+1250\right) r \rg^2 
+444 \rg^3
\Bigr) 
\\ \nonumber
&+ d_9  \frac{\rg^5}{2 r^9 (r-\rg)^2}
\Bigl(&&
-144 r^6 \omega^2
+342 r^5 \rg \omega ^2
+4 r^4 \Bigl(180 \left(J^2-2\right)-47 \rg^2 \omega^2 \Bigr) \\ \nonumber
&&&
+18 \bigl(281-129 J^2\bigr) r^3 \rg 
+ \bigl(2479 J^2-6334\bigr) r^2 \rg^2\\ \nonumber
&&&
+\left(3314-877 J^2\right) r \rg^3
-588 \rg^4
\Bigr)\,, \\ \nonumber 
&&& \\
\alpha_{h_1^\prime}(r) =
&d_8 \frac{9 \rg^6}{{4 r^8 (r-\rg)}}
\Bigl(&&
34 r^2-61 r \rg+28 \rg^2
\Bigr) 
\\ \nonumber 
+&d_9 \frac{3 \rg^5}{{2 r^8 (r-\rg)}}
\Bigl(&&
-240 r^3+834 r^2 \rg-913 r \rg^2+324 \rg^3
\Bigr)\,,
\end{alignat}
}

\subsection{Even Perturbations\label{secAppBH:pereven}}

At the perturbed level, the functions defined in Section~\ref{secR3:even_pert} (especially Eq.~\eqref{eqR3:mastervariable_even}, Eq.~\eqref{eqR3:field_redef_even} and Eq.~\eqref{eqR3:perturbed_Ve}) for the even modes are
\begin{alignat}{1}
\delta f_K(r) =
&d_8 \frac{1}{r^6 ((J^2-2)r + 3 \rg)}
\Bigl(
6J^2 (J^2-2) \rg^5 r^2 + 3 (J^2 + 10)\rg^6 r - 36 \rg^7
\Bigr) 
\\ \nonumber
&+ d_9  \frac{1}{r^6 ((J^2-2)r + 3 \rg)}
\Bigl(
144 (J^2-2) \rg^5 r^2 + 24 (25-8J^2)\rg^6 r - 378 \rg^7
\Bigr) \,, \\ \nonumber 
& \\
\delta f_H(r) =
&d_8 \frac{1}{r^6 ((J^2-2)r + 3 \rg)}
\Bigl(
6J^2 (J^2-2) \rg^5 r^2 + 12 (2J^2 -1)\rg^6 r + 27 \rg^7
\Bigr) 
\\ \nonumber
&+ d_9  \frac{1}{r^6 ((J^2-2)r + 3 \rg)}
\Bigl(
144 (J^2-2) \rg^5 r^2 + 6 (82-23 J^2)\rg^6 r - 216 \rg^7
\Bigr) \,,
\end{alignat}

\begin{alignat}{1}
\chi(r) =
&d_8 \frac{3 \rg}{ 2r^6}
\Bigl(
r^5 + r^4 \rg + r^3 \rg^2 + r^2 \rg^3 + r \rg^4 - 4 \rg^5
\Bigr) 
\\ \nonumber
&+ d_9  \frac{\rg}{r^6}
\Bigl(
5r^5 + 5r^4 \rg + 5r^3 \rg^2 + 5r^2 \rg^3 + 5r \rg^4 - 22 \rg^5
\Bigr) \,, 
\end{alignat}
and
\begin{alignat}{1}
\delta V _{\me}(r) = 
&d_8 \frac{3 \rg}{2r^9((J^2-2)r+3\rg)^3}
\Bigl(
2 \left( J^2-2\right)^3 (J^2-1) r^9 \\ \nonumber
& +(2J^4+5J^2-24) r^8 \rg \\ \nonumber
&+ (J^2-2) (2J^6-3J^2-64) r^7 \rg^2
 + (2J^8-5J^6-6J^4 + 11J^2 -44) r^6 \rg^3 \\ \nonumber
&+ (J^2+1)^2 (2J^4-10J^2+9) r^5 \rg^4 \\ \nonumber 
&+ (-22J^8 + 347J^6 - 1560 J^4 + 2717 J^2 - 1672) r^4 \rg^5 \\ \nonumber 
& + 3(-106J^6+1221J^4 -3642 J^2 + 3221) r^3 \rg^6 \\ \nonumber
& + 9 (-238J^4 + 1569 J^2 - 2198) r^2 \rg^7 
 + 45 (-134J^2 + 391 )r\,  \rg^8
-5832 \rg^9 
\Bigr) 
\\ \nonumber
&+ d_9  \frac{ \rg}{r^9((J^2-2)r+3\rg)^3}
\Bigl(
10 \left( J^2-2\right)^3 (J^2-1) r^9 \\ \nonumber
& + 5 (J^2-2)^2 (2J^4+5J^2-24) r^8 \rg \\ \nonumber
&+ 5 (J^2-2) (2J^6-3J^2-64) r^7 \rg^2
 + 5 (2J^8-5J^6-6J^4 + 11J^2 -44) r^6 \rg^3 \\ \nonumber
&-5 (142J^8 - 1722 J^6 +6921 J^4 -11528J^2 + 6903) r^5 \rg^4 \\ \nonumber 
&+ (766 J^8 - 19961 J^6 + 113448 J^4 - 239087 J^2 + 171544) r^4 \rg^5 \\ \nonumber 
& + 3( 3970 J^6-42031 J^4+124654 J^2 - 113079) r^3 \rg^6 \\ \nonumber
& + 9 (5294 J^4 - 28891 J^2 + 36546) r^2 \rg^7 
 + 135 (498J^2 - 1127 )r\,  \rg^8
-25272 \rg^9 
\Bigr)\,.
\end{alignat}

\newpage
\listoffigures

\printbibliography[heading=bibintoc]

\end{document}